\documentclass{article}
\usepackage{ifthen}
\usepackage[dvips]{graphicx}

\usepackage{amsfonts,amsmath,amssymb}

\def\be{\begin{eqnarray}}
\def\ee{\end{eqnarray}}
\def\nn{\nonumber}

\def\circled{\underbrace}

\hoffset= - 1.2in         

\def\set#1{\left\{#1\right\}}

\def\bb#1{\mathbb{#1}}

\def\cl#1{\mathcal{#1}}

\def\tx#1{{\rm#1}\,}

\def\Ait#1{{\it#1}}

\newtheorem{example}{Example}

\long\outer\def\EXz#1#2{\begin{example}\
\ifthenelse{\equal{#1}{}}{}{{\bf(#1)}\newline}
{\sf #2}
\end{example}}

\def\DMS#1,#2,#3{\includegraphics[width=#1pt,height=#2pt]}
\newcommand\FIGEPS[4]{
\ifthenelse{\equal{#3}{}}
{\begin{figure}\begin{center}\includegraphics[width=200pt,height=200pt,draft]
{./pics/#1.eps}\caption{\footnotesize#2}\label{#1}\end{center}\end{figure}}
{\begin{figure}\begin{center}\DMS#3,0{./pics/#1.pcx}
\caption{\footnotesize#4}\label{#1}\end{center}\end{figure}} }

\newcommand{\PFig}[3]{}

\newcommand{\Fig}[3]{
\ifthenelse{\equal{#2}{}}
{\begin{figure}\begin{center}\includegraphics[width=200pt,height=200pt,draft]
{#1.eps}\caption{\footnotesize#3}\label{#1}\end{center}\end{figure}}
{\begin{figure}\begin{center}\DMS#2,0{#1.eps}
\caption{\footnotesize#3}\label{#1}\end{center}\end{figure}} }

\def\cl#1{\mathcal{#1}}
\def\bb#1{\mathbb{#1}}

\begin{document}


\hfill ITEP/TH-35/06

\bigskip

\centerline{\Large{Introduction to Non-Linear Algebra }}

\bigskip

\centerline{V.Dolotin\ and\ A.Morozov}

\bigskip

\centerline{\it ITEP, Moscow, Russia}

\bigskip

\centerline{ABSTRACT}

\bigskip

Concise introduction to a relatively new subject of non-linear
algebra: literal extension of text-book linear algebra to the case
of non-linear equations and maps. This powerful science is based
on the notions of discriminant (hyperdeterminant) and resultant,
which today can be effectively studied both analytically and by
modern computer facilities. The paper is mostly focused on
resultants of non-linear maps. First steps are described in
direction of Mandelbrot-set theory, which is direct extension of
the eigenvalue problem from linear algebra, and is related  by
renormalization group ideas to the theory of phase transitions and
dualities.

\bigskip

\bigskip

\tableofcontents


\section{Introduction}

\subsection{Formulation of the problem}

{\it Linear algebra} \cite{G} is one of the foundations of modern
natural science: wherever we are interested in {\it calculations},
from engineering to string theory, we use linear equations,
quadratic forms, matrices, linear maps and their cohomologies.
There is a widespread feeling that the {\it non-linear} world is
very different, and it is usually studied as a sophisticated
phenomenon of interpolation between different approximately-linear
regimes. In \cite{DM} we already explained that this feeling can
be wrong: non-linear world, with all its seeming complexity
including "chaotic structures" like Julia and Mandelbrot sets,
allows clear and accurate description in terms of ordinary
algebraic geometry. In this paper we extend this analysis to
generic multidimensional situation and show that non-linear
phenomena are direct generalizations of the linear ones, without
any approximations. The thing is that the theory of generic
tensors and associated multi-linear functions and non-linear maps
can be built {\it literally} repeating everything what is done
with matrices (tensors of rank $2$), as summarized in the table in
sec.\ref{cota}. It appears that the only essential difference is
the lack of "obvious" canonical representations (like sum of
squares for quadratic forms or Jordan cells for linear maps): one
can not immediately choose between different
possibilities.\footnote{ {\it Enumeration} of such representations
is one of the subjects of "catastrophe theory" \cite{cath}. } All
other ingredients of linear algebra and, most important, its main
"special function" -- determinant -- have direct (moreover,
literal) counterparts in non-linear case.

Of course, this kind of ideas is hardly new \cite{Cay}-\cite{ChD},
actually, they can be considered as one of constituents of the
string program \cite{UFN2}. However, for mysterious reasons --
given significance of non-linear phenomena -- the field remains
practically untouched and extremely poor in {\it formulas}. In
this paper we make one more attempt to convince scientists and
scholars (physicists, mathematicians and engineers) that {\it
non-linear algebra} is as good and as powerfull as the linear one,
and from this perspective we"ll see one day that the non-linear
world is as simple and transparent as the linear one. This world
is much bigger and more diverse: there are more "phases", more
"cohomologies", more "reshufflings" and "bifurcations", but they
all are just the same as in linear situation and adequate
mathematical formalism does exist and is essentially the same as
in linear algebra.

One of the possible explanations for delay in the formulation of
{\it non-linear algebra} is the lack of adequate computer
facilities even in the close past. As we explain below, not all
the calculations are easily done "by bare hands" even in the
simplest cases. Writing down explicit expression for the simplest
non-trivial resultant $R_{3|2}$ -- a non-linear generalization of
the usual determinant -- is similar to writing $12!$ terms of
explicit expression for determinant of a $12\times 12$ matrix:
both tedious and useless. What we need to know are {\it
properties} of the quantity and possibility to evaluate it in a
particular practical situation. For example, for particular cubic
form $\frac{1}{3}ax^3 + \frac{1}{3}by^3 + \frac{1}{3}cz^3 +
2\epsilon xyz$ the resultant is given by a simple and practical
expression: $R_{3|2} = abc(abc + 8\epsilon^3)^3$. Similarly, any
other particular case can be handled with modern computer
facilities, like MAPLE or Mathematica. A number of results below
are based on computer experiments.

\bigskip

At the heart of our approach to {\it quantitative} non-linear
algebra are special functions -- {\it discriminants} and {\it
resultants} -- generalizations of determinant in linear algebra.
Sometime, when they (rarely) appear in modern literature
\cite{GKZ,recent} these functions are called {\it
hyperdeterminants}, but since we define them in terms of
consistency (existence of common solutions) of systems of
non-linear equations, we prefer to use "discrminantal" terminology
\cite{D1,D2,D3}. At least at the present stage of developement the
use of such terminology is adequate in one more respect. One of
effective ways to evaluate discriminants and resultants exploits
the fact that they appear as certain irreducible factors in
various auxiliary problems, and constructive definitions express
them through iterative application of two operations: taking an
ordinary resultant of two functions of a {\it single} variable and
taking an irreducible factor. The first operation is
constructively defined (and well computerized) for polynomials
(see s.4.10 of \cite{DM} for directions of generalization to
arbitrary functions), and in this paper we restrict consideration
to non-linear, but polynomial equations, i.e. to the theory of
tensors of finite rank or -- in homogeneous coordinates -- of
functions and maps between projective spaces $P^{n-1}$. The second
operation -- extraction of irreducible factor, denoted ${\bf
irf}(\ldots)$ in what follows, -- is very clear conceptually and
very convenient for pure-science considerations, but it is a
typical $NP$ problem from calculational point of view. Moreover,
when we write, say, $D = {\bf irf}(R)$, this means that $D$ is
{\it a} divisor of $R$, but actually in most cases we mean more:
that it is ${\it the}$ divisor, the somehow {\it distinguished}
irreducible factor in $R$ (in some cases $D$ can be divisor of
$R$, where $R$ can be obtained in slightly different ways, for
example by different sequences of iterations, -- then $D$ is a
{\it common} divisor of all such $R$). Therefore, at least for
practical purposes, it is important to look for "direct"
definitions/representations of discriminants and resultants (e.g.
like row and column decompositions of ordinary determinants), even
if aestetically disappealing, they are practically usefull, and --
no less important -- they provide concrete definition of ${\bf
irf}$ operattion. Such representations were suggested already in
XIX century \cite{Cay,Sylv}, and the idea was to associate with
original non-linear system some {\it linear-algebra} problem
(typically, a set of maps between some vector spaces), which
degenerates simulteneously with the original system. Then
discriminantal space acquires a linear-algebra representation and
can be studied by the methods of homological algebra \cite{Man}.
First steps along these lines are described in \cite{GKZ}, see
also s.\ref{Koshul} and s.\ref{polydet} in the present paper.
Another option is Feynman-style diagram technique \cite{CK,GMS},
capturing the structure of convolutions in tensor algebra with
only two kinds of invariant tensors involved: the unity
$\delta^i_j$ and totally antisymmetric $\epsilon_{i_1\ldots i_n}$.
Diagrams provide all kinds of invariants, made from original
tensor, of which discriminant or resultant is just one.
Unfortunately, enumeration and classification of diagrams is
somewhat tedious and adequate technique needs to be found for
calculation of appropriate generating functions.

\bigskip

Distinction between discriminants and resultants in this paper
refers to two essentially different types of objects: functions
(analogues of quadratic forms in linear algebra) and maps. From
tensorial point of view this is distinction between pure covariant
tensors and those with both contravariant and covariant indices
(we mostly consider the case of a single contravariant index). The
difference is two-fold. One difference -- giving rise to the usual
definition of co- and contra-variant indices --
is in transformation properties: 
under {\it linear} transformation $U$ of homogeneous coordinates
(rotation of $P^{n-1}$) the co- and contra-variant indices
transform with the help of $U$ and $U^{-1}$ respectively. The
second difference is that maps can be composed (contravariant
indices can be converted with covariant ones), and this opens a
whole variety of new possibilities. We associate {\it
discriminants} with {\it functions} (or forms or pure covariant
tensors) and {\it resultants} with {\it maps}. While closely
related (for example, in linear algebra discriminant of quadratic
form and resultant of a linear map are both determinants of
associated square matrices), they are completely different from
the point of view of questions to address: behaviour under
compositions and eigenvalue (orbit) problems for resultants and
reduction properties for tensors with various symmetries (like
$\det = {\rm Pfaff}^2$ for antisymmetric forms) in the case of
discriminants. Also, diagram technique, invariants and associated
group theory are different.

\bigskip

We begin our presentation -- even before discussion of relevant
definitions -- from two comparative tables:

One in s.\ref{cota} is comparison between notions and theorems of
linear and non-linear algebras, with the goal to demonstrate thet
{\it entire} linear algebra has {\it literal} non-linear
counterpart, as soon as one introduces the notions of discriminant
and resultant.

Another table in s.\ref{tedity} is comparison between the
structures of non-linear algebra, associated with different kinds
of tensors.

Both tables assume that discriminants and resultants are given.
Indeed, these are objectively existing functions (of coefficients
of the corresponding tensors), which can be constructively
evaluated in variety of ways in every particular case. Thus the
subject of non-linear algebra, making use of these quantities, is
well defined, irrespective of concrete discussion of the
quantities themselves, which takes the biggest part of the present
paper.

\bigskip

Despite certain efforts, the paper is not an easy reading.
Discussion is far from being complete and satisfactory. Some
results are obtained empirically and not all proofs are presented.
Organization of material is also far from perfect: some pieces of
discussion are repeated in different places, sometime even notions
are used before they are introduced in full detail. At least
partly this is because the subject is new and no traditions are
yet established of its optimal presentation. To emphasize
analogies we mainly follow traditional logic of linear algebra,
which in the future should be modified, according to the new
insghts provided by generic non-linear approach. The text may seem
overloaded with notions and details, but in fact this is because
it is too concise: actually every briefly-mentioned detail
deserves entire chapter and gives rise to a separate branch of
non-linear science. Most important, the set of results and
examples, exposed below is unsatisfactory small: this paper is
only one of the first steps in constructing the temple of
non-linear algebra. Still the subject is already well established,
ready to use, it deserves all possible attention and intense
application.

\subsection{Comparison of linear and non-linear algebra
\label{cota}}

{\bf Linear algebra} \cite{G} is the theory of {\bf matrices}
(tensors of rank $2$), {\bf non-linear algebra}
\cite{GKZ}-\cite{ChD} is the theory of generic tensors.

\bigskip \noindent

The four main chapters of {\it linear algebra},

$\bullet$ Solutions of systems of linear equations;

$\bullet$ Theory of linear operators (linear maps, symmetries of
linear equations), their eigenspaces and Jordan matrices;

$\bullet$ Linear maps between different linear spaces (theory of
rectangular matrices, Plukker relations etc);

$\bullet$ Theory of quadratic and bilinear functions, symmetric
and antisymmetric;

\noindent possess straightforward generalizations to {\it
non-linear algebra}, as shown in comparative table below.

\bigskip

Non-linear algebra is naturally split into two branches: theories
of solutions to {\it non-linear} and {\it poly-linear} equations.
Accordingly the main special function of linear algebra -- {\bf
determinant} -- is generalized respectively to {\bf resultants}
and {\bf discriminants}. Actually, discriminants are expressible
through resultants and vice versa -- resultants through
discriminants. Immediate applications "at the boundary" of
(non)-linear algebra concern the theories of $SL(N)$ invariants
\cite{D2}, of homogeneous integrals \cite{D3} and of algebraic
$\tau$-functions \cite{HR}.

Another kind of splitting -- into the theories of linear operators
and quadratic functions -- is generalized to distinction between
tensors with different numbers of covariant and contravariant
indices, i.e. transforming with the help of operators $U^{\otimes
r_1} \otimes \left(U^{-1}\right)^{\otimes (r-r_1)}$ with different
$r_1$. Like in linear algebra, the orbits of non-linear
$U$-transformations on the space of tensors depend significantly
on $r_1$ and in every case one can study canonical forms,
stability subgroups and their reshufflings (bifurcations). The
theory of eigenvectors and Jordan cells grows into a  deep theory
of {\it orbits} of non-linear transformations and {\it Universal
Mandelbrot set} \cite{DM}. Already in the simplest single-variable
case this is a profoundly rich subject \cite{DM} with non-trivial
physical applications \cite{UFN2,MN}.

\bigskip

\vspace*{-1.5cm} \hspace{-1.0cm}
\begin{tabular}{|c|c|}
\hline\hline
&\\
{\Large{\bf Linear algebra}} & {\Large {\bf Non-linear Algebra}}\\
&\\
\hline\hline\hline
&\\
{\bf SYSTEMS of linear equations}
& {\bf SYSTEMS of non-linear equations:}\\
and their {\bf DETERMINANTS}: & and their {\bf RESULTANTS}:\\
&\\
\hline\hline
&\\
{\it Homogeneous} & \\
&\\
\hline
&\\
$Az=0$ & $A(z) = 0$\\
&\\
$\sum_{j=1}^n A_i^j z_j = 0, \ \ \ i = 1,\ldots,n$ &
$A_i(z)=\sum_{j_1,\ldots, j_{s_i}=1}^n A_i^{j_1\ldots
j_{s_i}}z_{j_1}\ldots z_{j_{s_i}}=0,
\ \ \ i = 1,\ldots,n$ \\
&\\
Solvability condition: & Solvability condition: \\
&\\
${\det}_{1\leq i,j\leq n} A_i^j =
\left|\left| \begin{array}{ccc} A_1^1 &\ldots& A_1^n\\
&\ldots& \\ A_n^1 &\ldots & A_n^n\end{array}\right|\right| = 0$
&$\begin{array}{c}
{\cal R}_{s_1,\ldots,s_n}\{A_1,\ldots,A_n\} = 0 \\
{\rm or}\ \ \ {\cal R}_{n|s}\{A_1,\ldots,A_n\} = 0\ \ \
{\rm if\ all}\ \ \ s_1=\ldots=s_n=s \end{array}$ \\
& $d_{s_1,\ldots,s_n} \equiv {\rm deg}_A {\cal R}_{s_1,\ldots,s_n}
= \sum_{i=1}^n \Big(\prod_{j\neq i}^n s_j\Big)$,\ \
$d_{n|s} \equiv {\rm deg}_A {\cal R}_{n|s} = ns^{n-1}$ \\
&\\
Solution: & \\
&\\
$Z_j = \sum_{k=1}^n \check A_j^kC_k$ &  \\
where $\sum_{j=1}^n A_i^j\check A_j^k = \delta_i^k \det A$ &\\
&\\
Dimension of solutions space & \\
for homogeneous equation:  &\\
(the number of independent choices of $\{C_k\}$)&\\
${\rm dim}_{n|1} = {\rm corank}\{A\}$,\ \ typically\ \ ${\rm
dim}_{n|1} =1$ &
typically\ \ ${\rm dim}_{n|s} = 1$\\
&\\
\hline
&\\
{\it Non-homogeneous} & \\
&\\
\hline
&\\
$Az=a$ & $A(z) = a(z)$\\
&\\
$\sum_{j=1}^n A_i^j z_j = a_i$, & $\begin{array}{c}
\sum_{j_1,\ldots,j_s=1}^n A_i^{j_1\ldots j_s}z_{j_1}\ldots
z_{j_s}= \sum_{0\leq s'<s} \left(\sum_{j_1,\ldots,j_{s'}=1}^n
a_i^{j_1\ldots j_{s'}} z_{j_1}\ldots z_{j_{s'}}\right)
\end{array}$\\
&\\
$i = 1,\ldots,n$ & $i = 1,\ldots,n$ \\
&\\
Solution (Craemer rule): & Solution (generalized Craemer rule): \\
&\\
$Z_k$ is defined from a linear equation: &
$Z_k$ is defined from a single algebraic equation:\\
&\\
$\left|\left| \begin{array}{ccccc}
A_1^1 &\ldots& (A_1^kZ_k-a_1)&\ldots& A_1^n\\
&&\ldots&& \\
A_n^1 &\ldots &(A_n^kZ_k-a_n)&\ldots & A_n^n
\end{array}\right|\right| = 0$ &
${\cal R}_{n|s}\{A^{(k)}(Z_k)\} = 0$, \\
& where $A^{(k)}(Z_k)$ is obtained by substitutions \\
& $z_k \longrightarrow z_kZ_k$ and $a^{(s')} \rightarrow
z_k^{s-s'}a^{(s')}$ \\
&\\
$Z_k$ is expressed through principal minors
& the set of solutions $Z_k$ satisfies Vieta formula \\
$Z_k \det A = \check A_k^l a_l$ &
and its further generalizations, s.\ref{CraVie} \\
&\\
$\#$ of solutions of non-homogeneous equation:&\\
$\#_{n|1} = 1$ & $\#_{s_1,\ldots,s_n} = \prod_{i=1}^n s_i$, \ \ in
particular \
$\#_{n|s} = s^n$ \\
&\\
\hline\hline
\end{tabular}
\newpage
\hspace{-1.6cm}
\begin{tabular}{|c|c|}
\hline\hline
&\\
{\bf OPERATORS made from matrices: Linear maps} &
{\bf OPERATORS made from tensors: Poly-linear maps} \\
(symmetries of systems of linear equations):&
(symmetries of systems of non-linear equations):\\
&\\
\hline\hline
&\\
$z \rightarrow Vz$,  & $z \rightarrow V(z)$ of degree $d$,\\
 $A \rightarrow UA$: & $A \rightarrow U(A)$ of degree $d'$:\\
$(UA)_i(Vz) = \sum_{j,k,l=1}^n U_i^jA_j^kV_k^lz_l$ &
$U\left(A\left(V(z)\right)\right) = \sum^N
U \Big(A \left(V z^d\right)^s \Big)^{d'}$ (somewhat symbolically)\\
&\\
\hline
&\\
Multiplicativity of determinants w.r.t. compositions &
Multiplicativity of resultants w.r.t. compositions \\
of {\it linear} maps: & of {\it non-linear} homogeneous maps \\
&\\
for linear transforms $A \rightarrow UA$ and $z \rightarrow Vz$&
for transforms $A \rightarrow U(A)$ of degree $d'$ \\
& and $z \rightarrow V(z)$ of degree $d$ \\
$\det \Big((UA)(Vz)\Big) = \det U \det A \det V$& ${\cal
R}_{n|sdd'}\{ UA(Vz)\} ={\cal  R}_{n|d'}\{U\}^{(sd')^{n-1}}
{\cal R}_{n|s}\{A\}^{d^n{d'}^{n-1}} {\cal R}_{n|d}\{V\}^{(sd)^n} $ \\
&\\
\hline\hline
&\\
Eigenvectors (invariant subspaces) of linear transform $A$
& Orbits (invariant sets) of non-linear homogeneous transform $A$ \\
&\\
\hline
&\\
Orbits of transformations with $U = V^{-1}$ &
Orbits of transformations with $U = V^{-1}$\\
in the space of {\it linear} operators $A$: &
in the space of {\it non-linear} operators $A$:\\
&\\
generic orbit (diagonalizable $A$'s)  &
non-singular $A$'s  \\
and & and \\
$A$'s with coincident eigenvalues,    &
$A$'s with coincident orbits, \\
reducible to Jordan form &
belonging to the Universal Mandelbrot set \cite{DM}  \\
& \\
Invariance subgroup of $U = V^{-1}$: &
Invariance subgroup of $U = V^{-1}$ \\
a product of Abelian groups &  \\
&\\
$A_i^je_{j\mu} = \lambda_\mu e_{i\mu}$ & $A_i^{j_1\ldots
j_s}e_{j_1\mu}\ldots e_{j_s\mu} =
\Lambda_\mu e_{i\mu}$ \\
&\\
& or $A_i(\vec e_\mu) = \lambda_\mu(\vec e_\mu)e_{i\mu} =
I_i^{\lambda_\mu}(\vec e_\mu)$  \\
& \\
$A_i^j = \sum_{\mu=1}^n e_{i\mu} \lambda_\mu e^j_\mu$,\ \
$e_{i\mu}e^i_\nu = \delta_{\mu\nu}$ & $A_i^\alpha =
\sum_{\mu=1}^{M_{n|s}} e_{i\mu}\Lambda_\mu E^\alpha_{\mu} =
\sum_{\mu=1}^{M_{n|s}} \check e_{i\mu}
\check E^\alpha_{\mu}$, \ \ $\Lambda_\mu = \lambda(e_\mu)$\\
& $\{E^\alpha_\mu\} = ({\rm max.minor\ of\ }E)^{-1}$,\
$E_{\alpha\mu} = e_{j_1\mu}\ldots e_{j_s\mu}$,\ \
$\alpha = (j_1,\ldots, j_s)$ \\
&\\
\hline\hline
&\\
{\bf RECTANGULAR $m \times n$ matrices ($m<n$):} &
{\bf "RECTANGULAR" tensors of size} $n_1\leq\ldots\leq n_r$:\\
&\\
\hline
&\\
discriminantal condition: \ \ ${\rm rank}(T) < {\rm min}(m,n)$ &
${\cal D}^{(\kappa)}_{n_1\times\ldots\times n_r}(T)=0$,
\ \ $1\leq \kappa \leq N_{n_1\times\ldots\times n_r}$\\
&\\
\hline
&\\
Classification by ranks $\leq m$ &
Classification by "ranks" $\leq n_1$ \\
Plukker relations between $m\times m$ minors, &
Plukker relations between $n_1^{\times r}$ resultants, \\
Grassmannians $G(n,m)$, & PolyGrassmannians \\
Relations between Plukker relations & \\
(syzigies, Koshul complexes etc) & \\
Hirota relations for algebraic $\tau$-functions \cite{HR}&\\
&\\
\hline\hline
\end{tabular}
\newpage
\vspace*{-1.5cm} \hspace{-1.0cm}
\begin{tabular}{|c|c|}
\hline\hline
&\\
{\bf FUNCTIONS ("forms") made from matrices:} &
{\bf FUNCTIONS ("forms") made from tensors:}\\
&\\
\hline\hline
&\\
Bilinear forms & {\it Polylinear} forms \\
&\\
\hline
&\\
$T(\vec x,\vec y) = \sum_{i,j=1}^{n} T^{ij} x_iy_j$ & $T(\vec
x_1,\ldots,\vec x_r) = \sum_{i_1,i_2,\ldots, i_r=1}^n T^{i_1\ldots
i_r}
x_{1,i_1}x_{2,i_2}\ldots x_{r,i_r}$\\
&\\
Form is degenerate, if the system & Form is degenerate, if $dT=0$,
i.e. if the system \ \
$\frac{\partial T}{\partial x_{k,i_k}} =$ \\
$\left\{\begin{array}{c}
\sum_{j=1}^n T^{ij}y_j = \frac{\partial T}{\partial x_i} = 0 \\ \\
\sum_{i=1}^n T^{ij}x_i = \frac{\partial T}{\partial y_j} = 0
\end{array}\right.$ &
$\begin{array}{c} =\sum_{i_1,\ldots, \check i_k, \ldots, i_r=1}^n
T^{i_1\ldots i_r}
x_{1,i_1}\ldots x_{k-1,i_{k-1}}x_{k+1,i_{k+1}}\ldots x_{r,i_r} = 0\\
\ \ k=1,\ldots,r, \ \ \ i_k=1,\ldots,n \end{array}$\\
has non-vanishing solution & has non-vanishing solution \\
&\\
Degeneracy criterium: & \\
${\det}_{1\leq i,j \leq n} T^{ij} = 0$ &
${\cal D}_{n^{\times r}}(T) = 0$ \\
&\\
\hline
&\\
& for $r\geq 3$\ and \  $N = n(r-2)$ \ \\
&${\cal D}_{n^{\times r}}(T) = {\bf
irf}\left({R}_{N|N-1}\left\{\partial_I \Big(\det_{1\leq j,k \leq
n} \frac{\partial^2 T}{\partial x_{1j}\partial
x_{2k}}\Big)\right\}
\right)$\\
&\\
& $\partial_I = \left\{\frac{\partial}{\partial x_{ki_k}},
\ \ k=3,\ldots,r,\  \  i_k = 1,\ldots, n\right\}$ \\
&\\
\hline\hline
&\\
Quadratic forms: & {\it Symmetric} forms of rank $r=s+1$: \\
&\\
\hline
&\\
$S(\vec z) = \sum_{i,j=1}^n S^{ij} z_iz_j$ & $S(\vec z) =
\sum_{i_0,i_1\ldots,i_s=1}^n
S^{i_0i_1\ldots i_s} z_{i_0}z_{i_1}\ldots z_{i_s}$\\
&\\
Form is degenerate, if the system &
Form is degenerate, if $dS=0$, i.e. if the system \\
$\sum_{j=1}^n S^{ij}z_j =0$ & $\frac{1}{r!}\frac{\partial
S}{\partial z_i} = \sum_{i_1,\ldots,i_s=1}^n
S^{ii_1\ldots i_s} z_{i_1}\ldots z_{i_s} = 0$ \\
has non-vanishing solution & has non-vanishing solution\\
&\\
Degeneracy criterium: & \\
${\det}_{1\leq i,j \leq n} S^{ij} = 0$ & ${D}_{n|r}(S) = {\cal
D}_{n^{\times r}}(T=S) =
{R}_{n|r-1}\{\partial_i S(\vec z)\}$ \\
&\\
\hline
&\\
Orbits of transformations with $U = V$: &
Orbits of transformations with $U = V$ \\
diagonal quadratic forms,
classified by signature &  \\
&\\
Invariance subgroup of $U=V$: &
Invariance subgroup of $U=V$ \\
orthogonal and unitary transformations& \\
&\\
\hline\hline
&\\
Skew  forms
& Skew forms (totally antisymmetric rep) exist for $r\leq n$ \\
&  Forms in other representations of the \\
& permutation group $\sigma_r$ and braid group $B_r$ (anyons)\\
&\\
\hline
&\\
Stability subgroup: & Stability subgroup
\\
symplectic transformations & depends on the Young diagram
\\
&\\
Decomposition property: &
${\cal D}(T_{red})$ decomposes into irreducible factors,\\
$\det C = \Big({\rm Pfaff} A\Big)^2$ &
among them is appropriate reduced discriminant\\
& \\
\hline \hline
\end{tabular}

\subsection{Quantities, associated with tensors of different types
\label{tedity}}

\subsubsection{A word of caution}

We formulate {\it non-linear algebra} in terms of {\it tensors}.
This makes {\it linear algebra} a base of the whole construction,
not just a one of many particular cases. Still, at least at the
present stage of development, this is a natural formulation,
allowing to make direct contact with existing formalism of quantum
field theory (while its full string/brane version remains
under-\-investigated) and with other kinds of developed
intuitions. Therefore, it does not come as surprise that some
non-generic elements will play unjustly big role below. Especially
important will be: representations of tensors as poly-matrices
with indices (instead of consideration of arbitrary functions),
linear transformations of coordinates (instead of generic
non-linear maps), Feynman-diagrams in the form of graphs (instead
of generic symplicial complexes and manifolds). Of course,
generalizations towards the right directions will be mentioned,
but presentation will be starting from linear-\-algebra-\-related
constructions and will be formulated as {\it generalization}. One
day inverse logic will be used, starting from generalities and
going down to particular {\it specifications} (examples), with
linear algebra just one of many, but this requires -- at least --
developed and {\it generally accepted} notation and nomenclature
of notions in non-linear science (string theory), and time for
this type of presentation did not come yet.

\subsubsection{Tensors}

See s.IV of \cite{G} for detailed introduction of tensors. We
remind just a few definitions and theses.

$\bullet$ $V_n$ is $n$-dimensional vector space,\footnote{ We
assume that the underlying {\it number field} is $C$, though
various elements of non-linear algebra are defined for other
fields: the structure of tensor algebra requires nothing from the
field, though we actually assume commutativity and associativity
to avoid overloading by inessential details; in discussion of
solutions of polynomial equations we assume for the same reason
that the field is algebraically closed (that a polynomial of
degree $r$ of a single variable always has $r$ roots, i.e. Besout
theorem is true). Generalizations to other fields are
straightforward and often interesting, but we leave them beyond
the scope of the present text. } $V^*_n$ is its dual. Elements of
these spaces (vectors and covectors) can be denoted as $\vec v$
and $\vec v^*$ or $v_i$ and $v^i$, $i = 1,\ldots,n$. The last
notation -- more convenient in case of generic tensors -- implies
that vectors are written in some basis (not obligatory
orthonormal, no metric structure is introduced in $V_n$ and
$V^*_n$). We call lower indices (sub-scripts) {\it covariant} and
upper indices (super-scripts) -- {\it contravariant}.

$\bullet$ {\it Linear} changes of basises result into linear
transformations of vectors and covectors, $v_i \rightarrow
(U^{-1})_i^jv_j$,\ $v^i \rightarrow U^{i}_j v^j$ (summation over
repeated sub- and super-scripts is implied). Thus contravariant
and covariant indices are transformed with the help of $U$ and
$U^{-1}$ respectively. Here $U$ belongs to the {\it structure
group} of invertible linear transformations, $U \in GL(n)$, in
many cases it is more convenient to restrict it to $SL(n)$ (to
avoid writing down obvious factors of $\det U$), when group
averages are used, a compact subgroup $U(n)$ or $SU(n)$ is
relevant. Since choice of the group is always obvious from the
context, we often do not mention it explicitly. Sometime we also
write a hat over $U$ to emphasize that it is a {\it
transformation} (a {\it map}), not just a tensor.

$\bullet$ Tensor $T$ of the type $n_1,\ldots,n_p;m_1,\ldots m_q$
is an element from $V^*_{n_1}\otimes\ldots\otimes V^*_{n_p}\otimes
V_{m_1}\otimes\ldots \otimes V_{m_q}$, or simply $T^{i_{1}\ldots
i_{p}}_{j_{1}\ldots j_{q}}$, with $i_{k} = 1,\ldots, n_k$ and $j_k
= 1,\ldots,m_k$, transformed according to $T \longrightarrow
\Big(U_{n_1}\otimes\ldots\otimes U_{n_p}\Big) T
\Big(U^{-1}_{m_1}\otimes\ldots\otimes U^{-1}_{m_q}\Big)$ with
$U_{n_k} \in SL(n_k)$ and $U^{-1}_{m_k} \in SL^*(m_k)$ (notation
$SL^*$ signals that transformations are made with inverse
matrices). Pictorially such tensor can be represented by a vertex
with $p$ {\it sorts}\footnote{ In modern physical language one
would say that indices $i$ and $j$ label {\it colors}, and our
tensor is representation of a color group
$SL(n_1)\times\ldots\times SL^*(m_q)$. Unfortunately there is no
generally accepted term for parameter which distinguishes between
different groups in this product. We use "sort" for exactly this
parameter, it takes $r=p+q$ values. (Photons, $W/Z$-bosons and
gluons are three different "sorts" from this point of view. {\it
Sort} is the GUT-color modulo low-energy colors). } of  incoming
and $q$ {\it sorts} of outgoing  lines. We call $\prod_{k=1}^p
SL(n_k) \prod_{l=1}^q SL^*(m_l)$ the {\it structure group}.

$\bullet$ The number of sorts is {\it a priori} equal to the rank
$r = p+q$. However, if among the numbers $n_1,\ldots, m_q$ there
are equal, an {\it option} appears to identify the corresponding
spaces $V$, or identify sorts. Depending on the choice of this
option, we get different classes of tensors (for example, an
$n\times n$ matrix $T^{ij}$ can be considered as representation of
$SL(n)\times SL(n)$, $T^{ij} \rightarrow (U_1)^i_k(U_2)^j_lT^{kl}$
with independent $U_1$ and $U_2$, or as representation of a single
(diagonal) $SL(n)$, $T^{ij} \rightarrow U^i_kU^j_l T^{kl}$, and
diagram technique, invariants and representation theory will be
very different in these two cases). If any identification of this
kind is made, we call the emerging tensors {\it reduced}.
Symmetric and antisymmetric tensors are particular examples of
such reductions. There are no reductions in generic case, when all
the $p+q$ numbers $n_1,\ldots,m_q$ are different, but reduced
tensors are very interesting in applications, and {\it non-linear}
algebra is largely about reduced tensors, generic case is rather
{\it poly-linear}. Still, with no surprise, non-linear algebra is
naturally and efficiently embedded into poly-linear one.

$\bullet$ Tensors are associated with functions on the dual spaces
in an obvious way. Generic tensor is associated with an $r$-linear
function \be T(\vec v_1,\ldots,\vec v_p;\vec u^*_1,\ldots,\vec
u^*_q) = \sum_{\stackrel{1\leq i_k \leq n_k}{1\leq j_k \leq m_k}}
T^{i_1\ldots i_p}_{j_1\ldots j_q}\ \ v_{1i_1} \ldots v_{p\,i_p}
u_1^{j_1}\ldots u_q^{j_q} \ee In what follows  we mostly consider
pure contravariant tensors with the corresponding $r$-linear
functions $T(\vec v_1,\ldots,\vec v_r) = T^{i_1\ldots i_r}
v_{1i_1}\ldots v_{r\,i_r}$ and (non-linear) {\it maps} $V_n\to
V_n,\ A_i(\vec v) = A_i^{i_1\ldots i_s} v_{i_1}\ldots v_{i_s}$
(symmetric tensors with additional covariant index). It will be
important not to confuse upper indices with powers.

$\bullet$ {\it Reduced} tensors can be related to non-linear
functions (forms): for example, the {\it hypercubic} (i.e. with
all equal $n_1=\ldots=n_r=n$) contravariant tensor $T^{i_1\ldots
i_r}$, associated in above way with an $r-linear$ form $T(\vec
v_1,\ldots,\vec v_r) =  T^{i_1\ldots i_r} v_{1i_1}\ldots
v_{r\,i_r}$ can be reduced to symmetric tensor, associated with
$r$-form of power $r$ in a single vector, $S(\vec v) =
\sum_{i_1,\ldots,i_r=1}^n S^{i_1\ldots i_r}v_{i_1}\ldots v_{i_r}$.
For totally antisymmetric hypercubic tensor, we can write the same
formula with {\it anti-commuting} $\vec v$, but if only reduction
is made, with no special symmetry under permutation group
$\sigma_r$  specified, the better notation is simply $T_{n|r}(\vec
v) = T(\vec v,\ldots,\vec v) = T^{i_1\ldots i_r}\
v_{i_1}\otimes\ldots \otimes v_{i_r}$. In this sense tensors are
associated with functions on a huge tensor product of vector
spaces (Fock space) and only in special situations (like symmetric
reductions) they can be considered as ordinary functions. From now
on the label $n|r$ means that hypercubic tensor of rank $r$ is
reduced in above way, while polylinear covariant tensors will be
labeled by $n_1\times\ldots\times n_r$, or simply $n^{\times r}$
in hypercubic case: $T_{n|r}$ is the maximal reduction of
$T_{n^{\times r}}$, with all $r$ sorts identified and structure
group reduced from $SL(n)^{\times r}$ to its diagonal $SL(n)$.

\subsubsection{Tensor algebra}

\noindent

$\bullet$ Tensors can be added, multiplied and contracted. {\it
Addition} is defined for tensors of the same type
$n_1,\ldots,n_p;$ $m_1,\ldots,m_q$ and results in a tensor of the
same type. Associative, but non-commutative(!) {\it tensor
product} of two tensors of two arbitrary types results into a new
tensor of type $n_1,\ldots,n_p,n'_1,\ldots.n'_{p'};$
$m_1,\ldots,m_q,m'_1,\ldots,m'_{q'}$. Tensor products can also be
accompagnied by permutations of indices within the sets $\{n,n'\}$
and $\{m,m'\}$. {\it Contraction} requires identification of two
{\it sorts}: associated with one covariant and one contravariant
indices (allowed if some of $n$'s coincides with some of $m$'s,
say, $n_p=m_q=n$) and decreases both $p$ and $q$ by one: \be
T^{i_1\ldots i_{p-1}}_{j_1\ldots j_{q-1}} = \sum_{l=1}^n
T^{i_1\ldots i_{p-1}l}_{j_1\ldots j_{q-1}l} \label{convo} \ee Of
course, one can take for the tensor $T$ in (\ref{convo}) a tensor
product and thus obtain a contraction of two or more different
tensors. $k$ pairs of indices can be contracted simultaneously,
{\it multiplication} is a particular case of contraction for
$k=0$. Pictorially (see Fig.\ref{F1}) contractions are represented
by lines, connecting contracted indices, with sorts and arrows
respected: only indices of the same {\it sort} can be connected,
and incoming line (i.e. attached to a covariant index) can be
connected with an outgoing one (attached to a contravariant
index).

In order to avoid overloading diagrams with arrows, in what
follows we use slightly different notation (Fig.~\ref{diaver}):
denote covariant indices by white and contravariant ones by black,
so that arrows would go from black to white and we do not need to
show them explicitly. Tensors with some indices covariant and some
contravariant are denoted by semi-filled (mixed white-black)
circles (see Fig.~\ref{ta20}.B).

\Fig{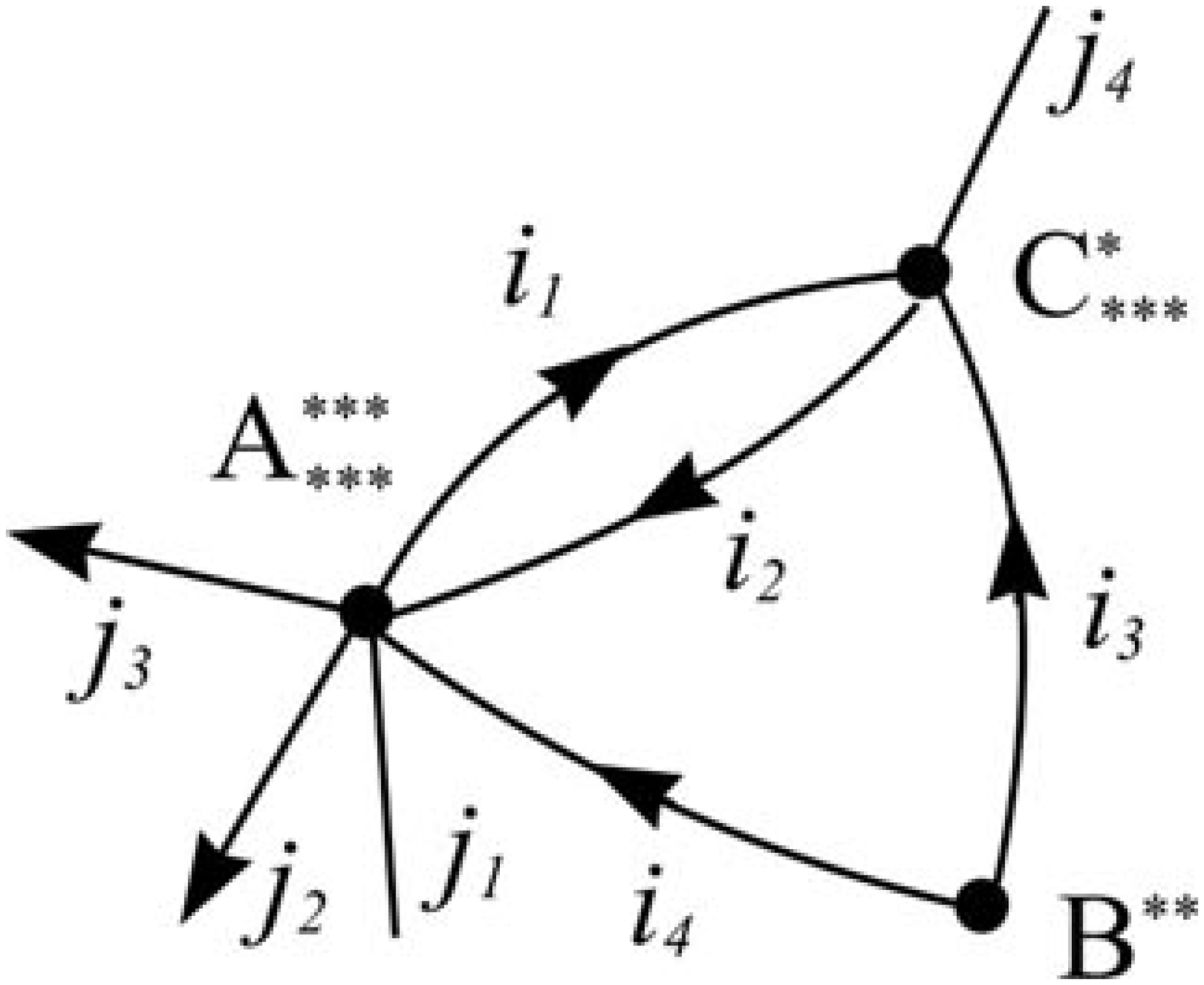} {206,167} {Example of diagram, describing particular
contraction of three tensors: $A$ of the type
$(i_1,j_2,j_3;i_1,i_4,j_1)$, $B$ of the type $(i_3,i_4;\empty)$
and $C$ of the type $(i_2;i_1,i_3,j_4)$. Sorts of lines are not
shown explicitly.}

\Fig{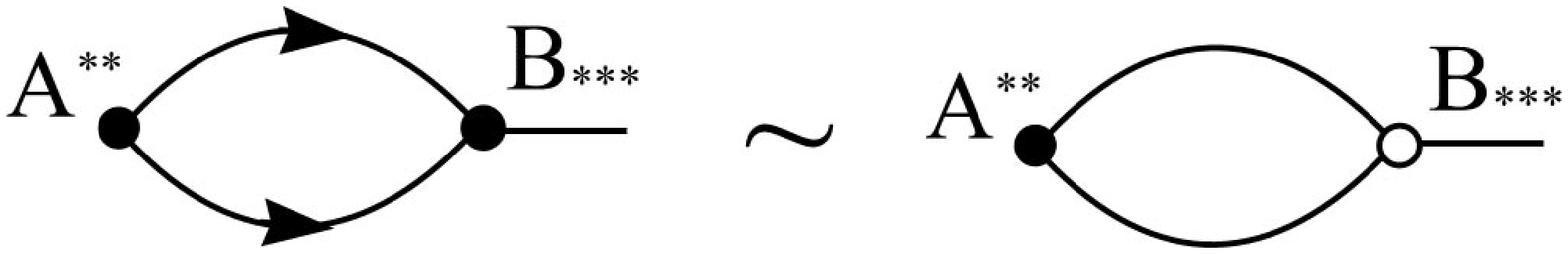} {346,54} {Contraction
$\sum_{i_1i_2}A^{i_1i_2}B_{i_1i_2j}$ in two different notations:
with arrows and with black/white vertices.}

$\bullet$ Given a structure group, one can define {\it invariant
tensors}. Existence of contraction can be ascribed to invariance
of the unit tensor $\delta^i_j$. The other tensors, invariant
under $SL(n)$, are totally antisymmetric covariant
$\epsilon_{i_1\ldots i_n}$ and contravariant $\epsilon^{i_1\ldots
i_n}$, they can be also considered as generating the
one-dimensional invariant subspaces w.r.t. enlarged structure
group $GL(n)$. These $\epsilon$-tensors can be represented by
$n$-valent diamond and crossed vertices respectively, all of the
same {\it sort}, see Fig.~\ref{ta20}.A. Reductions of structure
group increase the set of invariant tensors. Above-mentioned
reductions (which do not break $SL(n)$'s themselves, i.e. preserve
colors) just identify some sorts, i.e. add some sort-mixing
$\epsilon$-tensors.

\Fig{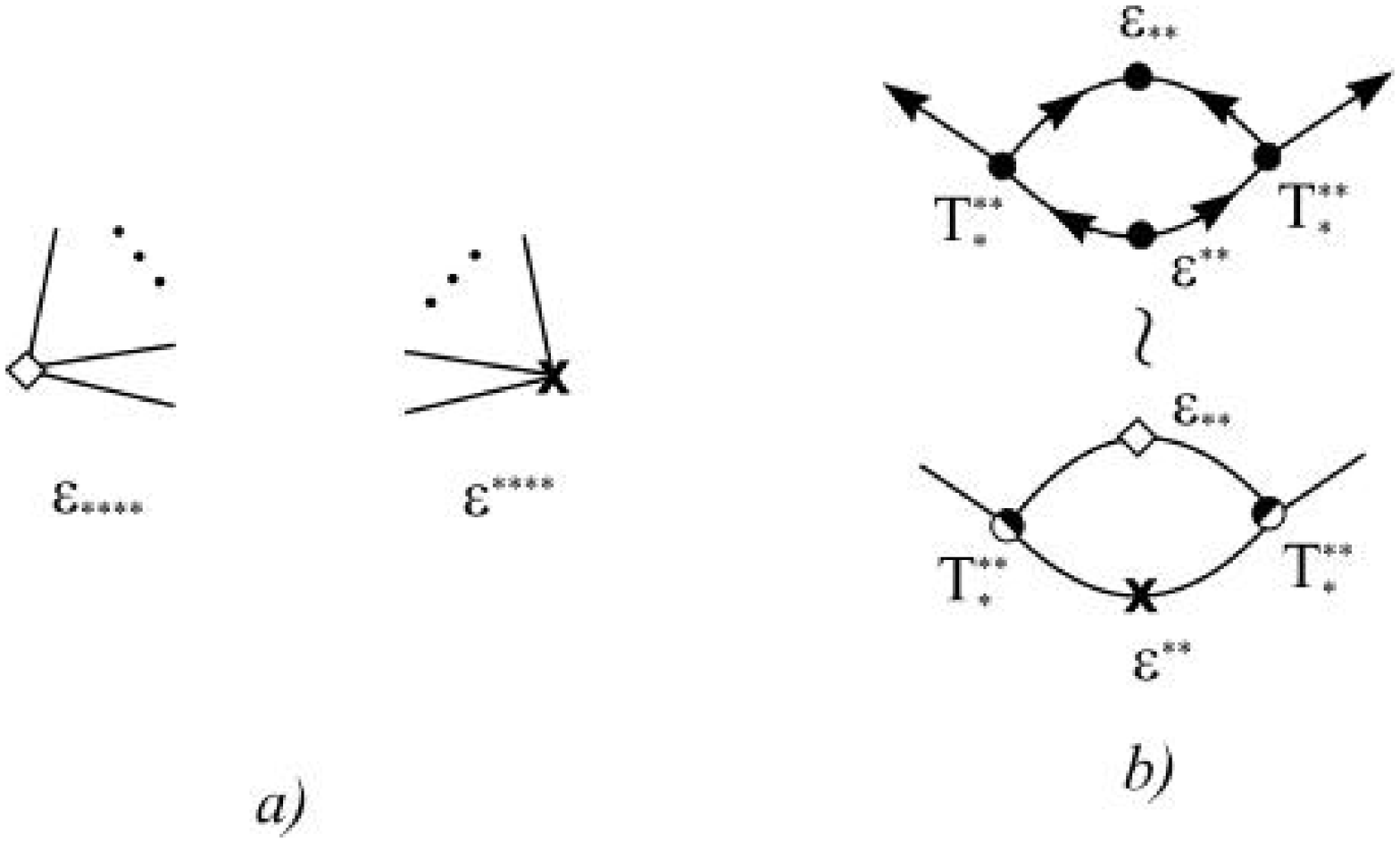} {457,273} {{\bf A.} Pictorial notation for covariant
and contravariant $\epsilon$-tensors.\ \ {\bf B.} Example of
diagram, constructed from a tensor $T$ with the help of
$\epsilon$'s. All possible diagrams made from any number of $T$'s
and $\epsilon$'s form the tensor algebra ${\cal T}(T,\epsilon)$ or
simply ${\cal T}(T)$.}

$\bullet$ Diagrams (see Fig.~\ref{ta20}.B), where all vertices
contain either the tensor $T$ or invariant $\epsilon$-tensors,
form ${\cal T}(T)$: the {\it tensor algebra}, generated by $T$.
Diagrams without external legs are homogeneous polynomials of
coefficients of $T$, invariant under the structure group. They
form a {\it ring of invariants} (or invariant's ring) ${\rm
Inv}_{{\cal T}(T)}$ of ${\cal T}(T)$. Diagrams with external legs
are representations of the structure group, specified by the
number and type of external legs.

$\bullet$ Invariants can be also obtained by taking an average of
any function of $T$ over the maximal compact subgroup of the
structure group.

$\bullet$ ${\cal T}(T)$ is an essentially non-linear object, in
all senses. It is much better represented {\it pictorially} than
{\it formally}: by necessity formulas have some {\it linear}
(line) structure, unnatural for ${\cal T}(T)$. However, pictorial
representation, while very good for qualitative analysis and
observation of relevant structures, is not very practical for
calculations. The compromise is provided by string theory methods:
diagrams can be converted to formulas with the help of
Feynman-like functional integrals. However, there is always a
problem to separate a particular diagram: functional integrals
normally describe certain sums over diagrams, and the depth of
separation can be increased by enlarging the number of different
couplings (actually, by passing from ${\cal T}(T)$ to ${\cal
T}(T,T',\ldots)$ with additional tensors $T',\ldots$); but -- as a
manifestation of complementarity principle -- the bigger the set
of couplings, the harder it is to handle the integral. Still a
clever increase in the number of couplings reveals new structures
in the integral \cite{UFN3} -- they are known as {\it integrable},
and there is more than just a game of words here, "integrability"
means deep relation to the Lie group theory. Lie structure is very
useful, because it is relatively simple and well studied, but the
real symmetry of integrals is that of the tensor algebras -- of
which the Lie algebras are very special example.

$\bullet$ ${\cal T}(T)$ can be considered as generated by a
"functional integral": \be \Big< \exp_\otimes\Big(T^{i_1\ldots
i_r}\phi_{1i_1}\otimes\ldots \otimes\phi_{ri_r}\Big), \ \
\otimes_{k=1}^r \exp_\otimes \Big(\epsilon_{i_1\ldots
i_{n_k}}\phi_k^{i_1}\otimes\dots\otimes\phi_k^{i_{n_k}}\Big) \Big>
\label{eq13} \ee e.g. for $k=2$ and $\phi_1 = \phi$, $\phi_2=\chi$
by \be \Big<\exp_\otimes\Big(T^{ij}\phi_i\chi_j\Big), \ \
\exp_\otimes \Big(\epsilon_{ij}\phi^i\phi^j\Big)\ \exp_\otimes
\Big(\epsilon_{ij}\chi^i\chi^j\Big)\Big> \ee The sign $\otimes$ is
used to separate (distinguish between) elements of different
vector spaces in $\ldots \otimes V\otimes V\otimes \ldots$
(actually, any other sign, e.g. comma, could be used instead).
Dealing with the average (\ref{eq13}), one substitutes all $\phi
\rightarrow \phi + \varphi$ and eliminates "quantum fields"
$\varphi$ with the help of the Wick rule: \be \langle
\varphi_{kj_1}\otimes\ldots\otimes \varphi_{kj_s},
\varphi_k^{i_1}\otimes\ldots\otimes\varphi_k^{i_s}\rangle
 = \delta^{i_1}_{j_1}\ldots \delta^{i_s}_{j_s}
\ee without summation over permutations(!), i.e. in the $k=2$ case
\be \langle \varphi_{j_1}\otimes\ldots\otimes \varphi_{j_s},\
\varphi^{i_1}\otimes\ldots\otimes\varphi^{i_s}\rangle\ =\ \langle
\chi_{j_1}\otimes\ldots\otimes \chi_{j_s},\
\chi^{i_1}\otimes\ldots\otimes\chi^{i_s}\rangle
 = \delta^{i_1}_{j_1}\ldots \delta^{i_s}_{j_s}
\ee Fields $\phi_k$ with different sorts $k$ are assumed
commuting. All quantities are assumed lifted to entire Fock space
by the obvious {\it comultiplication}, $\phi \rightarrow \ldots
\otimes \phi \otimes 0 \otimes  \ldots \ + \ \ldots \otimes 0
\otimes \phi \otimes  \ldots$ with (infinite) summation over all
possible positions of $\phi$. The language of tensor categories is
not very illuminating for many, fortunately it is enough to think
and speak in terms of diagrams. Formulation of {\it constructive}
functional-integral representation for ${\cal T}(T)$ remains an
interesting and important problem, as usual, integral can involve
additional structures, and (in)dependence of these structures
should provide nice reformulations of the main properties of
${\cal T}(T)$.

$\bullet$ Particular useful example of a functional integral,
associated with ${\cal T}(T)$ for a $n_1\times\ldots \times n_r$
contravariant tensor $T^{i_1\ldots i_r}$ with $i_k=1,\ldots,n_k$,
is given by
$$
Z(T) = \left\{\prod_{k=1}^r \left(\prod_{i=1}^{n_k} \int\int {\cal
D}x_{ki}(t){\cal D}\bar x^i_k(t) e^{\int x_{ki}(t)\bar
x^i_k(t)dt}\right)\cdot\right.
$$\vspace{-0.5cm}
\be \left.\cdot\exp {\int\ldots\int_{t_1<\ldots <t_{n_k}}
\epsilon_{i_1\ldots i_{n_k}} \bar x^{i_1}_k(t_1)\ldots \bar
x^{i_{n_k}}_k(t_n)dt_1\ldots dt_n} \right\} \exp {\int
T^{i_1\ldots i_r}x_{1i_1}(t)\ldots x_{ri_r}(t)dt} \label{fint} \ee
Here $t$ can be either continuous or discrete variable, and
integral depends on the choice of this integration domain. The
"fields" $x$ and $\bar x$ can be bosonic or fermionic
(Grassmannian). In operator formalism $Z(T)$ is associated with an
application of operators \be
\hat E_k = 
\exp \left(\epsilon_{i_1\ldots i_n}
\frac{\partial}{\partial x_{ki_1}}
\otimes \ldots \otimes \frac{\partial}{\partial x_{ki_1}} \right)
\ee with different $k$ ($n$ can depend on $k$) to various powers
of $(\oplus T)^m = \Big(\oplus T^{i_1\ldots i_r} x_{1i_1}\ldots
x_{ri_r}\Big)^m$: \be Z(T) = \left.\left(\prod_{k=1}^r \hat
E_k\right) \sum_{m=0}^\infty \frac{({\oplus T})^m}{m!} \
\right|_{\ {\rm all}\ x=0} \label{opint} \ee Generalizations to
many tensors $T$ and to $T$'s which are not pure contravariant are
straightforward.

In the simplest $2\times 2$ example ($r=2$, $n_1=n_2=2$), one can
switch to a more transparent notation: $x_{1i}(t) = x_i(t)$,
$x_{2i}(t)=y_i(t)$ and \be Z(T) = \left.\exp \sum_{t<t'}
\left(\epsilon_{ij} \frac{\partial^2}{\partial x_i(t)\partial
x_j(t')} + \epsilon_{ij} \frac{\partial^2}{\partial y_i(t)\partial
y_j(t')}\right) \exp \sum_t T^{ij}x_i(t)y_j(t)\ \right|_{\ {\rm
all}\ x,y=0}= \nn \\ = \int Dx_i(t) Dy_i(t) D\bar x^i(t) D\bar
y^i(t) \prod_t e^{x_i\bar x^i(t) + y_i\bar y^i(t) +
T^{ij}x_iy_i(t)} \prod_{t<t'} e^{\epsilon_{ij} \left(\bar
x^i(t)\bar x^j(t') + \bar y^i(t)\bar y^j(t')\right)}
\label{fint22} \ee (as usual $\int Dx(t) \equiv \prod_t \int
dx(t)$). In this particular case (all $n=2$) integral is Gaussian
and can be evaluated exactly (see s.\ref{sypo22}). For generic
$n>2$ non-Gaussian is the last factor with $\epsilon$-tensors.

$\bullet$ Above generating function $Z(T)$ is by no means unique.
We mention just a few lines of generalization.

Note, for example, that\ $({\oplus T})^m = T^m\otimes 0 \otimes
\ldots \otimes 0 \ + \ mT^{m-1}\otimes T \otimes \ldots \otimes 0\
+ \ \ldots$\ is not quite the same as\ $T^{\otimes m} = T\otimes T
\otimes \ldots \otimes T$. From $T^{\otimes m} = \Big(T^{i_1\ldots
i_r} x_{1i_1}\ldots x_{ri_r}\Big)^{\otimes m}$ one can build \be
\tilde Z(T) = \left.\left(\prod_{k=1}^r \hat E_k\right)
\sum_{m=0}^\infty \frac{T^{\otimes m}}{m!} \ \right|_{\ {\rm all}\
x=0} \label{opinttilde} \ee -- an analogue of $Z(T)$ with a more
sophisticated integral representation.

Both $Z(T)$ and $\tilde Z(T)$ respect {\it sorts} of the lines:
operator $\hat E_k$ carries the sort index $k$ and does not mix
different sorts. One can of course change this property and
consider integrals, where diagrams with sort-mixing are
contributing.

One can also introduce non-trivial totally antisymmetric weight
functions $h(t_1,\ldots,t_n)$ into the terms with $\epsilon$'s and
obtain new interesting types of integrals, associated with the
same ${\cal T}(T)$. An important example is provided by the {\it
nearest neighbors} weight, which in the limit of continuous $t$
gives rise to a {\it local} action
$$
\epsilon_{i_1\ldots i_n} \int x^{i_1}_k(t)\ d_tx^{i_2}_k(t)\
\ldots \ d^{n-1}_t x^{i_n}_k(t) dt.
$$

\subsubsection{Solutions to poly-linear and non-linear equations}

\noindent

$\bullet$ Tensors can be used to define systems of algebraic
equations, poly-linear and non-linear. These equations are
automatically projective in each of the {\it sorts} and we are
interested in solutions modulo all these projective
transformations. If the number of independent homogeneous
variables $N^{var}$ is smaller than the number of projective
equations $N^{eq}$, solutions exist only if at least $N^{con} =
N^{eq}-N^{var}$ constraints are imposed on the coefficients. If
this restriction is saturated, projectively-independent solutions
are discrete, otherwise we get $\Big(N^{sol} = N^{con} + N^{var} -
N^{eq}\Big)$-parametric continuous families of solutions (which
can form a discrete set of intersecting branches).

$\bullet$ Since the action of structure group converts solutions
into solutions, the $N^{con}$ constraints form a representation of
structure group. If $N^{con}=1$, then this single constraint is a
singlet respresentation, i.e. {\it invariant}, called {\it
discriminant} or {\it resultant} of the system of equations,
poly-linear and non-linear respectively.

$\bullet$ Resultant vanishes when the system of homogeneous
equations becomes {\it resolvable}. Equations define a map from
the space of variables and ask what points are mapped into zero.
Generically homogeneous map converts a projective space $P^{n-1}$
onto itself, so that zero of homogeneous coordinates on the target
space, which does not belong to $P^{n-1}$, has no pre-image
(except for original zero). Moreover, for non-linear maps each
point of the target space has several pre-images: we call their
number the {\it index} of the map at the point. For some maps,
however, the index is smaller than in generic case: this happens
exactly because some points from the image move into zero and
disappear from the target $P^{n-1}$. These are exactly the maps
with vanishing resultant.

When index is bigger than one, all points of the target $P^{n-1}$
stay in the image even when resultant vanishes: just the number of
pre-images drops down by one. However, if index already was one,
then the index of maps with vanishing resultant drops down to zero
at all points beyond some subvariety of codimension one, so that
most of points have no pre-images, and this means that the {\it
dimension} of image decreased. Still this phenomenon is nothing
but a particular case of the general one: decrease of the image
dimension is particular case of decrease of the index, occuring
when original index was unity.

The best known example is degeneration of {\it linear} maps: $C^n
\rightarrow C^n:\ z_i \rightarrow \sum_{j=1}^n A_i^j z_j$ \
usually maps vector space $C^n$ {\it onto} itself, but for some
$n\times n$ matrices $A_{ij}$ the image is $C^{n-k}$: has
non-vanishing codimension $k$ in $C^n$. This happens when matrix
$A_{ij}$ has rank $n-k<n$, and a necessary condition is vanishing
of its resultant, which for matrices is just a determinant, ${\cal
R}_{n|1}\{A\} \equiv \det A = 0$ (for $k>1$ also minors of smaller
sizes, up to $n+1-k$ should vanish).

The second, equally well known, example of the {\it same}
phenomenon is degeneration of {\it non-linear} maps, but only of
two homogeneous (or one projective) variables: $C^2 \rightarrow
C^2:  \ (x,y) \rightarrow \Big(P_{s_1}(x,y), P_{s_2}(x,y)\Big)$
with two homogeneous polynomials $P(x,y)$ of degrees $s_1$ and
$s_2$. Normally the image of this map is $s_1s_2$-fold covering of
$C^2$, i.e. has index $s_1s_2$. As a map $P^1 \rightarrow P^1$ it
has a lower index, \ ${\rm max}(s_1,s_2)$). When the two
polynomials, considered as functions of projective variable
$\xi=x/y$ have a common root, the latter index decreases by one.
Condition for this coincidence is again the vanishing of the
resultant: ${\rm Res}_\xi\Big(P_{s_1}(\xi,1), P_{s_2}(\xi,1)\Big)
= 0$.

To summarize, for linear maps the vanishing of resultant implies
that {\it dimension} of the image decreases. However, in
non-linear situation this does not need to happen: the map remains
a surjection, what decreases is not dimension of the image, but
the number of branches of inverse map. This number -- {\it index}
-- is appropriate non-linear generalization of quantities like
kernel dimensions in the case of linear maps, and it should be
used in construction of non-linear complexes and non-linear
cohomologies (ordinary linear complexes can also help in
non-linear studies, see, for example, s.\ref{Koshul} below).

$\bullet$ Thus ordinary determinant and ordinary resultant are
particular examples of generic quantity, which measures
degeneration of arbitrary maps, and which is called {\it
resultant} in the present paper. {\it Discriminant} is its
analogue for poly-linear functions, in above examples it is the
same determinant for the linear case and the ordinary discriminant
(condition that the two roots of a single function coincide) in
the polynomial case.

$\bullet$ Throughout the text we freely convert between
homogeneous and projective coordinates. Homogeneous coordinates
$\vec z = \{z_i, \ i=1,\ldots,n\}$ span a vector space $V_n$ Its
dual $V_n^*$ is the vector space of all linear functions of $n$
variables. Projectivization factorizes\ $V_n-0$\ (i.e. $V_n$ with
zero excluded) w.r.t. the common rescalings of all $n$
coordinates: $P^{n-1} = \Big\{ \vec z \sim \lambda \vec z, \
\forall \lambda\neq 0 \ {\rm and}\ \vec z \neq 0\Big\}$.
Projectivization is well defined for homogeneous polynomials of a
given degree and for homogeneous equations, where all items have
the same power in the variable $\vec z$. Any polynomial equation
can be easily made homogeneous by adding an auxiliary homogeneous
variable and putting it in appropriate places, e.g. \ $ax+b
\longrightarrow ax+by$,\ $ax^2+bx+c \longrightarrow ax^2 + bxy +
cy^2$ etc:
$$
\Big\{ {\rm the\ space\ of\ arbitrary\ polynomials\ of\ degrees}
\leq s\ {\rm of}\ n-1\ {\rm variables}\Big\} =
$$ $$
= \Big\{ {\rm the\ space\ of\ homogeneous\ polynomials\ of\
degree}\ s\ {\rm of}\ \ n\ \ {\rm variables}\Big\}
$$
A system of $n-1$ non-homogeneous equations on $n-1$ variables is
equivalent to a system of $n-1$ homogeneous equations, but of $n$
variables. The latter system has continuous one-parametric set of
solutions, differing by the value of the added auxiliary variable.
If this value is fixed, then in the section we normally get a
discrete set of points, describing solutions to the former system.
Of separate interest are the special cases when the one-parametric
set is tangent to the section at intersection point.

Projective coordinates can be introduced only in particular
charts, e.g. $\xi_k = z_k/z_n$, $k = 1,\ldots,n-1$. A system of
linear equations, $\sum_{j=1}^n A_i^jz_j = 0$, defines a map of
projective spaces \ \ $P^{n-1} \rightarrow P^{n-1}:\ \ \ z_i
\rightarrow \sum_{j=1}^n A_i^jz_j,\ i,j=1,\ldots,n,$\ which in
particular chart looks like a rational map
$$
\xi_i \rightarrow \frac{\sum_{j=1}^{n-1}A_i^j\xi_j + A_i^n}
{\sum_{j=1}^{n-1}A_n^j\xi_j + A_n^n},\ \ i,j=1,\ldots,n-1.
$$
However, the equations themselves has zero at the r.h.s., which
does not look like a point of $P^{n-1}$. And indeed, for
non-degenerate matrix $A$ equation does not have non-vanishing
solutions, i.e. no point of $P^{n-1}$ is mapped into zero, i.e.
$P^{n-1}$ is indeed mapped into $P^{n-1}$. In fact, this is a map
{\it onto}, since non-degenerate $A$ is invertible and every point
of the target $P^{n-1}$ has a pre-image. If $A$ is degenerate,
$\det A = 0$, the map still exists, just its image has codimension
one in $P^{n-1}$, but the seeming zero -- if properly treated --
belongs to this diminished image. For example, for $n=2$ we have
$\left(\begin{array}{c} x \\ y \end{array}\right)
\longrightarrow \left(\begin{array}{c} ax+by \\
cx+dy \end{array}\right)$ or $\xi \rightarrow
\frac{a\xi+b}{c\xi+d}$. If the map is degenerate, i.e. $ad=bc$,
then this ratio turns into constant: $ \xi \rightarrow
\frac{a}{c}$, i.e. entire $P^1$ is mapped into a single point
$a/c$ of the target $P^1$. By continuity this happens also to the
point $x/y = \xi = -c/d = -a/b$, which is the non-trivial solution
of the system
$\left\{\begin{array}{c} ax+by = 0\\
cx+dy = 0 \end{array}\right.$ Thus a kind of a l'Hopital rule
allows one to treat homogeneous equations in terms of projective
spaces. Of course, this happens not only for linear, but also for
generic non-linear and poly-linear equations (at least
polynomial): entire theory has equivalent homogeneous and
projective formulations and they will be used on equal footing
below without further comments.

\bigskip

\hspace{-2.27cm}
\begin{tabular}{|c|c|c|}
\hline && \\
Tensor & Relevant quantities & Typical results \\
&& \\ \hline &&\\
Generic rank-$r$ rectangular tensor & Discriminant (Cayley
hyperdeterminant) & $\bullet$\
${\rm deg}_T(D)$ -- see s.\ref{degdi} \\
of the type $n_1\times \ldots \times n_r$: & ${\cal D}_{n_1\times
\ldots \times n_r}(T)$ & $\bullet$\  itedisc (iteration in $r$):
\\
$T^{i_1\ldots i_r}$,\ \ $1\leq i_k \leq n_k$ & &${\cal
D}_{n_1\times\ldots\times n_r\times n_{r+1}} (T^{i_1\ldots
i_ri_{r+1}})=$
\\
or a function of $r$ $n_k$-component vectors & ${\cal D}(T)=0$ is
consistency condition &$= {\bf irf} \Big({\cal D}^{(t)}_{n_{r+1}|
{\rm deg}({\cal D}_{n_1\times\ldots\times n_r})}$
\\
$T(\vec x_1,\ldots,\vec x_k) = \sum_{\stackrel{1\leq i_k \leq
n_k}{1\leq k \leq r}} T^{i_1\ldots i_r} x_{1i_1}\ldots x_{ri_r}$
&(existence of solution with all $\vec x_k \neq \vec 0$)
&$\left({\cal D}_{n_1\times\ldots\times n_r} (T^{i_1\ldots
i_ri_{r+1}}t_{i_{r+1}})\right)\Big)$
\\
Coefficients $T_{i_1\ldots i_r}$ are placed at points &for the
system\ $\frac{\partial T}{\partial\vec x_k} = 0$\ \ (i.e.
$\frac{\partial T(x)}{\partial x_{ki_k}} = 0$) &$\bullet$\
Additive decomposition \cite{D1}
\\
of the $n_1\times \ldots \times n_r$ hyperparallepiped of
&&$\Big(D_{n^{\times r}}(T)\times\ {\rm sub-discriminants} \Big)$
\\
&& \\ \hline &&\\
Totally hypercubic symmetric rank-$r$ tensor, & Symmetric
discriminant &
\\
$\begin{array}{c}{\rm i.e.\ all}\ n_k = n\ {\rm and} \\
{\rm for\ any\ permutation}\ P \in \sigma_n\end{array}$ &
$D_{n|r}(S) = {\bf irf} \Big( {\cal
D}_{\stackrel{\underbrace{n\times \ldots \times n}} {r\ {\rm
times}}}(S)\Big)$ & $\bullet$ $D_{n|r}(S) = {\cal
R}_{n|r-1}\{\vec\partial S\}$
\\
$S^{i_1\ldots i_r} = S^{i_{P(1)}\ldots i_{P(r)}}$ & (an
irreducible factor in the full discriminant,
& $\bullet$\ ${\rm deg}_S{D}_{n|r}(S) = n(r-1)^{n-1}$ \\
or a function ($r$-form) of a single vector $\vec x$ & emerging
for {\it hypercube} and total symmetry); &
\\
$S(\vec x) = \sum_{\stackrel{1\leq i_k \leq n}{1\leq k \leq r}}
S^{i_1\ldots i_r}x_{i_1}\ldots x_{i_r}$ & ${D}(S) = 0$ is
consistency condition for $\frac{\partial S}{\partial \vec x} = 0$
&
\\
&& \\ \hline &&\\
Totally antisymmetric tensor (all $n_k = n$) & HyperPfaffian
&\\
$C^{i_1\ldots i_r} = (-)^P C^{i_{P(1)}\ldots i_{P(r)}}$ &
$\Big({\cal PF}_{n|r}(C)\Big)^{\nu} = {\bf irf} \Big( {\cal
D}_{\stackrel{\underbrace{n\times \ldots \times n}} {r\ {\rm
times}}}(C)\Big)$
&\\
for any permutation $P \in \sigma_n$ &for some power $\nu$
&\\
&& \\ \hline &&\\
Homogeneous map $V_n \rightarrow V_n$ of degree $s$ & Resultant
${\cal R}_{n|s}\{\vec A\} = {\bf irf} \Big({\cal
D}_{\stackrel{\underbrace{n\times\ldots\times n}} {s+1\ {\rm
times}}}(A_{i\alpha})\Big)$ & $\bullet$\ ${\rm deg}({\cal
R}_{n|s}) = ns^{n-1}$
\\
defined by a tensor of rank $r=s+1$, & ${\cal R}=0$ is consistency
condition & $\bullet$\ iteres (iteration in $n$):
\\
totally symmetric in the last $s$ indices & (existence of
non-vanishing solution $\vec z\neq 0$) & ${\cal
R}_{n+1|s}\left\{A_1(\vec z),\ldots, A_{n+1}(\vec z)\right\} = $
\\
$A_i^\alpha = A_i^{j_1\ldots j_s}$ & for the homogeneous system
$\vec A(\vec z)=0$ & $Res_{z_{n+1}}\Big({\cal
R}_{n|s}\left\{A_1(\vec z), \ldots,A_{n}(\vec z)\right\}$,
\\
with totally symmetric multi-index && ${\cal
R}_{n|s}\left\{A_2(\vec z),\ldots,A_{n+1}(\vec z)\right\}
\Big)$ \\
$\alpha = \{j_1\ldots,j_s\}$, & Eigenvectors and order-$p$
periodic orbits &
$\bullet$ Composition law \\
which takes $M_{n|s} = \frac{(n+s-1)!}{(n-1)!s!}$ values &
 of $\vec A(\vec z)$
(sol's $\vec z = \vec e_{\mu}^{(p)}$ to $A^{\circ p}_i(\vec z) =
\lambda(\vec z) z_i$) & ${\cal R}_{n|s_As_B}(A\circ B) = {\cal
R}_{n|s_A}^{s_B^{n-1}}!(A) {\cal R}_{n|s_B}^{s_A^{n}}!(B)$
\\
or $n$-comp. vector $\vec A(\vec z) = \left\{\sum_{\alpha}
A_{i\alpha}z^\alpha\right\} =$ & & $\bullet$\ Additive expansion
\\
$= \left\{\sum_{j_1,\ldots,j_s=1}^n A_i^{j_1\ldots
j_s}z_{j_1}\ldots z_{j_s}\right\}$ & Mandelbrot set ${\cal
M}_{n|s}$: & in Plukker determinants
\\
formed by homegeneous pol's of degree $s$ & $\vec A(\vec z) \in
{\cal M}_{n|s}$ if some two orbits of ${\cal A}$ merge:
&$\bullet$\ Eigenvalue decomposition\\
& $\vec e_{\mu}^{(p)} = \vec e_{\nu}^{(q)}$ for some $(p,\mu)\neq
(q,\nu)$ &${\cal R}_{n|s}\{A\} \sim \prod_{\mu=1}^{c_{n|s}}
\lambda(\vec e_\mu^{(1)})$\\
&& \\ \hline &&\\
Tensors of other types &&
\\
&& \\ \hline\hline &&\\
Arbitrary non-linear map $V_n \rightarrow V_n$, i.e. & Resultant
of generic non-linear system & $\bullet$ ${\cal
R}_{n|s_1,\ldots,s_n} = {\bf irf}\left({\cal R}_{n|{\rm
max}(s_1,\ldots,s_n)}\right)$
\\
collection of $n$ symmetric tensors & ${\cal
R}_{n|s_1,\ldots,s_n}\left\{A_i^{\alpha_i} \vec
z_{\alpha_i}\right\}$
&\\
$A_i^{j_1\ldots j_{s_i}}$ of ranks $s_1,\ldots,s_n$ & ${\cal
R}\{\vec A\} = 0$ if the system $\vec A(\vec z) = 0$
&\\
$\vec A(\vec z) =\left\{\sum_{j_1,\ldots,j_{s_i}=1}^n
A_i^{j_1\ldots j_{s_i}}z_{j_1}\ldots z_{j_{s_i}}\right\}$ & has
non-vanishing solution $\vec z\neq 0$
&\\
&& \\ \hline
\end{tabular}

\setcounter{equation}{0}

\section{Solving equations. Resultants}

\subsection{Linear algebra (particular case of $s=1$)}

We begin with a short summary of the theory of {\it linear}
equations. The basic problem of {\it linear} algebra is solution
of a system of $n$ {\it linear} equations of $n$ variables, \be
\sum_{j=1}^n A_i^jz_j = a_i \label{noholi} \ee In what follows we
often imply summation over repeated indices and omit explicit
summation sign, e.g. $A_i^jz_j = \sum_{j=1}^n A_i^jz_j$. Also, to
avoid confusion between powers and superscripts we often write all
indices as {\it sub}scripts, even if they label contravariant
components.

\subsubsection{Homogeneous equations}

In general position the system of $n$ homogeneous equations for
$n$ variables, \be A_i^jz_j = 0 \label{holi} \ee has a single
solution: all $z_j = 0$. Non-vanishing solution exists only if the
$n^2$ coefficients $A_i^j$ satisfy {\it one} constraint: \be
\det_{n\times n} A_i^j = 0, \ee i.e. the certain homogeneous
polynomial of degree $n$ in the coefficients of the matrix
$A_{ij}$ vanishes.

If $\det A = 0$, the homogeneous system (\ref{holi}) has solutions
of the form (in fact this is a {\it single} solution, see below)
\be Z_j = \check A_j^kC_k, \label{solholi} \ee where $\check
A_j^k$ is a minor -- determinant of the $(n-1)\times(n-1)$ matrix,
obtained by deleting the $j$-th row and $k$-th column from the
$n\times n$ matrix $A$. It satisfies: \be A_i^j\check A_j^k =
\delta^k_i\det A, \ \ \ \check A_j^kA_k^i = \delta_j^i\det A \
\label{AcheckA} \ee and \be \delta\det A = \sum_{i,j=1}^n \check
A_j^i\delta A_i^j \ \label{AdetA} \ee Eq.(\ref{solholi}) solves
(\ref{holi}) for any choice of parameters $C_k$ as immediate
corrolary of (\ref{AcheckA}), provided $\det A = 0$. However,
because of the same (\ref{AcheckA}), the shift $C_k \rightarrow
C_k + A_k^lB_l$ with any $B_l$ does not change the solution
(\ref{solholi}), and actually there is a single-parametric family
of solutions (\ref{solholi}), different choices of $C_k$ provide
projectively equivalent $Z_j$.

If rank of $A$ is smaller than $n-1$ (${\rm corank}(A) >1$), then
(\ref{solholi}) vanishes, and non-vanishing solution is given by
\be Z_j = \check A_{jj_1}^{k_1k_2} C^{j_1}_{k_1k_2} \ \ {\rm if\
corank}(A)=2,
\nn \\
Z_j = \check A_{jj_1\ldots j_{q-1}}^{k_1\ldots k_q} C^{j_1\ldots
j_{q-1}}_{k_1\ldots k_q} \ \ {\rm if\ corank}(A)=q \ee $\check
A_{\{j\}}^{\{k\}}$ denotes minor of the $(n-q)\times(n-q)$ matrix,
obtained by deleting the set $\{j\}$ of rows and the set $\{k\}$
of columns from $A$. Again most of choices of parameters $C$ are
equivalent, and there is a $q$-dimensional space of solutions if
${\rm corank}(A)=q$.

\subsubsection{Non-homogeneous equations}

Solution to non-homogeneous system (\ref{noholi}) exists and is
unique when $\det A \neq 0$. Then it is given by the Craemer rule,
which we present in four different formulations.

As a corollary of (\ref{AcheckA}) \be {\bf{Craemer\ I:}}\ \ \ \ \
\ Z_j = \frac{\check A_j^ka_k}{\det A} =
\left(A^{-1}\right)_j^ka_k \label{liCraI} \ee With the help of
(\ref{AdetA}), this formula can be converted into \be
{\bf{Craemer\ II:}}\ \ \ \ \ Z_j = \frac{\partial \log\det
A}{\partial A^j_k}a_k = \frac{\partial {\rm Tr}\log A}{\partial
A^j_k}a_k \label{liCraII} \ee Given the $k-th$ component $Z_k$ of
the solution to non-homogeneous system (\ref{noholi}), one can
observe that the following {\it homogeneous} equation: \be
\sum_{j\neq k}^n A_i^jz_j + \Big(A_i^kZ_k - a_i\Big) z_k = 0
\label{liCraIIIa} \ee (no sum over $k$ in this case!) has a
solution: $z_j = Z_j$ for $j\neq k$ and $z_k = 1$. This means that
determinant of associated $n\times n$ matrix \be
[A^{(k)}]_i^j(Z_k) \equiv (1-\delta^j_k)A_i^j + \delta^j_k
(A_i^kZ_k - a_i) \ee vanishes. This implies that $Z^k$ is solution
of the equation \be {\bf{Craemer\ III:}}\ \ \ \ \ \det_{n\times n}
[A^{(k)}]_i^j(z) = 0 \label{liCraIII} \ee The l.h.s. is a actually
a linear function of $z$: \be \det_{n\times n} [A^{(k)}]_i^j(z) =
z \det A - \det A^{(k)}_{\vec a} \ee where $n\times n$ matrix
$A^{(k)}_{\vec a}$ is obtained by substituting of $\vec a$ for the
$k$-th column of $A$: $A_j^k \rightarrow a_j$. Thus we obtain from
(\ref{liCraIII}) the Craemer rule in its standard form: \be
{\bf{Craemer\ IV:}}\ \ \ \ \ Z_k = \frac{\det A^{(k)}_{\vec
a}}{\det A} \label{liCraIV} \ee

If $\det A = 0$ non-homogeneous system (\ref{noholi}) is
resolvable only if the vector $a_i$ is appropriately constrained.
It should belong to the image of the linear map $A(z)$, or, in the
language of formulas, \be \check A_j^ka_k = 0, \ee as obvious from
(\ref{AcheckA}).

\subsection{Non-linear equations}

Similarly, the basic problem of {\it non-linear} algebra is
solution of a system of $n$ {\it non-linear} equations of $n$
variables. As mentioned in the introduction, the problem is purely
{\it algebraic} if equations are polynomial, and in this paper we
restrict consideration to this case, though {\it analytic}
extention should be also available (see s.4.11.2 of \cite{DM} for
preliminary discussion of such generalizations).

\subsubsection{Homogeneous non-linear equations}

As in linear algebra, it is worth distinguishing between
homogeneous and non-homogeneous equations. In homogeneous
(projective) case non-vanishing solutions exist {\it iff} the
coefficients of all equations satisfy a single constraint,
$${\cal R}\{{\rm system\ of\ homogeneous\ eqs}\}= 0,$$
and solution to non-homogeneous system is algebraically expressed
through the ${\cal R}$-functions by an analogue of the Craemer
rule, see s.\ref{Crae}. ${\cal R}$-function is called the {\it
resultant} of the system. It is naturally labelled by two types of
parameters: the number of variables $n$ and the set of powers
$s_1,\ldots,s_n$. Namely, the homogeneous system consisting of $n$
polynomial equations of degrees $s_1,\ldots,s_n$ of $n$ variables
$\vec z = (z_1,\ldots,z_n)$, \be A_i(\vec z) =
\sum_{j_1,\ldots,j_s=1}^n A_i^{j_1\ldots j_{s_i}} z_{j_1}\ldots
z_{j_{s_i}} = 0 \label{Aivecz} \ee has non-vanishing solution
(i.e. at least one $z_j \neq 0$) {\it iff} \be {\cal
R}_{s_1,\ldots,s_n}\left\{A_i^{j_1\ldots j_{s_i}}\right\}=0. \ee
Resultant is a polynomial of the coefficients $A$ of degree \be
d_{s_1,\ldots,s_n} = {\rm deg}_A {\cal R}_{s_1,\ldots,s_n} =
\sum_{i=1}^n \Big(\prod_{j\neq i} s_j\Big) \ee When all degrees
coincide, $s_1=\ldots=s_n=s$, the resultant ${\cal R}_{n|s}$ of
degree $d_{n|s} = {\rm deg}_A {\cal R}_{n|s} = ns^{n-1}$ is
parameterized by just two parameters, $n$ and $s$. Generic ${\cal
R}_{s_1,\ldots, s_n}$ is straightforwardly reduced to ${\cal
R}_{n|s}$, because multiplying equations by appropriate powers of,
say, $z_n$, one makes all powers equal and adds new solutions
(with $z_n=0$) in a controllable way: they can be excluded by
obvious iterative procedure and ${\cal R}_{s_1,\ldots,s_n}$ is an
easily extractable irreducible factor ({\bf irf}) of ${\cal
R}_{n|{\rm max}(s_1,\ldots,s_n)}$.

$A_i(\vec z)$ in (\ref{Aivecz}) can be considered as a map
$P^{n-1} \rightarrow P^{n-1}$  of projective space on itself, and
${\cal R}_{n|s}$ is a functional on the space of such maps of
degree $s$. In such interpretation one distinguishes between
indices $i$ and $j_1,\ldots,j_s$ in $A_i(\vec z) = A_i^{j_1\ldots
j_s} z_{j_1}\ldots z_{j_s}$: $j$'s are contravariant, while $i$
covariant.

If considered as elements of projective space $P^{n-1}$,
one-parametric solutions of homogeneous equations (existing when
resultant vanishes, but resultants of the subsystems -- the
analogues of the minors do not), are {\it discrete} points. The
number of these points (i.e. of {\it barnches} of the original
solution) is \be \#_{s_1,\ldots,s_n} = \prod_{i=1}^n s_i. \ee

Of course, in the particular case of the linear maps (when all
$s=1$) the resultant coincides with the ordinary determinant: \be
{\cal R}_{n|1}\{A\} = \det_{n\times n} A. \ee

\bigskip

{\bf Examples:}

\bigskip

For $n=0$ there are no variables and we assume ${\cal
R}_{0|s}\equiv 1$.

For $n=1$ the homogeneous equation of one variable is $Az^s=0$ and
$R_{1|s} = A$.

In the simlest non-trivial case of $n=2$ the two homogeneous
variables can be named $x=z_1$ and $y=z_2$, and the system of two
equations is
$$
\left\{\begin{array}{c} A(x,y) = 0 \\ B(x,y)=0 \end{array}\right.
$$ $$
{\rm with} \ \ \ A(x,y)  = \sum_{k=0}^s a_kx^ky^{s-k} =
a_s\prod_{j=1}^s (x-\lambda_jy) = y^s\tilde A(t) \ \ \ {\rm and} \
\ \ B(x,y)  = \sum_{k=0}^s b_kx^ky^{s-k} = b_s\prod_{j=1}^s
(x-\mu_jy) = y^s\tilde B(t),
$$
where $t = x/y$. Its resultant is just the ordinary resultant
\cite{resultant} of two polynomials of a single variable $t$: \be
{\cal R}_{2|s}\{A,B\} = {\rm Res}_t(\tilde A,\tilde B) =
(a_sb_s)^s\prod_{i,j=1}^s (\lambda_i - \mu_j) =
(a_0b_0)^s\prod_{i,j=1}^s \left(\frac{1}{\mu_j} -
\frac{1}{\lambda_i}\right) = \nn \\ = {\det}_{2s\times
2s}\left(\begin{array}{cccccccccc}
a_s & a_{s-1} & a_{s-2} & \ldots & a_1 & a_0 & 0 & 0 & \ldots & 0 \\
&&&& &&&& \\
0 & a_s & a_{s-1} & \ldots & a_2 & a_1 & a_0 & 0 & \ldots & 0 \\
&&&& &&&& \\
&&&&\ldots &&&& \\
&&&& &&&& \\
0 & 0 & 0 & \ldots & a_s & a_{s-1} & a_{s-2} & a_{s-3}& \ldots & a_0 \\
&&&& &&&& \\
b_s & b_{s-1} & b_{s-2} & \ldots & b_1 & b_0 & 0 & 0 & \ldots & 0 \\
&&&& &&&& \\
0 & b_s & b_{s-1} & \ldots & b_2 & b_1 & b_0 & 0 & \ldots & 0 \\
&&&& &&&& \\
&&&&\ldots &&&& \\
&&&& &&&& \\
0 & 0 & 0 & \ldots & b_s & b_{s-1} & b_{s-2} & b_{s-3}& \ldots &
b_0
\end{array}\right)
\label{ordresdetrep} \ee (If powers $s_1$ and $s_2$ of the two
polynomials are different, the resultant is determinant of the
$(s_1+s_2)\times(s_1+s_2)$ matrix of the same form, with first
$s_2$ rows containing the coefficients of degree-$s_1$ polynomial
and the last $s_1$ rows containing the coefficients of
degree-$s_2$ polynomial. We return to a deeper description -- and
generalizations -- of this formula in s.\ref{Koshul} below.) This
justifies the name {\it resultant} for generic situation. In
particular case of linear map ($s=1$) eq.(\ref{ordresdetrep})
reduces to determinant of the $2\times 2$ matrix and ${\cal
R}_{2|1}\{A\} = {\rm Res}_t\Big(a_1t + a_0,\ b_1t+b_0\Big) =
\left|\left|\begin{array}{cc} a_1 & a_0 \\ b_1 & b_0 \end{array}
\right|\right| = \det_{2\times 2} A$.

\subsubsection{Solution of systems of non-homogeneous equations:
generalized Craemer rule \label{Crae}}

Though originally defined for {\it homogeneous} equations, the
notion of the resultant is sufficient to solving {\it
non-homogeneous} equations as well. More accurately, this problem
is reduced to solution of ordinary algebraic equations of a single
variable, which is non-linear generalization of the ordinary
Craemer rule in the formulation (\ref{liCraIII}). We begin from
particular example and then formulate the general prescription.

\bigskip

{\bf Example of $n=2$, $s=2$:}

\bigskip

Consider the system of two non-homogeneous equations on two
variables: \be \left\{\begin{array}{c}
q_{111}x^2 + q_{112}xy + q_{122}y^2 = \xi_1 x + \eta_1 y + \zeta_1 ,\\
q_{211}x^2 + q_{212}xy + q_{222}y^2 = \xi_2 x + \eta_2 y + \zeta_2
\end{array}\right.
\label{nohoquexa2} \ee Homogeneous equation (with all
$\xi_i,\eta_i,\zeta_i = 0$) is solvable whenever \be {\cal R}_2 =
\left|\left|\begin{array}{cccc}
q_{111} & q_{112} & q_{122} & 0\\
0 & q_{111} & q_{112} & q_{122} \\
q_{211} & q_{212} & q_{222} & 0 \\
0 & q_{211} & q_{212} & q_{222}
\end{array}\right|\right| = 0
\ee (double vertical lines denote determinant of the matrix). As
to non-homogeneous system, if $(X,Y)$ is its solution, then one
can make an analogue of the observation (\ref{liCraIIIa}): the
{\it homogeneous} systems \be \left\{\begin{array}{c}
\Big(q_{111}X^2 - \xi_1X - \zeta_1\Big)z^2 +
\Big(q_{112}X-\eta_1\Big)yz + q_{122}y^2 = 0,\\
\Big(q_{211}X^2 - \xi_2X - \zeta_2\Big)z^2 +
\Big(q_{212}X-\eta_2\Big)yz + q_{222}y^2 = 0
\end{array}\right.
\label{nohoquexa2a} \ee and \be \left\{\begin{array}{c} q_{111}x^2
+ \Big(q_{112}Y-\xi_1\Big)xz +
\Big(q_{122}Y^2 - \eta_1 Y - \zeta_1\Big)z^2 = 0 ,\\
q_{211}x^2 + \Big(q_{212}Y-\xi_2\Big)xz + \Big(q_{222}Y^2 - \eta_2
Y - \zeta_2\Big)z^2 = 0
\end{array}\right.
\label{nohoquexa2b} \ee have solutions $(z,y) = (1,Y)$ and $(x,z)
= (X,1)$ respectively. Like in the case of  (\ref{liCraIIIa}) this
implies that the corresponding resultants vanish, i.e. that $X$
satisfies \be \left|\left|\begin{array}{cccc}
q_{111}X^2 - \xi_1X - \zeta_1 & q_{112}X-\eta_1 & q_{122} & 0\\
0 & q_{111}X^2 - \xi_1X - \zeta_1 & q_{112}X-\eta_1 & q_{122} \\
q_{211}X^2 - \xi_2X - \zeta_2 & q_{212}X - \eta_2 & q_{222} & 0 \\
0 & q_{211}X^2 - \xi_2X - \zeta_2 & q_{212}X-\eta_2 & q_{222}
\end{array}\right|\right| = 0
\label{nohoquexa21} \ee while $Y$ satisfies \be
\left|\left|\begin{array}{cccc}
q_{111} & q_{112}Y-\xi_1 & q_{122}Y^2-\eta_1Y - \zeta_1 & 0\\
0 & q_{111} & q_{112}Y -\xi_1 & q_{122}Y^2-\eta_1Y - \zeta_1  \\
q_{211} & q_{212}Y-\xi_2 & q_{222}Y^2-\eta_2Y - \zeta_2  & 0 \\
0 & q_{211} & q_{212}Y-\xi_2 & q_{222}Y^2-\eta_2Y - \zeta_2
\end{array}\right|\right| = 0
\label{nohoquexa22} \ee Therefore variables got separated:
components $X$ and $Y$ of the solution can be defined from
separate algebraic equations: solution of the {\it system} of
non-linear equations is reduced to that of individual {\it
algebraic} equations. The algebro-geometric meaning of this
reduction deserves additional examination.

Though variables $X$ and $Y$ are separated in
eqs.(\ref{nohoquexa21}) and (\ref{nohoquexa22}), solutions are
actually {\it a little} correlated. Equations (\ref{nohoquexa21})
and (\ref{nohoquexa22}) are of the $4$-th power in $X$ and $Y$
respectively, but making a choice of one of four $X$'s one fixes
associated choice of $Y$. Thus the total number of solutions to
(\ref{nohoquexa2}) is $s^2=4$.

For small non-homogeneity we have: \be X^4 R_2 \sim q^3 \Big(X^2
O(\zeta) + X^3 O(\xi,\eta)\Big) \ee i.e. \be X \sim
\sqrt{\frac{q^2 O(\zeta)}{R_2\{Q\}}} \ee This asymptotic behavior
is obvious on dimensional grounds: dependence on free terms like
$\zeta$ should be $X\sim \zeta^{1/r}$, on $x-linear$ terms like
$\xi$ or $\eta$ -- $X\sim \xi^{1/(r-1)}$ etc.

\bigskip

{\bf Generic case:}

\bigskip

In general case the non-linear Craemer rule looks literally the
same as its linear counterpart (\ref{liCraIII}) with the obvious
substitution of resultant instead of determinant: the $k$-th
component $Z^k$ of the solution to non-homogeneous system
satisfies \be {\bf{non-linear\ Craemer\ rule\ III:}}\ \ \ \ \
{\cal R}_{s_1,\ldots,s_n}\left\{A^{(k)}(Z_k)\right\} = 0
\label{CraIII} \ee Tensor $[A^{(k)}(z)]_i^{j_1\ldots j_{s_i}}$ in
this formula is obtained by the following two-step procedure:

1) With the help of auxiliary homogeneous variable $z_0$ transform
original non-homogeneous system into a homogeneous one (by
inserting appropriate powers of $z_0$ into items with unsufficient
powers of other $z$-variables). At this stage we convert the
original system of $n$ non-homogeneous equations of $n$
homogeneous variables $\{z_1,\ldots,z_n\}$ into a system of $n$
homogeneous equations, but of $n+1$ homogeneous variables $\{z_0,
z_1, \ldots, z_n\}$. The $k$-th variable is in no way
distinguished at this stage.

2) Substitute instead of the $k$-th variable the product $z_k =
z_0z$ and treat $z$ as {\it parameter}, not a variable. We obtain
a system of $n$ homogeneous equations of $n$ homogeneous variables
$\{z_0,z_1,\ldots,z_{k-1},z_{k+1},\ldots, z_n\}$, but coefficients
of this system depend on $k$ and on $z$. If one now renames $z_0$
into $z_k$, the coefficients will form the tensor
$[A^{(k)}(z)]_i^{j_1\ldots j_{s_i}}$.

It remains to solve the equation (\ref{CraIII}) w.r.t. $z$ and
obtain $Z^k$. Its degree in $z$ can be lower than
$d_{s_1,\ldots,s_n} = \sum_{j=1}^n\prod_{i\neq j}^n s_j$, because
$z$ is not present in {\it all} the coefficients
$[A^{(k)}(z)]_i^{j_1\ldots j_{s_i}}$. Also, the choices from
discrete sets of solutions for $Z_k$ with different $k$ can be
correlated in order to form a solution for original system (see
s.\ref{orexfa} for related comments). The total number of
different solutions $\{Z_1,\ldots,Z_n\}$ is $\#_{s_1,\ldots,s_n} =
\prod_{i=1}^n s_i$.

In s.\ref{CraVie} one more rephrasing of this procedure is given:
in the context of non-linear algebra Craemer rule belongs to the
same family with Vieta formulas for polynomial's roots and
possesses further generalizations, which do not have given names
yet.

\setcounter{equation}{0}

\section{Evaluation of resultants and their properties}

\subsection{Summary of resultant theory}

In this subsection we show how all the familiar properties of
determinants are generalized to the resultants. To avoid
overloading the formulas we consider symmetric resultants ${\cal
R}_{n|s}$. Nothing new happens in generic case of ${\cal
R}_{s_1,\ldots,s_n}$.

\subsubsection{Tensors, possessing a resultant:
generalization of square matrices}

Resultant is defined for tensors $A_i^{j_1\ldots j_s}$ and
$G^{ij_1\ldots j_s}$, symmetric in the last $s$ contravariant
indices. Each index runs from $1$ to $n$. Index $i$ can be both
covariant and contravariant.
Such tensor has $nM_{n|s}$ independent coefficients with $M_{n|s}
= \frac{(n+s-1)!}{(n-1)!s!}$.

Tensor $A$ can be interpreted as
a map $V_n \rightarrow V_n$ of degree $s = s_A = |A| = {\rm deg}_z
A(z)$,
$$ A_i(\vec z) = \sum_{j_1,\ldots,j_s = 1}^n
A_i^{j_1\ldots j_s} z_{j_1}\ldots z_{j_s} $$ It takes values in
the same space $V_n$ as the argument $\vec z$.

Tensor $G$ maps vectors into covectors, $V_n\rightarrow V^*_n$,
all its indices are contravariant and can be treated on equal
footing. In particular, it can be {\it gradient}, i.e. $G^i(\vec
z) = \frac{\partial}{\partial z_i}S(\vec z)$ with a form
(homogeneous symmetric function) $S(\vec z)$ of $n$ variables
$z_1,\ldots,z_n$ of degree $r=s+1$. Gradient tensor $G$ is totally
symmetric in all its $s+1$ contravariant indices and the number of
its independent coefficients reduces to $M_{n|s+1} =
\frac{(n+s)!}{(n-1)!(s+1)!}$.

Important difference between the two maps is that only $A:\
V_n\rightarrow V_n$ can be {\it iterated}: composition of any
number of such maps is defined, while $G:\ V_n\rightarrow V^*_n$
admit compositions only with the maps of different types.

\subsubsection{Definition of the resultant: generalization of
condition \ $\det A = 0$\ \ for solvability of system of
homogeneous linear equations}

Vanishing resultant is the condition that the map $A_i(\vec z)$
has non-trivial kernel, i.e. is the solvability condition for the
system of non-linear equations:

\centerline{ {\bf system $\Big\{A_i(\vec z) = 0\Big\}$ has
non-vanishing solution $\vec z \neq 0$ iff ${\cal R}_{n|s}\{A\} =
0$}} \noindent Similarly, for the map $G^i(\vec z)$:

\centerline{ {\bf system $\Big\{G^i(\vec z) = 0\Big\}$ has
non-vanishing solution $\vec z \neq 0$ iff ${R}_{n|s}\{G\} = 0$}}
\noindent Though $A_i(\vec z)$ and $G^i(\vec z)$ are maps with
different target spaces, and for $n>2$ there is no distinguished
(say, basis-independent, i.e. $SL(n)$-invariant) isomorphism
between them, the resultants ${\cal R}\{A\}$ and $R\{G\}$ are
practically the same: to obtain $R\{G\}$ one can simply substitute
all components $A_i^{\ldots}$ in ${\cal R}\{A\}$ by $G^{i\ldots}$
-- the only thing that is not defined in this way is the $A$ and
$G$-independent normalization factor in front of the resultant,
which is irrelevant for most purposes. This factor reflects the
difference in transformation properties with respect to extended
structure group $GL(n)\times GL(n)$: while both ${\cal R}\{A\}$
and $R\{G\}$ are $SL(n)\times SL(n)$ invariants, they acquire
different factors $\det U^{\pm d_{n|s}} \det V^{sd_{n|s}}$ under
$A_i(\vec z) \rightarrow U^j_i A_j(\vec(Vz))$ and $B^i(\vec z)
\rightarrow (U^{-1})i_jB^j(\vec(Vz))$ These properties are
familiar from determinant theory in linear algebra. We shall
rarely distinguish between covariant and contravariant resultants
and restrict most considerations to the case of ${\cal R}\{A\}$.

\subsubsection{Degree of the resultant: generalization of \
$d_{n|1} = {\rm deg}_A (\det A) = n$\ for matrices}

Resultant ${\cal R}_{n|s}\{A\}$ has degree \be d_{n|s} = {\rm
deg}_A{\cal R}_{n|s}\{A\} = ns^{n-1} \label{dimAres} \ee in the
coefficients of $A$.

Iterated resultant $\tilde{\cal R}_{n|s}\{A\}$, see s.\ref{iteres}
below, has degree
$$ \tilde d_{n|s} = {\rm deg}_A\tilde {\cal R}_{n|s}\{A\} =
2^{n-1} s^{2^{n-1}-1}$$ Iterated resultant $\tilde{\cal
R}_{n|s}\{A\}$ depends not only on $A$, but also on the sequence
of iterations; we always use the sequence encoded by the triangle
graph, Fig.\ref{trianglegraph}.A.

\subsubsection{Multiplicativity w.r.t. composition:
generalization of \ $\det AB = \det A \det B$ for determinants}

For two maps $A(z)$ and $B(z)$ of degrees $s_A = {\rm deg}_z A(z)$
and $s_B = {\rm deg}_z B(z)$ the composition $(A\circ B)(z) =
A(B(z))$ has degree $s_{A\circ B} = |A\circ B| = s_As_B$. In more
detail
\be
&&(A\circ B)^i_{k_1\ldots k_{|A||B|}}\nn\\
&&\hspace{-3mm}=\sum_{j_1,\ldots,j_{|A|}=1}^n A^i_{j_1j_2\ldots j_{|A|}}
B^{j_1}_{k_1\ldots k_{|B|}} B^{j_2}_{k_{|B|+1}\ldots k_{2|B|}}
\ldots\ B^{j_{|A|}}_{k_{(|A|-1)|B|+1}\ldots k_{|A||B|}}{\ \ \ \ \ \
\ \ \ }
\ee
Multiplicativity property of resultant w.r.t. composition:
$$
{\cal R}_{n| s_As_B}(A\circ B) = \Big({\cal
R}_{n|s_A}(A)\Big)^{s_B^{n-1}} \Big({\cal
R}_{n|s_B}(B)\Big)^{s_A^{n}}.
$$

This formula is nicely consistent with that for $d_{n|s}$ and with
associativity of composition. We begin from associativity.
Denoting degrees of by $A, B, C$ by degrees $\alpha, \beta,
\gamma$, we get from
\be {\cal R}_{n|\alpha\beta} (A\circ B) =
{\cal R}_{n|\alpha}(A)^{\beta^{n-1}} {\cal
R}_{n|\beta}(B)^{\alpha^n}, \label{composres} \ee
\be
&R_{n|\alpha\beta\gamma}(A\circ B\circ C) = {\cal
R}_{n|\alpha\beta} (A\circ B)^{\gamma^{n-1}}
R_{n|\gamma}(C)^{(\alpha\beta)^n} \nn\\
&=R_{n|\alpha}(A)^{(\beta\gamma)^{n-1}}
R_{n|\beta}(B)^{\alpha^n\gamma^{n-1}}
R_{n|\gamma}(C)^{(\alpha\beta)^n}
\nn\ee
and
\be
&R_{n|\alpha\beta\gamma}(A\circ B\circ C) = {\cal R}_{n|\alpha}
(A)^{(\alpha\beta)^{n-1}} R_{n|\beta\gamma}(B\circ
C)^{\alpha^n}\nn\\
&=R_{n|\alpha}(A)^{(\beta\gamma)^{n-1}}
R_{n|\beta}(B)^{\alpha^n\gamma^{n-1}}
R_{n|\gamma}(C)^{(\alpha\beta)^n}
\nn\ee
Since the two answers coincide, associativity is respected: \be\hspace{-6mm}
R_{n|\alpha\beta\gamma}(A\circ B\circ C) =
R_{n|\alpha}(A)^{(\beta\gamma)^{n-1}}
R_{n|\beta}(B)^{\alpha^n\gamma^{n-1}}
R_{n|\gamma}(C)^{(\alpha\beta)^n} \ee

The next check is of consistency between (\ref{composres}) and
(\ref{dimAres}). According to~(\ref{dimAres})
$$R_{N|\alpha}(A) \sim   A^{d_{N|\alpha}}$$
and therefore the composition $(A\circ B)$ has power $\alpha\beta$
in $z$-variable and coefficients $\sim AB^\alpha$: $z \rightarrow
A(Bz^\beta)^\alpha$. Thus
$$R_{N|\alpha\beta}(A\circ B) \sim
(AB^\alpha)^{d_{N|\alpha\beta}}$$ If it is split into a product of
$R$'s, as in (\ref{composres}),
 then -- from power-counting in above expressions --
this should be equal to:
$$R_{N|\alpha}(A)^{\frac{d_{N|\alpha\beta}}{d_{N|\alpha}}}
R_{N|\beta}(B)^{\alpha\frac{d_{N|\alpha\beta}}{d_{N|\beta}}}$$ In
other words the powers in (\ref{composres}) are:
$$\frac{d_{N|\alpha\beta}}{d_{N|\alpha}}
= \frac{(\alpha\beta)^{N-1}}{\alpha^{N-1}} = \beta^{N-1} $$ and
$$\alpha\frac{d_{N|\alpha\beta}}{d_{N|\beta}}
= \alpha\frac{(\alpha\beta)^{N-1}}{\beta^{N-1}} = \alpha^N$$

\subsubsection{Resultant for diagonal maps:
generalization of \ $\det \Big({\rm diag} \  a_j^j\Big) =
\prod_{j=1}^n a_j^j$ for matrices}

We call maps of the special form $A_i(\vec z) = A_{i} z_i^s$ {\it
diagonal}. For diagonal map \be {\cal R}_{n|s}(A) =
\Big(\prod_{i=1}^n A_i\Big)^{s^{n-1}} \ee Indeed, for the system
$\Big\{ A_iz_i^s = 0 \Big\}$ (no summation over $i$ this time!) to
have non-vanishing solutions, at least one of the coefficients
$A_i$ should vanish: then the corresponding $z_i$ can provide
non-vanishing solution. After that the common power $s^{n-1}$ is
easily obtained from (\ref{dimAres}).

\subsubsection{Resultant for matrix-like maps:
a more interesting generalization of \ $\det \Big({\rm diag} \
a_j^j  \Big) = \prod_{j=1}^n a_j^j$ for matrices}

Diagonal maps posses further generalization, which still leaves
one within the theory of {\it matrices}. We call maps of the
special form $A_i(z) = \sum_{j=1}^n A_i^j z_j^s$ {\it
matrix-like}. They can be also parameterized as
$$
\left\{\begin{array}{c}
A_1(z) = \sum_{j=1}^n {a_j}z_j^s, \\ \\
A_2(z) = \sum_{j=1}^n {b_j}z_j^s, \\ \\
A_3(z) = \sum_{j=1}^N {c_j}z_j^s, \\
\ldots
\end{array}\right.
$$
For the matrix-like map
$$
{\cal R}_{n|s}(A) = \Big( {\det}_{ij} A_i^j \Big)^{s^{n-1}}
$$


Iterated resultant (see s.\ref{iteres} below for details) is
constructred with the help of the triangle graph,
Fig.\ref{trianglegraph}.A, and its multiplicative decompostion for
diagonal map is highly reducible (contains many more than two
factors), but explicit: somewhat symbolically
$$
\tilde{\cal R}_{n|s}(A) = \left( {\rm Det}_n(A) {\rm
Det}_{n-2}(A^{(n-2)}) \ldots \prod {\rm Det}_1(A^{(1)})
\right)^{s^{2^{n-1}-1}}
$$
Structure and notation is clear from the particular example, see
eq.(\ref{itereslin}) below: \be\tilde{\cal R}_{6|s} = \ee
{\footnotesize
$$
\hspace{-1.35cm} \left( \left|\left| \begin{array}{cccccc}
a_1 & a_2 & a_3 & a_4 & a_5 & a_6 \\
b_1 & b_2 & b_3 & b_4 & b_5 & b_6 \\
c_1 & c_2 & c_3 & c_4 & c_5 & c_6 \\
d_1 & d_2 & d_3 & d_4 & d_5 & d_6 \\
e_1 & e_2 & e_3 & e_4 & e_5 & e_6 \\
f_1 & f_2 & f_3 & f_4 & f_5 & f_6
\end{array} \right|\right|
\left|\left| \begin{array}{cccc}
b_1 & b_2 & b_3 & b_4 \\
c_1 & c_2 & c_3 & c_4 \\
d_1 & d_2 & d_3 & d_4 \\
e_1 & e_2 & e_3 & e_4
\end{array} \right|\right|
\left|\left| \begin{array}{ccc}
b_1 & b_2 & b_3 \\
c_1 & c_2 & c_3 \\
d_1 & c_3 & d_3
\end{array} \right|\right|
\left|\left| \begin{array}{ccc}
c_1 & c_2 & c_3 \\
d_1 & d_2 & d_3 \\
e_1 & d_3 & e_3
\end{array} \right|\right|
\left|\left| \begin{array}{cc}
b_1 & b_2 \\
c_1 & c_2
\end{array} \right|\right|
\left|\left| \begin{array}{cc}
c_1 & c_2 \\
d_1 & d_2
\end{array} \right|\right|^2
\left|\left| \begin{array}{cc}
d_1 & d_2 \\
e_1 & e_2
\end{array} \right|\right|
b_1 c_1^3 d_1^3 e_1\right)^{s^{31}}
$$
} The resultant itself is given by the first factor, but in
another power: $s^{n-1} = s^5$ out of total $s^{2^{n-1}-1} =
s^{31}$, \be {\cal R}_{6|s} = \left|\left| \begin{array}{cccccc}
a_1 & a_2 & a_3 & a_4 & a_5 & a_6 \\
b_1 & b_2 & b_3 & b_4 & b_5 & b_6 \\
c_1 & c_2 & c_3 & c_4 & c_5 & c_6 \\
d_1 & d_2 & d_3 & d_4 & d_5 & d_6 \\
e_1 & e_2 & e_3 & e_4 & e_5 & e_6 \\
f_1 & f_2 & f_3 & f_4 & f_5 & f_6
\end{array} \right|\right|^{s^5}
\ee

\subsubsection{Additive decomposition: generalization of
$\det A = \sum_\sigma (-)^\sigma \prod_i A_i^{\sigma(i)}$ for
determinants}

Like determinant is obtained from diagonal term $\prod_{i=1}^n
a_i^i$ by permutations, the resultant for generic $\vec A(\vec z)$
is obtained by adding to the {\it matrix-like} contribution \be
\Big(\det_{ij} a_i^{j\ldots j}\Big)^{s^{n-1}} \label{malico} \ee
numerous other terms, differing from (\ref{malico}) by certain
permutations of upper indices between $s^{n-1}$ determinants in
the product. This is clarified by the example (here and in
analogous examples with $n=2$ below we often denote $a_1^{\ldots}
= a^{\ldots}$ and $a_2^{\ldots} = b^{\ldots}$):
\be {\cal R}_{2|2}
= (a_1^{11}a_2^{22}-a_1^{22}a_2^{11})^2 - (a_1^{11}a_2^{12} -
a_1^{12}a_2^{11}) (a_1^{12}a_2^{22}-a_1^{22}a_2^{12})
\label{resR22} \nn\ee
\be
&=(a^{11}b^{22}-a^{22}b^{11})^2 - (a^{11}b^{12} - a^{12}b^{11})
(a^{12}b^{22} - a^{22}b^{12}) \nn\\
&= \left|\left| \begin{array}{cc}
a^{11} & a^{22} \\
b^{11} & b^{22}
\end{array} \right|\right|
\left|\left| \begin{array}{cc}
a^{11} & a^{22} \\
b^{11} & b^{22}
\end{array} \right|\right|
- \left|\left| \begin{array}{cc}
a^{11} & a^{12} \\
b^{11} & b^{12}
\end{array} \right|\right|
\left|\left| \begin{array}{cc}
a^{12} & a^{22} \\
b^{12} & b^{22}
\end{array} \right|\right|
\ee
The number of independent elementary determinants is
$\frac{M_{n|s}!}{n!(M_{n|s}-n)!}$ with $M_{n|s} =
\frac{(n+s-1)!}{(n-1)!s!}$, the sum is over various products of
$s^{n-1}$ such elementary determinants, some products do not
contribute, some enter with non-unit coefficients.

Elementary determinants can be conveniently parameterized by the
numbers of different indices:
$U_{\nu_1,\nu_2,\ldots,\nu_{n-1}}^{(\alpha)}$ denotes elementary
determinant with $\nu_1$ indices $1$, $\nu_2$ indices $2$ and so
on. $\nu_n$ is not independent because the total number of indices
is fixed: $\nu_1 + \nu_2 + \ldots + \nu_{n-1} + \nu_n = ns$. For
example, (\ref{resR22}) can be written as
\be
&{\cal R}_{2|2} = U_{2}^2 - U_3U_1\ee
with
\be \ \ U_1 =
\left|\left| \begin{array}{cc}
a^{12} & a^{22} \\
b^{12} & b^{22}
\end{array} \right|\right|, \ \
U_2 = \left|\left| \begin{array}{cc}
a^{11} & a^{22} \\
b^{11} & b^{22}
\end{array} \right|\right|, \ \
U_3 = \left|\left| \begin{array}{cc}
a^{11} & a^{12} \\
b^{11} & b^{12}
\end{array} \right|\right|
\nn\ee
For bigger $n$ and $s$ the set $\{\nu_1,\nu_2,\ldots,\nu_{n-1}\}$
does not define $U_{\nu_1,\ldots,\nu_{n-1}}$ unambiguously,
indices can be distributed differently and this is taken into
account by additional superscript $(\alpha)$ (in examples with
small $n$ and $s$ we instead use $U$ for $U^{(1)}$, $V$ for
$U^{(2)}$ etc.) In these terms we can write down the next example:
\be {\cal R}_{2|3} = U_3^3 - U_2U_3U_4 + U_2^2U_5 + U_1U_4^2 -
2U_1U_3U_5 -U_1V_3U_5 \label{resR23} \ee with
$\frac{M_{2|3}!}{2!(M_{2|3}-2)!} = \frac{4!}{2!2!} = 6$ ($M_{2|3}
= \frac{4!}{1!3!} = 4$) {\it linearly} independent elementary
determinants given by
\be
&U_1 = \left|\left| \begin{array}{cc}
a^{122} & a^{222} \\
b^{122} & b^{222}
\end{array} \right|\right|, \ \
U_2 = \left|\left| \begin{array}{cc}
a^{112} & a^{222} \\
b^{112} & b^{222}
\end{array} \right|\right|, \ \
U_3 = \left|\left| \begin{array}{cc}
a^{111} & a^{222} \\
b^{111} & b^{222}
\end{array} \right|\right|, \nn\\
&V_3 = \left|\left| \begin{array}{cc}
a^{112} & a^{122} \\
b^{112} & b^{122}
\end{array} \right|\right|, \ \
U_4 = \left|\left| \begin{array}{cc}
a^{111} & a^{122} \\
b^{111} & b^{122}
\end{array} \right|\right|, \ \
U_5 = \left|\left| \begin{array}{cc}
a^{111} & a^{112} \\
b^{111} & b^{112}
\end{array} \right|\right|
\nn\ee
Eq.\,\,(\ref{resR23}) can be written in different forms, because
there are $2$ {\it non-linear} relations between the $10$ cubic
combinations with the proper gradation number   (i.e. with the sum
of indices equal to $9$) of $6$ elementary determinants, depending
on only $8$ independent coefficients $a^{111}, a^{112}, a^{122},
a^{222}, b^{111}, b^{112}, b^{122}, b^{222}$. These two cubic
relations are obtained by multiplication by $U_3$ and $V_3$ from a
single quadratic one:
$$U_3V_3 - U_2U_4+U_1U_5 \equiv 0.$$
The next ${\cal R}_{2|4}$ is a linear combination of quartic
expression made from $10$ elementary determinants {\footnotesize
\be
&U_1 = \left|\left| \begin{array}{cc}
a^{1222} & a^{2222} \\
b^{1222} & b^{2222}
\end{array} \right|\right|, \
U_2 = \left|\left| \begin{array}{cc}
a^{1122} & a^{2222} \\
b^{1122} & b^{2222}
\end{array} \right|\right|, \nn\\
&U_3 = \left|\left| \begin{array}{cc}
a^{1112} & a^{2222} \\
b^{1112} & b^{2222}
\end{array} \right|\right|,  \
V_3 = \left|\left| \begin{array}{cc}
a^{1122} & a^{1222} \\
b^{1122} & b^{1222}
\end{array} \right|\right|, \nn\\
&U_4 = \left|\left| \begin{array}{cc}
a^{1111} & a^{2222} \\
b^{1111} & b^{2222}
\end{array} \right|\right|,
V_4 = \left|\left| \begin{array}{cc}
a^{1112} & a^{1222} \\
b^{1112} & b^{1222}
\end{array} \right|\right|, \ \
U_5 = \left|\left| \begin{array}{cc}
a^{1111} & a^{1222} \\
b^{1111} & b^{1222}
\end{array} \right|\right|,\nn\\
&V_5 = \left|\left| \begin{array}{cc}
a^{1112} & a^{1122} \\
b^{1112} & b^{1122}
\end{array} \right|\right|, \ \
U_6 = \left|\left| \begin{array}{cc}
a^{1111} & a^{1122} \\
b^{1111} & b^{1122}
\end{array} \right|\right|, \ \
U_7 = \left|\left| \begin{array}{cc}
a^{1111} & a^{1112} \\
b^{1111} & b^{1112}
\end{array} \right|\right|
\nn\ee}

In general there are $\frac{M_{n|s}!}{(2n)!(M_{n|s}-2n)!}$
quadratic Plucker relations between $n\times n$ elementary
determinants: for any set $\alpha_1,\ldots,\alpha_{2n}$ of
multi-indices (of length $s$)
$$
\frac{1}{2!(n!)^2} \sum_{P \in \sigma_{2n}} (-)^P
U_{P(\alpha_1)\ldots P(\alpha_n)}U_{P(\alpha_{n+1})\ldots
P(\alpha_{2n})} \equiv 0
$$

\subsubsection{Evaluation of resultants}

From different approaches to this problem we select three,
addressing it from positions of elementary algebra (theory of
polynomial roots), linear (homological) algebra and tensor algebra
(theory of Feynman diagrams) respectively:

-- Iterative procedure of taking ordinary resultants w.r.t. one of
the variables, then w.r.t. another and so on. In this way one
obtains a set of {\it iterated resultants}, associated with
various simplicial complexes and the resultant itself is a common
irreducible factor of all iterated resultants, see s.\ref{iteres}.

-- Resultant can be defined as determinant of Koshul differential
complex, it vanishes when Koshul complex fails to be exact and
acquires non-trivial cohomology, see s.\ref{Koshul}.

-- Resultant is an $SL(n)\times SL(n)$ invariant and can be
represented as a certain combination of Feynman-like diagrams.
Entire set of diagrams reflects the structure of the tensor
algebra, associated with the given tensor $A_i^{j_1\ldots j_s}$,
see s.\ref{resdia}.

\subsection{Iterated resultants and solvability
of systems of non-linear equations \label{iteres}}

\subsubsection{Definition of iterated resultant
$\tilde {\cal R}_{n|s}\{A\}$}

Let us consider a system of $n$ {\it homogeneous} equations \be
\left\{ \begin{array}{c}
A_1(\vec z) = 0 \\
\ldots  \\
A_n(\vec z) = 0
\end{array}\right.
\label{Qsys} \ee where $A_i(\vec z)$ are homogeneous polynomials
of $n$ variables $\vec z = (z_1,\ldots,z_n)$. This system is
overdefined and non-vanishing solutions exist only if {\it one}
constraint ${\cal R}\{A\} = 0$ is imposed on the coefficients of
the polynomials. The goal of this section is to formulate this
constraint through a sequence of iterated resultants.

Namely, let\ ${\rm Res}_{z_i} (A_1,A_2)$ denote the resultant of
two polynomials $A_1(\vec z)$ and $A_2(\vec z)$, considered as
polynomials of a single variable $z_i$ (all other $z_j$ enter the
coefficients of these polynomials as sterile parameters). Let us
now define $\tilde{\cal R}_k\{A_1,\ldots,A_k\}$ by the iterative
procedure: \be
\tilde{\cal R}_1\{A\} = A , \nn \\
\tilde{\cal R}_{k+1}\{A_1,\ldots,A_{k+1}\} = {\rm Res}_{z_k}\Big(
\tilde{\cal R}_k\{A_1,\ldots,A_k\}, \tilde{\cal
R}_k\{A_2,\ldots,A_{k+1}\}\Big) \ee The lowest entries of the
hierarchy are (see Fig.\ref{trianglegraph}.A): \be
\tilde{\cal R}_2\{A_1,A_2\} = {\rm Res}_{z_1}(A_1,A_2), \nn \\
\tilde{\cal R}_3\{A_1,A_2,A_3\} = {\rm Res}_{z_2}\Big({\rm
Res}_{z_1}(A_1,A_2),
{\rm Res}_{z_1}(A_2,A_3)\Big), \nn \\
\tilde{\cal R}_4\{A_1,A_2,A_3,A_4\} = {\rm Res}_{z_3} \left({\rm
Res}_{z_2}\Big({\rm Res}_{z_1}(A_1,A_2), {\rm
Res}_{z_1}(A_2,A_3)\Big)^{\phantom{5^5}}\ ,\ {\rm
Res}_{z_2}\Big({\rm Res}_{z_1}(A_2,A_3), {\rm
Res}_{z_1}(A_3,A_4)\Big)
\right), \nn \\
\label{triangleseq} \ldots \ee

Two polynomials $f(z)$ and $g(z)$ of a single variable have a
common root iff their ordinary resultant ${\rm Res}_z(f,g) = 0$.
From this it is obvious that for (\ref{Qsys}) to have
non-vanishing solutions one should have \be \tilde{\cal R}_n\{A\}
= 0. \label{tildeR=0} \ee However, inverse is not true:
(\ref{tildeR=0}) can have extra solutions, corresponding to
solvability of subsystems of (\ref{Qsys}) instead of entire
system. What we need is an {\it irreducible} component ${\cal
R}\{A\} \equiv {\bf irf} \Big(\tilde {\cal R}\{A\}\Big)$. In other
words, one can say that along with (\ref{tildeR=0}) many other
iterated resultants should vanish, which are obtained by
permutations of $z$-variables in the above procedure (i.e.
described by Fig.\ref{trianglegraph}.B etc instead of
Fig.\ref{trianglegraph}).A. Resultant ${\cal R}\{A\}$ is a {\it
common divisor} of all these iterated resultants.

Actually, analytical expressions look somewhat better for
Fig.\ref{trianglegraph}.B than for Fig.\ref{trianglegraph}.A, and
we use Fig.\ref{trianglegraph}.B in examples below.

\Fig{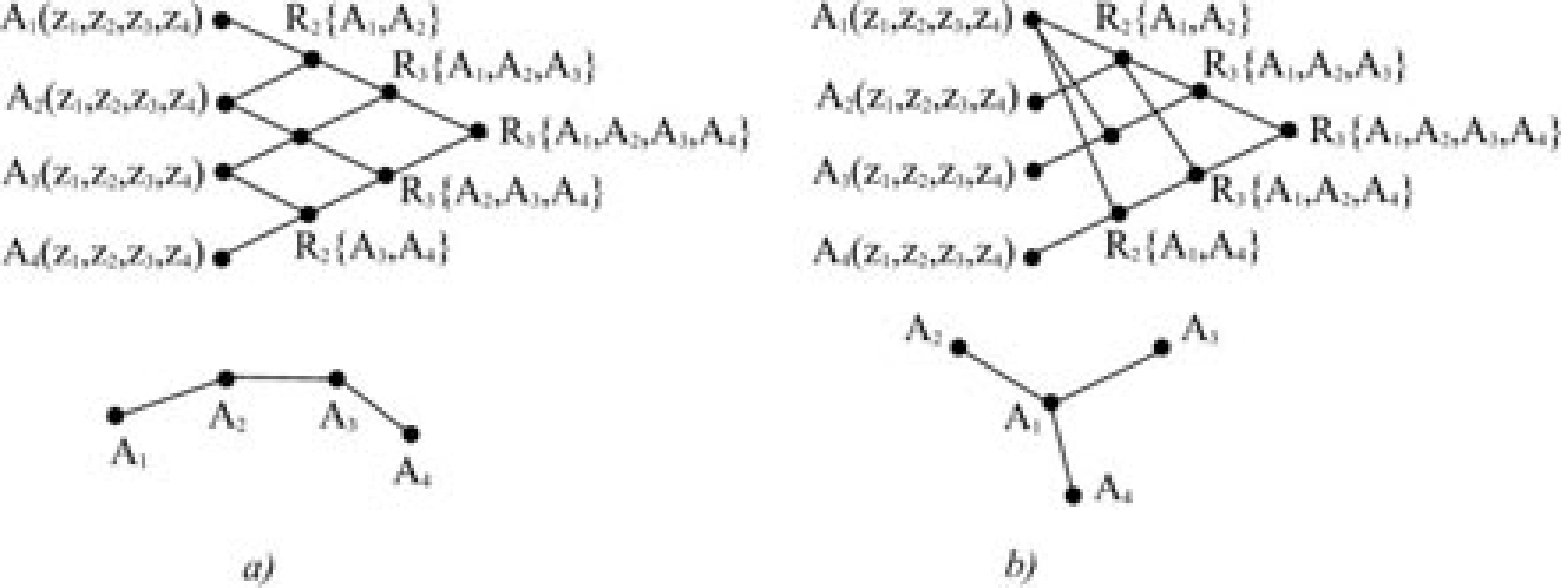} {423,159} {Sequences of iterations in the
definition of iterated resultants. \ \ A) Triangle graph, most
"ordered" from pictorial point of view and expressed by
eq.(\ref{triangleseq}). \ \ B) Another ordering, corresponding to
the "natural" iteration procedure, like in eqs.(\ref{natseq1}) and
(\ref{natseq2}). \ \ From these pictures it is clear that the
choice of the iteration sequence is in fact the choice of some
simplicial structure on the set of the equations. }

\subsubsection{Linear equations}

Let $A_i(\vec z) = \sum_{j=1}^n a_i^jz_j$. In this case the
solvability condition is nothing but $\det a_i^j = 0$.

Let us see now, how it arises in our iterated resultant
construction. For linear functions $A_i(\vec z)$ and $\tilde
a_i^k(\vec z) = \sum_{j=k}^n a_i^jz_j$ \be \tilde R_2\{A_1,A_2\} =
{\rm Res}_{z_1}\Big(a_1^1z_1 + \tilde a_1^2,\ a_2^1z_1 + \tilde
a_2^2\Big) = a_1^1\tilde a_2^2 - a_2^1\tilde a_1^2 \ee
(superscripts are indices, not powers!). Substituting now $\tilde
a_1^2 = a_1^2z_2 + \tilde a_1^3$ and $\tilde a_2^2 = a_2^2z_2 +
\tilde a_2^3$, we find \be \tilde R_3\{A_1,A_2,A_3\} = {\rm
Res}_{z_2}\Big(\tilde R_2\{A_1,A_2\}, \tilde R_2\{A_1,A_3\}\Big) =
\nn \\ = {\rm Res}_{z_2}\Big((a_1^1a_2^2-a_2^1a_1^2)z_2 +
(a_1^1\tilde a_2^3 - a_2^1\tilde a_1^3), \
(a_1^1a_3^2-a_3^1a_1^2)z_2 + (a_1^1\tilde a_3^3 - a_3^1\tilde
a_1^3)
\Big) = 
a_1^1\left|\left|\begin{array}{ccc}
a_1^1&a_1^2&\tilde a_1^3\\a_2^1&a_2^2&\tilde a_2^3\\
a_3^1&a_3^2&\tilde a_3^3
\end{array}\right|\right|
\label{natseq1} \ee The factor $a_1^1$ appears at the r.h.s.
because for $a_1^1=0$ both ${\rm Res}_{z_1}(A_1,A_2)$ and ${\rm
Res}_{z_1}(A_1,A_3)$ are proportional to $\tilde a_1^2 = a_1^2z_2
+ \tilde a_1^3$ and have a common root  $z_2 = -\tilde
a_1^3/a_1^2$, so that $\tilde{\cal R}_3$ vanishes; however, this
does not lead to non-trivial solution of entire system, since
$z_1$-roots of $A_2$ and $A_3$ are different unless the $3\times
3$ determinant also vanishes.

To make the next step, substitute $\tilde a_i^3 = a_i^3z_3 +
\tilde a_i^4$, and obtain \be \tilde R_4\{A_1,A_2,A_3,A_4\} =
\big(a_1^1\big)^2\Big(a_1^1a_2^2-a_1^2a_2^1\Big)
\left|\left|\begin{array}{cccc} a_1^1&a_1^2&a_1^3&\tilde
a_1^4\\a_2^1&a_2^2&a_2^3&\tilde a_2^4
\\a_3^1&a_3^2&a_3^3&\tilde a_3^4\\a_4^1&a_4^2&a_4^3&\tilde a_4^4
\end{array}\right|\right|
\label{natseq2} \ee and so on.

In general \be \tilde{\cal R}_n\{A_1,\ldots,A_n\} =
||a_1^1||^{2^{n-3}}\cdot \left|\left|\begin{array}{cc}
a_1^1&a_1^2\\a_2^1&a_2^2\end{array}\right|\right|^{2^{n-4}}\cdot
\left|\left|\begin{array}{ccc}
a_1^1&a_1^2&a_1^3\\a_2^1&a_2^2&a_2^3\\a_3^1&a_3^2&a_3^3
\end{array}\right|\right|^{2^{n-5}}\cdot \ldots = \nn \\ =
\prod_{k=1}^{n-2} \Big({\det}_{1\leq i,j\leq k}\
a_i^j\Big)^{2^{n-2-k}} \Big({\det}_{1\leq i,j \leq n}\ a_i^j\Big)
\label{prodfortildeR} \label{itereslin} \ee This $\tilde{\cal
R}_n$ is a homogeneous polynomial of power $n + \sum_{k=1}^{n-2}
2^{n-2-k}k = 2^{n-1}$ in $a$'s.

The irreducible resultant \be {\cal R}_n\{A_1,\ldots,A_n\} =
{\det}_{1\leq i,j \leq n}\ a_i^j, \ee providing the solvability
criterium of the system  of {\it linear} equations, is the last
factor in the product (\ref{prodfortildeR}). It can be obtained
from $\tilde {\cal R}$ by inverse iterative procedure: \be
{\cal R}_1 = \tilde{\cal R}_1, \nn \\
{\cal R}_2 = \tilde{\cal R}_2, \nn \\
{\cal R}_3 = \frac{\tilde{\cal R}_3}{\tilde{\cal R}_1}, \nn \\
{\cal R}_4 = \frac{\tilde{\cal R}_4}
{\tilde{\cal R}_2\tilde{\cal R}_1^2}, \nn \\
{\cal R}_5 = \frac{\tilde{\cal R}_5}
{\tilde{\cal R}_3\tilde{\cal R}_2^2\tilde{\cal R}_1^3}, \nn \\
\ldots, \nn \\
{\cal R}_n = \frac{\tilde{\cal R}_n}{\prod_{k=1}^{n-2} \tilde{\cal
R}_k^{n-1-k}} \ee

\subsubsection{On the origin of extra factors in $\tilde{\cal R}$
\label{orexfa}}

Though the linear example illustrates well the very fact that
$\tilde{\cal R}$ is reducible, the origin of the extra factors
$\tilde{\cal R}/{\cal R}$ is somewhat specific in this case.

Let $A_i(z)$ be polynomials of degree $s$ in their variables $z$.
Then vanishing of, say ${\cal R}_3\{A_1,A_2,A_3\}$ implies that
there exists a value $Y$ of $z_2$-variable, such that \be
\left.{\rm Res}_{z_1}(A_1,A_2)\right|_{z_2=Y} = 0, \nn \\
\left.{\rm Res}_{z_1}(A_1,A_3)\right|_{z_2=Y} = 0, \ee i.e. that
there are some values $X_2$ and $X_3$ of $z_1$-variable, such that
\be \left\{\begin{array}{c} A_1(X_2,Y) = 0, \\ A_2(X_2,Y) = 0
\end{array}\right.
\ee and \be \left\{\begin{array}{c} A_1(X_3,Y) = 0, \\ A_3(X_3,Y)
= 0
\end{array}\right.
\ee (all other $z_i$-variables with $i\geq 3$ are considered as
sterile parameters in this calculation). The solvability of the
$3-item$ subset \be \left\{\begin{array}{c} A_1(X,Y) = 0, \\
A_2(X,Y)=0, \\ A_3(X,Y) = 0
\end{array}\right.
\ee of original system requires that $X_2=X_3$, but this is not
necessarily the case.

{\it In general} the equation $A_1(x,Y)=0$ for $x$ has $s$
different roots $x=X_\mu(Y)$, $\mu=1,\ldots,s$, and our
$\tilde{\cal R}_3\{A_1,A_2,A_3\}$ gets contributions from common
solutions of \be
A_2\Big(X_\mu(Y), Y\Big) = 0, \nn \\
A_3\Big(X_\nu(Y), Y\Big) = 0 \label{Q23} \ee with all pairs of
$\mu$ and $\nu$. Actual multiplicity is smaller than $s^2$,
because individual (for given $\mu$ and $\nu$) $Y$-resultants of
the system (\ref{Q23}) are {\it not} polynomials of coefficients
of $A$'s: actually there are $s$ independent factors, and only one
of them, a product over all $s$ values of $\mu=\nu$, is our
irreducible resultant ${\cal R}_3\{A_1,A_2,A_3\}$.

Analysis is similar for higher ${\cal R}_k$.

A possibility to extract ${\cal R}$ from $\tilde{\cal R}$ is to
evaluate a set $\tilde{\cal R}$ with reordered (permuted)
$z$-variables and take their resultants as functions of some of
the coefficients of $A$. This is especially convenient if
resultant ${\cal R}$ is needed for a one-parametric family of
polynomials $A_i(z)$.

The linear case of $s=1$ is special: above analysis seems to imply
that no factors should occur in this case. The loophole is that
for $s=1$ the situation of {\it non-general position} has
codimension one: for {\it linear} function $A_1(x,Y)$ to have more
than one $x$-root (actually, infinitely many) it is enough to
impose a single condition $a_{11} = 0$. Similarly, for the system
of {\it linear} equations $A_i(z_1,\ldots,z_k,Z_{k+1})=0$,
$i=1,\ldots,k$, arising in analysis of $\tilde{\cal R}_l$ with
$l\geq k+2$, to have more than one (actually, infinitely many)
non-vanishing solution for $z_1=Z_1,\ldots,z_k=Z_k$, a single
condition is needed, ${\det}_{1\leq i,j\leq k} a_i^j = 0$. This is
the origin of extra factors in linear systems. For higher $s\geq
2$ such non-general positions have higher codimension and do not
affect the structure of the solvability constraints $\tilde{\cal
R}=0$.

\subsubsection{Quadratic equations}

Let now $A_i(z) = \sum_{j,k=1}^n a_i^{jk} z_jz_k$. Then \be \tilde
R_2\{A_{1},A_{2}\} = {\rm Res}_{z_1}\Big(a_1^{11}z_1^2 + \tilde
a_1^{12}z_1 + \tilde a_1^{22},\ \ a_2^{11}z_1^2 + \tilde
a_2^{12}z_1 + \tilde a_2^{22}\Big) = \nn \\= \Big(a_{1}^{11}\tilde
a_{2}^{22}-a_{2}^{11}\tilde a_{1}^{22}\Big)^2+ \Big(a_1^{11}\tilde
a_2^{12}-a_2^{11}\tilde a_1^{12}\Big) \Big(\tilde a_1^{22}\tilde
a_2^{12}- \tilde a_2^{22}\tilde a_1^{12}\Big) \ee Substituting now
\be
\tilde a_1^{12} = a_1^{12}z_2 + \tilde a_1^{13}, \nn \\
\tilde a_1^{22} = a_1^{22}z_2^2 + \tilde a_1^{23}z_2 +
\tilde a_1^{33},\nn \\
\tilde a_2^{12} = a_2^{12}z_2 + \tilde a_2^{13},\nn \\
\tilde a_2^{22} = a_2^{22}z_2^2 + \tilde a_2^{23}z_2 + \tilde
a_2^{33}, \ee we can find \be \tilde R_3\{A_1,A_2,A_3\} = {\rm
Res}_{z_2}\Big(\tilde R_2\{A_1,A_2\},\ \tilde R_2\{A_1,A_3\}\Big)
\ee Since $\tilde R_2$ are polynomials of degree $4$ in $z_2$,
$\tilde R_3$ will be a polynomial of degree $2\times 4 = 8$ in the
coefficients of $\tilde R_2$, which are themselves quartic in
$a$'s. Therefore total $a$-degree of $\tilde R_3$ should be
$8\times 4 = 32 = 12 + 20$. Symmetric resultant ${\cal
R}_3\{A_1,A_2,A_3\}$ is the irreducible factor of degree $12$.

\subsubsection{An example of cubic equation
\label{abc+e3}}

Take for a cubic form -- a cubic function of a single vector with
{\it three} components $x,y,z$ -- \be S(x,y,z) = \frac{1}{3}ax^3 +
\frac{1}{3}by^3 + \frac{1}{3}cz^3 + 2\epsilon xyz \label{cufexa1}
\ee i.e. non-vanishing elements $S^{ijk}$ are: \be S^{111} =
\frac{a}{3}, \ \ S^{222} = \frac{b}{3}, \ \ S^{333} = \frac{c}{3},
\ \ S^{123} = S^{132} = S^{213} = S^{231} = S^{312} = S^{321} =
\frac{\epsilon}{3} \ee

The resultant of the system $\vec\partial S = 0$, \be
\left\{\begin{array}{c}
S'_x= ax^2 + 2\epsilon yz = 0, \\
S'_y = by^2 + 2\epsilon xz = 0, \\
S'_z = cz^2 + 2\epsilon xy = 0
\end{array}\right.
\label{sys32Exa1} \ee is equal to degree-twelve polynomial of the
coefficients $S_{ijk}$, \be R_{3|2}(\vec\partial S) = abc(abc+
8\epsilon^3)^3 \label{R32exa1} \ee

Indeed, the typical resultants of {\it pairs} of equations in
(\ref{sys32Exa1}) are: \be
R_x(S'_x,S'_y) = ab^2y^4 + 8\epsilon^3yz^3, \nn \\
R_x(S'_y,S'_z) = \Big(2\epsilon (cz^3 - by^3)\Big)^3, \nn \\
R_y(S'_x,S'_y) = a^2b x^4 + 8\epsilon^3 xz^3, \nn \\
R_y(S'_y,S'_z) = bc^2z^4 + 8\epsilon^3x^3z, \ee so that \be
R_z\Big(R_x(S'_x,S'_y), R_x(S'_y,S'_z)\Big) \sim (abc +
8\epsilon^3)^3, \nn
\\
R_z\Big(R_y(S'_x,S'_y), R_y(S'_y,S'_z)\Big) \sim (a^2b^2c^2 -
64\epsilon^6)^3 \ee have (\ref{R32exa1}) as a common divisor.

\subsubsection{More examples of 1-parametric deformations}
Similarly, one can deduce resultants/discriminants in any particular case of interest.
We provide here just two examples of 1-parametric deformations of
diagonal tensors of $n=3$ variables:
\be
R_{3|r-1}(\vec\partial S)=D_{3|r}\left(S=\frac{ax^r+by^r+cz^r}{r}-\epsilon xy^{r-1}\right)\nn\\
=\left[a^{r-2}c^{r-1}(ab^{r-1}-(r-1)^{r-1}\epsilon^r)\right]^{r-1}
\label{DefDiag}\ee
Formula (\ref{R32exa1}) is a particular case of (\ref{DefDiag}) for $r=3$ and $\epsilon$ replaced
by $-2\epsilon$.

\subsubsection{Iterated resultant depends on symplicial structure}

As already mentioned, the iteration procedure, used to define
$\tilde{\cal R}$, depends on the choice of the iteration sequence.
Figs.\ref{trianglegraph} provide different iterated resultants
$\tilde{\cal R}$ for the same $A$. The difference between
Figs.\ref{trianglegraph}.A and \ref{trianglegraph}.B is the
difference between symplicial structures on the same set of
points. The resultant ${\cal R}\{A\}$ is a common divisor of all
$\tilde{\cal R}\{A|\Sigma\}$ for all possible symplicial
structures $\Sigma$.

\subsection{Resultants and Koszul complexes
\cite{Cay}-\cite{AG} \label{Koszul}\label{Koshul}}

In this approach evaluation of resultant ${\cal
R}_{s_1,\ldots,s_n}\{A\}$ is reduced to that of the ordinary
determinants of a few square matrices, which are synchronously
chosen minors of {\it rectangular} matrices made from (and linear
in) the coefficients of the tensor $A$. These rectangular matrices
describe the $A$-induced mappings between linear spaces of
polynomials of various degrees. Such mappings naturally arise if
one studies decomposition of arbitrary polynomial $P(\vec z)$ into
a linear combination of polynomials from the given set $\{
A_i(\vec z) \}$: $P(\vec z) = \sum_{i=1}^n P^i(\vec z)A_i(\vec z)$
with the coefficients $P^i(\vec z)$ which are also polynomials of
appropriate degrees, ${\rm deg}_z P^i = {\rm deg}_z P - s_i$, \
$s_i = \deg_z A_i$. Of course, the structure of decomposition
changes whenever different $A_i(\vec z)$ become "less
independent", namely when the resultant vanishes, ${\cal
R}_{s_1,\ldots,s_n}\{A_1,\ldots,A_n\} = 0$, and this opens the
possibility to study resultants with the help of complexes, i.e.
by methods of homological algebra.

Exact statement in this framework is that the resultant is {\it
determinant} of {\it Koszul complex}, associated with the set
$\{A_i(\vec z)\}$: \be {\cal R}\{A_i\} = {\rm DET} \Big(A_i(\vec
z)\frac{\partial}{\partial\theta_i}\Big) \label{KoresDET} \ee The
meaning of this formula will be explained below in the present
section. It should be kept in mind, that Koszul-complex method is
somewhat artificial from the resultant-theory point of view: it is
natural for decomposition-type problems (like Euclid decomposition
of integers into mutually prime components and its
algebro-geometric analogues), and it is a kind of lack that the
same technique can be applied to degeneracy problems (moreover, as
we shall see in the s.\ref{dico} below, applicability is at least
less straightforward in the case of poly-linear discriminants).
For above reason it is not surprising, that the r.h.s. of
(\ref{KoresDET}) involves additional structure, which is not
present at the l.h.s.: in (\ref{KoresDET}) it is Grassmannian
variable $\vec\theta = \{\theta_i\}$. Analogues of this formula
should exist with other kinds of auxiliary structures (see an
example in s.\ref{notKoszul} below). Of special interest would be
encoding of such structures in some functional-integral
representation of ${\cal R}$: then they would be put under control
and easily classified. Unfortunately, such interpretation of
(\ref{KoresDET}) and its analogues is not yet available.

It is worth mentioning that Koszul complexes start to appear in
physical literature: a certain deformation of Koszul complex plays
central role in the modern theory of superstring models
\cite{Berk}.

\subsubsection{Koszul complex. I. Definitions
\label{KoI}}

Let ${\cal P}_{n|s}$ be the linear space of {\it homogeneous}
polynomials of degree $s$ of $n$ variables. It has dimension
$M_{n|s} = \frac{(n+s-1)!}{(n-1)!s!}$, and convenient linear basis
is provided by collection of all monomials $z_{i_1}\ldots z_{i_n}$
with $1\leq i_1\leq \ldots \leq i_n\leq n$. Our set\footnote{
$A_i(\vec z)$ defines also {\it non-linear} map of projective
spaces, $P^{n-1} \rightarrow P^{n-1}$. This property will not be
exploited in the present section, but we continue to call the set
of polynomials a {\it map}. However, the theory of Koszul complex
remains literally the same, if the index $i$ is not covariant (as
required for a map), but contravariant (as it is in tensors $T$,
describing poly-linear forms and their reductions). The only
difference is that in this case one substitutes
$\partial/\partial\theta_i$ in the definition of the differential
$\hat d$ by $\theta_i$, in order to allow contraction with
contravariant $T^i$ instead of covariant $A_i$. To preserve the
structure of the complex one should simultaneously change in all
the vector spaces of the complex the $k$-th powers of $\theta$'s
for their dual $(n-k)$-th powers: $\theta_{j_1}\ldots\theta_{j_k}
\rightarrow \epsilon^{j_1\ldots j_k \tilde j_1\ldots \tilde
j_{n-k}} \theta_{\tilde j_1}\ldots \theta_{\tilde j_{n-k}}$ (this
operation preserves dimensions of the vector spaces). It is now
possible to substitute $\theta_i$ by anticommuting forms $dx^i$,
so that Koszul complex turns into that of the nilpotent operator
$\hat d = T^i(\vec x) dx_i$, acting by wedge multiplication in the
space of all differential forms, graded by their ranks. In
gradient case, $T^i(\vec x) = \partial^i T(\vec x)$, $\hat d =
dT$. \label{thetadx}} $\{A_i(\vec z)\} \in \oplus_{j=1}^n {\cal
P}_{n|s_j}$.

With a map $A_i(z)$ one can associate a nilpotent operator \be
\hat d = \sum_{i=1}^n A_i(z)\frac{\partial}{\partial \theta_i}, \
\ \ \hat d^2=0 \ee with auxiliary Grasmannian (anticommuting and
nilpotent) variables $\theta_i$,\ $\theta_i\theta_j +
\theta_j\theta_i = 0$. Given such $\hat d$, one can define Koszul
complex: \be 0 \longrightarrow \tilde H \theta_1\ldots\theta_n
\stackrel{\hat d}{\longrightarrow} \oplus_{i=1}^n \tilde H_i
\theta_1\ldots\theta_{i-1}\theta_{i+1} \ldots\theta_n
\stackrel{\hat d}{\longrightarrow} \ldots \stackrel{\hat
d}{\longrightarrow} \oplus_{i=1}^n H^i \theta_i \stackrel{\hat
d}{\longrightarrow} H \longrightarrow  0 \label{KoshI} \ee where
$H^{i_1\ldots i_k} = {\cal P}_{n|s(i_1,\ldots,i_k)}$. The set of
degrees is defined by those of original map, $A_i(z) \in {\cal
P}^{s_i}$, and by  degree of the end-point polynomials, $\deg_z
H(z) = p$. Then $s(i_1,\ldots,i_k) = p - s_{i_1} -\ldots -
s_{i_k}$.

The main property of Koszul complex is that it is {\it exact},
{\it unless} the resultant ${\cal
R}_{s_1,\ldots,s_n}\{A_1,\ldots,A_n\} = 0$, so that discriminantal
variety is associated with the cohomologies of Koszul complex.
Accordingly the resultant is given by {\it determinant} of this
complex. If basises are chosen in all the spaces $H^{i_1\ldots
i_k}$, the Koszul complex becomes a set of rectangular matrices.
Determinant is alternated product of maximal minors of these
matrices, invariant under linear transformations of basises.

In practice it is useful to adjust $m$ so that only a few terms in
the complex contribute -- as we demonstrate below this drastically
simplifies the calculation.\footnote{ We are indebted to
A.Gorodentsev for comments about this method.} If degree $m$ is
adjusted so that only the last of these matrices is non-vanishing
and square, the resultant is nothing but determinant of this
matrix. Unfortunately, such a deep adjustment is rarely possible,
though the two elementary cases, $s_1=\ldots =s_n=1$ (i.e.
ordinary $n\times n$ matrices) and $n=2$ (i.e. ordinary resultants
of two polynomials of degrees $s_1$ and $s_2$ of a single
variable), are exactly of this type. In more sophisticated
situations one still can adjust $m$ so that the number of
non-vanishing matrices is minimal (actually, one should take $m$
as small as possible), and extract the resultant from this data.

Now we consider a few examples and afterwards return to some more
details about the general case.

\subsubsection{Linear maps (the case of $s_1=\ldots=s_n=1$)}

\noindent

{\bf Minimal option, only one term in the complex is non-trivial:}

In this case $s_1=\ldots=s_n=1$, $A_i(z) = \sum_{j=1}^n A_i^jz_j$,
one can choose $p=1$ and then the entire complex reduces to the
last term:
$$
0 \longrightarrow \oplus_{i=1}^n {\cal P}_{n|0} \stackrel{\hat
d}{\longrightarrow} {\cal P}_{n|1} \longrightarrow 0
$$
i.e. the set of constants (polynomials of degree zero)
$\alpha_1,\ldots,\alpha_n$ (each from one of the $n$ copies of
${\cal P}_{n|0}$) is mapped into linear functions $\sum_{i=1}^n
\alpha^i A_i(\vec z) = \sum_{i,j=1}^n \alpha^i A_i^jz_j$. In the
basises $\{\alpha^i\}$ in the left space and $\{z_j\}$ in the
right space this map is given by the matrix $A_i^j$, and the
resultant is given by determinant of this matrix: \be {\cal
R}_{n|1}\{A_i^jz_j\} = \det_{1\leq i,j \leq n} A_i^j \ee

{\bf Two terms contributing, an example:}

If for the same collection of linear maps, $s_1=\ldots=s_n=1$,
$A_i(z) = \sum_{j=1}^n A_i^jz_j$, we take $p=2$ instead of $p=1$,
then the complex reduces to the last {\it two} terms:
$$
0 \longrightarrow \oplus_{1\leq i<j\leq n} {\cal P}_{0|n}
\stackrel{\hat d}{\longrightarrow} \oplus_{i=1}^n {\cal P}_{1|n}
\stackrel{\hat d}{\longrightarrow} {\cal P}_{2|n} \longrightarrow
0
$$
The first map takes the set of $\frac{n(n-1)}{2}$ constants
$\alpha^{ij} = -\alpha^{ji}$ into $\Big(A_k(\vec
z)\frac{\partial}{\partial \theta_k}\Big)
\alpha^{ij}\theta_i\theta_j = A_k(\vec z)(\alpha^{ki} -
\alpha^{ik})\theta_i$ i.e. into $-2\sum_{k=1}^n \alpha^{ik}
A_k(\vec z) = -2\sum_{j,k=1}^n \alpha^{ik}A_k^jz_j$, i.e.
described by rectangular $\frac{n(n-1)}{2}\times n^2$ matrix. The
second map takes the set of $n$ linear functions $\sum_{j=1}^n
\beta^{ij}z_j$ into the $\frac{n(n+1)}{2}$ linear space of
quadratic functions, $\Big(A_k(\vec z)\frac{\partial}{\partial
\theta_k}\Big) \beta^{ij}z_j\theta_i = \beta^{ij} A_i(\vec z)z_j =
\beta^{ij}A_i^kz_jz_k$ described by rectangular $n^2 \times
\frac{n(n+1)}{2}$ matrix.

For example, if $n=2$ the two matrices are $1\times 4$ and
$4\times 3$:
$$
\Big({\cal A}_a) = \Big( -A_2^1, -A_2^2, A_1^1, A_1^2\Big) \ \
{\rm and} \ \ \Big({\cal B}^a_j\Big) =
\left(\begin{array}{ccc} A_1^1 & A_1^2 & 0 \\ 0 & A_1^1 & A_1^2\\
A_2^1 & A_2^2 & 0 \\ 0 & A_2^1 & A_2^2 \end{array}\right)
$$
It is easy to check that $\sum_{a=1}^4 {\cal B}_j^a{\cal A}_a = 0$
(i.e. that $\hat d^2=0$) and
$$
\epsilon_{aa_1a_2a_3} {\cal B}^{a_1}_{j_1}{\cal
B}^{a_2}_{j_2}{\cal B}^{a_3}_{j_3} \epsilon^{j_1j_2j_3} = {\cal
A}_a {\cal R}_{2|1}\{A\}
$$
with ${\cal R}_{2|1}\{A\} = \det \left(\begin{array}{cc} A_1^1 &
A_1^2 \\ A_2^1 & A_2^2 \end{array}\right)$ and totally
antisymmetric  tensors $\epsilon_{a_1a_2a_3a_4}$ and
$\epsilon^{j_1j_2j_3}$ of ranks $3$ and $4$.


{\bf $m$ terms contributing:}

In general, for linear maps $s_1=\ldots = s_n=1$ the number of
non-trivial terms in Koszul complex is equal to $m$, and complex
is a collection of $m$ rectangular matrices of sizes
$\frac{M_{n|0} n!}{(n-m)!}\times \frac{M_{n|1} n!}{(n-m+1)!}$,
$\ldots$, $\frac{M_{n|k} n!}{(n-m+k)!}\times \frac{M_{n|k+1}
n!}{(n-m+k+1)!}$, $\ldots$, $M_{n|m-1}n \times M_{n|m}$. Still
alternated combinations of minors provides the same quantity
${\cal R}_{2|1}\{A\} = \det_{2\times 2}A$.

\subsubsection{A pair of polynomials (the case of $n=2$)}

\noindent

{\bf Minimal option, only one term in the complex is non-trivial:}

For $n=2$ $A^i(z)$ consists of two polynomials $f(z) =
\sum_{k=0}^{s_1} f_k z^k$, $g(z) = \sum_{k=0}^{s_2} g_k z^k$, and
the complex reduces to the last term,
$$
{\cal P}_{2|p-s_1}\oplus {\cal P}_{2|p-s_2} \stackrel{\hat
d}{\longrightarrow} {\cal P}_{2|p}
$$
if $p$ is adjusted to make the matrix square: $M_{2|p-s_1} +
M_{2|p-s_2} = M_{2|p}$, i.e. $(p-s_1+1) + (p-s_2+1) = p+1$ or $p =
s_1+s_2-1$. With this choice $p-s_1=s_2-1 < s_2$ and $p-s_2=s_1-1
< s_1$, so that the preceding term in the complex would involve
negative degrees and thus does not contribute.

With this choice of $p$ the map $\hat d$ is:
$$
\Big(\sum_{i=0}^{p-s_1} \alpha_i z^i \Big) \Big(\sum_{k=0}^{s_1}
f_k z^k\Big) + \Big(\sum_{i=0}^{p-s_2} \beta_i z^i \Big)
\Big(\sum_{k=0}^{s_2} g_k z^k\Big)
$$
and in the basis $\{\alpha_k,\beta_k\}$ in the left space and
$\{z^0,\ldots,z^m\}$ in the right space the matrix looks \be
\left(\begin{array}{cccccc}
f_0 & f_1 & \ldots & f_{s_1} & 0 & \ldots \\
0 & f_0 & \ldots & f_{s_1-1} & f_{s_1} & \ldots \\
&&&\ldots && \\
g_0 & g_1 & \ldots & g_{s_1} & g_{s_1+1} & \ldots \\
0 & g_0 & \ldots & g_{s_1-1} & g_{s_1} & \ldots
\end{array}\right)
\ee and determinant of this $(s_1+s_2)\times(s_1+s_2)$ matrix
(where $s_1+s_2 = M_{2|s_1+s_2-1}$) is exactly the resultant of
$f(z)$ and $g(z)$, cited in \ref{ordresdetrep}.

%
%

\subsubsection{A triple of polynomials (the case of $n=3$)
\label{abc+e3fromK}}

For three polynomials $f_1(\vec z), f_2(\vec z), f_3(\vec z)$ of
three homogeneous variables of degrees $s_1$, $s_2$ and $s_3$ the
last two terms of Koszul complex are:
$$
\ldots \stackrel{\hat d}{\longrightarrow} \oplus_{i=1}^3 {\cal
P}_{3|p-s_j-s_k} \stackrel{\hat d}{\longrightarrow} \oplus_{i=1}^3
{\cal P}_{3|p-s_i} \stackrel{\hat d}{\longrightarrow} {\cal
P}_{3|p}
$$
(as usual, $i,j,k$ denote ordered triples: if $i=1$, then $j=2$,
$k=3$ and so on). Dimensions of the three linear spaces are
$\sum_{i=1}^3 M_{3|p-s_j-s_k} = \frac{1}{2} \sum_{i=1}^3
(p-s_j-s_k+1)(p-s_j-s_k+2)$, $\sum_{i=1}^3 M_{3|p-s_i} =
\frac{1}{2}\sum_{i=1}^3 (p-s_i+1)(p-s_i+2)$ and $M_{3|p} =
\frac{(p+1)(p+2)}{2}$. The middle dimension equals the sum of the
two others, $\sum_{i=1}^3 (p-s_i+1)(p-s_i+2) = (p+1)(p+2) +
\sum_{i=1}^3 (p-s_j-s_k+1)(p-s_j-s_k+2)$, if either $p =
s_1+s_2+s_3 - 2$ or $p = s_1+s_2+s_3 - 1$. Then the middle space
is a direct sum of the two other spaces, and the ratio of
determinants of emerging square matrices is the resultant ${\cal
R}_{s_1,s_2,s_3}\{f_1,f_2,f_3\}$.

\vspace{2mm}
{\bf Example:} Let $s_1=s_2=s_3=2$ and take a special family of
maps: $f_i = a_i z_i^2 + 2\epsilon z_jz_k$ ($a_1=a$, $a_2=b$,
$a_3=c$, $z_1 = x$, $z_2 = y$, $z_3 = z$). The first map
{\be
&(\alpha,\beta,\gamma) \stackrel{d}{\longrightarrow} \Big(
\beta(cz^2+2\epsilon xy) - \gamma(by^2 + 2\epsilon xz),\ \nn\\
&\hskip-4mm\gamma(ax^2+2\epsilon yz) - \alpha(cz^2+2\epsilon xy),\
\alpha(by^2 + 2\epsilon xz) - \beta(ax^2+2\epsilon yz)\Big)
\ee
}
is described by the $3\times 18$ matrix:
\vskip2mm
$$
\Big({\cal A}^I_a\Big) = \left(\begin{array}{cccccc|cccccc|cccccc}
x^2 & y^2 & z^2 & xy & yz & zx & x^2 & y^2 & z^2 & xy & yz & zx &
x^2 & y^2 & z^2 & xy & yz & zx \\
&&&&&&&&&&&&&&&&&\\
\hline
&&&&&&&&&&&&&&&&&\\
&&&&&&&& -c & -2\epsilon &&&& b &&&&2\epsilon \\
&& c & 2\epsilon &&&&&&&&& -a &&&& -2\epsilon & \\
& -b &&&& -2\epsilon & a &&&& 2\epsilon &&&&&&& \\
&&&&&&&&&&&&&&&&&
\end{array}\right)
$$\vskip2mm
(all other entries are zeroes). The second map,
$$\Big(\xi_1(\vec z), \xi_2(\vec z), \xi_3(\vec z)\Big)
\stackrel{d}{\longrightarrow} \xi_1(\vec z)f_1(\vec z) +
\xi_2(\vec z)f_2(\vec z) + \xi_3(\vec z)f_3(\vec z),$$
-- by the
$18\times 15$ matrix\vskip1mm
{\footnotesize$$\hspace{-5mm}
\Big({\cal B}^a_i\Big) =
\left(\begin{array}{c||ccc|cccccc|ccc|ccc} & x^4 & y^4 & z^4 &
x^3y &x^3z & y^3x & y^3z & z^3x & z^3y & x^2y^2 &
y^2z^2 & z^2x^2 & x^2yz & y^2xz & z^2xy \\
\hline
x^2 & a &&&&&&&&&&&&2\epsilon &&\\
y^2 & &&&&& & 2\epsilon &&& a &&&&&\\
z^2 & &&&    &&&&&2\epsilon&   &&a&  &&\\
xy & &&&  a &&&&&&   &&&  &2\epsilon &\\
yz & &&&   &&&&&&   &2\epsilon&& a &&\\
zx & &&&   &a&&&&&   &&&  &&2\epsilon\\
\hline
x^2 & &&&   &2\epsilon &&&&& b  &&&  &&\\
y^2 & &b&&   &&&&&&   &&&  &2\epsilon&\\
z^2 & &&&   &&&&2\epsilon&&   &b&&  &&\\
xy & &&&   &&b&&&&   &&&  2\epsilon &&\\
yz & &&&   &&&b&&&   &&&  &&2\epsilon\\
zx & &&&   &&&&&&   &&2\epsilon&  &b&\\
\hline
x^2 & &&&  2\epsilon &&&&&&   &&c&  &&\\
y^2 & &&&   &&2\epsilon &&&&   &c&&  &&\\
z^2 & &&c&   && &&&&   &&&  &&2\epsilon\\
xy & &&&   &&&&&&  2\epsilon &&&  &&c\\
yz & &&&   &&&&&c&   &&&  &2\epsilon&\\
zx & &&&   &&&&c&&   &&&  2\epsilon &&\\
&&&&&&&&&&&&&&&\\
\end{array}\right)
$$}
\vskip3mm\noindent
Then for any triple of indices $1\leq\tilde a_1 < \tilde a_2 <
\tilde a_3\leq 18$ we have \be {\cal R}_{3|2}\cdot
\epsilon_{I_1I_2I_3} {\cal A}^{I_1}_{\tilde a_1}{\cal
A}^{I_2}_{\tilde a_2} {\cal A}^{I_3}_{\tilde a_3}=
\epsilon_{\tilde a_1\tilde a_2\tilde a_3 a_1\ldots a_{15}} {\cal
B}^{a_1}_{i_1}\ldots {\cal B}^{a_{15}}_{i_{15}}
\epsilon^{i_1\ldots i_{15}} \label{KoresR32} \ee If choice is such
that the $3\times 3$ minor at the l.h.s. is non-vanishing, this
provides explicit formula for the resultant ${\cal R}_{3|2} =
abc\Big(abc + 8\epsilon^3\Big)^3$, which was already found by
direct computation in s.\ref{abc+e3}.

Eq.\,\,(\ref{KoresR32}) is in fact an explicit formula for generic
${\cal R}_{3|2}\{A\}$: for a given set $\{A^i(z)\}$ one
straightforwardly builds the $3\times 18$ and $18\times 15$
matrices ${\cal A}$ and ${\cal B}$, then choose any three columns
in the $3\times 18$ matrix, take the complementary 15 lines in the
$18\times 15$ matrix, and divide $15\times 15$ determinant by the
$3\times 3$ one. The ratio does not depend on the choice of the
three lines and is equal to ${\cal R}_{3|2}\{A\}$.

%
%

\subsubsection{Koszul complex. II. Explicit expression for
determinant of exact complex \label{KoII}}

Now, after examples are considered, we can return to generic case.

\bigskip

{\bf The case of a two-term complex:}

We have two rectangular matrices, $p\times q$ and $q\times r$,
${\cal A}^i_a$ and ${\cal B}^a_\alpha$.

The relation is as follows: \be \epsilon_{i_1\ldots i_p} {\cal
A}^{i_1}_{\tilde a_1}\ldots {\cal A}^{i_p}_{\tilde a_p}  =
\epsilon^{\alpha_1\ldots \alpha_r} {\cal B}^{a_1}_{\alpha_1}
\ldots {\cal B}^{a_r}_{\alpha_r} \epsilon_{\tilde a_1\ldots \tilde
a_p a_1\ldots a_r} \cdot{\cal R}^{\pm 1} \label{Kores2term} \ee
Here the last epsilon has exactly $p+r=q$ indices and {\cal R} is
our resultant. It appears at the r.h.s. in power ${\rm
sign}(p-r)=\pm 1$. If $p=r$, then ${\cal R}=1$.

From (\ref{Kores2term}) it follows that the resultant can be
considered as irreducible factor of {\it any} maximal-size minor
of matrix $A$ (if $p>r$) of $B$ (if $p<r$). However,
eq.(\ref{Kores2term}) -- and its analogues for multi-term Koszul
complexes -- explicitly identifes also the other factors (as the
maximal-size minors of another matrix -- $B$ or $A$
respectively,-- they depend on the choice of the original minor).

\bigskip

{\bf The case of 3-term complex:}

For 3-term {\it exact} complex (see Fig.\ref{23ntermcomplex}.B)
$$
0 \longrightarrow \begin{array}{c} k\\+\\0 \end{array}
\longrightarrow \begin{array}{c} l\\+\\k \end{array}
\longrightarrow \begin{array}{c} m\\+\\l \end{array}
\longrightarrow \begin{array}{c} 0\\+\\m \end{array}
\longrightarrow 0
$$
with three rectangular matrices ${\cal A}^I_a$, ${\cal B}^a_i$ and
${\cal C}^i_\mu$ of sizes $k\times (k+l)$, $(k+l)\times (l+m)$ and
$(l+m)\times m$ we have:
$$
\epsilon_{I_1\ldots I_k}{\cal A}^{I_1}_{\tilde a_1}\ldots {\cal
A}^{I_k}_{\tilde a_k} \epsilon^{\mu_1\ldots\mu_m} {\cal C}^{\tilde
i_1}_{\mu_1}\ldots {\cal C}^{\tilde i_m}_{\mu_m} =
\epsilon_{\tilde a_1\ldots \tilde a_k a_1\ldots a_l}
\epsilon^{\tilde i_1\ldots \tilde i_m i_1\ldots i_l} {\cal
B}^{a_1}_{i_1}\ldots {\cal B}^{a_l}_{i_l} \cdot {\cal R}^{\pm 1}
$$
Dimensions $k$, $l+k$, $m+l$, $m$ are made from the relevant
$M_{n|s}$. Resultant at the r.h.s. appears in power ${\rm
sign(k+m-l)} = \pm 1$. If $k+m=l$, ${\cal R}=1$.


\Fig{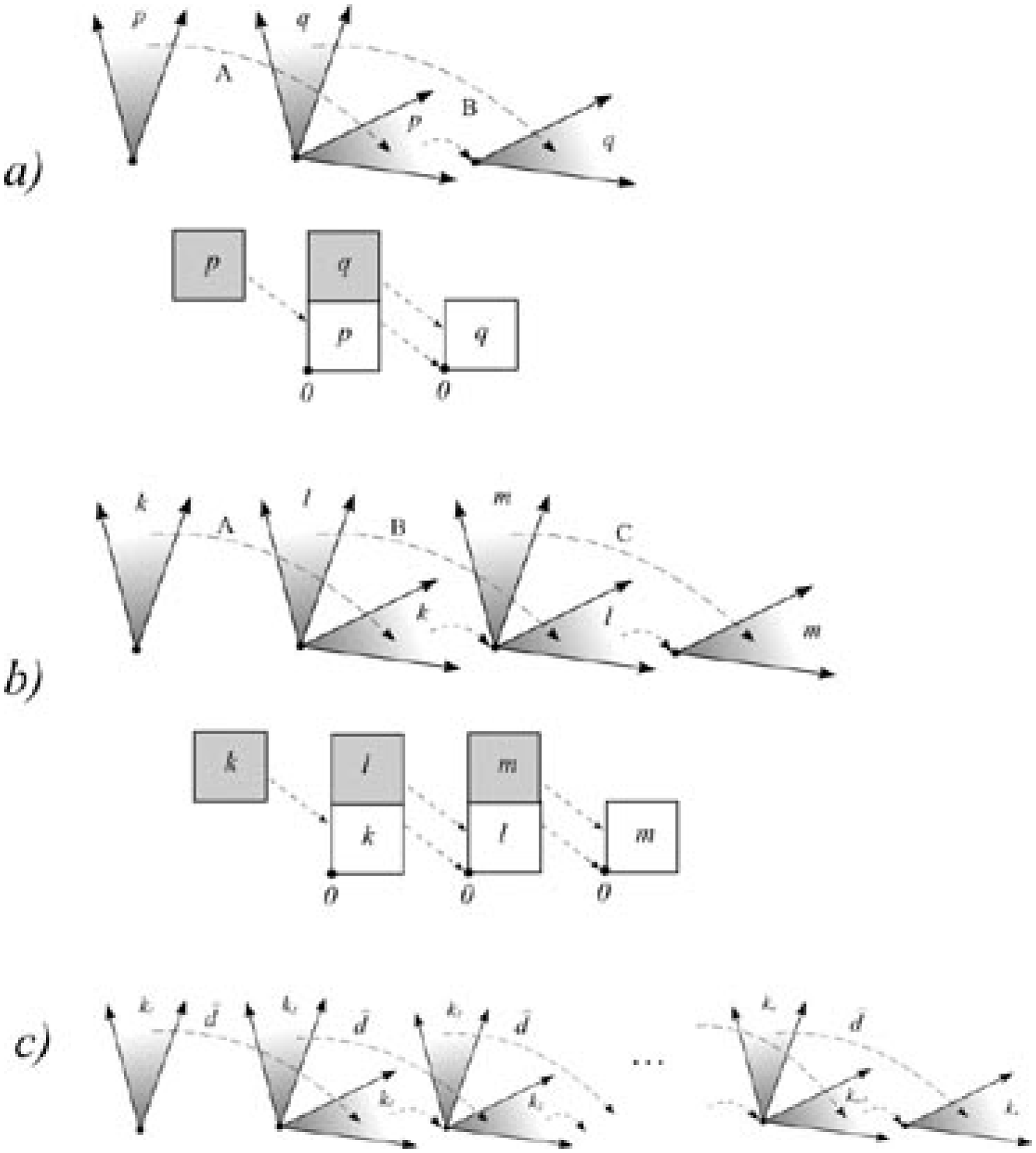} {442,493} {Typical form of {\it exact} linear
complexes, i.e. such that the image of each map coincides with the
kernel of the next map. This imposes constraints on dimensions of
vector spaces: they are equal to the sum of kernel dimensions $d_j
= k_{j} + k_{j+1}$. \ \ A) The two-term complex (two maps between
three vector spaces). \ \ B) The three-term complex (three maps
between four vector spaces). \ \ C) The $n$-term complex. \ \ Note
that numeration of vector spaces in complex starts from $0$, this
helps to slightlu simplify the formulas.}

{\bf Generic case:}

The $n$ terms of exact sequence are formed by maps between $n+1$
linear spaces of dimensions $d_j = k_{j} + k_{j+1}$,\ $j =
0,\ldots,n$: (Fig.\ref{23ntermcomplex}.C)
$$
0 \longrightarrow \begin{array}{c} k_1\\+\\0 \end{array}
\stackrel{\hat d}{\longrightarrow}
\begin{array}{c} k_2\\+\\k_1 \end{array}
\stackrel{\hat d}{\longrightarrow} \ldots \stackrel{\hat
d}{\longrightarrow}
\begin{array}{c} k_j\\+\\k_{j-1} \end{array}
\stackrel{\hat d}{\longrightarrow} \ldots \stackrel{\hat
d}{\longrightarrow}
\begin{array}{c} k_{n}\\+\\k_{n-1} \end{array}
\stackrel{\hat d}{\longrightarrow}
\begin{array}{c} 0\\+\\k_{n} \end{array}
\longrightarrow 0
$$
Assume for simplicity that all $s_1=\ldots=s_n=s$ are the same.

If the starting ($j=0$) vector space is ${\cal P}_{n|q}$, i.e. is
formed by homogeneous polynomials of degree $q$, then $d_0 = k_1 =
M_{n|q}$.

The next space consists of $n$ copies (because there are $n$
different products of $n-1$ out of $n$ parameters $\theta$) of
${\cal P}_{n|q+s}$ and has dimension $d_1 = k_1+k_2 = nM_{n|q+s}$.

The $j$-th space consists of $\frac{n!}{j!(n-j)!}$ copies of
${\cal P}_{n|q+js}$ and has dimension $d_j = k_j + k_{j+1} =
\frac{n!}{j!(n-j)!}M_{n|q+js}$.

The final ($j=n$) space is a single copy of ${\cal P}_{n|q+ns}$
(this means that $p = q+ns$) and has dimension $d_n = k_n =
M_{n|q+ns}$ (note, that $k_{n+1} = 0$ if the complex is indeed
{\it exact}).

Above is described a recurrent procedure, defining kernel
dimensions $k_j$ for all $j$ through $q$ and $s$: \be
k_1 = M_{n|q}, \nn \\
k_2 = nM_{n|q+s} - M_{n|q}, \nn \\
k_3 = \frac{n(n-1)}{2}M_{n|q+2s} - nM_{n|q+s} + M_{n|q}, \nn \\
k_4 = \frac{n(n-1)(n-2)}{6}M_{n|q+3s} -
    \frac{n(n-1)}{2}M_{n|q+2s} + nM_{n|q+s} - M_{n|q}, \nn \\
k_5 = \frac{n(n-1)(n-2)(n-3)}{24}M_{n|q+4s} -
    \frac{n(n-1)(n-2)}{6}M_{n|q+3s} +
    \frac{n(n-1)}{2}M_{n|q+2s} - nM_{n|q+s} + M_{n|q}, \nn \\
\ldots \nn \ee We remind that $M_{n|p} =
\frac{(n+p-1)!}{(n-1)!p!}$. Consistency check is that emerging
$k_{n+1}$ vanishes for any $q$.
Indeed, \\ $k_1$ vanishes for $n=0$;\\
for $n=1$\ $k_2 = M_{1|q+s}-M_{1|q} =
\frac{(q+s)!}{(q+s)!} - \frac{q!}{q!} = 0$;\\
for $n=2$\ $k_3 = M_{2|q+2s} - 2M_{2|q+s} + M_{2|q} =
\frac{(q+2s+1)!}{(q+2s)!} - 2\frac{(q+s+1)!}{(q+s)!} +
\frac{(q+1)!}{q!} = 0$; \\
and so on. Criterium $k_{n+1}=0$ of the complex exactness
continues to hold for negative values of $q$, provided the
formulas for dimensions remain true. However, $d_1 = k_1 =
M_{n|q}$ vanishes -- as it should -- for $q<0$ only for \be q\geq
1-n, \label{q+n} \ee and this is condition of validity of the
resultant formulas, provided by the theory of Koszul complexes.
One can easily check that examples with {\it truncated} complexes
from s.\ref{KoI} satisfy (\ref{q+n}). If one puts $q=1-n$, then
the first non-trivial space for linear maps with $s=1$ is the one
with $q+m=0$, i.e. $m = n-1$, so that the complex reduced to a
single map between spaces with $m=n-1$ and $m=n$. Similarly, for
the case of $n=2$ one can start from $q=-1$, and only two of the
three vector spaces remain non-trivial. Finally, in the example of
$n=3$ and $s=2$ in s.\ref{abc+e3fromK} the actual choice was
$q=-2$: then only three out of four vector spaces are non-trivial.

The matrix, describing the map between $j$-th and $j+1$-st spaces
has dimension $d_j\times d_{j+1}$, we denote it by ${\cal
A}_{d_j}^{d_{j+1}}$. Determinant of the complex -- which gives the
resultant of original tensor -- is defined from
$$
\Big({\cal A}_{k_1}^{k_1+k_2}\Big)^{\wedge k_1} \Big({\cal
A}_{k_2+k_3}^{k_3+k_4}\Big)^{\wedge k_3}\ldots \Big({\cal
A}_{k_{2i}+k_{2i+1}}^{k_{2i+1}+ k_{2i+2}}\Big)^{\wedge k_{2i+1}}
\ldots = $$ $$ = \Big({\cal A}_{k_1+k_2}^{k_2+k_3}\Big)^{\wedge
k_2} \Big({\cal A}_{k_3+k_4}^{k_4+k_5}\Big)^{\wedge k_4} \ldots
\Big({\cal A}_{k_{2i+1}+k_{2i+2}}^{k_{2i+2}+
k_{2i+3}}\Big)^{\wedge k_{2i+2}} \ldots\cdot {\cal R}
$$
Wedge tensor powers here denote contractions with
$\epsilon$-tensors in each of the spaces. $\epsilon$ actring in
the space of dimension $d_j = k_j + k_{j+1}$ has $d_j$ indices:
one can easily check that indeed the sum of numbers of covariant
indices in this space at one side of equality plus the sum of
contravariant indices in the same space at another side is equal
$d_j$. Since all matrices ${\cal A}$ are {\it linear} in original
tensor $A$ (which defines the differential $\hat d$, of which
${\cal A}$ are various representations), the total power of $A$ --
which should be equal to the $A$-degree of the resultant -- is \be
k_1 - k_2 + k_3 - \ldots - (-)^n k_n = ns^{n-1} \ee This
reproduces correct answer for $d_{n|s} = {\rm deg}_A ({\cal R}) =
ns^{n-1}$.

Generalization is straightforward to the case of arbitrary
$s_1,\ldots,s_n$.

\subsubsection{Koszul complex. III. Bicomplex structure}

In this section we add a little more details to description of
Koszul complex and also introduce additional -- {\it bicomplex} --
structure, unifying infinitely many Koszul complexes for a given
map $A^i(z)$ into a finite set of bi-complexes, labeled by $p\
{\rm mod}(\sum s_i)$.

Koszul complex (\ref{KoshI}) depends on the choice of the map
$A_i(\vec z)$, which defines the differential \be \hat d =
A_i(\vec z)\frac{\partial}{\partial \theta_i} = \sum_i^n \sum_{
\{j\} }^n A_i^{j_1\ldots j_{s_i}}z_{j_1}\ldots z_{j_{s_i}}
\frac{\partial}{\partial\theta_i} \ee and of degree $p$ of the
polynomials in the end-point space. Another differential \be
\hat\delta = \epsilon_{ii_1\ldots i_{n-1}}A_i(\vec z)
\theta_{i_1}\ldots\theta_{i_{n-1}} = \sum_{\{i\},\{j\}}^n
\epsilon^{ii_1\ldots i_{n-1}} A_i^{j_1\ldots j_{s_i}}z_{j_1}\ldots
z_{j_{s_i}} \theta_{i_1}\ldots\theta_{i_{n-1}} \ee with the
properties \be \hat d^2 = \hat\delta ^2 = 0, \ \ \ \hat
d\hat\delta + \hat\delta\hat d = 0 \ee can be used to connect
complexes with parameters $p$ and $p + S \equiv p + \sum_{i=1}^n
s_i$ and form a peculiar {\it bicomplex} \be
\begin{array}{ccccccccccccc}
&&&&& \ldots &&&&&&& \\
&&&&\hat\delta&&\hat\delta&&&&&& \\
&&&&\swarrow&&\swarrow&&&&&& \\
0 & \stackrel{\hat d}{\rightarrow} & V^{(n|p)} & \stackrel{\hat
d}{\rightarrow} & V^{(n-1|p)} & \stackrel{\hat d}{\rightarrow} &
\ldots & \stackrel{\hat d}{\rightarrow} & V^{(1|p)} &
\stackrel{\hat d}{\rightarrow} & V^{(0|p)}
& \rightarrow & 0 \\
&&&&\hat\delta&&\hat\delta&&&&&& \\
&&&&\swarrow&&\swarrow&&&&&& \\
&&&&&&&&&&&& \\
0 & \stackrel{\hat d}{\rightarrow} & V^{(n|p+S)} & \stackrel{\hat
d}{\rightarrow} & V^{(n-1|p+S)} & \stackrel{\hat d}{\rightarrow} &
\ldots & \stackrel{\hat d}{\rightarrow} & V^{(1|p+S)} &
\stackrel{\hat d}{\rightarrow} & V^{(0|p+S)}
& \rightarrow & 0 \\
&&&&\hat\delta&&\hat\delta&&&&&& \\
&&&&\swarrow&&\swarrow&&&&&& \\
&&&&& \ldots &&&&&&& \\
\end{array}
\ee where \be V^{(k|p)} = \left\{ \sum_{\{i\}}^n \sum_{\{j\}}^n
H^{i_1\ldots i_k|j_1\ldots j_{s(p|i_1,\ldots,i_k)
}} \theta_{i_1}\ldots\theta_{i_k} z_{j_1}\ldots
z_{j_{s(p|i_1,\ldots,i_k)
}} \right\} \ee and $s(p|i_1,\ldots,i_k) = p-s_{i_1} - \ldots -
s_{i_k}$ (or simply $p-ks$ if all $s_1=\ldots=s_n=s$, note that in
this case $V^{(k|p)}$ consists of polynomials of degree $p-ks$,
not $p$, in $z$, in general the $z$-degree depends on the set of
antisymmetric indices $\{i\}$). Linear dimensions of these vector
spaces are \be {\rm dim} V^{(k|p)} = \sum_{1\leq i_1 < \ldots <i_k
\leq n} \frac{\left(p-s_{i_1} - \ldots - s_{i_k}+ n-1 \right)!}
{(n-1)!\left(p-s_{i_1} - \ldots - s_{i_k}\right)!}, \ee the sum
over $i$'s contains $\frac{n!}{k!(n-k)!}$ terms. Operators
$\hat\delta$ map spaces $V^{(1|p)}$ and $V^{(0|p)}$ into
$V^{(n|p+S)}$ and $V^{(n-1|p+S)}$ respectively and annihilate all
other $V^{(k|p)}$ with $k>1$.

\subsubsection{Koszul complex. IV. Formulation through
$\epsilon$-tensors}

In the mixed tensor algebra $\cl{T}(V)\otimes \cl{T}^*(V)$ of an
$n$-dimensional linear space $V$, there are two distinguished
1-dimensional $GL_n$-invariant subspaces spanned by
$$\epsilon:=\wedge^nx:=x_1\wedge\dots\wedge x_n:=
\sum_{\sigma\in S_n}(-1)^{|\sigma|}x_{\sigma(1)}\dots
x_{\sigma(n)}\in \cl{T}^n$$ and its contravariant dual
$$\epsilon^*:=\wedge^nx^*\in (\cl{T}^{*})^{n}$$
This gives an $n$-tuple of operators $\Lambda^k,\ k=1,\dots,n$
being tensors from $\cl{T}_{k}^{k}=\cl{T}^{k}\otimes
(\cl{T}^*)^{k}$ defined by mutual contractions between $\epsilon$
and $\epsilon^*$: \be \Lambda_{i_1\dots i_k}^{j_1\dots j_k}=
\epsilon^{j_1\dots j_kl_{1}\dots l_{n-k}} \epsilon_{i_1\dots
i_kl_{1}\dots l_{n-k}} \ee represented by the diagram on
Fig.~\ref{eek0}.

\Fig{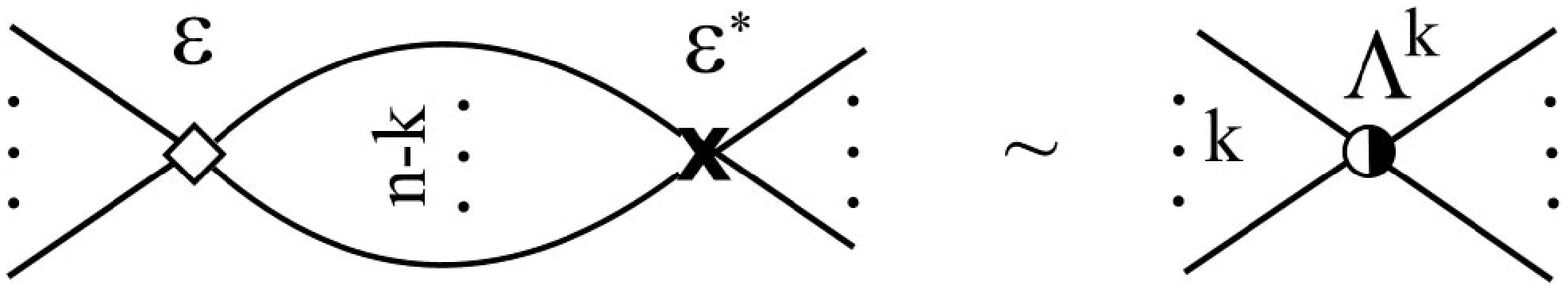} {333,58} {Diagrammatic representation of the operator
$\Lambda^k$}

Then for each $\cl{T}^k$ and a partition $p_1+\dots+p_r=k$ we have
the operator $\Lambda^{p_1\dots p_r}:\cl{T}^k\to \cl{T}^k$
represented by the diagram in Fig.~\ref{eer0}.

\Fig{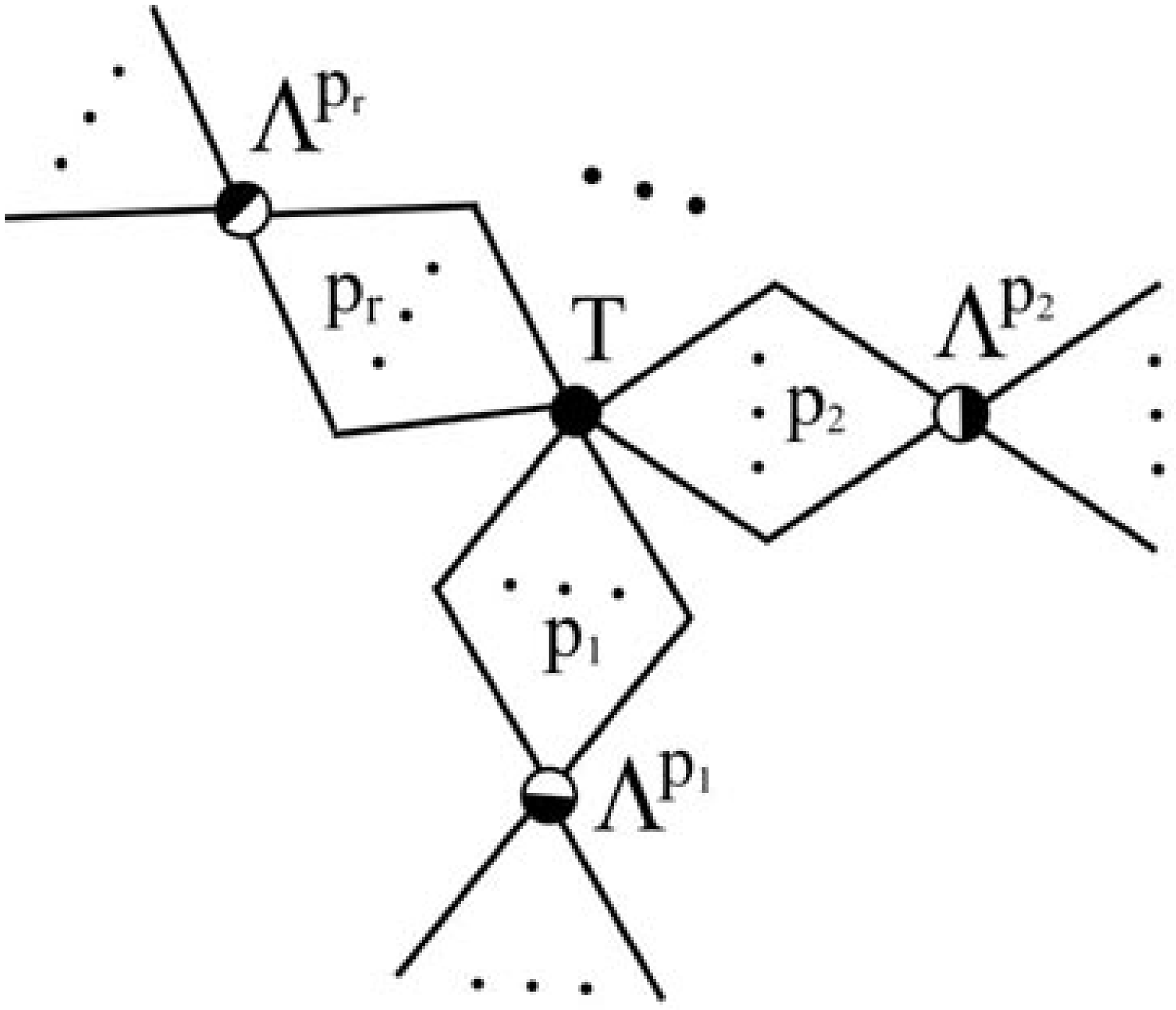} {233,199} {Diagram, representing the action of
$\Lambda^{p_1,\dots,p_r}$ on a contravariant tensor $T$ of rank
$p_1+\ldots+p_r$.}

The image of $\Lambda^{P}$ is a $GL_n$-invariant subspace in
$\cl{T}^k$. If we have a set of $p_{i_1}=\dots=p_{i_s}=p$ in the
partition $P$, then $\Lambda^P$ containes an invariant subspace
obtained by symmetrization ${\rm S}$ with respect to permutations
of groups of indices corresponding to those $p_i=p$.

{\bf Example:} {For $P=(1,\dots,1)$ the $\Lambda^P$ is the
identity operator and space ${\rm S}\Lambda^P\cl{T}^k={\rm
S}\cl{T}^k$ is just the space of symmetric tensors.}

The spaces ${\rm S}\,\Lambda^{P}(\cl{T}^k)\subset \cl{T}^k$
correspond to irreducible representations of $GL_n$.

\bigskip

{\bf Exterior product}

In particular, the operator $\Lambda^k$ (in this case also denoted
by $\Lambda$) maps $\cl{T}^k$ into a $C_n^k$-dimensional
$GL_n$-invariant subspace $\Lambda^kV\subset \cl{T}^k(V)$, spanned
by $\sum_{\sigma\in S_k}x_{\sigma(i_1)}\wedge\dots\wedge
x_{\sigma(i_k)},\ i_1<\dots<i_k\in C^k_n$. Its elements are called
skew- or anti-symmetric tensors.

The composition of a tensor product $\otimes_\tau$ with the
projection operator $\Lambda$ is called {\it exterior product},
$a\wedge b:=\Lambda(a\otimes_\tau b)$.

{\bf Claim:} For any $T\in \cl{T}$ and a tensor product
$\otimes_\tau$ we have $T\wedge T\wedge\cdot\equiv 0$ (see
Fig.~\ref{dtree}) or \be (\Lambda\, T\otimes_\tau)^{\circ 2}\equiv
0 \label{nilpot} \ee Thus any $T\in \cl{T}$ defines the structure
of complex on $\cl{T}$ with the differential $\delta_T:=\Lambda\,
T\otimes_\tau$.

\Fig{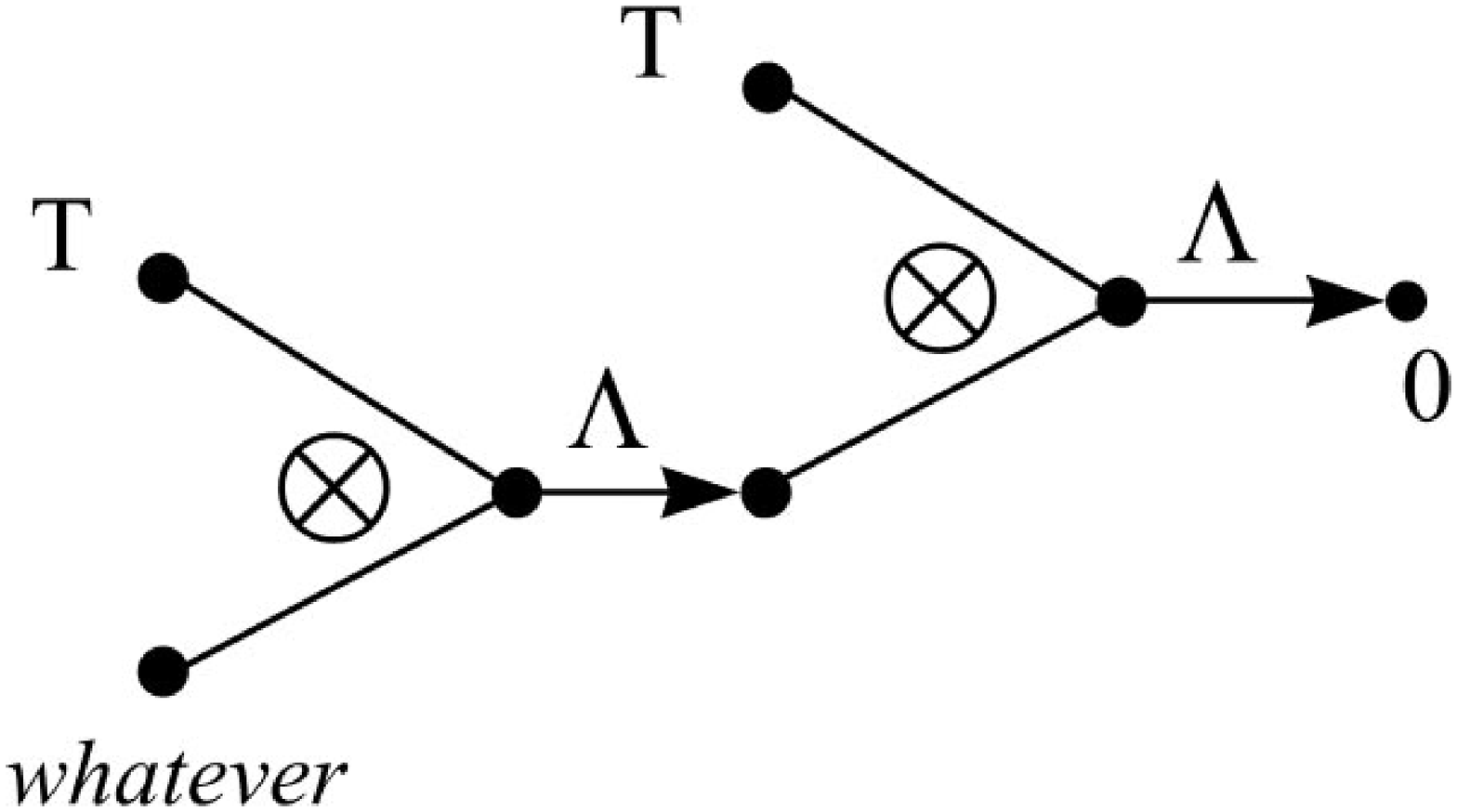} {200,111} {Schematic presentation of the nilpotency
property (\ref{nilpot}) of the $\Lambda\, T\otimes_\tau$
operation.}

\bigskip

{\bf Koszul complex}

For $W\in \cl{T}(U)\otimes \cl{T}(V)$ and a tensor product
$W\otimes_{\tau',\tau''}$ we have a set of operators
$\delta_{P'P''}:=(\wedge^{P'}_{U},
\wedge^{P''}_{V})(W\otimes_{\tau',\tau''}\cdot)$.

{\bf Example:} { Let $\dim U=\dim V=2$, $W=b_{ij}\in U\otimes V$,
$P'=P''=(2)$. Then for $\delta_{P'P''}:\cl{T}^mU\otimes
\cl{T}^nV\to \cl{T}^{m+1}U\otimes \cl{T}^{n+1}V$ we have
$\delta^2$ acting from scalars to $\wedge^2U\otimes\wedge^2 V$ as
follows:
$$\delta^2:\lambda\mapsto|b_{ij}|\lambda\epsilon_U\otimes\epsilon_V$$}

{\bf Claim:} { For $P'=(1,\dots,1)$ and $P''=(\dim V)$ we have
$\delta_{P'P''}^2\equiv 0$.}

The corresponding structure of the complex on $\cl{T}(U)\otimes
\cl{T}(V)$ is called {\it Koszul complex}.

Other operators $\delta_{P'P''}$, not obligatory nilpotent,
generate other important structures, not obligatory complexes.

\subsubsection{Not only Koszul and not only complexes
\label{notKoszul}}

As already mentioned in the beginning of this section, the idea of
linear-algebra approach to the resultant theory is to find some
{\it linear} mapping between some auxiliary vector spaces, which
is somehow induced by non-linear map $A^i(\vec z)$ of projective
spaces and degenerates simultaneously with $A^i(\vec z)$. Koszul
complexes provide explicit realization of this idea, but by no
means unique and even distinguished. Moreover, even the structure
of complex, i.e. the use of {\it nilpotent} mapping, is
unnecessary. It deserves noting, that the conventional constraint
$d^2=0$ on the external derivative, which was one of the original
motivations for the complex construction,
is in no way distinguished: for
example, on simplicial complexes there are natural
operations $d_n$, satisfying
$d_n^n=0$ with any integer $n$ \cite{DMS2}.
Some interesting phenomenon occurs every time, when
dimension of an image of any map drops for $A^i(\vec z)$ of a
special form. This can happen when $A$ belongs to discriminantal
subspace, $A \in {\rm Disc}$, -- then one can extract information
about ${\cal R}\{A\}$. However, this is not the only, and not even
generic, possibility.

We begin with examples of the first type: when degeneration
condition provides information about ${\cal R}\{A\}$.

\bigskip

{\bf Example of linear maps:}

\be {\cal O} = A_i^j z_j \frac{\partial}{\partial z_i} \ee This is
an operator of degree zero in $x$, it maps every space ${\cal
P}_{n|p}$ on itself and this action is represented by square
$M_{n|p}\times M_{n|p}$ matrix ${\cal O}_p$. It turns out that for
all $n$ and all $p>0$ determinant of this matrix is proportional
to $\det A$: \be {\cal R}_{n|1}\{A\} = \det A = {\bf irf}\Big(\det
{\cal O}_p\Big) \ee

If $p=0$, the matrix vanishes.

If $p=1$, the matrix coincides with $A_i^j$ and its determinant is
just $\det A$.

The first non-trivial case is $p=2$. Then \be \det {\cal O}_2 =
\det_{ij,kl}\Big(A^k_i\delta_j^l + A^l_i\delta^k_j +
A^k_j\delta^l_i + A^l_j\delta^k_i\Big) \label{O2detAdelta} \ee For
diagonal matrix $A_i^j = a_i\delta_i^j$ determinant
$\det {\cal O}_2 \sim \det A \cdot \prod_{i<j} (a_i+a_j)$.\\
For $n=1$ the r.h.s. of (\ref{O2detAdelta})
equals $A_1^1 = {\cal R}_{1|1}\{A\}$. \\
For $n=2$  we get determinant of the $3\times 3$ ($3 = M_{2|2}$)
\be
&\left|\left|\begin{array}{ccc} 4A^1_1 & 2A^2_1 & 0 \\
2A^1_2 & A^1_1+A^2_2 & 2A^2_1 \\ 0 & 2A^1_2 & 4A^2_2
\end{array}\right|\right| =
16(A^1_1+A^2_2)(A^1_1A^2_2 - A^1_2A^2_1) \nn\\
&= 16\cdot{\rm tr} A \cdot
\det A \sim \det A = {\cal R}_{2|1}\{A\}.
\ee
For $n=3$ we get in a similar way
$$
\det {\cal O}_2 \sim \Big(({\rm tr} A)^3 - {\rm tr} A^3\Big) \det
A \sim \det A = {\cal R}_{3|1}\{A\}
$$

Consideration for higher $p$ is similar. Always in this example we
get $\det_{M_{n|p}\times M_{n|p}}{\cal O}_p \sim \det_{n\times n}
A = {\cal R}_{n|1}\{A\}$. What is important, operator ${\cal O}$
is not nilpotent and does not generate anything like exact
complex. Still, like in the case of Koszul complex, this operator
degenerates whenever original map does, and thus the study of this
operator provides information about ${\cal R}\{A\}$. However, the
lack of the structure of complex makes difficult interpretation of
additional factors, appearing in explicit expressions for $\det
{\cal O}$ and allows defining ${\cal R}$ only as an {\bf irf} in
this determinant.

For some non-linear examples see s.\ref{Exacov} below.

\bigskip

However, determinant of generic operator is in no way related to
the resultant of original tensor: degeneration of operators is
{\it not} related to degeneration of the generator of the tensor
algebra ${\cal T}(A)$. Generic degeneration condition is described
by some element of invariants ring ${\rm Inv}_{{\cal T}(A)}$,
which is adequately described in terms of diagram technique to be
considered in the next subsection.

Before closing this subsection, we give a couple of examples,
where operators degenerate {\it outside} the discriminantal domain
of original tensor $A$.

\bigskip

{\bf Example of a linear differential operator, which is not {\it
obligatory} related to a resultant.}

Let $T \in {\cal M}_{2^{\times 4}}$, i.e. $n=2$ and tensor $T$ has
$4$ contravariant indices. Then one can form a linear operator
(Fig.\,\,\ref{337}.A) ${\cal O}_i^j = T_{iklm}T^{jklm} =
\epsilon_{ii'}\epsilon_{kk'}\epsilon_{ll'}\epsilon_{mm'}
T^{i'k'l'm'}T^{jklm}$ or $\hat{\cal O} = {\cal O}_i^j
x_j\frac{\partial}{\partial x_i}$. It maps any space ${\cal
P}_{2|p}$ with $p\geq 1$ onto itself, and {\it this} map
degenerates whenever $\det_{M_{2|p}\times M_{2|p}} \hat{\cal O}_p
= 0$.

\bigskip

\Fig{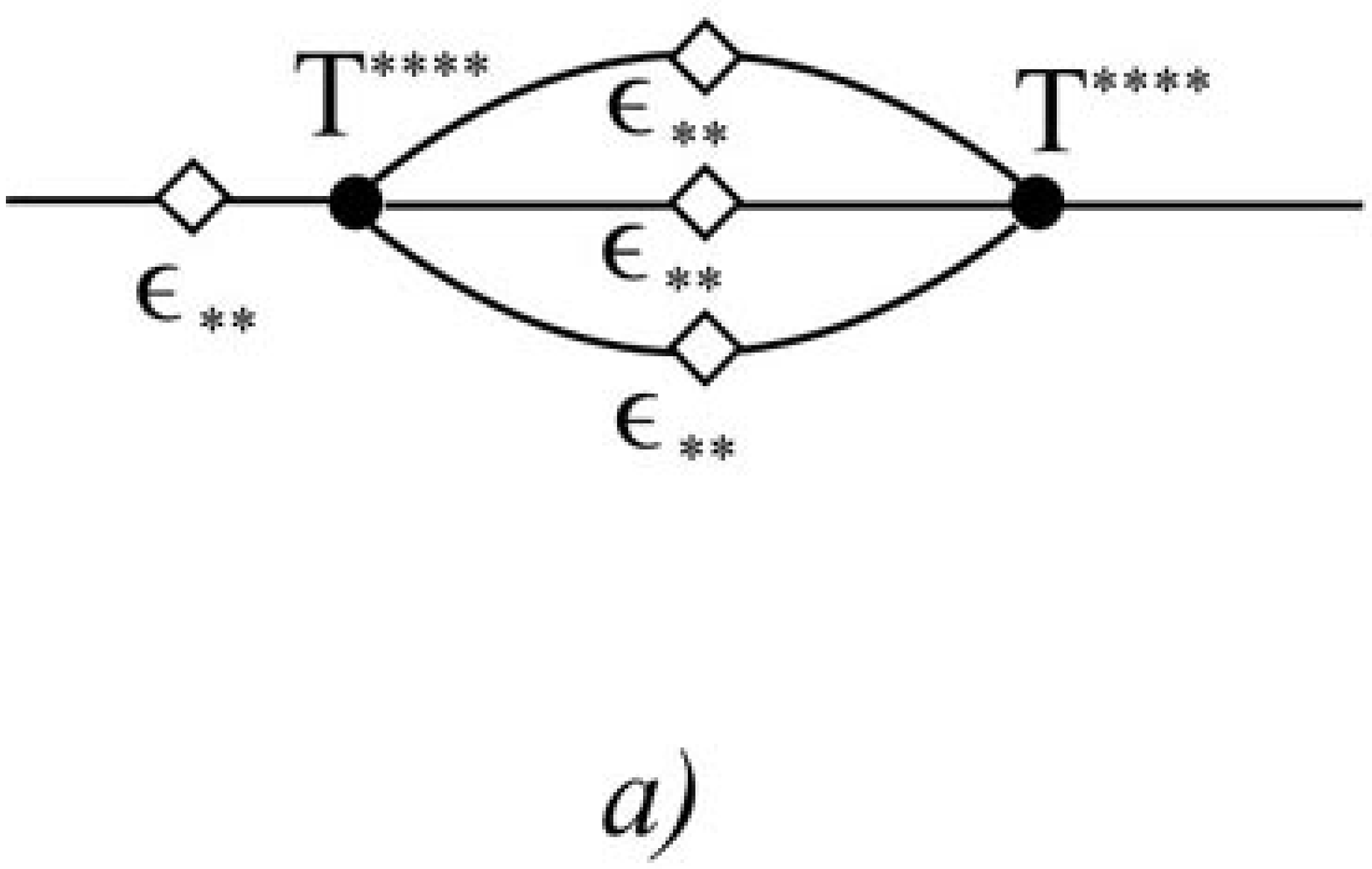}
{320,148}
{Examples of linear operators, with
determinants {\it non-related} to resultants.}

{\bf Example of a quadratic differential operator, which is not
{\it obligatory} related to a resultant.}

In order for an operator to map ${\cal P}_{n|p}$ on itself, it
should have degree zero in $z$. As an example we consider an
operator, which is quadratic in $x$-derivatives. Example will be
associated with the case $n|s = 2|3$, i.e. with cubic map
$A_i(\vec z) = a_i^{jkl}z_jz_kz_l$ of two variables
(Fig.\,\,\ref{337}.B):
\be
{\cal O} &=& \frac{1}{2}a_i^{klm}a_j^{k'l'n}
\epsilon_{kk'}\epsilon_{ll'} z_mz_n\frac{\partial}{\partial
z_i}\frac{\partial}{\partial z_j}\nn\\
&=&\Big((a^{111}x+a^{112}y)(a^{122}x+a^{222}y) -
(a^{112}x+a^{122}y)^2\Big)\frac{\partial^2}{\partial x^2}\nn\\
&+&\Big((b^{111}x+b^{112}y)(b^{122}x+b^{222}y) -
(b^{112}x+b^{122}y)^2\Big)\frac{\partial^2}{\partial y^2}\nn\\
&-&\Big((a^{111}x + a^{112}y)(b^{122}x+b^{222}y) -
2(a^{112}x + a^{122}y)(b^{112}x+b^{122}y) \nn\\
&+&(a^{122}x+a^{222}y)(b^{111}x+b^{112}y)\Big)
\frac{\partial^2}{\partial x\partial y}
\ee
in obvious notation. The
spaces ${\cal P}_{2|0}$ and ${\cal P}_{2|1}$ of constant and
linear functions are annihilated by this ${\cal O}$. When acting
in the space ${\cal P}_{2|2}$ of quadratic polynomials, ${\cal
O}_2$ is represented by a $3\times 3$ matrix {\footnotesize
$$\left(\begin{array}{c|cccc}
&& x^2 & xy & y^2 \\
&&&&\\
\hline
& \  &&&\\
x^2 && 2\Big(a^{111}a^{122} - (a^{112})^2\Big) &
2\Big(a^{111}a^{222}-a^{112}a^{122}\Big) &
2\Big(a^{112}a^{222} - (a^{122})^2\Big) \\ &&&& \\
xy && -a^{111}b^{122} + 2a^{112}b^{112}  &
 -a^{111}b^{222} + a^{112}b^{122}  &
-a^{112}b^{222} + 2a^{122}b^{122}  \\
&&- a^{122}b^{111}&+ a^{122}b^{112} - a^{222}b^{111}& - a^{222}b^{112}\\
&&&& \\
y^2 && 2\Big(b^{111}b^{122}-(b^{112})^2\Big) &
2\Big(b^{111}b^{222}-b^{112}b^{122}\Big)&
2\Big(b^{112}b^{222}-(b^{122})^2\Big)
\end{array} \right)
$$ }
Its determinant is {\it not} equal to the resultant
\be
&{\cal R}_{2|3}\{A\}
= {\rm Res}_t\Big(a^{111}t^3 + 3a^{112}t^2 + 3a^{122}t +
a^{222},\nn\\
&b^{111}t^3 + 3b^{112}t^2+ 3b^{112}t + b^{222}\Big),\ee
which is also a polynomial of degree $6$ in the coefficients $a,b$.
The lack of coincidence is not surprising, because in this case $n|s
= 2|3$ the space of degree-$6$ invariants is two dimensional,
represented by linear combinations of ${\rm Inv}_2^3$ and ${\rm
Inv}_3^2$, see s.\ref{23resulta} below.

\subsection{Resultants and diagram representation of tensor algebra
\label{resdia}}

Feynman-like diagrams provide powerfull technique for study of
generic invariants and representations of the structure group.
Invariants are represented by "vacuum" diagrams, non-trivial
representations -- by those with external legs. Application of
diagram technique to resultant/discriminant theory is less
straightforward: no reliable criteria are known to select
diagrams, contributing to the resultant or discriminant, i.e. it
is yet unclear how degeneracy of original tensor affects
particular diagrams and their linear combinations. This problem is
closely related to integral representation of diagram technique:
it is easy to write down an integral, sensitive to degenerations,
but not all such integrals are reduced to a sum of finitely many
diagrams. Here the subject touches the theory of localization and
topological theories.

\subsubsection{Tensor algebras ${\cal T}(A)$ and ${\cal T}(T)$,
generated by $A^i_I$ and $T$ \cite{GMS} }

Conceptually tensor algebra ${\cal T}(A,B,\ldots)$ is formed by
all tensors which can be constructed from the given set of
generating tensors $A, B, \ldots$. Allowed operations include
summation, multiplication, contractions between $A,B,\ldots$ and
invariant totally antisymmetric $\epsilon$-tensors. These
$\epsilon$-tensors are distinguished for the theory of
$SL(n)$-tensors and thus we do not mention them explicitly in the
notation ${\cal T}(A,B,\ldots) = {\cal T}(\epsilon,A,B,\ldots)$. A
convenient description of a linear basis in ${\cal T}(A,B,\ldots)$
is provided by Feynman-like diagrams. We now proceed to a more
detailed definition.

Let $T^{i_1\ldots i_r}$ be a tensor of the type
$n_1\times\ldots\times n_r$, i.e. with $r$ ordered indices, and
$k$-th index taking values $i_k = 1,\ldots,n_k$. Such generic
tensors are relevant for considerations of {\it poly-linear}
equations, see s.\ref{polydet} for more details. Non-linear map
$A_i(\vec z) = A_i^{j_1\ldots j_s}z_{j_1}\ldots z_{j_s}$, relevant
for consideratons of {\it non-linear} equations, is built from a
tensor $A_i^\alpha$ of special type: with all equal
$n_0=n_1=\ldots=n_s = n$, moreover $A$ is symmetric in the last
$s$ indices (so that one can introduce a multi-index $\alpha =
{\rm symm}(j_1,\ldots,j_s)$, taking $M_{n|s} =
\frac{(n+s-1)!}{(n-1)!s!}$ different values) and the very first
index $i$ is covariant, not contravariant. Tensor $T$ can also
possess additional symmetries, be totally symmetric or
antisymmetric, have a symmetry of any $r$-cell Young diagram and
even more sophisticated (parastatistic or braid) invariances.
However, for the purposes of this subsection concrete symmetry
properties and even the difference between $T$ and $A$ is not
essential, and everything told about $A$ can be immediately
extended to the case of $T$, see s.\ref{polydet} below.

Following \cite{GMS}, we can associate with any tensor, say, $A$
or $T$, a set ${\cal T}(A)$ or ${\cal T}(T)$ of all graphs
(Feynman diagrams), with vertices of two types. Vertices of the
first type contain $A$ (or $T$) and have valence $r=s+1$. In the
case of $T$ there are $r$ different types ({\it sorts}) of lines,
associated with particular vector spaces $V_{n_k}$, and $k$-th leg
of the $T$-vertex is of the $k$-th sort. Lines are directed (carry
arrows), say, to covariant index from a contravariant: in the case
of $T$ vertices all lines are incoming, in the case of $A$ one
(associated with the only contravariant index $i$) is outgoing.
$T$ vertices can not be connected by such lines, or, in other
words, pure contravariant tensors can not be contracted.
Contractions are provided by invariant $\epsilon$
tensors\footnote{ For the purposes of ref.\cite{GMS} symmetric
(metric) tensors $G_{ij}$ and $G^{ij}$ were used instead of
totally antisymmetric $\epsilon$ tensors. The corresponding tensor
algebra ${\cal T}(G,A)$ posesses also a simple Feynman-{\it
integral} interpretation, which is somewhat more obscure for
${\cal T}(\epsilon,A)$. However, despite of many advantages,
conceptual and technical, provided by existence of explicit
integral representation, the tensor algebra ${\cal T}$ and
Feynman-like diagrams can be well defined with no reference to it.
} $\epsilon_{i_1\ldots i_{n_k}}$, which  -- together with
$\epsilon^{i_1\ldots i_{n_k}}$ -- provide vertices of the second
type:  of valences $n_k$ with all $n_k$ lines belonging to the
same sort and either all incoming (if covariant $\epsilon$ stands
in the vertex) or all outgoing (if $\epsilon$ is contravariant).
In the case of $A$ there are only two sorts of epsilons,
associated with indices $j$ and $i$, and even these can be often
identified. In this case one can also express combinations of
$\epsilon$-tensors in terms of multi-indices $I$, but this is a
somewhat sophisticated group-theory exercise, involving routine of
Clebsh-Gordan coefficients for symmetric representations.

The set ${\cal T}(A)$ or ${\cal T}(T)$ is naturally called {\it
tensor algebra}, associated with (generated by) original tensor
$A$ or $T$.

\subsubsection{Operators}

Of certain convenience in applications is the use of particular
diagrams, constructed from {\it operators}, because they can be
expressed through {\it matrices} and handled by means of the
elementary linear algebra. An example of such kind is provided by
one-particle reducible diagrams, but this is only the simplest
possibility, related to operators in the simplest (first
fundamental) representation of the structure group.

We define {\it linear operators} as represented by {\it square}
matrices, acting in some vector space, for example, in the space
${\cal P}_{n|p} = {\cal S}^p\Big({\cal X}_n\Big)$ of homogeneous
polynomials of certain degree $p$ or ${\cal X}_{n_1}\otimes \ldots
\otimes {\cal X}_{n_p}$ of poly-linear functions of certain degree
(rank) $p$. Tensor algebra ${\cal T}(A)$ or ${\cal T}(T)$,
generated by original tensor like $A_i^\alpha = A_i^{j_1\ldots
j_s}$ or $T^{i_1\ldots i_r}$ contains many such operators,
represented by diagrams with equal number ($p$) of incoming and
outgoing legs, i.e. with $p$ covariant and $p$ contravariant
indices.

Operators can be multiplied, commuted, they have traces and
determinants. Various invariants (some or all) of the tensor
algebra ${\cal T}(A)$ or ${\cal T}(T)$, associated with non-linear
map $A^i(\vec z)$ or with a poly-linear function $T(\vec
x_1,\ldots, \vec x_r)$ can be represented as traces of various
operators, belonging to ${\cal T}(A)$ and ${\cal T}(T)$.

The archetypical example of such approach is generation of
invariants ring of the universal enveloping  algebra, ${\cal
T}(A)$, associated with an $n\times n$ matrix (linear map)
$A^i_j$, by $n$ traces $t_k^* = {\rm tr} A^k$, $k = 1,\ldots,n$.
Determinant and higher traces are non-linearly expressed through
these lowest $t_k^*$ through the celebrated formula \be
\det_{n\times n} (I - \lambda A) = \exp \Big\{{\rm tr}\log
(I-\lambda A)\Big\} = \exp \left\{ - \sum_{k=1}^\infty
\frac{\lambda^k}{k}{\rm tr}A^k\right\} \label{dettra1} \ee
Relations follow from the fact that the l.h.s. is polynomial of
degree $n$ in $\lambda$, therefore all coefficients in front of
$\lambda^k$ with $k>n$ at the r.h.s. should vanish, and this
recursively expresses ${\rm tr} A^k$ with $k>n$ through the first
$n$ traces. Equating the coefficients in font of $\lambda^n$ at
both sides, one obtains also an expression for determinant: \be
\det_{n\times n} A = (-)^n \oint \frac{d\lambda}{\lambda^{n+1}}
\exp \left\{ - \sum_{k=1}^\infty \frac{\lambda^k}{k}{\rm
tr}A^k\right\} = \frac{1}{n!}\Big({\rm tr} A\Big)^n - \ldots +
(-)^n \sum_{k=1}^{n-1} \frac{{\rm tr} A^k\ {\rm tr}
A^{n-k}}{2k(n-k)} + \frac{(-)^{n+1}}{n}{\rm tr} A^n,
\label{dettra2} \ee for example \be
\begin{array}{ccc}
n=1: &\ \ \ & \det_{1\times 1} A = {\rm tr} A \\
n=2: && \det_{2\times 2} A =
\frac{1}{2}\left[ ({\rm tr} A)^2 - {\rm tr} A^2\right] \\
n=3: && \det_{3\times 3} A = \frac{1}{6}({\rm tr} A)^3 -
\frac{1}{2} {\rm tr} A\ {\rm tr} A^2 + \frac{1}{3}{\rm tr} A^3 \\
& \ldots &
\end{array}
\label{dettra3} \ee This particular example is somewhat special,
because $\epsilon$-vertices play no role: identities like
(\ref{dettra2}) allow one to build ${\cal T}(A)$ from $A$-vertices
only. Of course, this is not true in the general case (it is
enough to remind that no non-trivial diagrams can be made from
pure covariant tensors $T$ without the use of $\epsilon$'s):
relations like (\ref{dettra1})-(\ref{dettra3}) still exist (and
are very interesting to find and explore), but they involve
"traces", made with the help of $\epsilon$-tensors.

Such reformulation of eqs.(\ref{dettra3}) can be given in terms of
diagonal Plukker minors \be I_d[i_1,\ldots,i_k] =
I[i_1,\ldots,i_k|i_1,\ldots,i_k] = \det_{1\leq r,s\leq k}
A^{i_s}_{i_r}, \ee for example, $I_d[ij] = A^i_iA^j_j-A^i_jA^j_i$
and $I_d[ijk] = \left|\left|\begin{array}{ccc}
A^i_i & A^j_i & A^k_i \\ A^i_j & A^j_j & A^k_j \\
A^i_k & A^j_k & A^k_k \end{array} \right|\right|$. Conbinations of
traces at the r,.h.s. of (\ref{dettra3}) are in fact equal to \be
I_k(A) = \sum_{1\leq i_1<\ldots i_k\leq n} I_d[i_1,\ldots, i_k],
\ee for example, \be
{\rm tr} A = \sum_{i=1}^n A^i_i = I_1(A), \nn \\
\frac{1}{2}\left[ ({\rm tr} A)^2 - {\rm tr} A^2\right] =
\sum_{1\leq i<j\leq n} \Big(A^i_iA^j_j - A^i_jA^j_i\big)
= I_2(A), \nn \\
\ldots \ee Expressed in these terms, eq.(\ref{dettra2}) is
obvious.

\subsubsection{Rectangular tensors and linear maps}

A linear map between linear spaces of two different dimensions,
$A:\ V \rightarrow W$, ${\rm dim} V = n$, ${\rm dim W} = m$ with
fixed basises $\{e_j\} \subset V$ and $\{f_i\} \subset W$ is given
by an $n\times m$ rectangular matrix $A^i_j$, where the element
$A^i_j$ is the $i$-th coordinate of the $A$-image of $e_j$.

If ${\rm rank}\ A = m < n$ then the kernel of $A$ is an
$m$-dimensional subspace ${\rm ker}\ A \subset V$, i.e. a point of
Grassmannian $Gr_{m,n}(W) \subset \Lambda^m V$ with Plukker
coordinates $(-)^{\sigma(\alpha)}\Delta_\alpha(A)$, where
$\Delta_\alpha$ is an $m\times m$ minor of $A$ built on the
complement to the combination $\alpha \in C_n^{n-m}$ of columns of
$A$, and $\sigma(\alpha)$ is the signature of this combination.

For $n=m+1$ the kernel of $A$ is a one-dimensional subspace,
spanned by the vector $\Big(\Delta_1, -\Delta_2, \ldots,
(-)^m\Delta_{m+1}\Big) \in V$ or a point of the projective space
$PV$. If $\Delta_{m+1}\neq 0$ then it may be written as an element
\be \left(\frac{\Delta_1}{\Delta_{m+1}},\ldots,
\frac{(-)^{m-1}\Delta_m}{\Delta_{m+1}}\right) \ee in the chart
$V^{(m+1)} \subset PV$ with coordinates $\tilde x_i =
\frac{x_i}{x_{m+1}}$. This formula is nothing but the Craemer rule
for solution of non-homogeneous linear system \be \sum_{j=1}^m
A_i^j \tilde x_j = A_i^{m+1} \ee

{\bf Example:} For $n=2$ and $m=1$ the kernel of the operator
$(a,b): \ V\rightarrow W$\ is given by the one-dimensional
subspace $\lambda(be_1 - ae_2) \subset V$ with Plukker coordinates
$(\Delta_1, -\Delta_2) = (b,-a)$.

\subsubsection{Generalized Vieta formula
for solutions of non-homogeneous equations
\label{CraVie}}


Symmetric combinations of solutions to non-homogeneous equations
are expressed through the coefficients of the equations. The well
known Craemer formula for inverse matrix and Vieta formulas
involving the roots of a polynomial are particular examples of the
general formula, appearing in the cases of $n|s=n|1$ and $n|s=2|s$
respectively. The general formula includes a sum over diagrams
with coefficients which are known only hypothetically.

{\bf The structure of general relation}

\noindent

Consider the system of $n-1$ equations of $n$ homogeneous
variables: \be \sum_{j_1,\ldots, j_{S_I} = 1}^n A_I^{j_1\ldots
j_{s_I}} x_{j_1}\ldots x_{j_{s_I}} = 0, \ \ \ \ I = 1,\ldots,n-1
\label{nonhomprob} \ee Any non-homogeneous system, consisting of
$n-1$ non-homogeneous equations of $n-1$ non-homogeneous
variables, can be rewritten in this form at expense of introducing
one more variable, see examples below. Note that degrees $s_I$ of
different equations can be different: $s$ depends on $I$.
Generically such system has {\it discrete} projectively
non-equivalent solutions (no moduli except for common rescalings),
and their number ${\cal N} = \prod_{i=1}^n s_i$. We denote label
{\it solutions} by index $\mu = 1,\ldots,{\cal N}$ and denote them
through $X^{(\mu)}_j$. Then {\it generalized Vieta formula} has
the form: \be {\rm symm} \left\{ \prod_{\mu = 1}^{{\cal N}}
X^{(\mu)}_{i_\mu}\right\}\ \sim\ \prod_{\mu}^{{\cal N}}
\epsilon_{i_\mu k_{\mu,1}\ldots k_{\mu,n-1}} \left\{ \sum_{{\rm
diagrams}} c_{{\rm diag}} \otimes_{I=1}^{n-1} A_I^{\otimes {\cal
N}/s_I} \right\}^{k_{1,1}\ \ldots\ k_{{\cal N},n-1}}
\label{genviaf} \ee At the l.h.s. of this formula symmetrization
is performed over permutations of $i$-indices: \be {\rm symm}
\left\{ \prod_{\mu = 1}^{{\cal N}} X^{(\mu)}_{i_\mu}\right\}
\equiv \sum_{P\in\sigma_{{\cal N}}} \prod_{\mu=1}^{{\cal N}}
X^{P(\mu)}_{i_\mu} \label{symmsol} \ee The r.h.s. of
(\ref{genviaf}) contains each of $n-1$ tensors $A_I$ in
$\prod_{J\neq I}^n s_J = {\cal N}/s_I$ copies, and the total
number of their upper $k$-indices is $\sum_{I=1}^{n-1}
s_I\frac{{\cal N}}{s_I}$ i.e. indeed equals to the number
$(n-1){\cal N}$ of lower $k$-indices in the set of
$\epsilon$-tensors. These upper $k$-indices can be distributed
over $A$-tensors in various ways, which can be conveniently
represented by diagrams. In many cases the weight $c_{{\rm diag}}$
is unity for all possible diagrams. If some $k$ of degrees $s_I$
coincide, the problem (\ref{nonhomprob}) acquires an extra $SL(k)$
symmetry factor in the structure group, and associated sub-scripts
of $A_I$'s at the r.h.s. (\ref{genviaf}) should be contracted with
the help of the corresponding rank-$k$ tensors $\varepsilon$.
These contractions will be denoted in diagrams below by wavy lines
with crosses at vertices of valence $k$. The sign $\sim$ denotes
projective equivalence, i.e. solutions $X^{(\mu)}$ can be
normalized so that eq.(\ref{genviaf}) becomes and exact equality
(note that the r.h.s. of (\ref{genviaf}) is polynomials in the
coefficients $A$).

{\bf The case of $n=2$: the ordinary Vieta formula}

\noindent

This is the case of a single non-homogeneous equation of a single
variable, defining the roots of a single polynomial of degree $s$.
In homogeneous coordinates $x_1,x_2 = x,y$ it can be written in
several differently-looking forms: \be \sum_{j_1,\ldots, j_s=1}^2
A^{j_1\ldots j_s}x_{j_1}\ldots x_{j_s} = \sum_{l=0}^s a_l x^l
y^{s-l} = \prod_{\mu=1}^{s} \Big(xY^{(\mu)} - yX^{(\mu)}\Big) = 0
\ee It has exactly ${\cal N} = s$ projectively non-equivalent
solutions, and the $s$ roots of original polynomial are ratios
$z_\mu = X^{(\mu)}/Y^{(\mu)}$. In eq.(\ref{genviaf}) there are no
$I$-indices, ${\cal N} = s$, the $A$-tensor appears at the r.h.s.
only once ${\cal N}/s=1$, and the role of ${\cal N}=s$ rank-$2$
$\epsilon$-symbols is to convert upper indices into lower ones:
\be {\rm symm} \left\{ \prod_{\mu = 1}^{{\cal N}}
X^{(\mu)}_{i_\mu}\right\} \ \sim\ \epsilon_{i_1j_1}\ldots
\epsilon_{i_sj_s} A^{j_1\ldots j_\mu} = A_{i_1\ldots i_\mu}
\label{viaf} \ee Fig.\ref{Vietdia} shows the single relevant
diagram.

\Fig{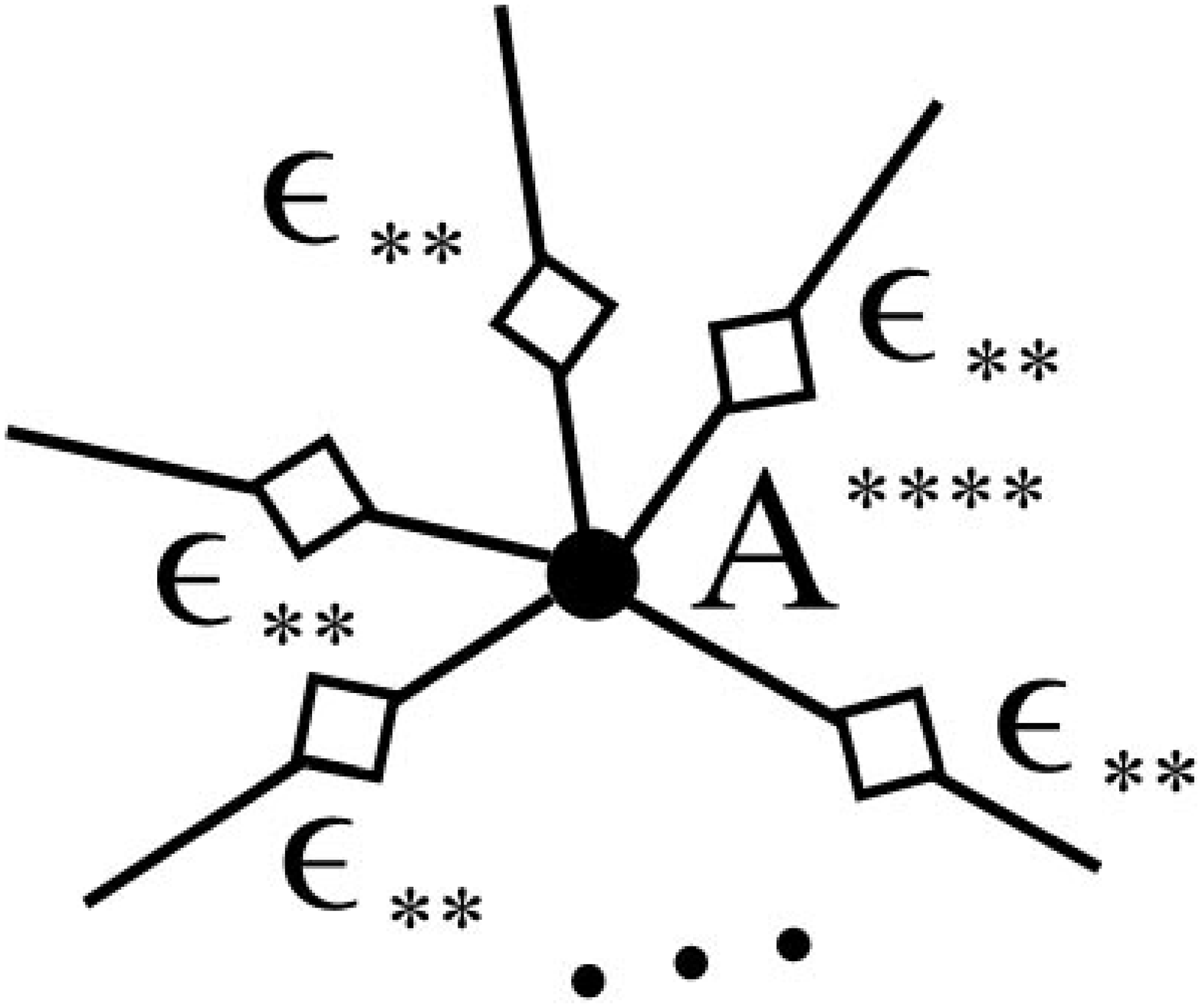} {118,102} {Pictorial representation of Vieta formula
for $n=2$. Dots at valence-$s$ vertices represent tensor
$A^{j_1\ldots j_s}$ of rank $s$. Diamonds at valence-$2$ vertices
represent $\epsilon$-tensors of rank $n=2$.}

Represented in terms of coefficients $a_l = C^l_s A^{1\ldots
12\ldots 2}$ with $l$ appearances of $1$, $s-l$ appearances of $2$
and with binomial coefficients $C^l_s = \frac{s!}{l!(s-l)!}$,
eq.(\ref{viaf}) turns into conventional Vieta formula: \be \sum
z_{\mu_1}\ldots z_{\mu_l} = (-)^l \frac{a_l}{a_0} \ee where the
sum at the r.h.s. is over all $C^l_s$ subsets
$\{\mu_1,\ldots,\mu_l\}$ of the size $l$ of the set of all the $s$
roots of the polynomial.

For example, for $s=2$ we get the usual quadratic equation,
$A^{11}z^2 + 2A^{12}z + A^{22} = 0$ for $z = x_1/x_2=x/y$ with two
roots, which are projectively equivalent to
$$ \begin{array}{c} X^\pm= X_1^\pm = -A^{12} \pm \sqrt{D},\\
Y^\pm= X_2^\pm = 2A^{11}, \end{array}$$ $D =
(A^{12})^2-4A^{11}A^{22}$. Vieta formula (\ref{viaf}) states:
$$
\left(\begin{array}{cc} A_{11} & A_{12} \\ A_{12} &
A_{22}\end{array}\right) = \left(\begin{array}{cc} A^{22} &
-A^{12} \\ -A^{12} & A^{11}\end{array}\right) \cong
\frac{1}{2}\left\{\left(\begin{array}{c}X^+\\
Y^+\end{array}\right) \otimes \Big( X^-\ \  Y^-\Big) +
\left(\begin{array}{c}X^-\\ Y^-\end{array}\right) \otimes \Big(
X^+\ \  Y^+\Big)\right\} = $$ \be = \left(\begin{array}{cc} X^+X^-
&
\frac{1}{2}(X^+Y^- +X^-Y^+) \\
 \frac{1}{2}(X^+Y^- +X^-Y^+) &
 Y^+Y^-\end{array}\right)
\ee or simply $A^{22}/A^{11} = z_+z_-$, $-A^{12}/A^{22} =
\frac{1}{2}\big(z_+ + z_-\big)$.

{\bf The case of $n=3$ with $s_1=1$ and $s_2=2$:}

\noindent

In this example we have a system of two equations, one linear,
another quadratic, for $3$ homogeneous variables
$x_1,x_2,x_3=x,y,z$: \be
\left\{ \begin{array}{c} A^i x_i = 0, \\
B^{ij} x_ix_j = 0.  \end{array} \right. \ee Resolving the first
equation with respect to, say, $z$, and substituting it into the
second equation, we reduce the system to a single quadratic
equation. The number of projectively non-equivalent solutions is
${\cal N} = s_1s_2 = 2$ and (\ref{genviaf}) states that \be {\rm
symm} \left\{ X^+_iX^-_l\right\} \sim \epsilon_{ijk}\epsilon_{lmn}
A^jA^mB^{kn} \label{viaf12} \ee One easily checks that this is
indeed the case. Only one diagram in Fig.\ref{Vietdia12}
contributes.

\Fig{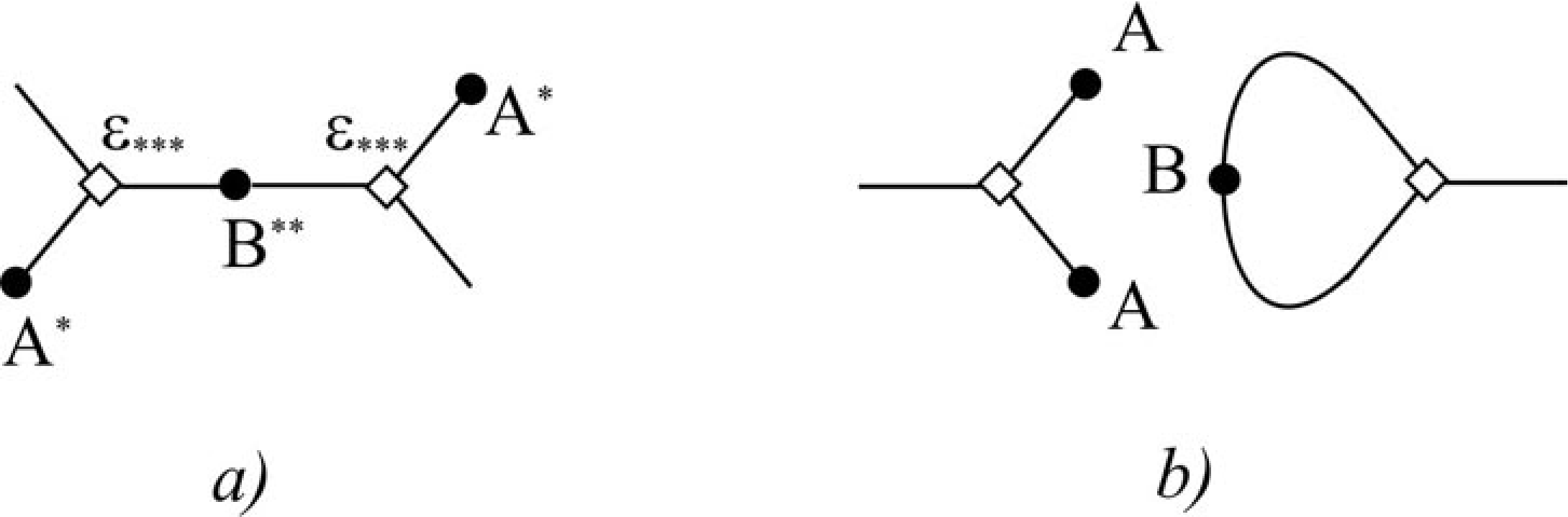} {400,131} {{\bf A}. Pictorial representation of
generalized Vieta formula (\ref{viaf12}) for $n=3$ and $s_1=1$,
$s_2=2$. Black circles at valence-$1$ vertices represent
contravariant tensor $A^j$ of rank $s_1=1$, black circle at the
valence-$2$ vertex represents symmetric contravariant tensor
$B^{jk}$ of rank $s_2=2$. White diamonds at valence-$3$ vertices
represent $\epsilon$-tensors of rank $n=3$. \ {\bf B}. Another
thinkable diagram vanishes for symmetry reasons: in fact both
components vanish separately, one because $\epsilon_{ijk}A^jA^k =
0$, another because $B^{mn}$ is symmetric and
$\epsilon_{lmn}B^{mn} = 0$.}

{\bf The case of arbitrary $n$, but with $s_1=\ldots =s_{n-2}=1$
and only $s_{n-1}=s$:}

\noindent

This is immediate generalization of the previous example, only the
$A$-vector now acquires additional label $\tilde I$, which runs
from $1$ to $n-2$ (tilde reminds that it is not quite the index
$I$, enumerating the equations in the system: there is one more
equation of degree $s$): \be \left\{ \begin{array}{cc} A^i_{\tilde
I} x_i = 0&
\tilde I = 1,\ldots,n-2, \\
B^{j_1\ldots j_s} x_{j_1}\ldots x_{j_s} = 0. &  \end{array}
\right. \ee This is the first of our examples where the wavy lines
and $\varepsilon$ tensors of the rank $n-2$ appear: see
Fig.\ref{Vietdia11s}. Still a single diagram contributes to the
generalized Vieta formula (\ref{genviaf}): \be {\rm symm} \left\{
\prod_{\mu = 1}^s X^{(\mu)}_{i_\mu}\right\} \sim \left\{\prod_{\mu
= 1}^s \Big( \epsilon_{i_\mu j_1\ldots j_{n-2}k_\mu}
A^{j_1}_{\tilde I_1}\ldots A^{j_{n-2}}_{\tilde I_{n-2}}
\varepsilon^{\tilde I_1\ldots \tilde I_{n-2}}\Big) \right\}
B^{k_1\ldots k_s} \label{viaf1...1s} \ee

\Fig{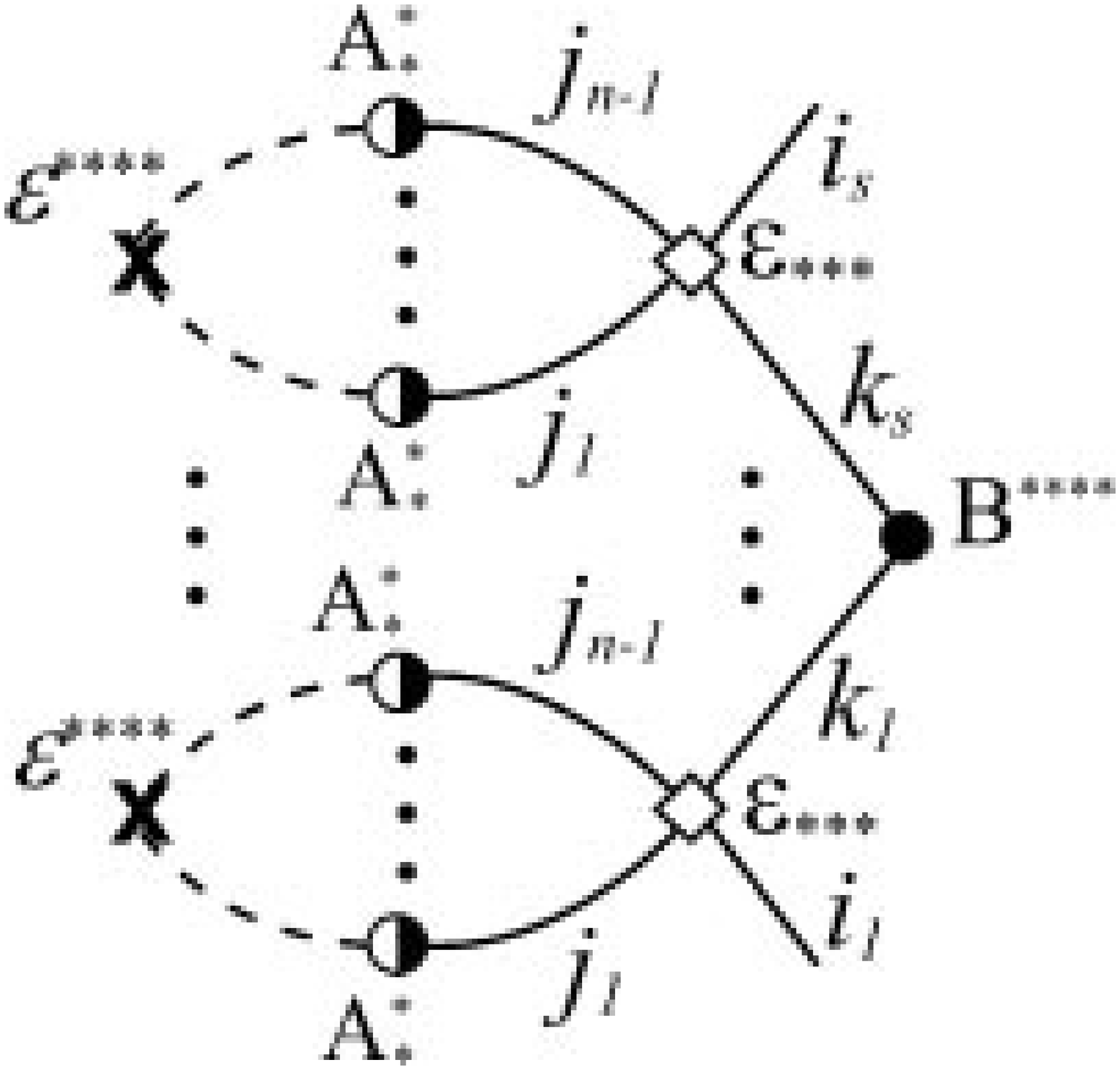} {202,193} {Pictorial representation of
generalized Vieta formula (\ref{viaf1...1s}) for $n=3$ and
$s_1=\ldots=s_{n-2}=1$, $s_{n-1}=s$. White circles  represent
tensor $A^i_{\tilde I}$ with one solid (carrying the $i$-index)
and one wavy (carrying the $\tilde I$-index) line. Black circle at
valence-$s$ vertex represents symmetric tensor $B^{k_1\ldots k_s}$
of rank $s_{n-1}=s$. Crosses at valence-$n$ vertices with all
lines solid represent $\epsilon$-tensors of rank $n$ with
covariant (lower) indices, while those at valence-$n-2$ vertices
with all lines wavy represent $\varepsilon$ tensors of rank $n-2$
with contravariant (upper) indices.}

{\bf The case of $n=3$ with $s_1=s_2=2$:}

\noindent

In this example we have a system of two quadratic equations, for
$3$ homogeneous variables $x_1,x_2,x_3=x,y,z$: \be
\left\{ \begin{array}{c} A^{ij} x_ix_j = A_1^{ij} x_ix_j =0, \\
B^{ij} x_ix_j = A_2^{ij} x_ix_j =0.  \end{array} \right.
\label{viet22prob} \ee It is instructive to look at it in two
ways: as defined by the two rank-$2$ tensors $A^{ij}$ and $B^{ij}$
and by a single $2\times 3\times 3$ tensor $A_I^{ij}$. This is our
first example where several diagrams contribute, but since the two
degrees coincide, $s_1=s_2$, the number of independent diagrams
with wavy lines and rank $n-1=2$ tensors $\varepsilon^{IJ}$ is
actually smaller than it may seem if one uses the $A-B$
formulation, see Fig.\,\,\ref{Vietdia22} and eq.\,\,(\ref{viaf22}) below.

The number of projectively non-equivalent solutions of the system
(\ref{viet22prob}) is ${\cal N} = s_1s_2=4$, and generalized Vieta
formula (\ref{genviaf}) states:
\be
&{\rm symm} \left\{ X^{(1)}_{i_1}X^{(2)}_{i_2}
X^{(3)}_{i_3}X^{(4)}_{i_4}\right\} \sim 
\epsilon_{i_1j_1k_1}\epsilon_{i_2j_2k_2}
\epsilon_{i_3j_3k_3}\epsilon_{i_4j_4k_4}\nn\\
&\Big\{A^{j_1j_2}A^{j_3j_4} \Big(c_A B^{k_1k_2}B^{k_3k_4} + c_B
B^{k_1k_3}B^{k_2k_4} + c_C B^{k_1k_4}B^{k_2k_3} \Big)\nn\\
&+ c_DA^{j_1j_2}A^{k_1k_2} B^{j_3j_4}B^{k_3k_4} + c_E
A^{j_1j_2}A^{k_1j_3} B^{k_2k_4}B^{k_3j_4} \Big\}\nn\\
&=\epsilon_{i_1j_1k_1}\epsilon_{i_2j_2k_2}
\epsilon_{i_3j_3k_3}\epsilon_{i_4j_4k_4} A_I^{j_1j_2}A_J^{j_3j_4}
A_K^{k_1k_2}A_L^{k_3k_4}
\Big(c_F \varepsilon^{IK}\varepsilon^{JL} + c_G
\varepsilon^{IJ}\varepsilon^{KL}\Big) \label{viaf22} \nn\\
\ee
Five coefficients $c_A-c_E$ are actually expressed through $c_F-c_K$,
associated with the diagrams, respecting the total $SL(2)\times
SL(3)$-symmetry.

For example, if $A_I$ are {\it matrix-like}, i.e. $A_I^{ij}x_ix_j
= \sum_{i=1}^3 a_I^i x_i^2$, then solutions of the system
(\ref{viet22prob}) are given by $X_i^2 = \epsilon_{ijk} a^j_J
a^k_K\varepsilon^{JK}$ and ${\cal N} = s_1s_2 = 4$ different
solutions correspond to the $4$ choices of relative signs when
square roots of these expressions are taken to obtain
$X_i^{\mu)}$. For, say, component with $i_1,i_2,i_3,i_4 = 1,1,1,1$
the eighte diagrams are equal to:
$$
\begin{array}{|c|cccc|}
\hline
{\bf A}&(a^2b^3+a^3b^2)^2&&&\\
{\bf B}&(a^2b^3+a^3b^2)^2&&&\\
{\bf C}&(a^2b^3+a^3b^2)^2&&&\\
{\bf D}&4a_2a_3b_2b_3&&&\\
{\bf E}&4a_2a_3b_2b_3&&&\\
\hline
{\bf F}&(a^2b^3-a^3b^2)^2&&&\\
{\bf G}&(a^2b^3-a^3b^2)^2&&&\\
{\bf H}&(a^2b^3-a^3b^2)^2&&&\\
{\bf K}&(a^2b^3-a^3b^2)^2&&&\\
\hline
\end{array}
$$
and only their appropriate linear combinations, respecting $SL(2)$
symmetry provide correct expressions.

\Fig{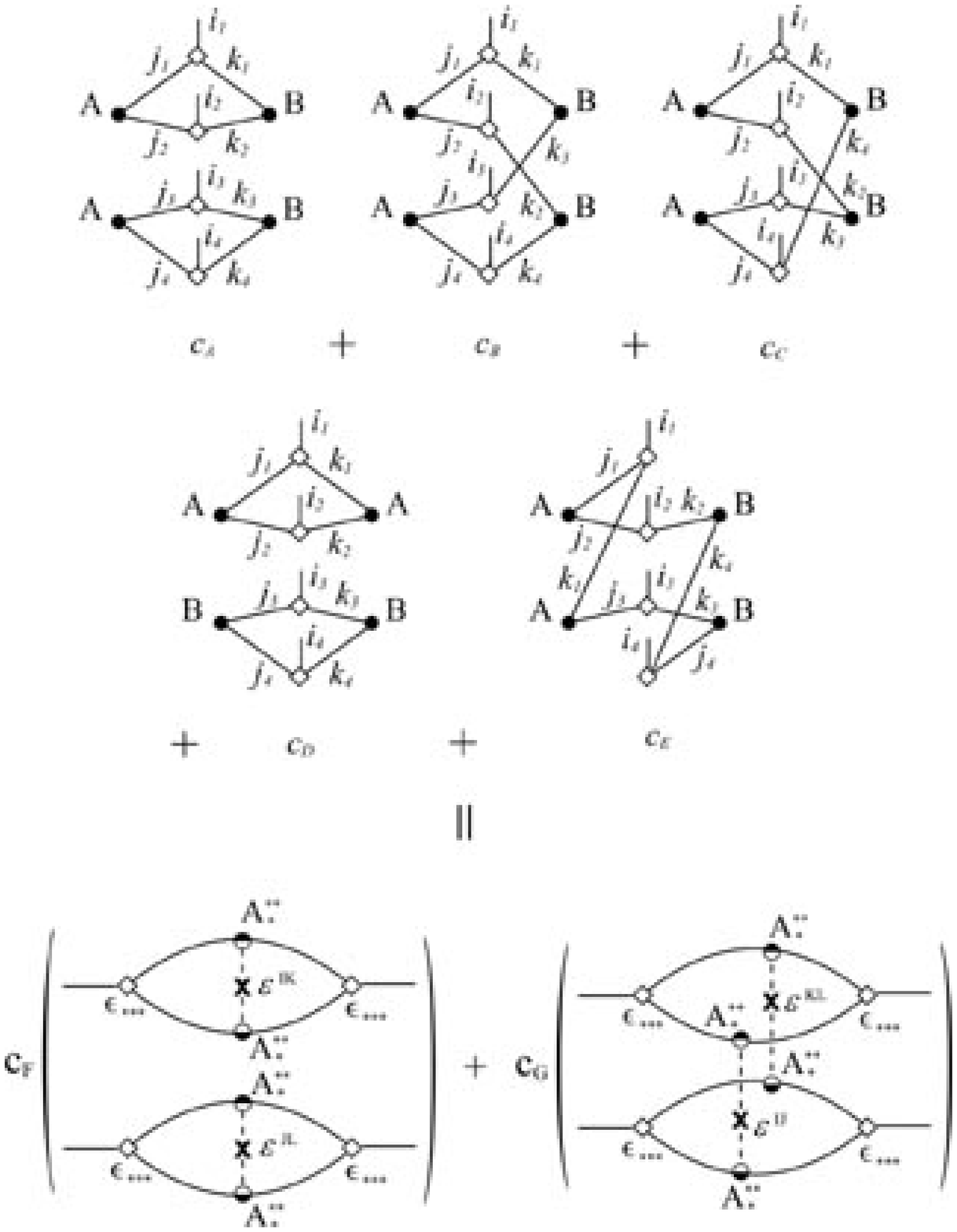}
{320,370}
{Pictorial representation of generalized
Vieta formula (\ref{viaf22}) for $n=3$ and $s_1=s_2=2$. White
circles at valence-$2$ vertices represent tensor $A^{jk}$ of rank
$s_1=2$, black circles at valence-$2$ vertices represent symmetric
tensor $B^{jk}$ of rank $s_2=2$. Crosses at valence-$3$ vertices
represent $\epsilon$-tensors of rank $n=3$. Equation
(\ref{viaf22}) is a sum with weights $\pm 1$ of five diagrams,
{\bf A-E}, of which two, {\bf B} and {\bf C}, are actually
equivalent. This bi-tensor representation takes into account only
the $SL(n)=SL(3)$ symmetry of the problem. If $SL(2)$ component of
the structure group $SL(2)\times SL(3)$ is taken into account,
then relevant are diagrams {\bf F-K} -- certain linear
combinations of {\bf A-E}. They involve wavy ($I$-index) lines and
rank-$2$ tensors $\varepsilon$. }

{\bf Cramer rule and inverse matrix: any $n$, but all $s_1=\ldots
s_{n-1}=1$}

\noindent

At least one more example deserves mentioning in the present text.
It is a non-homogeneous system of linear equations \be A_I^Jz_J =
a_J, \ \ \ I,J=1\ldots n-1, \ee which in homogeneous coordinates
turns into \be A_I^i x_i = 0, \ \ \ i = 1,\ldots,n \ee with
$A_I^{n} = a_I$ and $z_I = x_I/x_n$. Its single (since ${\cal N} =
s_1\ldots s_{n-1} = 1$) solution, as given by generalized Vieta
formula (\ref{genviaf}), is shown in Fig.\,\,\ref{liCradia}: \be X_i
\sim \epsilon_{ij_1\ldots j_{n-1}} A_{I_1}^{j_1}\ldots
A_{I_{n-1}}^{j_{n-1}} \varepsilon^{I_1\ldots I_{n-1}}
\label{linsol} \ee After conversion to non-homogeneous coordinates
in particular chart $z_I = x_I/x_n$, it turns into a usual Cramer
rule. Indeed,
\be
&X_{n} \sim \sum_{j_1,\ldots,j_{n-1}=1}^{n}\epsilon_{nj_1\ldots
j_{n-1}} A_{I_1}^{j_1}\ldots A_{I_{n-1}}^{j_{n-1}}
\varepsilon^{I_1\ldots I_{n-1}} \nn\\
&=\sum_{J_1,\ldots,J_{n-1}=1}^{n-1} \varepsilon_{J_1\ldots J_{n-1}}
A_{1}^{J_1}\ldots A_{n-1}^{J_{n-1}}\varepsilon^{I_1\ldots I_{n-1}}
= \det_{(n-1)\times (n-1)} A,
\nn\ee
while
$$
X_J \sim \sum_{j_1,\ldots,j_{n-1}=1}^{n} \epsilon_{Jj_1\ldots
j_{n-1}}A_{I_1}^{j_1}\ldots
A_{I_{n-1}}^{j_{n-1}}\varepsilon^{I_1\ldots I_{n-1}}$$
\be =
\sum_{K=1}^{n-1} A_K^n \sum_{J_1,\ldots,J_{n-2}=1}^{n-1}
\varepsilon_{JJ_1\ldots J_{n-2}} A_{I_1}^{J_1}\ldots
A_{I_{n-2}}^{J_{n-2}} \varepsilon^{I_1\ldots I_{n-2}K} \nn\\
=\sum_{K=1}^{n-1} A_n^K \det_{(n-2)\times(n-2)} \check
A^{(J,n)}_{(K)} = \det_{(n-1)\times (n-1)}\check
A^{(J)}\hspace{5mm}
\ee
where $(n-2)\times (n-2)$ matrix $\check A^{(k;J,n)}$ is obtained
from rectangular $A_I^j$ by deleting the $k$-th row and columns
with numbers $J$ and $n$. Summation over $k$ provides determinant
of the $(n-1)\times (n-1)$ matrix $\check A^{(J)}$, obtained by
omission of the $J$-th column only.

Comparing this with the standard formulation (\ref{liCraI}) of the
Cramer rule,
$$x_J = \left(A^{-1}\right)_J^KA_K^{n}$$
we obtain the usual expression for inverse matrix: \be
\left(A^{-1}\right)_J^K \det A  = \varepsilon_{JJ_1\ldots
J_{n-2}}\varepsilon^{KI_1\ldots I_{n-2}} A_{I_1}^{J_1}\ldots
A_{I_{n-2}}^{J_{n-2}} \label{linsolinv} \ee -- and it is extracted
from projective diagram (\ref{linsol}). Thus we see that {\bf the
Cramer rule and expression for inverse-matrix are particular
examples of the generalized Vieta formula}.

\bigskip

\Fig{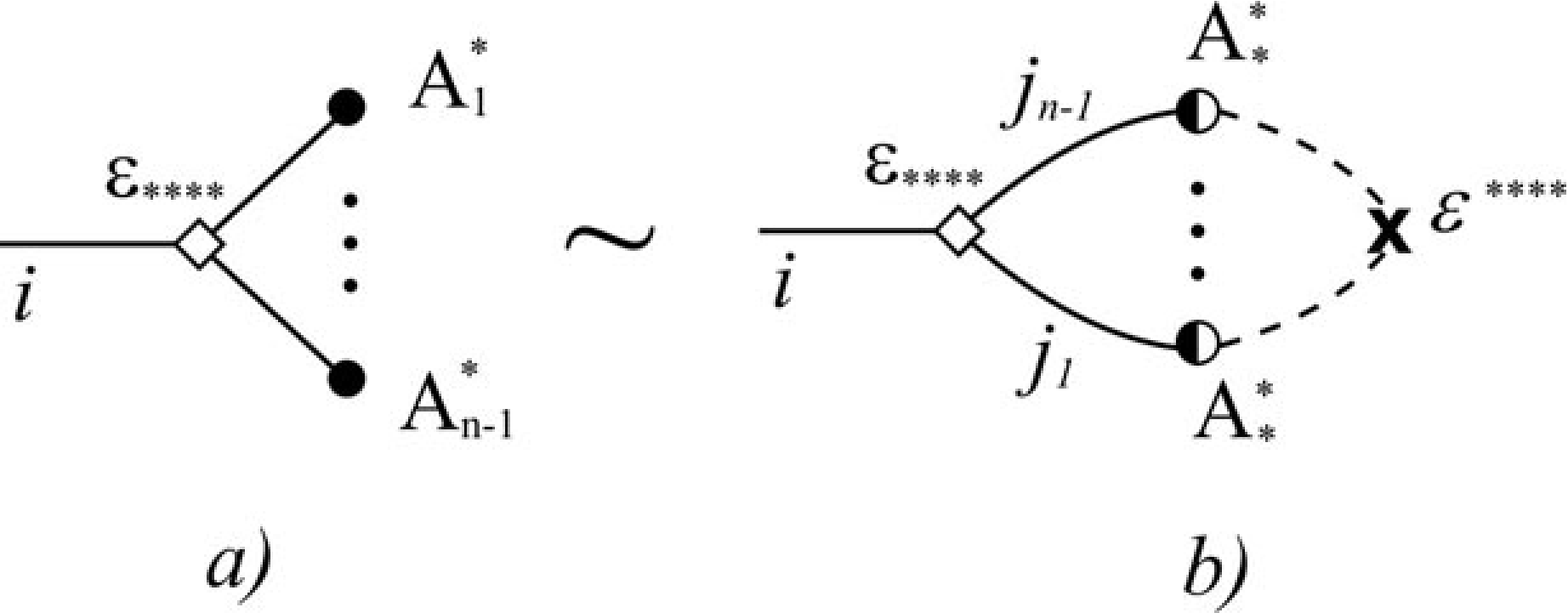}
{320,124}
{{\bf A.} Pictorial representation of
Cramer rule. Dots at valence-$1$ vertices represent $n-1$ tensors
$A_I^j$ of rank $1$. Cross at the valence-$n$ vertex represents
$\epsilon$-tensors $\epsilon_{i_1\ldots i_n}$ of rank $n$. {\bf
B.} In the present case one can also associate valence-$2$
vertices with the $A_I^j$, introduce wavy $I$-lines with
$I=1,\ldots,n+1$ ($i$-lines with $i=1,\ldots,n$ are shown solid,
as in the case {\bf A}) and a wavy cross for associated
$\epsilon$-tensor $\varepsilon^{I_1\ldots I_{n-1}}$ of rank
$n-1$.}

\subsubsection{Coinciding solutions of non-homogeneous equations:
generalized discriminantal varieties \label{hidiva}}

\noindent

When some two or more solutions $X^{(\mu)}$ projectively coincide
(i.e. the corresponding homogeneous vectors become collinear), the
l.h.s. of (\ref{genviaf}) becomes of special type. For example, in
the case of ${\cal N}=2$, if the two vectors $X^+_i$ and $X^-_i$
become collinear, determinant of the l.h.s. vanishes:
$\det_{n\times n}\Big( {\rm symm}\left\{ X^+_iX^-_j \right\}\Big)
= 0$. In this way we can introduce and describe various
discriminantal conditions for generic non-homogeneous system. The
problem is badly investigated and interesting already in the case
of a single polynomial equation, $n=2$. See eq.(\ref{Bij23}) and
the very end of s.\ref{24resulta} below for the simplest examples
and \cite{Sha} for more profound results in this
direction.\footnote{ We acknowledge fruitful discussions with
I.Gyakhramanov, Sh.Shakirov and A.Smirnov on the problem of higher
discriminantal varieties.}

If $n>2$ {\it collinearity} of two solutions is no longer a
condition of codimension one. Of codimension one is rather a {\it
complanarity} constraint $\Delta^{\mu_1\ldots\mu_n} = 0$ where \be
\Delta^{\mu_1\ldots\mu_n} = \epsilon_{i_1\ldots i_n}
X_{i_1}^{\mu_1}\ldots X_{i_n}^{\mu_n} \ee For $n=2$ the product
\be \prod_{\mu\neq \nu}^{{\cal N}} \Delta^{\mu\nu} = D_{2^{\times
s}} \ee is discriminant of the polynomial, and higher
discriminantal constraints can be also expressed through
$\Delta^{\mu\nu}$ \cite{Sha}. Similarly, for arbitrary $n$
determinants $\Delta^{\mu_1\ldots\mu_n}$ provide the building
blocks for discriminantal constraints.

\setcounter{equation}{0}

\section{
Discriminants of polylinear forms
\label{polydis} \label{polydet} }

After a brief survey of resultants we now switch to {\it
discriminant} -- a quantity, somewhat more natural from tensorial
point of view, because it is defined for arbtrary tensors (while
resultant is associated only with $A_i^{j_1\ldots j_s}$, which are
{\it symmetric} in the last $s$ indices). However, discriminant is
also defined from consistency condition of a system of homogeneous
equations, and {\it in this sense} is a quantity, derived from
resultant theory -- this is why it was natural to begin from
resultants. Of course, resultants in their turn are part of
discriminant theory, associated with reduction to tensors with
particular symmetry. This peculiar russian-doll embedding pattern
of different subjects is very typical for the structure of
non-linear algebra.

\subsection{Definitions}

\subsubsection{Tensors and polylinear forms}

With a  rank-$r$ covariant tensor of the type $n_1\times \ldots
\times n_r$ (rectangular polymatrix) $T^{i_1 \ldots i_r}$ we
associate an $r$-linear combination of $r$ vectors $\vec
x_1,\ldots,\vec x_r$, probably with different numbers of
components $n_1,\ldots,n_r$: \be T\{\vec x\} = T(\vec
x_1,\ldots,\vec x_r) = \sum_{\stackrel{1\leq i_k \leq n_k}{1\leq k
\leq r}} T^{i_1 \ldots i_r}x_{1,i_1}\ldots x_{r,i_r}
\label{tensorT} \ee
Notation $n_1\times\ldots\times n_r$ refers to a matrix-style
representation of tensor $T$ by an $n_1\times\ldots\times n_r$
hyper-parallelogram, with item $T^{i_1\ldots i_r}$ placed at the
point of an integer lattice $Z^{\otimes r}$ with coordinates
$i_1,\ldots,i_r$.

\subsubsection{Discriminantal tensors}

We call $T\{\vec x\}$ degenerate or discriminantal, if a system of
polylinear equations \be T^{k,i_k}\{\vec x\}=\frac{\partial
T\{\vec x\}}{\partial x_{k,i_k}} = \sum_{i_1\ldots \check i_k
\ldots i_r} T^{i_1 \ldots i \ldots i_r}x_{1,i_1}\ldots
x_{k-1,i_{k-1}}x_{k+1,i_{k+1}}\ldots x_{r,i_r} = 0
\label{degcondforT} \ee has a solution  with all vectors
non-vanishing (i.e. for each $k$ at least one component $x_{k,i_k}
\neq 0$). Discriminantal tensors form a subspace ${\cal
D}_{n_1\times\ldots\times n_r}$ in a {\it moduli space} ${\cal
M}_{n_1\times\ldots\times n_r}$ of all tensors of the given type.
It has codimension $1$ if, for example, all $n_k$ are equal, in
general it exceeds unity and we denote the codimension by
$N_{n_1\times\ldots\times n_r}$.

Discriminantal subspace is algebraic and can be described as the
space of zeroes of \be {\cal D}_{n_1\times\ldots\times
n_r}^{(\kappa)}(T) = 0, \ \ \ \kappa = 1,\ldots,
N_{n_1\times\ldots\times n_r} \label{manydisc} \ee (we use the
same letter to denote the space and the constraints, this should
not cause confusion). For {\it hypercubic} tensors, when all $n_k$
coincide, $n_1 = \ldots = n_r = n$, the codimension $N_{n^{\times
r}}=1$ and we call the {\it single} function \be {\cal D}_{n|r}(T)
= {\cal D}_{n^{\times r}}(T) \ee discriminant of the hypercubic
$T$. In rectangular case \be N_{n_1\times\ldots\times n_r}=1\ \
{\rm iff}\ \ n_k-1 \leq \sum_{j\neq k}^r (n_j-1) \ \ \forall k,
\label{onecompdisc} \ee and only under this condition there is a
single discriminant. When it exists, discriminant is a polynomial
in the coefficients of $T$ of degree $deg_{n_1\times n_r}  \equiv
{\rm deg}_T{\cal D}_{n_1\times n_r}(T)$ or $deg_{n|r} \equiv {\rm
deg}_T{\cal D}_{n|r}(T)$ (in generic rectangular case, when
(\ref{onecompdisc}) is not true, all ${\cal D}^{(\kappa)}$ in
(\ref{manydisc}) can have different degrees in $T$).

\subsubsection{Degree of discriminant \label{degdi}}

According to \cite{GKZ} degrees of discriminants are described by
the following generating function: \be {DEG}_r(t_1,\ldots,t_r)
\equiv \sum_{n_1,\ldots,n_r\geq 1} {\rm deg}_T\Big({\cal
D}_{n_1\times\ldots\times n_r}(T)\Big)\ t_1^{n_1-1}\ldots
t_n^{n_r-1} = \frac{1}{\Big(\prod_{i=1}^n (1 + t_i) - \sum_{i=1}^n
t_i \prod_{j\neq i}^n(1+t_j)\Big)^2} \label{degdigkz} \ee This
formula is informative only when discriminantal subspace is of
codimension one. When (\ref{onecompdisc}) is not true and there is
more then one discriminantal constraint, eq.(\ref{degdigkz}) puts
${\rm deg}_T({\cal D}) = 0$.

\bigskip

{\bf Important particular cases:}

\bigskip

$\bullet$ $r=1:$ Discriminantal constraints for $T(z) =
\sum_{i=1}^n T_iz_i$ require that $T_i = 0$ for all $i=1,\ldots,n$
at once. This set reduces to a single condition only for $n=1$,
and this condition, $T_1=0$ is of degree one in $T$. Accordingly,
\be {DEG}_1(t) = 1 \label{r1DEG} \ee

\bigskip

$\bullet$ $r=2:$ This is the case of $n_1\times n_2$ matrices,
discriminantal constraint is given by a single function when
$n_1=n_2=n$, i.e. matrix is square (in rectangular case
discriminantal constraint requires vanishing of all the principal
minors and thus is multi-component), accordingly \be
{DEG}_2(t_1,t_2) = (1-t_1t_2)^{-2} = \sum_{n\geq 1} n(t_1t_2)^n
\ee i.e. ${\rm deg}_T({\cal D}_{n_1\times n_2} =
n_1\delta_{n_1,n_2}$, Discriminant of the square $n\times n$
matrix is its determinant and it indeed is a polynomial of degree
$n$ of the matrix coefficients.

\bigskip

$\bullet$ $r=3:$ This is the first non-trivial family, \be
{DEG}_3(t_1,t_2,t_3) = (1-t_1t_2-t_2t_2-t_3t_1-2t_1t_2t_3)^{-2}
\ee is already a non-trivial function, and no convenient
representation exists for its generic term. Discriminantal
constraint is one-component whenever $n_i-1 \leq (n_j-1) +
(n_k-1)$ for all the $6$ triples $(ijk) = (123),$\ $(132),$\
$(213),$\ $(231),$\ $(312),$\ $(321)$ The first term of the
hypercubic sequence $n^{\times 3}$ are \cite{GKZ}: \be
\begin{array}{|c|ccccccc|}
\hline
&&&&&&&\\
n & 1 & 2 & 3 & 4 & \ldots & n & \ldots \\
&&&&&&&\\
\hline
&&&&&&&\\
{\rm deg}_T({\cal D}_{n\times n\times n}) & 1 & 4 & 36 & 272 & &
\sum_{0\leq j \leq \frac{n-1}{2}}
2^{n-2j-1}\frac{(n+j)!}{(j!)^3(n-2j-1)!}&\\
&&&&&&&\\
\hline
\end{array}
\label{dimsDncube} \ee

$\bullet$ Another interesting hypercubic sequence is $2^{\times
r}$. Appropriate generating function in this case is simple
\cite{GKZ}: \be \sum_{r=0}^\infty {\rm deg}_T\left({\cal
D}_{2^{\times r}}\right) \frac{\kappa^r}{r!} =
\frac{e^{-2\kappa}}{(1-\kappa)^2} \ee and \be
\begin{array}{|c|ccccccc|}
\hline
&&&&&&&\\
r & 1 & 2 & 3 & 4 & 5 & 6 & \ldots \\
&&&&&&&\\
\hline
&&&&&&&\\
{\rm deg}_T\left({\cal D}_{2^{\times r}}\right) &
0 & 2 & 4 & 24 & 128 & 880 & \\
&&&&&&&\\
\hline
\end{array}
\label{dimD2timesr} \ee (for $r=1$ discriminantal constraint is
two-component, see eq.(\ref{r1DEG} above, thus zero in the table)

\bigskip

$\bullet$ In the {\it boundary} case, when the condition
(\ref{onecompdisc}) for the single-componentness of discriminant
is saturated \cite{GKZ} \be {\rm deg}_T({\cal
D}_{n_1\times\ldots\times n_r}) =
\frac{n_1!}{(n_2-1)!\ldots(n_r-1)!} \ \ \ {\rm for} \ \ \ n_1-1 =
(n_2-1) + \ldots + (n_r-1) \ee

\subsubsection{Discriminant as an
$\prod_{k=1}^r SL(n_k)$ invariant}

Discriminantal subspace in ${\cal M}_{n_1\times\ldots\times n_r}$
does not change under {\it linear} transformations of the tensor,
i.e. is invariant of the {\it structure group} $\prod_{k=1}^r
SL(n_k)$. Therefore the set of discriminantal constraints
(\ref{manydisc}) form a linear representation of this group, and
if $N_{n_1\times\ldots\times n_r} = 1$ the single discriminant is
its invariant. $GL$ transformations multiply discriminant by
determinants. Non-linear transformations result in appearance of
resultant factors and raise discriminant to some powers.

When some $n_k$ coincide one can also produce invariants of
smaller groups, with identical $SL(n_k)$ factors excluded. They
are related to tensors with additional symmetry properties:
symmetric or antisymmetric in some indices or having more
compicated symmetries, not seen at the level of matrices (of
$r=2$). In this case, in addition to continuous $SL$-symmetries,
discriminantal conditions (\ref{degcondforT}) are also invariant
under discrete transmutations of indices: while tensor $T$ does
not need to have this symmetry,-- in general $T^{i_1\ldots i_r}
\neq T^{i_{P(1)}\ldots i_{P(r)}}$ for permutations $P \in
\sigma_r$ even if all $n_k$ are the same,-- such permutations take
one solution $\vec X_k$ of (\ref{degcondforT}) into another, $\vec
X_{P(k)}$.

When $T$ has special symmetry properties, discriminant functions
$D^{(\kappa)}(T)$ become reducible. Especially interesting are
associated multiplicative decompositions of the single
discriminant $D(T)$ when $N_{n_1\times\ldots n_r} = 1$ (for
example, in hypercubic case).

\subsubsection{Diagram technique for the $\prod_{k=1}^r SL(n_k)$
invariants}

All $SL(n)$ invariants are made from the coefficients of covariant
tensors $T$ by contraction with $SL(n)$-invariant
$\epsilon$-tensors $\epsilon_{i_1\ldots i_n}$. Since all $SL(n_k)$
subgroups act independently, there are separate $\epsilon$-tensors
for all $n_k$. This means that invariants are represented by
graphs (diagrams) with lines (edges, propagators) of $r$ different
{\it sorts}. Vertices are either $r$-valent, with $T^{i_1\ldots
i_r}$, where $r$ different {\it sorts} merge, or $n_k$-valent with
$\epsilon_{i_1\ldots i_{n_k}}$ and lines of only one {\it sort}.
In dual representation fixed are the numbers and sorts of faces of
every symplex: allowed are $r$-symplices with all $r$ faces
different and $T$ in the center and $n_k$-simplices with all $n_k$
faces of the same color $k$ ($k = 1,\ldots,r$) and the
corresponding $\epsilon$ in the center.

Of course, particular diagrams do not respect the discrete
$\sigma_r$ invariance. As we shall see in examples in
s.\ref{Exacov} below, {\it expressions} for some diagrams possess
this enhanced symmetry, even if the {\it diagram} itself does not.
This means that $\sigma_r$-non-symmetric diagrams {\it can}
contribute to expressions for $\sigma_r$-symmetric discriminant,
and this makes diagram representation less powerful in application
to discriminant theory.

This diagram technique can be expressed in terms of the Fock
space: the infinite tensor product $\otimes_{-\infty}^\infty
\Big(V_{n_1}\otimes\ldots \otimes V_{n_k}\Big)$. Here $V_n$ is a
linear space of dimension $n$. An operator acting on one $V$, can
be raised to entire Fock space by a trivial comultiplication: $ V
\longrightarrow  \ldots \ + \ \ldots\otimes 0 \otimes V \otimes 0
\otimes 0 \otimes \ldots \ +\ \ldots\otimes 0 \otimes 0 \otimes V
\otimes 0 \otimes\ldots \ + \ \ldots $ -- the sum is over all
possible positions of $V$. Similarly, an operator acting on a pair
$V_1\otimes V_2$ us raised to an infinite double sum and so on. In
obvious way an exponential raise to the Fock space $\exp_\otimes$
is defined. For a scalar $\lambda$ $\exp_\otimes (\lambda) =
e^\lambda \Big(\ldots \otimes I \otimes I \otimes \ldots\Big)$,
where $I$ is the unit operator. Invariants ${\rm Inv}_p(T)$ are
generated by \be \exp_\otimes \Big(\sum_p{\rm Inv}_p(T) \Big) =
\prod_{k=1}^r \exp_\otimes\left( \epsilon_{i_1\ldots i_{n_k}}
\frac{\partial}{\partial x_{k,i_1}}\otimes\ldots\otimes
\frac{\partial}{\partial x_{k,i_{n_k}}} \right)      \exp_\otimes
\left( T^{i_1\ldots i_r}x_{1,i_1}\ldots x_{r,i_r}\right) \ee Of
course, to invariant of a given power of $T$ only finite number of
terms contribute.

As usual, Fock space representation can be encoded in terms of a
field-theory integral: a topological theory of fields
$\phi_{k,i_k}(s)$ (actually, $x_{k,i_k}$) with the action of the
kind \be \phi^{k,i_k}\partial_s \phi_{k,i_k} + T^{i_1\ldots i_r}
\phi_{1,i_1}\partial_s \phi_{2,i_2}\ldots
\partial_s^{r-1}\phi_{r,i_r}
+ \sum_{k=1}^r \epsilon_{i_1\ldots i_{n_k}} \phi^{k,i_1}\partial_s
\phi^{k,i_2}\ldots
\partial_s^{n_k-1}\phi^{k,i_{n_k}}
\ee The only role of $s$-variable is to avoid vanishing of
antisymmetric combinations in the action, at the end of the day,
all $\partial_s$ in propagators and vertices are substituted by
unities.

If some $n_k$ coincide, and we are interested of invariants of the
smaller group, with some of coincident $SL(n_k)$ factors excluded,
one adds new $\epsilon$-vertices, which mix colors, see examples
in s.\ref{Exacov} below.

\subsubsection{Symmetric, diagonal and other specific  tensors}

If tensor $T$ is {\it hypercubic} (all $n_k = n$), one can
restrict it on diagonal of the space $V_n^{\otimes r}$. Then we
obtain {\it symmetric hypercubic} tensor, such that $S^{i_1\ldots
i_r} = S^{i_{P(1)}\ldots i_{P(r)}}$ for all $r!$ permutations $P
\in \sigma_r$. When restricted to symmetric tensors, the $\sigma_r
\times SL(n)^{\times r}$-symmetric discriminant ${\cal
D}_{n|r}(S)$, which is irreducible on entire moduli space ${\cal
M}_{n|r}$ of hypercubic tensors, decomposes into irreducible
factors. One of them, describing consistency condition for the
system (\ref{degcondforT}) in the case of totally symmetric $T =
S$, is symmetric discriminant $D_{n|r}(S) = {\bf irf}\Big({\cal
D}_{n|r}(T)\Big)$. This discriminant is invariant of a single
$SL(n)$ (though in some particular cases its symmetry can be
occasionally enhanced, see examples in s.\ref{Exacov}). Its degree
$d_{n|r} \equiv {\rm deg}_S D_{n|r}(S)$ is typically smaller than
$deg_{n|r} \equiv {\rm deg}_T {\cal D}_{n|r}(T)$ and can be extracted from
the generating function
$$
\sum_nd_{n|r}t^n=\frac{1}{(1-(r-1)t)^2}
.$$
i.e. $d_{n|r}=n(r-1)^{n-1}$. For a tensor symmetric in $k$ groups of variables of dimensions
$n_1,\dots,n_k$ and degrees $r_1,\dots,r_k$ correspondingly (for example
$\sum\limits_{i,j=1}^n\sum\limits_{\alpha,\beta,\gamma=1}^mS_{ij\,\alpha\beta\gamma}x_ix_j
y_\alpha y_\beta y_\gamma$ is of the type $n,m|2,3$\,) the generating function is \cite{GKZ}
\be
F_{r_1\dots r_k}(t_1,\dots,t_k)=\sum_{n_1\dots n_k}d_{n_1\dots n_k|r_1\dots r_k}
t_1^{n_1}\dots t_k^{n_k}\nn\\
=\frac{1}{(\prod_j(1+t_j)-\sum_jr_jt_j\prod_{i\ne j}(1+t_j))^2}
.\ee

We call hypercubic  symmetric tensors of the form
$$T^{i_1\ldots i_r} =
\sum_{i=1}^n a_i\delta^{i_1}_i\ldots \delta^{i_r}_i$$ {\it
diagonal}. They are direct generalizations of diagonal matrices.
Discriminant of diagonal tensor is a power of $(a_1\ldots a_n)$,
the power depends on what is considered: the  $r$-linear
discriminant ${\cal D}_{n|r} = (a_1\ldots a_n)^{deg_{n|r}/n}$ or
the symmetric $r$-form one $D_{n|r} = (a_1\ldots
a_n)^{d_{n|r}/n}$.

For $r=2$, when any square matrix can be diagonalized, determinant
is always reduced to the one of diagonal matrices. However,
diagonalization procedure is in fact a property of {\it maps}, not
forms, and thus is generalized to the case of resultants, not
discriminants. Moreover, as we shall see in s.\ref{eige}, even for
resultants {\it canonical} diagonalization is somewhat subtle when
tensors have rank $r>2$.

\bigskip

One can of course consider tensors with other symmetries, not
obligatory totally symmetric. They can be symmetric in {\it some}
indices only (and then only some $n_k$ should coincide). They can
be {\it anti}symmetric in all (in hypercubic case with $r\leq n$)
or some indices. They can have symmetries of particular Young
diagrams. They can be representations of braid group (i.e. have
anionic-type symmetries). They can have parastatistic-type
symmetries. In all cases generic ${\cal D}_{n_1\times\ldots\times
n_r}(T)$, when restricted to the corresponding subspace in ${\cal
M}_{n_1\times\ldots\times n_r}$ decomposes into irreducible
factors with lower symmetries, and among these factors the one
with appropriate transformation properties becomes the
corresponding specific discriminant. This is a far-going
generalization of emergency of Pfaffians in the case of
antisymmetric matrices.

\PFig{Johann-Friedrich-Pfaff}
{150,176}
{Johann Friedrich Pfaff (1765 -- 1825)}

\subsubsection{Invariants from group averages}

Like $SO(2)$ invariant can be obtained by averaging of any
monomial over the {\it compact} group,
$$ a^2 \rightarrow \oint
(a\cos\phi + b\sin\phi)^2d\phi \sim a^2 + b^2, $$ we have for an
$SU(2) \subset SL(2|C)$-invariant
$$ ad \rightarrow \oint
(a\cos\phi + c\sin\phi)(-b\sin\phi + d\cos\phi)d\phi \sim ad-bc
$$
In variance with, say, $ab$,  $ad$ itself is invariant under
$$\left(\begin{array}{cc} a & b \\ c & d \end{array}\right)
\rightarrow \left(\begin{array}{cc} e^{i\varphi} & 0 \\ 0 &
e^{-i\varphi} \end{array}\right) \left(\begin{array}{cc} a & b \\
c & d \end{array}\right)
$$
This averaging over maximal compact subgroup is one of the obvious
ways to produce and study invariants of the structure group.

\subsubsection{Relation to resultants}

\noindent

$\bullet$ By definition, the vanishing of discriminant implies
that a resultant of a huge system (\ref{degcondforT}) of $M \equiv
\sum_{k=1}^r n_k$ equations with $M$ variables vanishes. However,
since the $A$-tensor associated with this system is of a special
(polylinear) form, this resultant can decompose into irreducible
factors, and discriminant constraints are only some of them. In
particular, if (\ref{onecompdisc}) is true, \be {\cal
D}_{n_1\times\ldots\times n_r}(T) = {\bf irf} \left({R}_{M|r-1}
\left\{\frac{\partial T}{\partial x_{k,i_k}}\right\} \right)
\label{disres0} \ee Often the resultant at the r.h.s. is just a
power of discriminant at the l.h.s.

\bigskip

$\bullet$ Relation (\ref{disres0}) simplifies significantly for
particular case of {\it symmetric hypercubic} tensors. Then the
number of $x$-variables is $n$ rather than $M=nr$, and we get
instead of (\ref{disres0}): \be
D_{n|r}(S) =  
{R}_{n|r-1} \left\{\frac{\partial S}{\partial x_{i}}\right\}
\label{disres1} \ee The argument of the resultant at the r.h.s. is
not {\it quite} generic: it is a {\it gradient} tensor. However,
resultant of generic gradient tensor appears irreducible, and
(\ref{disres1}) is actually exact relation, without need to
extract irreducible factors, like in the poly-linear case of
(\ref{disres0}).

\bigskip

$\bullet$
Another interesting combination from the first derivatives
of $T$ is \cite{Sham}
\be
\det_{1\leq i,k \leq n} \frac{\partial T}{\partial x_{k,i}}
= A_{I_1\ldots I_n} (\vec x_1)^I_{r-1}\ldots (\vec x_n)^I_{r-1}
\label{TnA}
\ee
At the r.h.s. $\Big\{(\vec x)_s^I\Big\}$ is a collection of
monomials of degree $s$ made out of $n$ components
$x_1,\ldots,x_n$ of the vector $\vec x$.
There are $N_{n|s} = \#(I) = \frac{(n+s-1)!}{(n-1)!s!}$
such monomials, and $A$ is an antisymmetric tensor of
the type $N_{n|r-1}|n$, i.e. it has rank $n$ and depends on
$N_{n|r-1}$ variables.
The coefficients of $A$ are polynomials of degree $n$
in the coefficients of original $T$.

If discriminant ${\cal D}_{n^{\times r}}(T) = 0$,
this means that
$n$ equations $\frac{\partial T}{\partial x_{1,i}}=0$
possess a common solution
$\{\vec x_2^{(0)},\ldots,\vec x_n^{(0)}\}$.
On this solution determinant at the l.h.s. of (\ref{TnA})
vanishes and this implies that the same happens with
the r.h.s. for any $\vec x_1$, i.e.
$\{\vec x_2^{(0)},\ldots,\vec x_n^{(0)}\}$ is also a common
solution to ${\cal N}_{r-1|n}$ equations
$A_{I_1\ldots I_n} (\vec x_2)^I_{r-1}\ldots (\vec x_n)^I_{r-1}
= 0$. Repeating the same argument for all $\vec x_k$ at the
place of $\vec x_1$ we conclude that whenever
${\cal D}_{n^{\times r}}(T)$ the same happens to discriminant
$D^{asy}_{N_{r-1|n}|n}(A)$, i.e.
\be
{\cal D}_{n^{\times r}}(T) =
{\bf irf}\Big\{D^{asy}_{N_{r-1|n}|n}(A)\Big\}
 = {\bf irf}\Big\{ {\cal D}_{N_{r-1|n}^{\times n}}(A)\Big\}
\label{didirela}
\ee
Inverse is not true: a zero-mode of $A$ need not imply the
existence of a zero-mode of $T$.
This is because  a generic vector $\{z^I\}$ does not have
a form of $\Big\{(\vec x)_s^I\Big\}$: its $N_{n|s}$ components
are not obligatory $s$-linear combinations of just $n$
independent variables.
In (\ref{didirela})
$D^{asy}_{N|s}(A) = {\bf irf}\Big\{ {\cal D}_{N^s}(A)\Big\}$
is discriminant of a
totally antisymmetric rank-$s$ tensor of $N$ variables.
In the simplest case of $s=2$ this is just a Pfaffian:
$D^{asy}_{N|2}(A) = {\rm Pfaff}(A) = \det^{1/2} A$.

Relation (\ref{didirela}) allows to
move parameter $n$ from the first to the second position
(from the number of variables to rank), and this allows to
decrease the rank in the problem
(the degree of non-linearity) at expense of increasing
the number of variables.

In particular case of the rank-$2$ tensor $T$ the l.h.s.
of (\ref{TnA}) is divisible by $\det_{1\leq i,k \leq n} x_{k,i}$,
the ratio is nothing but determinant of $T$,
and (\ref{didirela}) can be strengthened to
${\cal D}_{n^2}(T) = \det_{n\times n}(T)$, see s.5.2.
A similar phenomenon occurs in the case of homogeneous
polynomials $n=2$, see \cite{Sham} and eq.\,\,(\ref{sydirel})
below.

\bigskip

$\bullet$ In what follows we shall also use peculiar $x$-dependent
{\it matrices}, associated with the tensor $T\{\vec x\}$: \be \hat
T[x]:\ \ \ T[x]^{ij} = \frac{\partial T\{\vec x\}}{\partial
x_{1,i}\partial x_{2,j}} = \sum_{i_3\ldots i_r} T^{ij i_3 \ldots
i_r}x_{3,i_3}\ldots x_{r,i_r} \label{hatTmatrix} \ee Instead of
$1$ and $2$ one can choose any other pair of ``sorts".

If all $n_1=\ldots=n_r=n$, then one can show, see s.\ref{diandre}
and eq.\,\,(\ref{polydi}) below, that \be {\cal D}_{n^{\times r}}(T) =
{\bf irf}\left( {R}_{N|N-1}\left\{\partial_I \det_{n\times n} \hat
T\right\} \right) \label{disres2} \ee where $N = n(r-2)$ is the
number of $x$ variables of the matrix $\hat T(x)$, and $I =
1,\ldots,N$ is used to enumerate these variables.

\subsection{Discrminants and resultants: Degeneracy condition
\label{diandre}}

\subsubsection{Direct solution to discriminantal constraints}

In this subsection we express degeneracy condition for hypercubic
(all $n_k = n$) $T$ in terms of the square matrix $T_{ij}[z]$, see
(\ref{hatTmatrix}), where $z$ denote the set of $r-2$ vectors
$\vec x_3,\ldots, \vec x_r$, while $\vec x_1 = \vec x$ and $\vec
x_2 = \vec y$. Then the $n\times r$ degeneracy equations are: \be
x_i T^{ij}[z] = 0, \nn \\
T^{ij}[z] y_j = 0, \nn \\
x_i dT^{ij}[z]y_j = 0, \label{degcon} \ee where $dT$ in the last
equation includes derivatives w.r.t. all of the $N = n\cdot(r-2)$
components of $z$ (so that the third line in (\ref{degcon}) is
actually a set of $N$ different equations).

From the first two of the equations (\ref{degcon}) we have \be
x_i = u^l \check T_{li}[z], \nn\\
y_j = \check T_{jm}[z] v^m, \label{xyvstildeA} \ee provided \be
\det \ T[z] = 0, \ee where $\check T_{ij}$ is the complement minor
of the matrix element $T_{ij}$: \be T^{ij}\check T_{jk} =
\delta^i_k \det \hat T \ee and inverse matrix $T^{-1}_{ij} = (\det
T)^{-1} \check T_{ij}$. $u^l$ and $v^m$ are two arbitrary
covectors, the one-dimensional zero-eigenvalue spaces containing
$x_i$ and $y_j$ do not depend on the choice of $u$ and $v$, what
follows from the elementary identity \be \check T_{ik}\check
T_{jl} - \check T_{il}\check T_{jk}  = \check T_{ij,kl} \det \hat
T \label{Aijkl} \ee Eqs.(\ref{xyvstildeA}) assume that $rank(\hat
T) = n-1$, if it is lower, than all $\check T_{ij}=0$ and $x_i$,
$y_j$ are expressed through smaller minors (and more solutions
appear, because  the zero-eigenvalue space can have more
dimensions).

After substitution of (\ref{xyvstildeA}) the third of the
equations (\ref{degcon}) becomes: \be x_idT^{ij}[z] y_j =
u^l\check T_{li}dT^{ij} \check T_{jm} v^m = (\det \hat T)^2 u^l
d\left(T^{-1}\right)_{lm} v^m = 0 \label{idefordeg} \ee
Homogeneity of $\hat T$ implies that $z_k\frac{\partial \hat
T^{-1}}{\partial z_k} = -(r-2)\hat T^{-1}$, therefore such linear
combination of equations in (\ref{idefordeg}) gives $-(det \hat
T)^2 \hat T^{-1} = (2-r)\det \hat T \cdot \check T = 0$, i.e.
(\ref{idefordeg}) {\it implies} that $\det \hat T = 0$.

\subsubsection{Degeneracy condition in terms of $\det \hat T$}

Conditions (\ref{idefordeg}), with re-labeled indices, \be \check
T_{ik} dT^{kl} \check T_{lj} = 0 \label{idefordeg'} \ee can be
rewritten in another form: \be (\det \hat T)\ d\check T_{ij} =
\check T_{ij}\ d(\det \hat T) \label{tildeAprime} \ee Indeed,
since $d(\det \hat T) = \check T_{lk} dT^{kl}$ and, similarly,
$d\check T_{ij} = \check T_{il,jk}dT^{kl}$, eq.(\ref{tildeAprime})
states that \be \Big( \check T_{lk} \check T_{ij} - \check
T_{il,jk}\det \hat T\Big)dT^{kl}\ \stackrel{(\ref{Aijkl})}{=}\
\check T_{lj}\check T_{ik}dT^{kl} = 0, \ee i.e. reproduces
(\ref{idefordeg'}).

Now, since we already know that $\det  T[z]=0$ is a corollary of
homogeneity and (\ref{idefordeg'}) or,equivalently,
(\ref{tildeAprime}), we conclude that degeneracy condition is
nothing but the set of $N = n(r-2)$ identities \be dP \equiv
d(\det  T) = 0 \label{detcon} \ee In other words, starting from
$T\{x\}$ we proceed to the matrix $T^{ij}[z]$, take its
determinant, \be P[z] \equiv \det_{ij} T^{ij}[z] = \sum_{I_k=1}^N
\alpha^{I_1,\dots I_N} z_{I_1}\ldots z_{I_N} \label{defP[z]} \ee
and impose $N$ constraints (\ref{detcon}). These are $N$ equations
for $N$ variables $z_I = \{x_{3,1},\ldots,x_{3,n};\ldots;
x_{n,1},\ldots,x_{n,n}\}$, but homogeneity of $P[z]$ implies that

the number of independent equations exceeds that of independent
variables by one, so that (\ref{detcon}) actually imposes {\it
one} constraint on the coefficients $\alpha$, which, in turn, are
made from the coefficients of $T$.

\subsubsection{Constraint on $P[z]$}

This constraint can be expressed by in iterative procedure through
the resultants of polynomials $dP_I = \partial P[z]/\partial z_I$.

The degeneracy condition for $T$ with $r\geq 3$ is \be {\cal
D}_{n^{\times r}}(T) = {\bf irf} \Big( {\cal
R}_{N|N-1}\{dP_1,\ldots,dP_N\}\Big) = 0 \label{polydi} \ee Because
of the homogeneity of $P[z]$ the l.h.s. actually contains a factor
of $z_N$ in a high power, which can be easily eliminated.

Of course, above arguments guarantee only that discriminant at the
l.h.s. is a divisor of the resultant at the r.h.s.: for
discriminant to vanish, the constraints (\ref{detcon}) should be
fulfilled, but inverse is not obligatory true. In fact degree of
the r.h.s. in the coefficients of $dP$ is equal to $N(N-1)^{N-1}$
and $dP$ itself has degree $n$ in $T$, so that the total
$T$-degree of the r.h.s. is $nN(N-1)^{N-1}$. Already for $r=3$,
when $N=n(r-2)=n$ this sequence $n^2(n-1)^{n-1} =
\{1,4,36,432,\ldots\}$ deviates from (\ref{dimsDncube}), starting
from $n=4$ (of course, when deviation occurs, resultant in
(\ref{polydi}) has higher degree than discriminant).

\subsubsection{Example}

For example, for $n=2$ and $\hat T[z] = \left(\begin{array}{cc}
a[z] & b[z] \\ c[z] & d[z]
\end{array}\right)$ eq.(\ref{idefordeg}) states:
\be (ad-bc)^2 \ d\left[ \frac{\left(\begin{array}{cc} d & -b \\ -c
& a
\end{array}\right)}{ad-bc}\right] = 0,
\ee or \be
a'(ad-bc) = a(ab-cd)', \nn\\
b'(ad-bc) = b(ab-cd)', \nn\\
c'(ad-bc) = c(ab-cd)', \nn\\
d'(ad-bc) = d(ab-cd)' \label{degconn=2} \ee where prime denotes
variation, $a' = da$. If $r=3$, then $a = a_1z_1 + a_2z_2$ etc,
\be ad-bc = (a_1d_1-b_1c_1)z_1^2 + (a_1d_2+a_2d_1 -
b_1c_2-b_2c_1)z_1z_2 + (a_2d_2-b_2c_2)z_2^2 \equiv \alpha z_1^2 +
2\beta z_1z_2 + \gamma z_2^2, \ee while prime can denote
derivative w.r.t. $z_1$ or $z_2$, so that (\ref{degconn=2})
becomes: \be (a_1z_1+a_2z_2)(2\alpha z_1 + 2\beta z_2) -
a_1(\alpha z_1^2 + 2\beta z_1z_2 +\gamma z_2^2) = a_1\alpha z_1^2
+ 2a_2\alpha z_1z_2 +
(2a_2\beta - a_1\gamma)z_2^2 = 0,\nn \\
(a_1z_1+a_2z_2)(2\beta z_1 + 2\gamma z_2) - a_2(\alpha z_1^2 +
2\beta z_1z_2 +\gamma z_2^2) = (2a_1\beta - a_2\alpha) z_1^2 +
2a_1\gamma z_1z_2 + a_2\gamma z_2^2 = 0, \label{propro} \ee plus
analogous equations, obtained by substitutions $a \rightarrow
b,c,d$. These look like $8$ homogeneous equations for two
variables homogeneous variables $z_1$ and $z_2$. However,
according to (\ref{detcon}) they are all equivalent to just two
equations $d\det T = 0$, where
\be \det T = \alpha z_1^2 + 2\beta z_1z_2 +\gamma z_2^2 \ee
Indeed, $d\det T = 0$ implies that \be
\left\{\begin{array}{c} \alpha z_1 + \beta z_2 = 0 \\
\beta z_1 + \gamma z_2 = 0 \end{array}\right. \label{detpro} \ee
while the r.h.s. first equation in (\ref{propro}) is $a_1(\alpha
z_1^2-\gamma z_2^2) + 2a_2z_2(\alpha z_1+\beta z_2)$. The second
item vanishes by (\ref{detpro}), in the first item $\alpha
z_1^2-\gamma z_2^2 = (\alpha z_1 + \beta z_2)z_1 - (\beta z_1 +
\gamma z_2)z_2$ also vanishes due to (\ref{detpro}). Note also
that (\ref{detpro}) implies that $\det T = (\alpha z_1 + \beta
z_2)z_1 + (\beta z_1 + \gamma z_2)z_2 = 0$, as expected. Thus
descriminantal condition reduces to the solvability of the system
(\ref{detpro}), i.e. \be {\cal R}_{2\times 2\times 2} =
\alpha\gamma - \beta^2  \sim (a_1d_2+a_2d_1 - b_1c_2-b_2c_1)^2 -
4(a_1d_1-b_1c_1)(a_2d_2-b_2c_2) \ee We shall see in s.\ref{cayh}
below that this is indeed the correct expression for Cayley
hyperdeterminant.

\subsubsection{Degeneracy of the product}

If $A_{ij}[z,w] = B_{ik}[z]C_{kl}[w]$, then $A$ is always
degenerate, because $z$ and $w$ can be chosen so that {\it both}
$\det B[z]=0$ and $\det C[w] = 0$, and if prime in
(\ref{idefordeg}) denotes $z$ or $w$ derivatives, then the l.h.s.
of (\ref{idefordeg}) vanishes automatically: \be \check A\ dA\
\check A = \check C \check B \big(dB\ C + B dC\big) \check C
\check B = \check C (\check B \ dB\ \check B) \det C + (\check C\
dC\ \check C) \check B \det B = 0 \ee though the combinations in
brackets can remain non-vanishing, i.e. $B$ and $C$ can be
non-degenerate.

\subsubsection{An example of consistency between
(\ref{disres1}) and (\ref{disres2})}

For $n=3$, $r=3$ we have $N=n(r-2)=3$, and \be {D}_{n|r}(S)\
\stackrel{(\ref{disres2})}{=}\ {\bf irf}\left(
{R}_{N|N-1}\Big(\partial_I (\det \partial^2 S)\Big)\right) \
\stackrel{(\ref{disres1})}{=}\
 {\bf irf}\Big({R}_{n|r-1}(\partial_i S)\Big)
\label{diviconexa0} \ee turns into relation between two resultants
${\cal R}_{3|2}$: \be {D}_{3|3}(S) = {\bf irf}\left(
{R}_{3|2}\Big(\vec\partial (\det \partial^2 S)\Big)\right) = {\bf
irf}\Big(R_{3|2}(\vec\partial S)\Big) \label{diviconexa1} \ee

\bigskip

Take for a cubic form -- a cubic function of a single vector with
{\it three} components $x,y,z$ -- \be S(x,y,z) = \frac{1}{3}ax^3 +
\frac{1}{3}by^3 + \frac{1}{3}cz^3 + 2\epsilon xyz \label{cufexa2}
\ee As we know from ss.\ref{abc+e3} and \ref{abc+e3fromK} the
resultant of the system $\vec\partial S = 0$, is equal to
degree-twelve polynomial of the coefficients of $S$, \be
R_{3|2}(\vec\partial S) = abc(abc+ 8\epsilon^3)^3 \label{R32exa2}
\ee

Consider now a 3-linear form, made from the same tensor $S_{ijk}$:
it is a function $S(\vec u,\vec v,\vec w) = \sum_{i,j,k=1}^3
S_{ijk}u_iv_jw_k$ of {\it nine} homogeneous variables
$u_i,v_i,w_i$. Determinant of the matrix of its second derivatives
w.r.t. $u_i$ and $v_j$ is a cubic form -- a cubic function of a
single vector with {\it three} components $w_1,w_2,w_3$: \be \det
\partial^2 S (\vec w) = \frac{2}{6^3} \left|\left|
\begin{array}{ccc}
aw_1 & \epsilon w_3 & \epsilon w_2 \\
\epsilon w_3 & bw_2 & \epsilon w_1 \\
\epsilon w_2 & \epsilon w_1 & cw_3
\end{array}\right|\right| = \nn \\ =
\frac{2}{6^3} \Big((abc + 2\epsilon^3) w_1w_2w_3 - \epsilon^2
(aw_1^3 + bw_2^3 + cw_3^3)\Big) \ee i.e. has exactly the same
shape as original anzatz (\ref{cufexa2}), only with different
value of parameter \be \epsilon \rightarrow \varepsilon =
-\frac{abc + 2\epsilon^3}{6\epsilon^2} = \epsilon -
\frac{1}{6\epsilon^2}(abc + 8\epsilon^3) \label{epvarep} \ee Thus
its resultant -- which is the discriminant of the three-linear
form $Q(\vec u,\vec v,\vec w)$ is given by the same formula
(\ref{R32exa2}), only with the substitution (\ref{epvarep}). Thus
we have: \be {\cal D}_{3|3} (S) = {\cal R}_{3|2}\Big(\vec\partial
(\det \partial^2 S)\Big) \sim abc (abc+8\varepsilon^3)^3 \sim abc
\left(abc - \left(\frac{abc +
2\epsilon^3}{3\epsilon^2}\right)^3\right)^3 \ee and, in accordance
with (\ref{diviconexa1}), it is indeed divisible by
$R_{3|2}(\vec\partial S) = abc(abc+ 8\epsilon^3)^3$, because at
$(abc + 8\epsilon^3) = 0$ $\varepsilon = \epsilon$ and
$2\varepsilon = -\frac{abc + 2\epsilon^3}{3\epsilon^2} =
2\epsilon$, see (\ref{epvarep}).

An simpler example of the same relation (\ref{diviconexa0}), for
$n|r = 2|3$ and $N=2$, is given below in s.\ref{cayh}.

\subsection{Discriminants and complexes
\label{dico}}

\subsubsection{Koshul complexes, associated with
poly-linear and symmetric functions}

\noindent

$\bullet$ Making use of the footnote \ref{thetadx} in s.\ref{KoI}
we can associate with the tensor $T\{\vec x_1,\ldots,\vec x_r\}$ a
Koshul complex with nilpotent differential
$$
\hat d = dT = \sum_{k=1}^r \sum_{i_k=1}^{n_k} \frac{\partial
T}{\partial x_{k,i_k}}\ dx_{k,i_k}
$$
acting by wedge multiplication in the space of differential forms.
If all forms are taken into account, this complex is {\it exact},
unless its determinant -- which is nothing but the resultant
${R}_{M|r-1}\left\{\partial T/\partial x_{k,i_k} \right\}$ with $M
= \sum_{k=1}^r n_k$ -- vanishes. If we want to relate determinant
of this complex with {\it discriminant} ${\cal
D}_{n_1\times\ldots\times n_r}(T)$ of $T$, we reproduce
eq.(\ref{disres0}). Already for the $3$-linear tensor of $3$
variables the resultant has degree $M(r-1)^{M-1} = 9\cdot 2^8$ in
the coefficients of $T$, which is $2^6 = 64$ times bigger than\
${\rm deg}_T {\cal D}_{3\times 3\times 3}(T) = 36$, see
(\ref{dimsDncube}): in this particular case we actually get the
$64$-th power of discriminant.

Situation is better for symmetric tensors $S(\vec x)$: then
determinant of Koshul complex provides ${R}_{n|r-1}\Big\{\partial
S/\partial\vec x\Big\}$, and we can reproduce eq.(\ref{disres1}).
For $n|r = 3|3$ the resultant has degree $3\cdot 2^2 = 12$ in $S$,
which is exactly equal to ${\rm deg}_S D_{3|3}(S) = 12$, so that
in this case Koshul complex provides correct answer for
discriminant.

$\bullet$ Above Koshul complex in the poly-linear case is huge:
for $n_1\times\ldots\times n_r = 3\times 3\times 3$ it consists of
$M=9$ maps between $M+1=10$ vector spaces of dimensions
$\frac{M!}{m!(M-m)!}M_{M|q+2m}$, which even for the smallest
possible $q=1-n=-2$ (when Koshul complex is truncated by one map,
see s.\ref{KoII}) can be as big as $\frac{24!}{8!16!}$.

In symmetric case for $n|r = 3|3$ there are only $n=3$ maps and
$4$ vector spaces with relatively modest dimensions $d_m =
\frac{3(q+2m+1)(q+2m+2)}{m!(3-m)!}$:
$$
\begin{array}{|c|c|| c|c|c|c|c|}
\hline
&&&&&&\\
& & q=-2 & q=-1 & q=0 & q=1 & \ldots \\
&&&&&&\\
\hline
&&&&&&\\
d_0 = k_1 & \frac{(q+1)(q+2)}{2}& 0 & 0 & 1 & 3 & \\
&&&&&&\\
\hline
&&&&&&\\
d_1=k_1+k_2 & \frac{(3(q+3)(q+4)}{2} &
   3 & 9 & 18 & 30 & \\
k_2 & q^2+9q+17 & 3 & 9 & 17 & 27 & \\
&&&&&&\\
\hline
&&&&&&\\
d_2 = k_2+k_3 & \frac{(3(q+5)(q+6)}{2} &
  18 & 30 & 45 & 63 & \\
k_3 & \frac{q^2+15q + 56}{2} & 15 & 21 & 28 & 36 & \\
&&&&&&\\
\hline
&&&&&&\\
d_3=k_3+k_4 & \frac{(q+7)(q+8)}{2} &
15 & 21 & 28 & 36 & \\
&&&&&&\\
\hline\hline
&&&&&&\\
{\rm deg}_S D_{3|3}(S) & k_1-k_2+k_3 &
  12 & 12 & 12 & 12 & \\
&&&&&&\\
\hline
\end{array}
$$
\bigskip
It is clear from the table that $k_4=0$, as required for {\it
exact} complex, and discriminant can be evaluated for complexes
with arbitrary $q\geq 1-n = -2$:\ $S$-degree is stably $12$.

\subsubsection{Reductions of Koshul complex for
poly-linear tensor}

The situation in poly-linear case can be significantly improved,
if involved vector spaces are cleverly reduced. However, reduction
can be somewhat tricky. It is based on poly-linear gradation: the
action of $\hat d= dT$ not only increases the overall $x$-power by
$r$ (if powers of $dx$ are also counted), it changes the power of
every particular $\vec x_k$ by one. This means that when acting,
for example, on constants, $\hat d$ produces not the entire space
of $x$-polynomials of degree $r-1$, but a much smaller subspace
with the linear basis $\left(\prod_{k\neq l}^r
x_{k,i_k}\right)dx_{l,i_l}$: monomials like $x_{1i_1}\ldots
x_{1i_{r-1}}dx_{1i_r}$ and many other cannot appear. This
gradation splits Koszul complex into independent sub-complexes.

\bigskip

{\bf The $1\times 1$ example:}

\bigskip

\noindent

$\bullet$ The first example to illustrate the story is trivial:
the $n_1\times n_2 = 1\times 1$ bilinear tensor $Txy$. The
simplest version ($q=0$) of reduced complex has two terms, and the
three vector spaces are: $1$-dimensional space of constants,
$2$-dimensional space with linear basis $\{xdy,\ ydx\}$ and
$1$-dimensional space of two-forms $\sim xydx\wedge dy$. The two
matrices, describing the action of $\hat d = T(xdy + ydx)$ between
these three spaces are: \be {\cal A}_a = \Big( T\ \ T\Big)\ \ \
{\rm and} \ \ \ {\cal B}^a = \left(\begin{array}{c} T \\ -T
\end{array}\right) \label{ABvsTrK} \ee This complex is obviously
exact, but its determinant is identically unit (independent of
$T$): \be 1\cdot {\cal A}_a = \epsilon_{ab} {\cal B}^b
\label{unitdisc} \ee and thus does {\it not} produce discriminant
${\cal D}_{1\times 1}(T) = T$.

After reduction to symmetric case, $Txy \longrightarrow Sx^2$, the
second vector spaces shrinks to $1$-dimensional $\sim xdx$, while
the third space simply disappears, the remaining $1\times 1$
matrix ${\cal A}$ now reproduces ${\cal A} = S = D_{1|2}(S)$.

$\bullet$ Let us now return to bilinear case, but consider more
carefully what happens if Koszul complex is {\it not} reduced: the
second vector is left $4$-dimensional (with $xdx$ and $ydy$ left
in addition to $xdy$ and $ydx$) and the the third space is left
$3$-dimensional (with $x^2dx\wedge dy$ and $y^2dx\wedge dy$ left
along with $xydx\wedge dy$). It is instructive also to extend the
tensor $T$ to symmetric ${\cal S}(x,y) = \frac{1}{2}\alpha x^2 +
Txy + \frac{1}{2}\beta y^2$, so that the original bilinear $Txy$
arises for $\alpha=\beta=0$. Now the two $\hat d$-matrices,
describing the wedge multiplication by $d{\cal S} = (\alpha x +
Ty)dx + (Tx + \beta y)dy$, are
\be
{\cal A}_a &=& \left( \begin{array}{cccc}
xdx & ydx & xdy & ydy \\
&&&\\
\hline
&&&\\
\alpha & {\bf T} & {\bf T} & \beta
\end{array}\right) \ \ \nn\\
{\rm and} \ \ \
{\cal B}^a_i &=& \left(\begin{array}{c|ccc}
& x^2dx\wedge dy & xy dx\wedge dy & y^2dx\wedge dy \\
&&&\\
\hline
&&&\\
xdx & T & \beta & 0 \\ ydx & 0 & {\bf T} & \beta \\
xdy & -\alpha & {\bf -T} & 0 \\ ydy & 0 & -\alpha & -T
\end{array}\right)
\ee
Boldfaced are the two matrices, $1\times 2$ and $2\times 1$,
familiar from (\ref{ABvsTrK}), arising after reduction of the
complex. The problem is that the formula \be D_{2|2}({\cal
S})\cdot {\cal A}_a = \epsilon_{abcd} {\cal B}^b_i{\cal
B}^c_j{\cal B}^d_k \epsilon^{ijk} \label{D2|2fromK} \ee for \be
D_{2|2}({\cal S}) = T^2-\alpha\beta \label{D2|2fromKc} \ee does
not reduce in desirable way: as we saw in (\ref{unitdisc}),
instead of ${\cal D}_{1\times 1}(T) = T$ it produces unity.

$\bullet$ Fig.\,\,\ref{Koshul1x1dia} shows the way out. Instead of
reducing Koszul complex to ``symmetric" part in the center (inside
an oval), one can take another reduction to the upper or lower
fragments. Such reduced complex has one term, and $xdx
\stackrel{\hat d}{\longrightarrow} x^2ydx\wedge dy$ is represented
by the $1\times 1$ matrix of multiplication by $T$. In other
words, determinant of {\it so} reduced Koszul complex is {\it
exactly} our ${\cal D}_{1\times 1}(T) = T$ -- with no need to
extract irreducible factors, taking roots etc.

\Fig{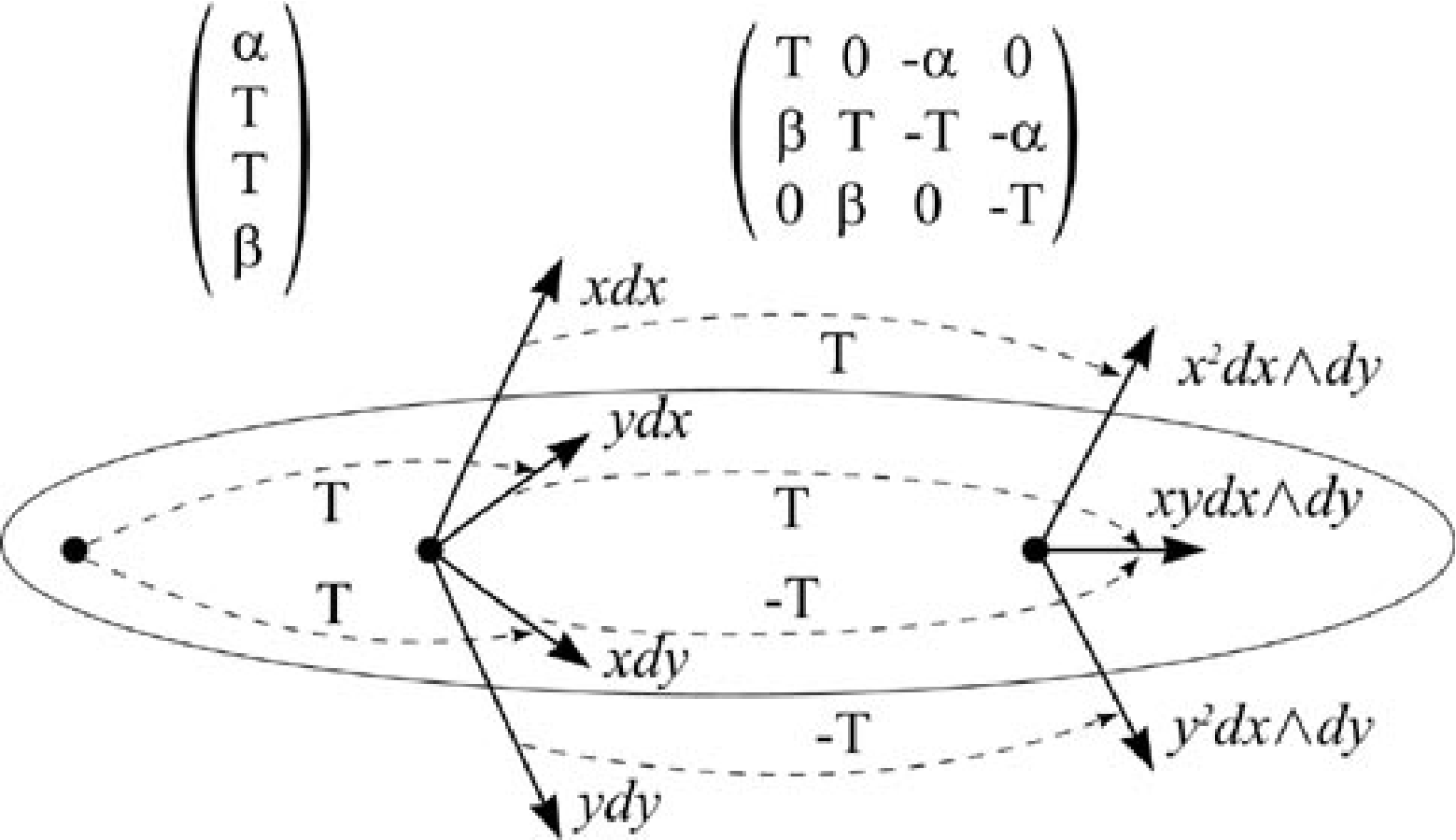}
{320,184}
{The structure of extended
(non-reduced) Koszul complex and its different sub-complexes. }

$\bullet$ One can of course make use of $D_{2|2}({\cal S})$ in
(\ref{D2|2fromKc}): when $\alpha=\beta=0$ it is a full square, and
taking the square root one obtains ${\cal D}_{1\times 1}(T) = T$.
This is approach through the resultant theory, with the use of
relations (\ref{disres0})-(\ref{disres2}).

\bigskip

{\bf The $2\times 2$ example:}

\bigskip

\noindent

$\bullet$ Things work exactly the same way for the next example:
of $2\times 2$ bilinear tensor $T^{11}x_1y_1 + T^{12}x_1y_2 +
T^{21}x_2y_1 + T^{22}x_2y_2$ (as usual, superscripts are indices,
not powers). Koszul complex with $q=0$ now has $4$ terms. In
``symmetrically reduced" version of the complex the first of the
$5$ vector spaces has dimension $1$ (constants), the second one --
with the linear basis $\{x_idy_j,\ y_jdx_i\}$ -- has dimension
$4+4=8$, the third one -- with the basis $\Big\{x_ix_jdy_1\wedge
dy_2,\ x_iy_kdx_j\wedge dy_l,\ y_ky_ldx_1\wedge dx_2 \Big\}$ --
has dimension $3+16+3 = 22$, the fourth -- with the basis $\Big\{
x_ix_jy_kdx_l\wedge dy_1\wedge dy_2,\ x_iy_jy_kdy_l\wedge
dx_1\wedge dx_2\Big\}$ -- is $12+12=24$-dimensional, and the fifth
space -- of $4$-forms $\Big\{x_ix_jy_ky_ldx_1\wedge dx_2 \wedge
dy_1\wedge dy_2\Big\}$ -- is $9$-dimensional. Thus the sequence of
dimensions is
$$
\begin{array}{ccccc}
d_0 & d_1 & d_2 & d_3 & d_4 \\
1   &  8  &  22 &  24 &  9  \\
\hline
k_1 & k_2 & k_3 & k_4 & k_5 \\
1  &   7  &  15 &   9 &  0  \\
\end{array}
$$
Since $k_5=0$, this complex can be exact, but vanishing of the
alternated sum $k_1 - k_2 + k_3 - k_4 = 0$ suggests that
determinant of the complex will be identical unity, rather than
${\cal D}_{2\times 2}(T)$.

$\bullet$ Like in the previous example, we now consider the
asymmetric reduction of Koszul complex, which starts from the
$1$-dimensional space of $2$-forms $\sim x_1x_2dx_1\wedge dx_2$
(such form is not $\hat d$-images of any 1-form). The next two
vector spaces are: $4$-dimensional space with the basis $\Big\{
x_idy_j\wedge( x_1x_2dx_1\wedge dx_2)\Big\}$ and $3$-dimensional
space with the basis $\Big\{x_ix_j dy_1\wedge dy_2 \wedge(
x_1x_2dx_1\wedge dx_2)\Big\}$. Thus reduced complex has two terms
and the two matrices, describing the wedge-multiplication by
$(T^{11}x_1 + T^{21}x_2)dy_1 + (T^{12}x_1 + T^{22}x_2)dy_2$ (the
other two items with $dx_1$ and $dx_2$ annihilate all the three
spaces) are: \be {\cal A}_a = \Big( T^{11}\ \ T^{21} \ \ T^{12} \
\ T^{22}\Big) \ \ \ {\rm and}\ \ \ {\cal B}^a_i =
\left(\begin{array}{ccc}
T^{12} & T^{22} & 0 \\ 0 & T^{12} & T^{22} \\
-T^{11} & - T^{21} & 0 \\ 0 & -T^{11} & -T^{21}
\end{array}\right)
\ee and from \be {\cal D}_{2\times 2}(T) \cdot {\cal A}_a =
\epsilon_{abcd} {\cal B}^b_i{\cal B}^c_j {\cal B}^d_k
\epsilon^{ijk} \label{D2t2fromK} \ee the bilinear discriminant is
\be {\cal D}_{2\times 2}(T) \sim T^{11}T^{22} - T^{12}T^{21} =
\det_{2\times 2} T^{ij} \ee Note that despite the obvious
similarity of formulas (\ref{D2|2fromK}) and (\ref{D2t2fromK})
they are written for different complexes and provide different
discriminants for different types of tensors: symmetric and
bilinear respectively.

\subsubsection{Reduced complex for generic bilinear
$n\times n$ tensor: discriminant is determinant of the square
matrix}

If $\vec x$ and $\vec y$ are $n$ component vectors, then the
relevant asymmetrically reduced complex starts from
$\wedge_{i=1}^n dx_i$ and has $n$ terms. $\hat d$ actually reduced
to $\hat d_{red} = d_yT = T^{ij}x_idy_j$, and the corresponding
vector spaces are formed by polynomials of increasing degree of
$x$-variables multiplied by wedge polynomials of $dy$-forms. These
spaces have dimensions \be d_m = C^n_m M_{n|m} =
\frac{n!}{m!(n-m)!}\frac{(n+m-1)!}{(n-1)!m!} = \frac{n}{(m!)^2}
\prod_{l=n-m+1}^{n+m-1} l = \prod_{j=0}^m \frac{(n^2-(j-1)^2}{j^2}
\ee (the product for $\prod_0^{-1}$ for $m=0$ by definition is
unity). The corresponding kernel dimensions $d_m = k_m+k_{m+1}$
are \be k_m = \prod_{j=1}^{m-1} \frac{n^2-j^2}{j^2} \ee From this
formula it follows that $k_{n+1}=0$ -- as needed for the complex
to be exact,-- and alternated sum, which defines $T$-dimension of
determinant of the complex, is \be \sum_{m=1}^n (-)^{n-m}k_m =
\sum_{m=1}^n (-)^{n-m} \prod_{j=1}^{m-1} \frac{n^2-j^2}{j^2} = n
\ee -- as required for \be {\cal D}_{n\times n}(T) = DET\Big(\hat
d = d_yT\Big) = \det_{n\times n} T^{ij} \ee The following table
list dimensions $d_m$ and $k_{m}$ for the first few $n$:

\bigskip

{\footnotesize
$$
\begin{array}{c|cccccc|c}
& m=0 & m=1 & m=2 & m=3 & m=4 & \ldots & \\
&&&&&&&k_n-k_{n-1} + \ldots \\
& d_0=1 & d_1=n^2 & d_2 = \frac{n^2(n^2-1)}{4} & d_3 =
\frac{n^2(n^2-1)(n^2-4)}{36} &
d_4 = \frac{n^2(n^2-1)(n^2-4)(n^2-9)}{24^2} &&\\
& k_1=1 & k_2=n^2-1 & k_3 = \frac{(n^2-1)(n^2-4)}{4} & k_4 =
\frac{(n^2-1)(n^2-4)(n^2-9)}{36} &k_5=
\frac{(n^2-1)(n^2-4)(n^2-9)(n^2-25)}{24^2}&&\\
&&&&&&&\\
\hline
&&&&&&&\\
& 1 & 1 &&&&&\\
n=1&&&&&&& 1\\
& 1 & 0 &&&&&\\
&&&&&&&\\
\hline
&&&&&&&\\
& 1 & 4 &3&&&&\\
n=2&&&&&&& 2\\
& 1 & 3 &0&&&&\\
&&&&&&&\\
\hline
&&&&&&&\\
& 1 & 9 &18&10&&&\\
n=3&&&&&&& 3\\
& 1 & 8 &10&0&&&\\
&&&&&&&\\
\hline
&&&&&&&\\
& 1 & 16 &60&80&35&&\\
n=4&&&&&&& 4\\
& 1 & 15 &45&35&0&&\\
&&&&&&&\\
\hline
\end{array}
$$}

\subsubsection{Complex for generic symmetric discriminant}

For $S(\vec x) = \sum_{j_1\ldots j_r=1}^n S^{j_1\ldots j_r}
x_{j_1}\ldots x_{j_r}$ the relevant Koshul complex is formed by
$n+1$ vector spaces, the $m$-th space, $0\leq m \leq n$, consists
of $m$-forms $dx_{i_1}\wedge\ldots \wedge dx_{i_m}$ with
coefficients, which are polynomials of degree $q+ms$ of $n$
variables $x^i$. Nilpotent operator $\hat d = dS$ acts by wedge
multiplication and increases $m$ by one. Dimensions of the vector
spaces are: \be d_m = C^n_m M_{n|q+ms} = \frac{n!}{m!(n-m)!}
\frac{(q+ms+n-1)!}{(n-1)!(q+ms)!} \ee Here $s=r-1$.
In what follows we put $q=0$. 
The values of $d_m$, the corresponding kernel dimensions $k_m$,
$d_m = k_m+k_{m+1}$ and alternated sums $\sum_{m=1}^n (-)^{n-m}
k_m$, defining the $S$-degree of determinant of the complex, are
listed in the table:

\bigskip
{\footnotesize
$$
\begin{array}{c|cccccc|c}
& m=0 & m=1 & m=2 & m=3 & m=4 & \ldots & \\
&&&&&&&k_n-k_{n-1} + \ldots \\
& d_0=P_{n|0}=1 & d_1=nP_{n|s} & d_2 = C^n_2P_{n|2s} &
d_3 = C^n_3 P_{n|3s} & d_4 = C^n_4 P_{n|4s} &&\\
& k_1=d_0 & k_2 = d_1-k_1 & k_3 = d_2-k_2 &k_4 = d_3-k_3&
k_5=d_4-k_4&&\\
&&&&&&&\\
\hline
&&&&&&&\\
& 1 & 1 &&&&&\\
n=1&&&&&&& 1\\
& 1 & 0 &&&&&\\
&&&&&&&\\
\hline
&&&&&&&\\
&1&2(s+1)&2s+1&&&&\\
n=2 &&&&&&&2s\\
&1 &2s+1&0&&&&\\
&&&&&&&\\
\hline
&&&&&&&\\
&1&\frac{3(s+1)(s+2)}{2}&\frac{3(2s+1)(2s+2)}{2}&
             \frac{(3s+1)(3s+2)}{2}&&&\\
n=3&&&&&&&3s^2\\
&1&\frac{3s^2+9s+4}{2}&\frac{9s^2+9s+2}{2}&0&&&\\
&&&&&&&\\
\hline
&&&&&&&\\
&1&\frac{4(s+1)(s+2)(s+3)}{6}&
{(2s+1)(2s+2)(2s+3)}&\frac{4(3s+1)(3s+2)(3s+3)}{6}&
\frac{(4s+1)(4s+2)(4s+3)}{6}&&\\
n=4&&&&&&&4s^3\\
&1&\frac{2s^3+12s^2+22s+9}{3}&\frac{22s^3+60s^2+44s+9}{3}&
\frac{32s^3+48s^2+22s+3}{3}  &0&&\\
&&&&&&&\\
\hline
\hline
\end{array}
$$ }
We see, that $k_{n+1}=0$, as required for exact complex.
$S$-dimension of $DET(\hat d)$ is shown in the last column,
actually it does not depend on $q$ and coincides with the
$S$-dimensions  of symmetric discriminant \be {\rm deg}_S
D_{n|r}(S)\ \stackrel{(\ref{disres1})}{=}\ {\rm deg}_S
{R}_{n|s}\{\vec\partial S\} = ns^{n-1} \ee
Explicit formulas for determinant of the complex in terms of $\hat
d$-matrices were given in s.\ref{KoII}. In each particular case
(of given $n|r$ and/or $S(x)$ of particular shape) one easily
writes down concrete matrices and evaluate \be D_{n|r}(S) =
DET\Big(\hat d = dS\Big) \ee

\subsection{Other representations}

\subsubsection{Iterated discriminant}

For tensor $T^{i_1\ldots i_{r+1}}$ of the rank $r+1$ \be {\cal
D}_{n_1\times \ldots \times n_{r+1}}(T) = {\rm irf} \left\{
{D}^{(t)}_{n|d_{n|r}} \Big({\cal D}_{n_1\times\ldots\times n_r}
(T^{\ldots k}t_k)\Big)\right\}, \label{itediscor} \ee in other
words, ${\cal D}_{n_1\times \ldots \times n_{r+1}}
\Big(T^{i_1\ldots i_{r+1}}\Big)$  divides ${D}_{n|d_{n|r}}^{(t)}
\left( {\cal D}_{n_1\times \ldots \times n_{r}} \Big(\sum_j
T^{i_1,\ldots, i_rj}t_j\Big)\right)$. In these formulas at the
r.h.s. we have symmetric functions $S(t)$ of $t$-variables of
degree $d_{n|r}$, and ${D}_{n|d_{n|r}}^{(t)} (S) =
{R}_{n|d_{n|r}-1}(\vec\partial_t S)$.

Let us now give some examples of this hierarchy: \be \bullet\
{D}_{n|0}(T) = T \ee Put by definition \be \bullet\
{D}_{n|1}\Big( T^i\Big) = \sum_{i,j=1}^n g_{ij}T^iT^j, \ee then
\be {D}_{n|1}^{(t)} \left( {D}_{n|0}\Big(\sum_{i=1}^n T^i
t_i\Big)\right) = {D}_{n|1}^{(t)}\Big(\sum_{i=1}^n T^i t_i\Big) =
\sum_{i,j=1}^n g_{ij}T^iT^j \ee (in this case vanishing of
discriminant is not the same as the system of equations $\{T^i =
0\}$, at least for complex $T^i$). \be \bullet\ {D}_{n|2}
\Big(T^{ij}\Big) = \det_{1\leq i,j \leq n}  T^{ij}, \ee and \be
{D}_{n|2}^{(t)} \left( {D}_{n|1}\Big(\sum_{i=1}^n T^{ik}
t_k\Big)\right) = {D}_{n|2}^{(t)}
\left(g_{ij}T^{ik}t_kT^{jl}t_l\right)= {D}_{n|2} \left(
g_{ij}T^{ik}T^{jl} \right) = \nn \\ = \det_{k,l} \left(
g_{ij}T^{ik}T^{jl} \right) \det (TgT^{tr}) = \det g \cdot (\det
T)^2 \ee Similarly one expects \be \bullet\ {D}_{n|3}
\Big(T^{ijk}\Big)\  \in \ {D}_{n|n}^{(t)} \left( {D}_{n|2}
\Big(\sum_k T^{ijk} t_k\Big) \right) \ee

\subsubsection{Discriminant through paths}

For matrices we have the following ``path-integral" representation
\be {D}_{n|2}\Big( T^{ijk}t_k\Big) = {\det}_{ij}\Big(
T^{ijk}t_k\Big) = \sum_{paths} \Big({\det}_{path} T\Big)
t^{path}\hspace{14mm}\nn\\
= \sum_{{\rm paths}\ k(j)} \sum_P (-)^P \prod_{j=1}^n
\Big(T^{P(j)jk(j)} t_{k(j)}\Big) = \sum_{{\rm paths}\ k(j)}
\Big({\det}_{ij} T^{ijk(j)}\Big) \prod_{j=1}^n t_{k(j)}
\nn\ee
Similarly for ${D}_{n|r}$ with $r>2$ one should have: \be
{D}_{n|r}\Big(T^{i_1\ldots i_rk}t_k\Big) = \sum_{paths}
{D}_{n|r}^{path}(T) t^{path} \ee

\subsubsection{Discriminants from diagrams \label{difrodi}}

Diagrammatic description of discriminants is a big subject. As in
the case of resultants, diagrams are well suited for consideration
of entire space of representations of structure groups, with
vacuum diagrams describing invariants: the singlet
representations. Already the elementary examples are so numerous,
that we devote to them a special section \ref{Exacov}.

Reflection of discriminantal property in diagrammatic terms
remains an open issue, closely related to devising appropriate
integral (ordinary or functional) realizations. Sensitive to
degeneracy of tensor $T$ is, for example, the {\it quasiclassical
limit} of integral (\ref{fint}), irrespective of the choice of the
totally antisymmetric weight function $h(t_1,\ldots,t_n)$.
Equations of motion, with quasiclassical parameter $\hbar$
introduced, \be \left\{ \begin{array}{c} \bar x^i_k(t) =
\frac{1}{\hbar}
\frac{\partial T}{\partial x_{ki}(t)}, \\ \\
x_{ki}(t) = \epsilon_{ij_1\ldots j_{n-1}} \int
h(t,t_1,\ldots,t_{n-1}) \bar x^{j_1}_k(t_1)\ldots \bar
x^{j_{n-1}}_k(t_{n-1}) dt_1\ldots dt_{n-1},
\end{array}\right.
\label{eqmfint} \ee acquire non-trivial solution, when $D_{n|r}(T)
= 0$. If $X_{ki}$ is non-trivial root of $\frac{\partial
T}{\partial x_{ki}}(X) = 0$ (its existence is exactly the
condition that $D(T)=0$), then solution to (\ref{eqmfint}) can be
built as a series in non-negative powers of $\hbar$: for
$$
\left\{\begin{array}{c}
x_{ki}(t) = X_{ki} + \hbar\chi_{ki}(t) + O(\hbar^2),\\ \\
\bar x_k^i(t) = \bar\chi^i_k(t) + O(\hbar)
\end{array}\right.
$$
we have:
$$
\bar\chi^i_k(t) = \frac{\partial^2 T(X)}{\partial x_{ki}
\partial x_{lj}}\chi_{lj}(t)
$$
while $\chi_{lj}$ is defined from  the system (non-linear, if
$n>2$)
$$
\chi_{lj}(t) = \left(\epsilon_{ij_1\ldots j_{n-1}}
\frac{\partial^2 T(X)}{\partial x_{lj_1}\partial x_{m_1i_1}}
\ldots \frac{\partial^2 T(X)} {\partial x_{lj_{n-1}}\partial
x_{m_{n-1}i_{n-1}}}\right) \int h(t,t_1,\ldots, t_{n-1})
\chi_{m_1i_1}(t_1)\ldots \chi_{m_{n-1}i_{n-1}}(t_{n-1}) dt_1\ldots
dt_{n-1}
$$
It is nautral to assume therefore that the quasiclassical (i.e.
$\hbar \rightarrow 0$ or simply the large-$T$) asymptotics of
$Z(T)$ depends on $D(T)$ alone, and this will be indeed the case
in the simplest examples in s.\ref{fintsec} below.

\setcounter{equation}{0}

\section{Examples of resultants and discriminants
\label{Exacov}}

We are going to discuss spaces ${\cal M}$ of tensors of various
types and associated tensor algebras. Various associated
homogeneous equations will be considered and spaces ${\cal X}$ of
their solutions. The difference between the numbers $N^{var}$ of
independent variables and $N^{eq}$ of algebraically independent
equations defines the difference
$${\rm dim} {\cal X} - {\rm dim}\ {\cal D} =
N^{sol} - N^{con} = N^{var} - N^{eq}$$ between dimensions of the
space ${\cal X}$ of solutions and the space ${\cal D}$ of
discriminantal constraints imposed on the coefficients of original
tensors. All dimensions can be also counted modulo projective
transformations, in order to enumerate projectively-inequivalent
quantities. Of special interest for us will be situations when
either ${\rm dim} {\cal D} =1$ or ${\rm dim}{\cal X} = 1$. When
discriminantal space in ${\cal M}$ has non-unit codimension or
when the space of solutions is multi-dimensional they are
described in terms of non-trivial resolvents. Both discriminantal
and solution spaces are invariant spaces w.r.t. the action of the
structure group, can be effectively described with the help of
diagrams and can be associated with {\it linear} entities at
sufficiently high level of the tensor algebra.

\subsection{The case of rank $r=1$ (vectors)}

\noindent

$\bullet$ Rank-one tensor \be T(\vec x) = \sum_{i=1}^n T^ix_i =
\vec T\vec x. \label{tenr1} \ee Moduli space (the set of
projectively non-equivalent tensors of rank one) ${\cal M}_{n} =
P^{n-1}$ has dimension $n-1$.

$\bullet$ Associated solution space ${\cal X}_n$ consists of
projectively-\-non-\-equivalent solutions $x^i$ to the equation
\be T(\vec x) = 0 \label{diseqr1} \ee and has dimension ${\rm
dim}\ {\cal X}_n = N^{sol}_n = n-2$ ($n$ homogeneous coordinates
minus one rescaling minus one equation = $n-2$). Negative
dimension in the case of $n=1$ means that there are no solutions
unless the coefficients of $T$ satisfy $N^{con} = 1$ {\it
discriminantal} constraint: ${\cal D}_1(T)=0$ with ${\cal D}_1(T)
=T^1$.

$\bullet$ The space ${\cal X}_n$ of solutions to (\ref{diseqr1})
can be represented as Grassmann manifold $G_{n-1,n-2}$ of
codimension-$1$ hyperplanes $P^{n-2}$ in $P^{n-1}$ and possesses
description in Plukker coordinates.

$\bullet$ Solutions to (\ref{diseqr1}) form vector representation
of the structure group $SL(n)$. Diagrammatically
(Fig.\ref{diseqr1dia}) it can be obtained from contraction of the
$\epsilon$-tensor with $n$ legs with the single-leg $T$ and some
antisymmetric tensor $C^{(1)}$ of valence (rank) $n-2$: \be x_i =
\epsilon_{ij k_1\ldots k_{n-2}} T^{j}C_{(1)}^{k_1\ldots k_{n-2}}
\label{53eq} \ee However, different $C_{(1)}$ can define the same
solution $x$. Equivalent are two $C_{(1)}$, differing by $\delta
C_{(1)}^{k_1\ldots k_{n-2}} = \sum_{l=1}^{n-2} (-)^l T^{k_l}
C_{(2)}^{k_1\ldots\check k_l\ldots k_{n-2}}$ with antisymmetric
tensor $C^{(2)}$ of rank $n-3$. In turn, $C_{(2)}$ are equivalent
modulo antisymmetric tensors of rank $n-4$ and so on. Since there
are $C^n_m = n!/m!(n-m)!$ antisymmetric tensors of rank $m$, we
get for dimension of solution space $N^{sol}_n  = -1 + C^n_{n-2} -
C^n_{n-3} + C^n_{n-4} - \ldots + (-)^nC^n_0 = -1 + (1-1)^n - 1 +
C^n_{n-1} = n - 2 + \delta_{n,0}.$ This is the simplest example of
{\it resolvent} construction, describing solutions of algebraic
equations.

\Fig{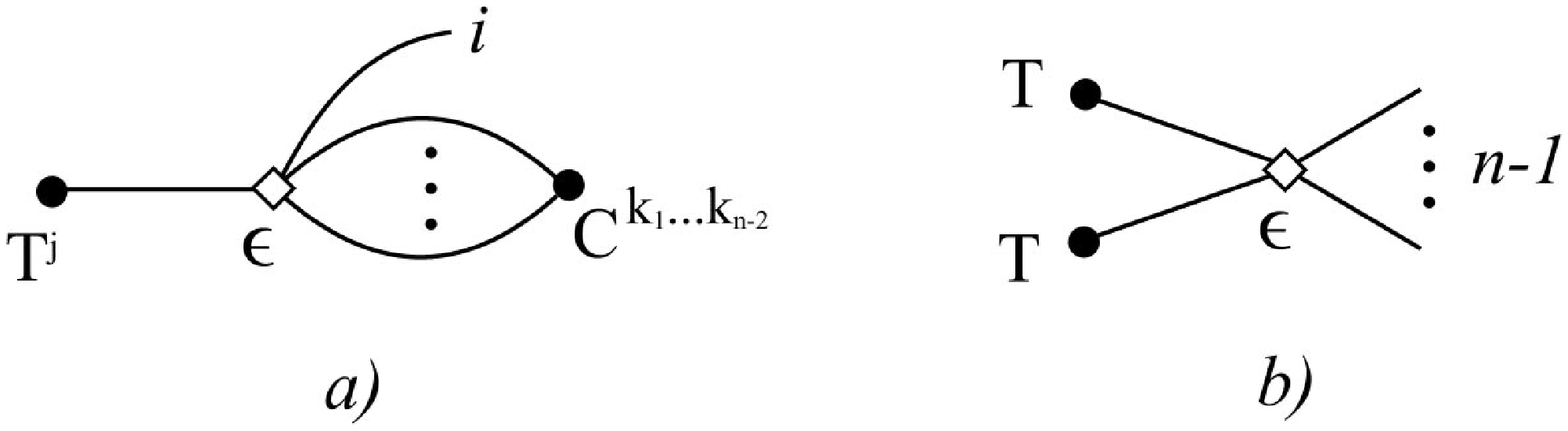} {400,108} {{\bf A.} Generic solution $x^i$ to the
equation (\ref{diseqr1}), labeled by arbitrary tensor $C$, see
(\ref{53eq}). Actually, many different choices of $C$ produce the
same $x^i$, because diagram {\bf B} vanishes identically.}

$\bullet$ The set of {\it projectively non-equivalent} solutions
is discrete, $N^{sol}_n=0$, for $n=2$. In this case the resolvent
is trivial and \be x_i = \epsilon_{ij}T^j \label{soln2r1} \ee This
formula can be considered as a diagrammatic solution $(X_1,X_2)
\sim (T^2,-T^1)$ to the equation $T^1 X_1 + T^2X_2 = 0$. In
projective coordinate $x = X_1/X_2$, solution to $T^1x + T^2 = 0$
is a fraction $x = -T^2/T^1$, and transition to homogeneous
coordinates makes it polynomial. This property becomes very
convenient in general situation. Immediate generalizations of
eq.(\ref{soln2r1}) are expression for inverse matrix and Craemer
rule in linear algebra, Vieta formulas in the theory of
polynomials, further generalizations do not have commonly accepted
names yet.

\subsection{The case of rank $r=2$ (matrices) \label{matr|2}}

\noindent

$\bullet$ Covariant tensor of rank 2 is associated with bilinear
form: \be T(\vec x, \vec y) = \sum_{i=1}^n\sum_{j=1}^m
T^{ij}x_iy_j \in V^*_n\otimes V^*_m \ee and the structure group
$SL(2)\times SL(2)$ In the square case, $m=n$ it has two quadratic
{\it reductions} with the smaller structure group $SL(2)$: \be
{\rm symmetric:} \ \ \ \ S(\vec x) =  \sum_{i,j=1}^n S^{ij}x_ix_j
\ee and \be {\rm antisymmetric:} \ \ \ \ A(\vec x) = \sum_{1\leq
i<j \leq n} A^{ij} \Big(x_i\otimes x_j - x_j\otimes x_i\Big) \ee

$\bullet$ With $T^{ij}$ one can associate the equations \be
T^{ij}y_j =  \sum_{j=1}^m T^{ij}y_j = 0, \label{Tijyj} \ee \be
T^{ij}x_i = \sum_{i=1}^n T^{ij} x_i = 0 \label{Tijxi} \ee These
are $n+m$ equations on $n+m$ homogeneous variables $x$ and $y$,
and there is a single relation between all these equations, \be
T(\vec x,\vec y) = T^{ij}x_iy_j = 0 \label{Txy} \ee This gives for
the number of projectively non-equivalent solutions $N^{sol} = -2
+ (n+m) - (n+m) + 1 = -1$, what means that there is at least one
discriminantal constraint on $T^{ij}$ for non-trivial solution
(i.e. with both $x \neq 0$ and $y\neq 0$) to exist. For square
matrix with $n=m$ this is indeed a single constraint: \be {\cal
D}_{n\times n} (T) = \det_{n\times n} T = \frac{1}{n!}
\epsilon_{i_1\ldots i_n}\epsilon_{j_1\ldots j_n} T^{i_1j_1}\ldots
T^{i_nj_n} = 0 \ee -- {\bf determinant is particular case of
discriminant}. However, the story is more complicated for $m>n$:
the first system (\ref{Tijyj}) has $N_{n\times m} = m-n-1$
linearly independent solutions for arbitrary $T$, and for
(\ref{Tijxi}) to have non-vanishing solutions $m-n+1$
discriminantal constraints should be imposed on the coefficients
of matrix $T^{ij}$.

$\bullet$ Eq.(\ref{Tijyj}) is especially interesting in the case
$m=n+1$: then it is nothing but a non-homogeneous system of $n$
linear equations, written in homogeneous coordinates. From linear
algebra we know that such system has a single solution, given by
Craemer rule. Diagrammatically (Fig.\ref{soln2r1dia}) we can
immediately write it down, as direct generalization of
eq.(\ref{soln2r1}): \be Y_j = \varepsilon_{jj_1\ldots
j_n}\epsilon_{i_1\ldots i_n} T^{i_1j_1}\ldots T^{i_nj_n},\ \ \
i=1,\ldots,m=n+1 \label{512eq} \ee As usual, in homogeneous
coordinates solution is polynomial. It can be brought to
conventional form in a chart: for $y_j = Y_j/Y_{n+1}$ we have:
$$
Y_j = (n+1)!\varepsilon_{jj_1\ldots j_{n-1},n+1}
\epsilon_{i_1\ldots i_n}T^{i_1j_1} \ldots T^{i_{n-1}j_{n-1}}
T^{i_n,n+1} = $$ $$ = \Big(\epsilon_{jj_1\ldots
j_{n-1}}\epsilon_{ii_1\ldots i_{n-1}} T^{i_1j_1}\ldots
T^{i_{n-1}j_{n-1}}\Big) T^{i,n+1}(-)^n\frac{(n+1)!}{Y_{n+1}}
$$
(indices in $\epsilon$ run from $1$ to $n$, while they run from
$1$ to $n+1$ in $\varepsilon$) and
$$
Y_{n+1} = (-)^n(n+1)!\epsilon_{j_1\ldots j_n}\epsilon_{i_1\ldots
i_n} T^{i_1j_1}\ldots T^{i_nj_n} = (-)^n n \det_{n\times n} T.
$$
In particular, we read from this diagrammatic representation the
standard expression for inverse of the square matrix $T^{ij}$ with
$1\leq i,j \leq n$: \be (T^{-1})_{ji} = \frac{\epsilon_{jj_1\ldots
j_{n-1}} \epsilon_{ii_1\ldots i_{n-1}}T^{i_1j_1}\ldots
T^{i_{n-1}j_{n-1}}} {\det T} \equiv {\check T_{ji}}{\det T}^{-1},
\label{invema} \ee thus this formula is direct generalization of
eq.(\ref{soln2r1}), and it is further generalized to Vieta
formulas, as already explained in s.\ref{CraVie}. In what follows
we often mark various minors by check-sign, in particular check
over $T$ signals that expression is non-linear in $T$.

\Fig{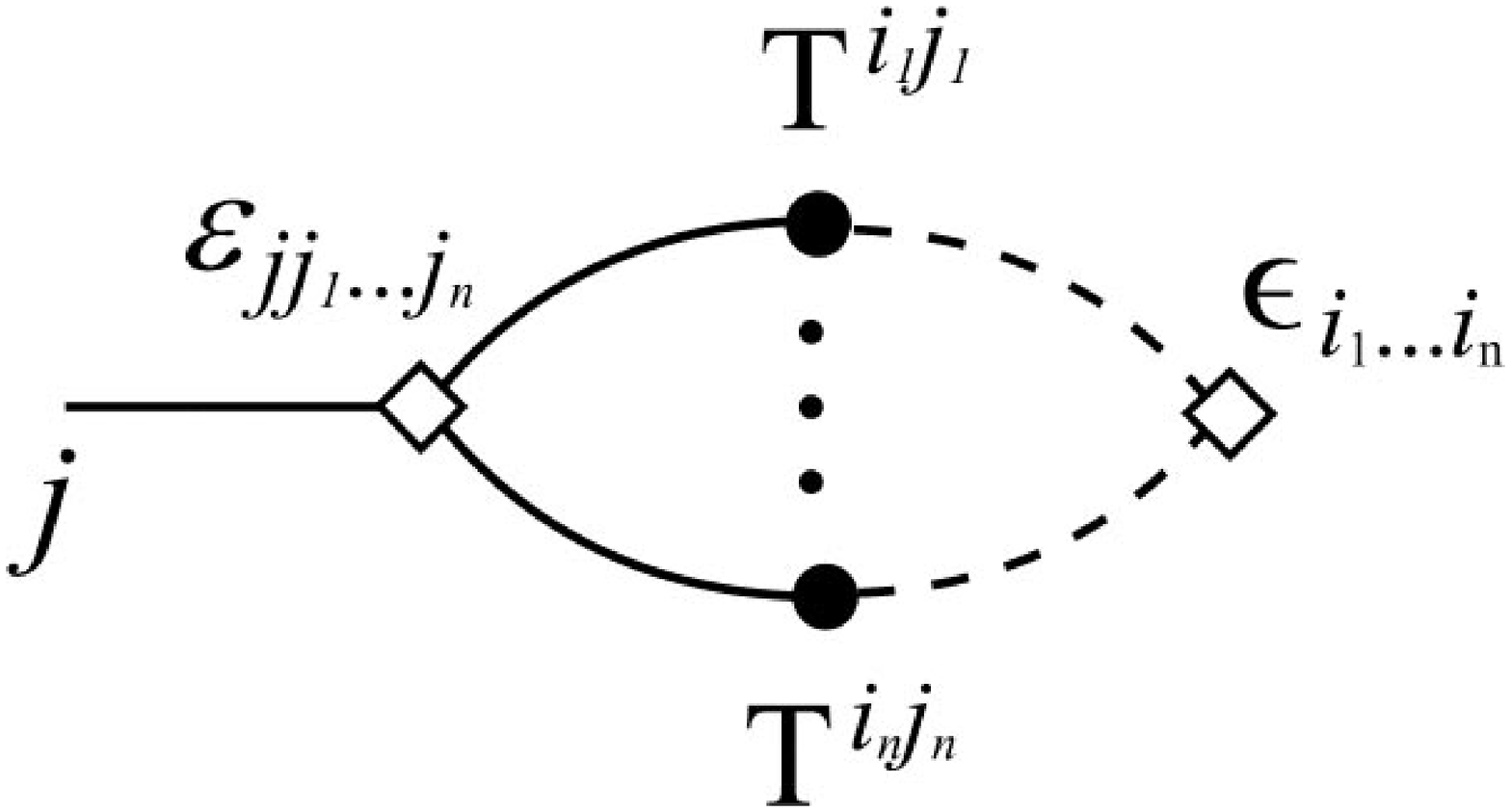} {209,107} {Pictorial representation of the
Craemer rule (\ref{512eq}). The number of rectangular
$n\times(n+1)$ tensors $T$ in the picture is $n$.}

$\bullet$ If $m>n$ the over-defined system (\ref{Tijxi}) is
resolvable iff all the principal minors \be \check T[j_1\ldots
j_n] = \epsilon_{i_1\ldots i_n} T^{i_1j_1}\ldots T^{i_nj_n}
\label{priminors} \ee for all the $\rho_{n+1}(m+1)$ $n$-ples
$1\leq j_1 < \ldots < j_n \leq m$ vanish simultaneously.
$\rho_{n+1}(m+1)$ is the number of representations of $m+1$ as a
sum of $n+1$ positive integers $j_1,j_2-j_1,\ldots,m+1-j_n$ and
$\check T[j_1\ldots j_n]$ can be also labeled by Young diagrams.
Minors (\ref{priminors}) are not algebraically independent: they
are constrained by $T\check T$ relations \be \sum_{j=1}^m T^{ij}
\check  T[jj_1\ldots j_{n-1}] = 0, \ \ \forall i = 1,\ldots,n;\
j_1,\ldots,j_{n-1} = 1,\ldots,m. \ee and their various
corollaries. Among those especially interesting are quadratic (in
$\tilde T$) Plukker relations \be \check T[ij\ldots] \check
T[kl\ldots] - \check T[ik\ldots] \check T[jl\ldots] + \check
T[il\ldots] \check T[jk\ldots] = 0 \ \ \ \forall i,j,k,l. \ee
These constraints in turn are not independent and satisfy
relations between relations and so on -- again entire resolvent
emerges.

$\bullet$ If in the square-matrix ($m=n$) case the discriminantal
constraint\  $\det T = 0$\ is satisfied, solution for the system
(\ref{Tijyj}) is given by Fig.\ref{soln2r0dia}: \be y_j =
\epsilon_{jj_1\ldots j_{n-1}}\epsilon_{ii_1\ldots i_{n-1}}
T^{i_1j_1}\ldots T^{i_{n-1}j_{n-1}}C^i = \check T_{ji}C^i, \ \ \
\forall C^i, \label{soln2r0} \ee but two vectors $C^i$ provide the
same solution $y_j$ whenever they differ by \be \delta C^i =
T^{ik}C^{(2)}_k \ee Resolvent is short in this case.

\Fig{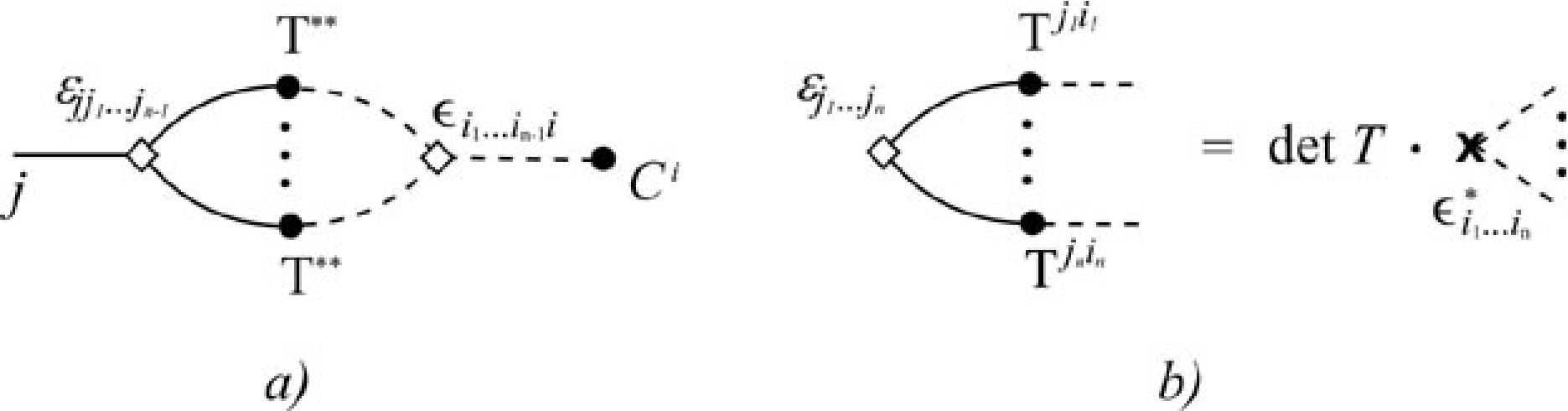} {261,114} {{\bf A.} Pictorial representation of
solution (\ref{soln2r0}) to the linear system $T(\vec x)=0$ with
the square matrix $T$. Such solution exists when $\det T = 0$, as
shown in {\bf B.} The numbers of $T$-tensors in pictures {\bf A}
and {\bf B} are $n-1$ and $n$ respectively.}

$\bullet$ If $n=2$, symmetric tensors of all ranks are
projectivizations of polynomials of a single variable, see
s.\ref{sypo} below for more details. In particular, when rank is
two and tensor is a matrix, $S(\vec x) = \sum_{i,j=1}^2
S^{ij}X_iX_j = 0$ is equivalent to the ordinary quadratic
equation, $S^{11}x^2 + 2S^{12}x + S^{22} = 0$ for $x = X_1/X_2$
with two roots, $X_1^\pm = -S^{12} \pm \sqrt{D}$, $X_2^{\pm} =
S^{11}$ and discriminant $D_{2|2}(S) = (S^{12})^2-S^{11}S^{22}$.
Discriminant $D_{2|2}(S) = 0$ (tensor $S$ is degenerate) when the
two roots coincide. Everywhere in the moduli space ${\cal
M}_{2|2}$ the Vieta formula holds, see (\ref{Vietaf}) in
s.\ref{CraVie}:
$$
\left(\begin{array}{cc} S_{11} & S_{12} \\ S_{12} & S_{22}
\end{array}\right) =
\left(\begin{array}{cc} S^{22} & -S^{12} \\ -S^{12} & S^{11}
\end{array}\right) \cong
\frac{1}{2}\left\{\left(\begin{array}{c}X_1^+\\
X_2^+\end{array}\right) \otimes \Big( X_1^-\ \  X_2^-) +
\left(\begin{array}{c}X_1^-\\ X_2^-\end{array}\right) \otimes
\Big( X_1^+\ \  X_2^+)\right\} = $$ $$ = \left(\begin{array}{cc}
X_1^+X_1^- &
\frac{1}{2}(X_1^+X_2^-+X_1^-X_2^+) \\
\frac{1}{2}(X_1^+X_2^-+X_1^-X_2^+) &  X_2^+X_2^-\end{array}\right)
$$
or simply $S^{22}/S^{11} = x_+x_-$, $-S^{12}/S^{22} =
\frac{1}{2}(x_+ +x_-)$.

$\bullet$ For antisymmetric square matrix discriminant
(determinant) is reduced to Pfaffian: \be \det_{n\times n} A =
\left\{ \begin{array}{cc}
0 & {\rm for\ odd}\ n \\
{\rm Pfaff}^2(A) & {\rm for\ even} \ n=2k
\end{array}
\right. \ee where (Fig.\ref{Pfaffdia}) \be {\rm Pfaff}(A) =
\epsilon_{i_1\ldots i_{2k}}A^{i_1i_2}\ldots A^{i_{2k-1}i_{2k}} \ee

\Fig{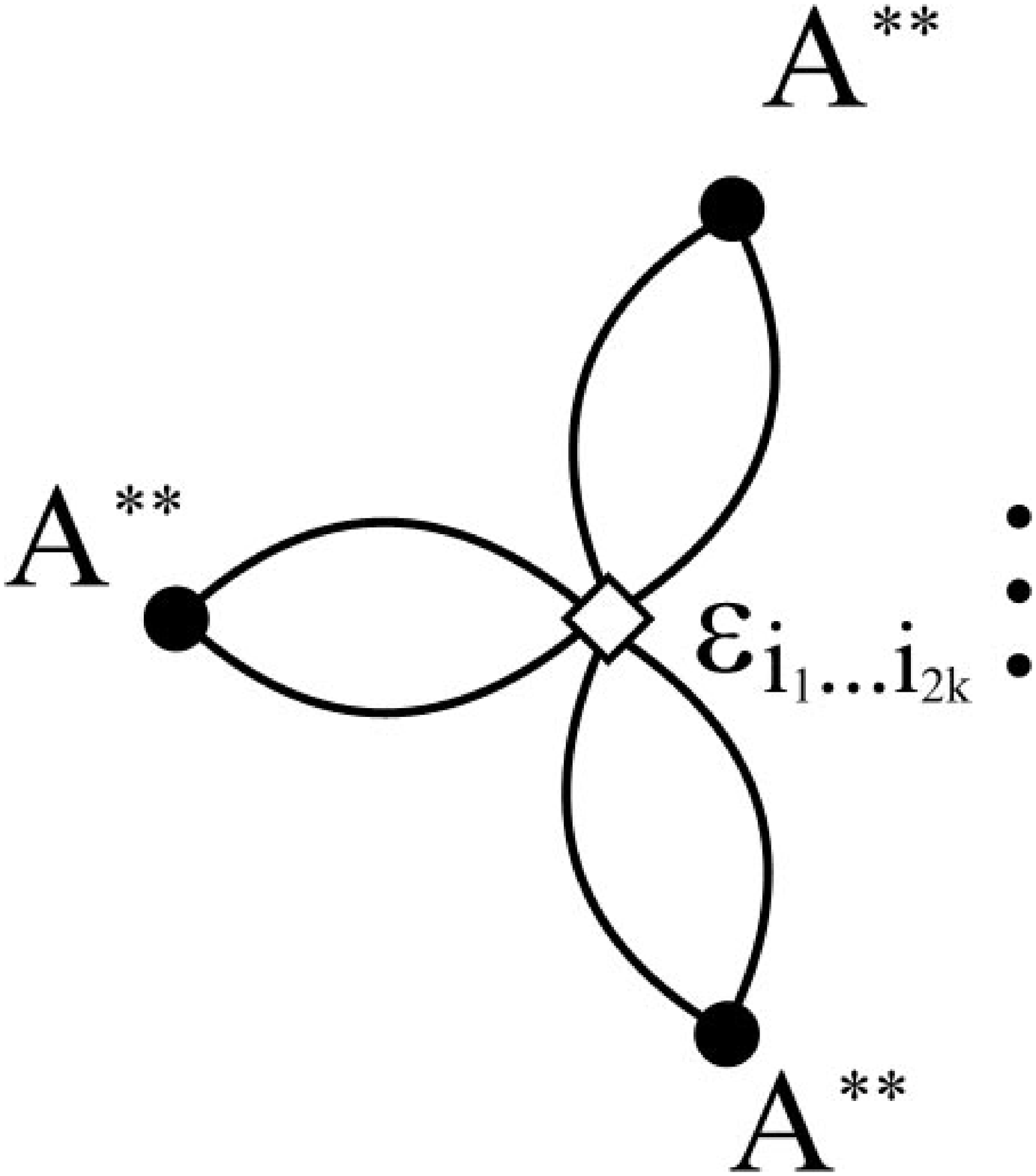} {144,168} {Pictorial representation of the Pfaffian
of a rank-two contravariant tensor $A$, Pfaffian vanishes for
symmetric $A$.}

$\bullet$ The simplest example ${n|r} = {2|2}$ can be used to
illustrate investigation of tensor algebra ${\cal T}(T)$ with the
help of operators. In this case the relevant operator is $\hat T =
T_i^j = \epsilon_{ik}T^{kj}$, and connected diagrams are exhausted
by two sequences: the closed (vacuum) loops\ ${\rm Tr}\ \hat T^m$\
without external legs and the trees (actually, lines)\ $\hat T^m$\
with two external legs. There are at most $4$ algebraically
independent quantities: the simplest option is to take for them
just $4$ components of $\hat T$. However, this choice does not
explain how the diagrams are interrelated.

One kind of relations between diagrams is given by the identity
$\log\det = {\rm Tr}\log$: \be \det (\hat I - \hat T) = \exp
\left(-\sum_{k=1}^\infty \frac{1}{k} {\rm Tr}\ \hat T^k\right)
\label{dettr22} \ee Since the l.h.s. is equal to
$$\det (\hat I - \hat T) = 1 - {\rm Tr}\ T + \det T$$
eq.(\ref{dettr22}) expresses $\det T = \frac{1}{2} \Big(({\rm Tr}\
T)^2 - {\rm Tr}\ T^2\Big)$ and all ${\rm Tr}\ T^m$ with $m\geq 3$
through two independent ${\rm Tr}\ \hat T = T^1_1 - T^2_2 = T_{21}
- T_{12}$ and ${\rm Tr}\ \hat T^2 = (T^1_1)^2 + 2T^1_2T^2_1 +
(T^2_2)^2 = T_{21}^2 + T_{12}^2 - 2T_{11}T_{22}$. These are the
two generators of the ring of the $SL(2)\otimes SL(2)$ invariants,
associated with this model.

$\hat T^m$ provide non-singlet spin-one-half representation of the
structure group. In generic position they can be parameterized by
two eigenvalues and two coefficients in the diagonalizing matrix
$U$. This parametrization becomes singular when $\det S= 0$.

$\bullet$ In the case of matrices ($r=2$) the integral
(\ref{fint}) is Gaussian in $x$-variables (but not in $\bar x$ if
$n>2$), and they can be explicitly integrated out. This converts
$Z_{r=2}(T)$ into
$$
Z_{r=2}(T) \sim \Big(\det T\Big)^{\mp\#(t)} \int D\bar x^i(t)
D\bar y^i(t) \exp \left\{ \int \bar x^i(t)T^{-1}_{ij}\bar y^j(t)
dt + \right.
$$
\be \left. + \int_{t_1<\ldots< t_n} \epsilon_{i_1\ldots i_n}
\Big(\bar x^{i_1}(t_1)\ldots \bar x^{i_n}(t_n) + \bar
y^{i_1}(t_1)\ldots \bar y^{i_n}(t_n)\Big) dt_1\ldots dt_n\right\}
\ee Here $\bar x^i(t) = \bar x_1^i(t)$, $\bar y^i(t) = \bar
x_2^i(t)$; $\#(t)$ is the number of points in the $t$-lattice and
$\mp$ correponds to the choice between bosonic (-) and fermionic
(+) integrations. For $n=2$ integrals over $\bar x$ and $\bar y$
are also Gaussian and can be performed explicitly, see the end of
s.\ref{sypo22} below.

\subsection{The $2\times 2\times 2$ case (Cayley
hyperdeterminant \cite{Cay}) \label{cayh} }

This example is the only well-known chapter of non-linear algebra,
open already in XIX century and slowly reappearing in physical
literature of the XXI century \cite{recent}. However, as all
examples with $n=2$ it reduces to ordinary resultants.

$\bullet$ The poly-linear $2\times 2\times 2$ tensor \be T(\vec
x,\vec y,\vec z) = \sum_{i,j,k=1}^2 T^{ijk}x_iy_jz_k \ee has\ \
${\rm dim}{\cal M}_{2\times 2\times 2} = 2^3-1=7$\ \
projectively-non-equivalent parameters. Its symmetric reduction
$S(\vec x) = T(\vec x,\vec x,\vec x)$ has\ \ ${\rm dim}{\cal
M}_{2|3} = 4-1 = 3$\ \ parameters.

$\bullet$ Discriminant of the polylinear tensor (Cayley
hyperdeterminant \cite{Cay}) is given by (Fig.\ref{Caydia}) \be
{\cal D}_{2\times 2\times 2}(T) =
T^{ijm}T^{i'j'n}T^{klm'}T^{k'l'n'}
\epsilon_{ii'}\epsilon_{jj'}\epsilon_{kk'}\epsilon_{ll'}
\epsilon_{mm'}\epsilon_{nn'} \equiv T^{ijm}T_{ij}^{\cdot\cdot n}
T^{kl}_{\cdot\cdot m} T_{kln} = \label{Cahyp} \ee
$$
= \Big(T^{111}T^{222} + T^{112}T^{221} - T^{121}T^{212} -
T^{122}T^{211}\Big)^2 - 4\Big(T^{111}T^{221}-T^{121}T^{211}\Big)
\Big(T^{112}T^{222} - T^{122}T^{212}\Big)
$$

\Fig{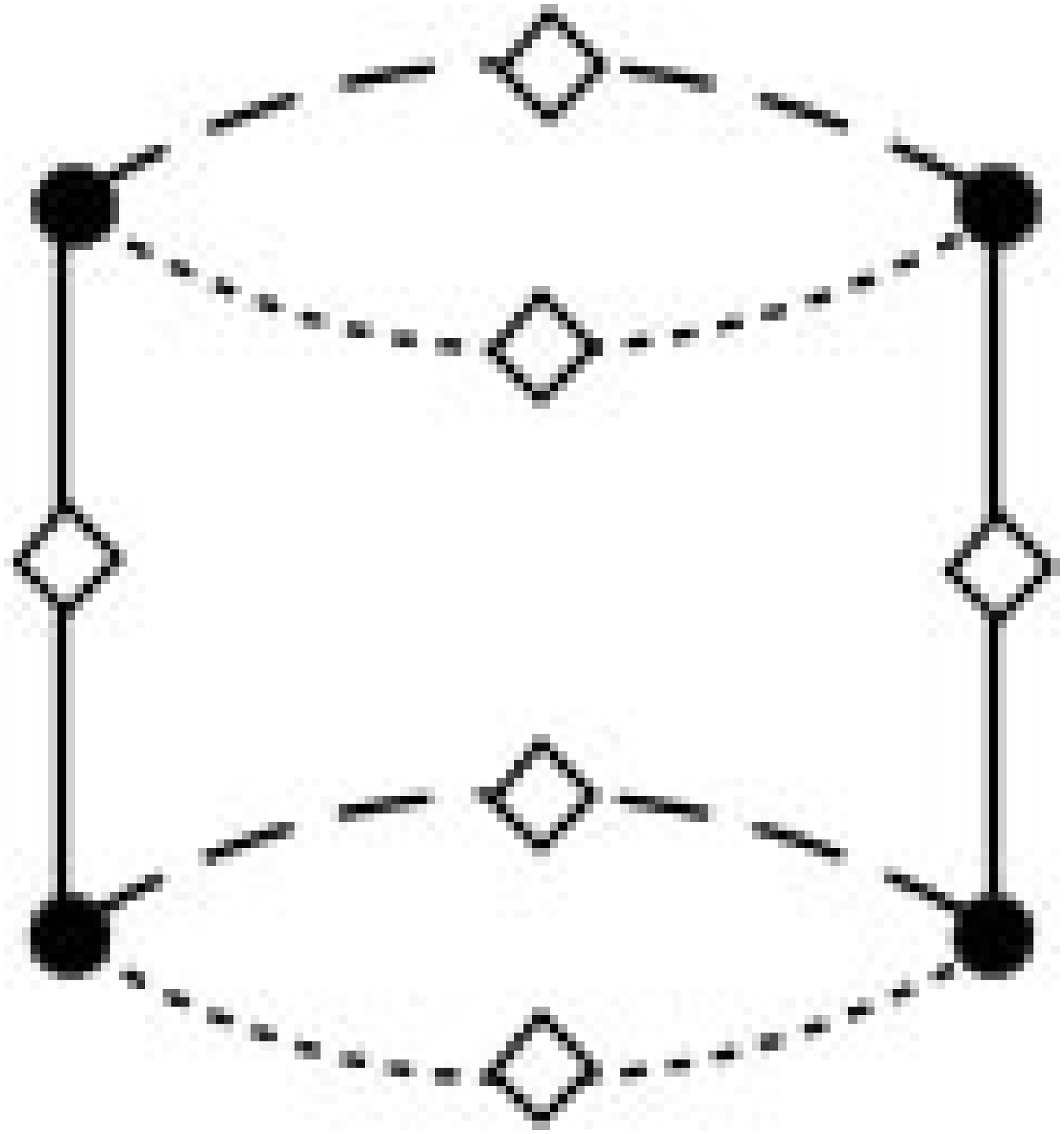} {120,128} {Pictorial representation of the Cayley
hyperdeterminant (\ref{Cahyp}). Three different {\it sorts} of
lines are explicitly shown.}

$\bullet$ The three diagrams in Fig.\ref{Cay3dia}, differing by
permutation of the lines {\it sorts} are actually equal:
$$T^{ijm}T_{ij}^{\cdot\cdot n} T^{kl}_{\cdot\cdot m} T_{kln} =
T^{imj}T_{i\cdot j}^{\cdot n\cdot} T^{k\cdot l}_{\cdot m\cdot}
T_{knl} = T^{mij}T_{\cdot ij}^{n\cdot\cdot} T^{\cdot
kl}_{m\cdot\cdot} T_{nkl}$$ (indices are raised with the help of
$\epsilon$-tensor $u^{i\ldots}_{\ldots} \equiv
\epsilon^{ii'}u^{\ldots}_{i'\ldots}$). The explanation of this
non-evident equality (not seen at the level of diagrams) is that
the action of the structure group $SL(2)^{\times 3}\times\sigma_3$
converts one diagram into another, while discriminant remains
intact.

\Fig{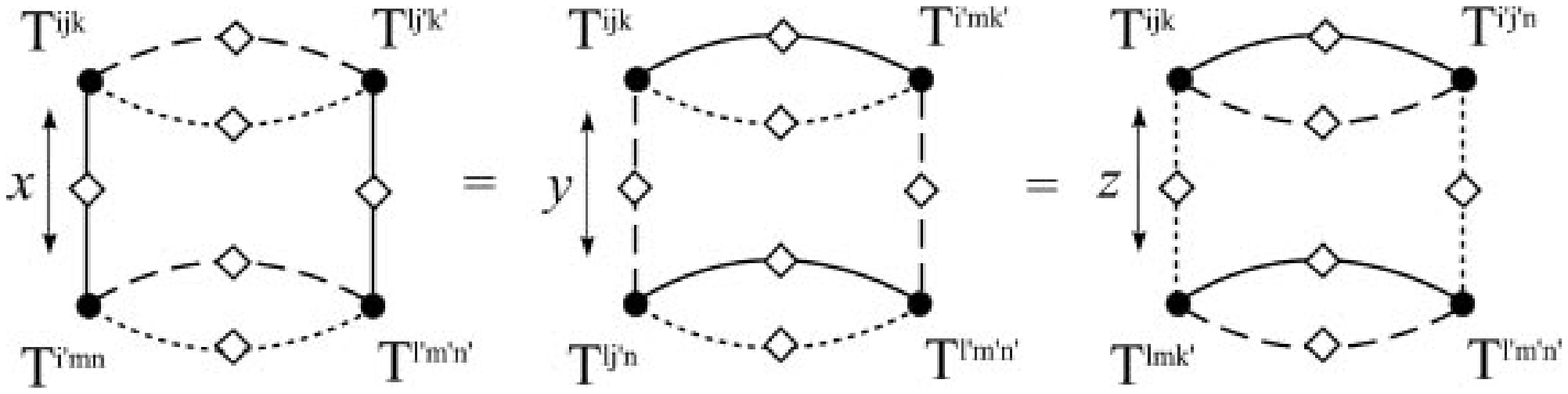} {460,113} {Equivalence of different representations
of Cayley hyperdeterminant.}

$\bullet$ Fig.\ref{Cayotherdia} shows some other closed (without
external legs) diagrams: one of the two order-two diagrams,
Fig.\ref{Cayotherdia}.A \be T_{ijk}T^{ijk} = 0 \ee is {\it
identically} zero, while Fig.\ref{Cayotherdia}.B \be T_{ij}^i
T_k^{kj}, \label{Catadp} \ee with the structure group reduced from
$SL(2)^{\times 3}$ to a single $SL(2)$, can be non-vanishing. Also
vanishing is the tetrahedron diagram in order four
(Fig.\ref{Cayotherdia}.C): it is obviously zero for {\it diagonal}
tensors, but actually it vanishes for arbitrary $T^{ijk}$. Higher
order diagrams are expressed through (\ref{Cahyp}) and tadpole
(\ref{Catadp}).

\Fig{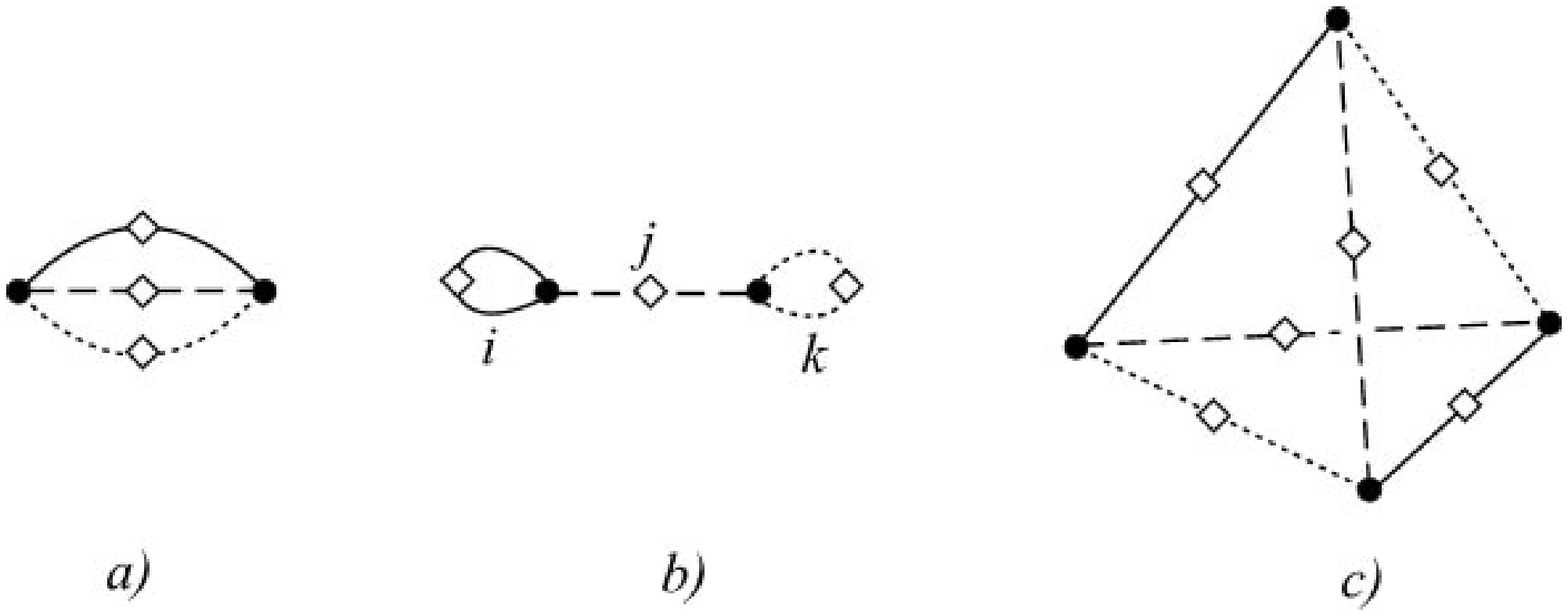} {441,171} {Examples of vacuum (without external
lines) diagrams from ${\cal T}(T)$ for a $2\times 2\times 2$
tensor $T$. {\bf A} and {\bf C} vanish identically, while {\bf B}
can be non-vanishing.}

$\bullet$ Symmetric reduction in the $2\times 2\times 2$ case does
not affect discriminant: symmetric discriminant is given by the
same expression (\ref{Cahyp}), only the three diagrams in
Fig.\ref{Cay3dia} are now equivalent already at the level of
pictures, not only expressions. Symmetric discriminant in this
case is nothing but the ordinary discriminant (well known as
$27p^3+4q^2$) of the cubic polynomial $S_3(t) = S^{111}t^3 +
3S^{112}t^2 + 3S^{122}t + S^{222} = s_3t^3 + s_2t^2 + s_1t + s_0$:
$$
D_{2|3}(S) = {\rm Disc}_t\Big( s_3t^3 + s_2t^2 + s_1t + s_0\Big) =
s_{3}^{-1} {\rm Res}_t\Big(S_3(t), S'_3(t)\Big) =
\frac{1}{s_3}\left|\left|\begin{array}{ccccc}
s_3 & s_2 & s_1 & s_0 & 0  \\
0 & s_3 & s_2 & s_1 & s_0  \\
3s_3 & 2s_2 & s_1 & 0 & 0 \\
0 & 3s_3 & 2s_2 & s_1 & 0 \\
0 & 0 & 3s_3 & 2s_2 & s_1 \end{array}\right|\right| = $$ $$ =
4s_1^3s_3 + 27 s_0^2s_3^2 + 4s_0s_2^3 - s_1^2s_2^2 - 18
s_0s_1s_2s_3 = $$ $$ = 27\Big(4S^{111}(S^{122})^3 + (S^{111})^2
(S^{222})^2 + 4S^{222}(S^{112})^3 - 3(S^{112})^2(S^{122})^2 - 6
S^{111}S^{112}S^{122}S^{222}\Big) =
$$
\be = 27\left(\Big(S^{111}S^{222}-S^{112}S^{122}\Big)^2 -
\Big(S^{111}S^{122}- (S^{112})^2\Big) \Big(S^{112}S^{222} -
(S^{122})^2\Big)\right) \label{DiscS3} \ee One easily see that
this is nothing but symmetrization of the Cayley hyperdeterminant
(\ref{Cahyp}). This is not a too big surprise: as we already saw,
for low values of $n$ and $r$ many quantities coincide, even if
generically they belong to different families.

$\bullet$ The simplest diagram with two external legs,
Fig.\ref{Caylegsdia}, a direct generalization of
eq.(\ref{soln2r1}), provides a bi-covector \be X^{ij} =
T^i_{mn}T^{jmn} \ee After symmetric reduction the three conditions
$X^{ij}(S) = 0$ define a codimension-two subspace in ${\cal
M}_{2|3}$, where all the {\it three} roots of associated cubic
polynomial $S_3(t)$ coincide (while discriminant (\ref{Cahyp})
vanishes when any {\it two} of these roots coincide, and this
happens in codimension {\it one} in ${\cal M}_{2|3}$).

\Fig{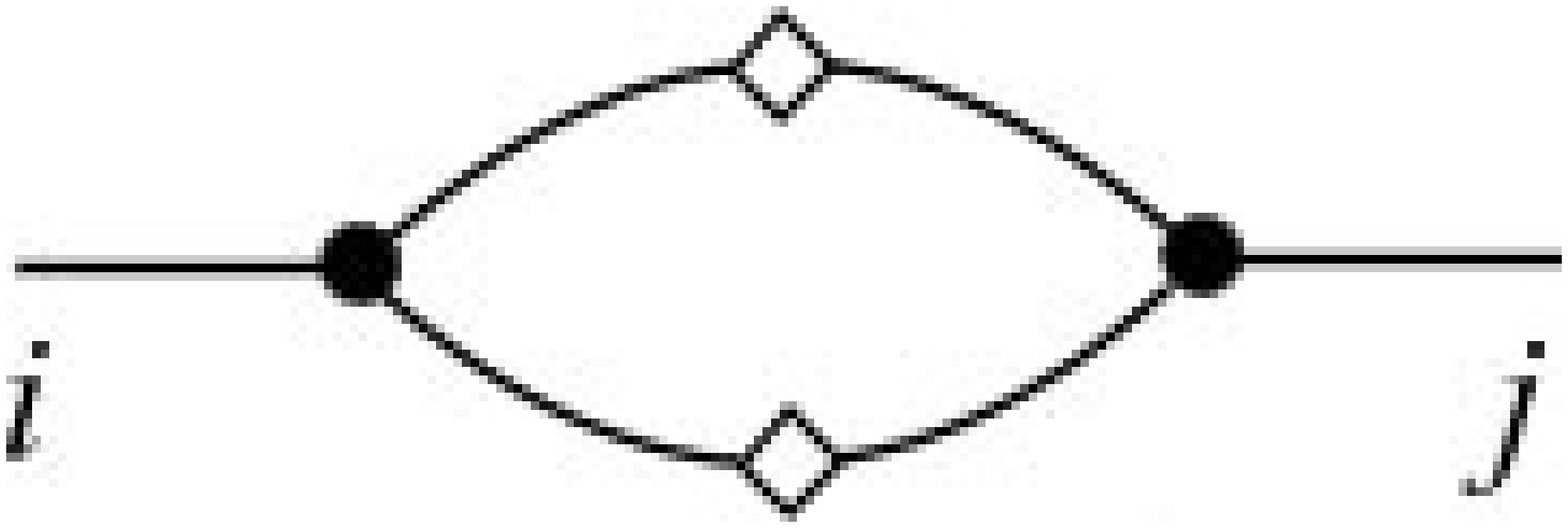} {181,60} {A diagram with two external legs from
the same ${\cal T}(T)$. In case of totally symmetric $T$ this
particular diagram plays role in description of discriminantal
spaces, see eq.(\ref{Bij23}) and s.\ref{hidiva}.}

$\bullet$ For ${n|r}={2|3}$ we have $N=n(r-2)=2$, and \be
{D}_{n|r}(S)\ \stackrel{(\ref{disres2})}{=}\ {\bf irf}\left(
{R}_{N|N-1}\Big(\partial_I (\det \partial^2 S)\Big)\right) \
\stackrel{(\ref{disres1})}{=}\
 {\bf irf}\Big({R}_{n|r-1}(\partial_i S)\Big)
\label{diviconexa01} \ee implies \be {D}_{2|3}(S) =
{R}_{2|1}\Big(\vec\partial (\det \partial^2 S)\Big) =
R_{2|2}(\vec\partial S) \label{diviconexa11} \ee As usual for low
$n$ and $r$, the resultants at the r.h.s. appear irreducible and
no ${\bf irf}$ operation is needed.

$\bullet$ The arguments of the resultant ${R}_{2|1}$ in
(\ref{diviconexa11}) are two linear functions: $x=z_1$ and $y=z_2$
derivatives of determinant ${\cal S}[z] \equiv \det \partial^2 S =
S^{11}[z]S^{22}[z] - S^{12}[z]^2$,\ $S^{ij}[z] = S^{ijk}z_k$. The
resultant of two linear functions is in turn a determinant of the
$n\times n = 2\times 2$ matrix:
$$
{R}_{2|1}\Big(\vec\partial (\det \partial^2 S)\Big) =
{R}_{2|1}\Big(\partial_1{\cal S}, \partial_2{\cal S}\Big) =
\left|\left|\begin{array}{cc}
\partial^2_{11}{\cal S} & \partial^2_{12}{\cal S} \\
\partial^2_{12}{\cal S} & \partial^2_{22}{\cal S}
\end{array}\right|\right| \sim
\left|\left|\begin{array}{cc}
2(S^{111}S^{122}-S^{112}S^{112}) & S^{111}S^{222}-S^{112}S^{122}\\
S^{111}S^{222}-S^{112}S^{122} & 2(S^{112}S^{222}-S^{122}S^{122})
\end{array}\right|\right| =
$$
\be = 4\Big(S^{111}S^{122}-(S^{112})^2\Big)
\Big(S^{112}S^{222}-(S^{122})^2\Big) -
\Big(S^{111}S^{222}-S^{112}S^{122}\Big)^2 \label{resdetS} \ee

$\bullet$ Resultant ${R}_{2|2}$ measures degeneracy of a pair of
quadratic equations, \be \left\{ \begin{array}{c}
A(x,y) = a_{11}x^2 + 2a_{12}xy + a_{22}y^2 = 0  \\
B(x,y) = b_{11}x^2 + 2b_{12}xy + b_{22}y^2 = 0
\end{array}\right.
\label{ABsys} \ee This system has non-vanishing solution {\it iff}
one of the roots of the first equation coincides with one of the
roots of the second: \be \frac{-a_{12} \pm
\sqrt{a_{12}^2-a_{11}a_{22}}}{a_{11}} = \frac{-b_{12} \pm
\sqrt{b_{12}^2-b_{11}b_{22}}}{b_{11}} \ee This equation can be
converted into polynomial form by double squaring, then it becomes
an equation of degree $4$ in $a$ and $b$ which is exactly the
condition ${R}_{2|2}\{A,B\} = 0$ with \be {R}_{2|2}\{A,B\}  =
\left|\left|\begin{array}{cccc}
a_{11} & 2a_{12} & a_{22} & 0 \\
0 & a_{11} & 2a_{12} & a_{22} \\
b_{11} & 2b_{12} & b_{22} & 0 \\
0 & b_{11} & 2b_{12} & b_{22} \end{array}\right|\right| =
\Big(a_{11}b_{22}-a_{22}b_{11}\Big)^2 + 4\Big(a_{11}b_{12} -
a_{12}b_{11}\Big) \Big(a_{22}b_{12}-a_{12}b_{22}\Big)
\label{resAB} \ee In the case of (\ref{diviconexa11}) the system
(\ref{ABsys}) is gradient: with $z_1=x$, $z_2=y$ and
$$S(z) = S^{111}x^3 + 3S^{112}x^2y + 3S^{122}xy^2 + S^{222}y^3$$
we have: \be A(x,y) = \partial_x S(z) =
3\Big(S^{111}x^2 + 2S^{112}xy + S^{122}y^2\Big)\nn \\
B(x,y) = \partial_y S(z) = 3\Big(S^{112}x^2 + 2S^{122}xy +
S^{222}y^2\Big) \nn \ee and \be {\cal R}(\vec\partial S) =
\Big(S^{111}S^{222}-S^{122}S^{112}\Big)^2 -
4\Big(S^{111}S^{122}-(S^{112})^2\Big) \Big(S^{112}S^{222} -
(S^{122})^2\Big) \ee Then (\ref{resAB}) obviously coincides (up to
an overall sign and other inessential numerical factors) with
(\ref{resdetS}) and with (\ref{DiscS3}), which is the
symmetrization of (\ref{Cahyp}), so that (\ref{diviconexa11}) is
indeed true.

$\bullet$ Operator approach in the case of ${n|r} = {2|3}$ works
as follows. Introduce \be \hat A = \left(\begin{array}{cccc}
S_1^{11} & S_1^{12} & S_1^{21} & S_1^{22} \\
S_2^{11} & S_2^{12} & S_2^{21} & S_2^{22}
\end{array}\right) = \left(\begin{array}{cccc}
s_2 & s_1 & s_1 & s_0 \\ -s_3 & -s_2 & -s_2 & -s_1
\end{array}\right)
\label{oper23.1} \ee \be \hat B = \left(\begin{array}{cc}
S^1_{11} & S^2_{11} \\ S^1_{12} & S^2_{12} \\
S^1_{21} & S^2_{21} \\ S^1_{22} & S^2_{22}
\end{array}\right) = \left(\begin{array}{cc}
s_1 & s_0 \\ -s_2 & -s_1 \\ -s_2 & -s_1 \\ s_3 & s_2
\end{array}\right)
\label{oper23.2} \ee Indices here are raised with the help of
$\epsilon$-tensor. From rectangular $\hat A$ and $\hat B$ one can
build two basic square matrices ({\it operators}): \be \hat C =
\hat A\hat B = \left(\begin{array}{cc} C^1_1 & C^1_2 \\ C^2_1 &
C^2_2 \end{array}\right) = \left(\begin{array}{cc}
s_3s_0-s_1s_2 & 2(s_0s_2-s_1^2) \\
2(s_2^2 - s_1s_3) & s_1s_2-s_0s_3
\end{array}\right)
\label{oper23.3} \ee \be \hat E = \hat B \hat A  =
\left(\begin{array}{cccc}
s_1s_2-s_0s_3 & s_1^2 - s_0s_2 & s_1^2-s_0s_2 & 0 \\
s_1s_3-s_2^2 & 0 & 0 & s_1^2-s_0s_2 \\
s_1s_3-s_2^2 & 0 & 0 & s_1^2-s_0s_2 \\
0 & s_1s_3-s_2^2 & s_1s_3-s_2^2 & s_0s_3-s_1s_2
\end{array}\right)
\label{oper23.4} \ee \be D_{2|3} = {\rm Tr} \hat C^2
\label{oper23.5} \ee


\subsection{Symmetric hypercubic tensors $2^{\times r}$
and polynomials of a single variable \label{sypo}}

\subsubsection{Generalities}

\noindent

$\bullet$ For $n=2$ symmetric hypercubic tensors are
projectivizations of ordinary polynomials of a single variable.
With a tensor of the type ${2|r}$ (i.e. symmetric $2^{\times r}$),
$\sum_{i_1,\ldots,i_r=1}^2 S^{i_1\ldots i_r}x_{i_1}\ldots
x_{i_r}$, we associate a polynomial of degree $r$:
\be
S_r(\xi) &=& S^{11\ldots 11}\xi^r + rS^{11\ldots 12}\xi^{r-1} +
\frac{r(r-1)}{2}S^{11\ldots 22} \xi^{r-2} + \ldots
 + rS^{12\ldots
22}\xi \nn\\
&+& S^{22\ldots 22}  = s_r\xi^r + s_{r-1}\xi^{r-1} +
\ldots + s_0 = s_r\prod_{k=1}^r(\xi-\xi_k)
\ee
where $\xi = x_1/x_2$. In this case discriminant reduces to the
ordinary discriminant from the elementary algebra \cite{resultant}
($\dot S$ denotes the $\xi$-derivative),
\be &D_{2^{\times r}}(S) =
D_{2|r}(S) = {\rm Disc}_\xi\Big(S_r(\xi)\Big) \nn\\
&= s_r^2 \prod_{k<l}
(\xi_k-\xi_l)^2 = s_r^{-1}{\rm Res}_\xi \Big(S(\xi),\dot
S(\xi)\Big) \nn\\
&= \frac{1}{s_r}\det\left(\begin{array}{cccccccccc} s_r & s_{r-1} &
s_{r-2} & \ldots & s_{1} & s_0 & 0 &
\ldots & 0 \\
0 & s_r & s_{r-1} & \ldots & s_{2} & s_{1} & s_0 &
\ldots & 0 \\
\ldots & 0 \\
&&&& \ldots &&&& \\
0 & 0 & 0 & \ldots & s_r & s_{r-1} & s_{r-2} & \ldots & s_0 \\
rs_r & (r-1)s_{r-1} & (r-2)s_{r-2} & \ldots & s_{1} & 0 & 0 &
\ldots & 0 \\
0 & rs_r & (r-1)s_{r-1} & \ldots & 2s_{2} & s_{1} & 0 &
\ldots & 0 \\
&&&& \ldots &&&&& \\
0 & 0 & 0 & \ldots & 0 & rs_r & (r-1)s_{r-1} & \ldots & s_1
\end{array} \right)
\label{discpol}\nn\\\ee
This is a $(2r-1)\times(2r-1)$ matrix and discriminant
$D_{2|r}(S)$ has degree $2r-2$ in coefficients of $S$ \be {\rm
deg}_S D_{2|r}(S) = 2r-2 \ee (while degree of discriminant of
generic -- without special symmetry -- tensor, ${\cal
D}_{2^{\times r}}(T)$, grows much faster with $r$, see
(\ref{dimD2timesr}): while ${\rm deg}_T {\cal D}_{2\times 2}(T) =
2$ ${\rm deg}_T {\cal D}_{2\times 2 \times 2}(T) = 4$, already
${\rm deg}_T {\cal D}_{2\times 2 \times 2\times 2}(T) = 24 > 6$
and the difference grows fast with the further increase of $r$).

$\bullet$ In accordance with (\ref{disres1}) discriminant
$D_{2|r}(S)$ of a polynomial $S_r(\xi)$ of degree $r$ coincides
with the resultant $R_{2|r-1}(S_1,S_2)$ of a pair of polynomials
of degree $r-1$, made from symmetric tensors $(S_i)^{k_1\ldots
k_{r-1}} = \epsilon_{ij}S^{jk_1\ldots k_{r-1}} =
\frac{1}{r!}\epsilon_{ij}\frac{\partial S(x)}{\partial x_j}$ with
$i=1$ and $i=2$. This is important property of homogeneous
polynomials: every such polynomial gives rise to {\it two}
ordinary ones, obtained by differentiation over $x_1$ and $x_2$
respectively.

$\bullet$
In application to polynomials the general relation
(\ref{didirela}) can be made substantially stronger
\cite{Sham}.
Given a homogeneous polynomial of $n=2$ variables
$x_1=x$ and $x_2=y$,
the skew product of its two derivatives $S^1=\partial_x S$
and $S^2 = \partial_y S$ is always divisible by
$\epsilon^{ij}x_i\tilde x_j$ and
\be
\frac{\epsilon_{ij}S^i(x)S^j(\tilde x)}
{\epsilon^{ij}x_i\tilde x_j} =
B_{IJ}(x)_{r-2}^I(\tilde x)_{r-2}^J
\ee
where this time $B_{IJ}$ is a symmetric $(r-1)\times(r-1)$
matrix ($N_{r-2|2} = r-1$),
so that the analogue of (\ref{didirela}) is now
\be
D_{2|r}(S) \sim D_{r-1|2}\Big(B(S)\Big) \sim
\det_{(r-1)\times(r-1)} B_{IJ}\label{e545}
\label{sydirel}
\ee
Such formula can be also obtained from (\ref{ordresdetrep}),
which can be used to represent discriminant of a
homogeneous polynomial $S(x,y)$
as a resultant of two polynomials $S^1 = \partial_x S$
and $S^2 = \partial_y S$ and provides the answer in the form
of determinant of an $2(r-1)\times 2(r-1)$ matrix.
This matrix can be decomposed into $2\times 2$ minors,
and this decomposition can be rewritten as determinant of
a $(r-1)\times (r-1)$ matrix $B(S)$.
If $corank(B) >1$, the corresponding $S$ belongs to a
higher discriminant variety: some $corank(B)+1$ roots of
$S$ merge \cite{Sham}.

For example, if $n=2$, $\vec x = (x,y)$ and $r=4$, then
\be
&S(\vec x) = ax^4+bx^3y+cx^2y^2+dxy^3+ey^4\nn\\
&= S^{1111}x^4 + 4S^{1112}x^3y + 6S^{1122}x^2y^2 +
4S^{1222}xy^3 + S^{2222}y^4,
\ee
and
$$
\begin{array}{c}
S^1(\vec x) = \frac{1}{4}\partial_x S(\vec x) =
S^{1111}x^3+3S^{1112}x^2y + 3S^{1122}xy^2 + S^{1222}y^3 \\
S^2(\vec x) = \frac{1}{4}\partial_y S(\vec x) =
S^{1112}x^3+3S^{1122}x^2y + 3S^{1222}xy^2 + S^{2222}y^3
\end{array}
$$
In these terms we have
$$
\frac{S^1(\vec x_1)S^2(\vec x_2) - S^2(\vec x_1)S^1(\vec x_2)}
{x_1y_2-x_2y_1} $$
\be
&=3\frac{x_1^3x_2^2y_2-x_1^2y_1x_2^3}{x_1y_2-x_2y_1}
\Big(S^{1111}S^{1122} - (S^{1112})^2\Big)\nn\\
&+\ 3\frac{x_1^3x_2y_2^2-x_1y_1^2x_2^3}{x_1y_2-x_2y_1}
\Big(S^{1111}S^{1222}-S^{1112}S^{1122}\Big)\nn\\
&+\
\frac{x_1^3y_2^3 - y_1^3x_2^3}{x_1y_2-x_2y_1}
\Big(S^{1111}S^{2222}-S^{1112}S^{1222}\Big)\nn\\
&+ \ 9\frac{x_1^2y_1x_2y_2^2 - x_1y_1^2x_2^2y_2}{x_1y_2-x_2y_1}
\Big(S^{1112}S^{1222} - (S^{1122})^2\Big)\nn\\
&+\ 3\frac{x_1^2y_1y_2^3 - y_1^3x_2^2y_2}{x_1y_2-x_2y_1}
\Big(S^{1112}S^{2222}-S^{1122}S^{1222}\Big) \nn\\
&+3\frac{x_1y_1^2y_2^3 - y_1^3x_2y_2^2}{x_1y_2-x_2y_1}
\Big(S^{1122}S^{2222}-(S^{1222})^2\Big) \nn\\
&= \frac{1}{16}\Big(x_1^2x_2^2 M_{12}
+\big(x_1^2(x_2y_2)+(x_1y_1)x_2^2\big) M_{13}\nn\\
&+\big(x_1^2y_2^2+(x_1y_1)(x_2y_2) + y_1^2x_2^2\big)M_{14}
+(x_1y_1)(x_2y_2)M_{23}\nn\\
&+\big((x_1y_1)y_2^2 + y_1^2(x_2y_2)\big)M_{24}
+y_1^2y_2^2 M_{34}\Big) =
\ee
\be
= \frac{1}{16}\ \Big(\begin{array}{ccc}
x_1^2 & x_1y_1 & y_1^2\end{array}\Big)
\left(\begin{array}{ccc}
M_{12} & M_{13} & M_{14} \\
M_{13} & M_{23}+M_{14} & M_{24} \\
M_{14} & M_{24} & M_{34}
\end{array}\right)
\left(\begin{array}{c}
x_2^2 \\ x_2y_2 \\ y_2^2
\end{array}
\right)
\label{disc4}
\ee
The $3\times 3$ matrix here is exactly $B(S)$, and
it is made from $M_{ij}$, $1\leq i<j\leq 4$,
$$
\begin{array}{c}
M_{12} = 16\cdot 3\Big(S^{1111}S^{1122} - (S^{1112})^2\Big) =
16\cdot 3\Big(\frac{ac}{6}-\frac{b^2}{16}\Big) = 8ac-3b^2\\
M_{13} = 16\cdot 3\Big(S^{1111}S^{1222}-S^{1112}S^{1122}\Big) =
16\cdot 3\left(\frac{ad}{4} - \frac{bc}{24}\right) = 12ad -2bc \\
M_{14} = 16\cdot\Big(S^{1111}S^{2222}-S^{1112}S^{1222}\Big) =
16\cdot \left(ae - \frac{bd}{16}\right) = 16ae-bd\\
M_{23} = 16\cdot9 \Big(S^{1112}S^{1222} - (S^{1122})^2\Big) =
16\cdot9\left(\frac{bd}{16} - \frac{c^2}{36}\right) = 9bd - 4c^2\\
M_{24} = 16\cdot 3\Big(S^{1112}S^{2222}-S^{1122}S^{1222}\Big) =
16\cdot 3\left(\frac{be}{4} - \frac{cd}{24}\right) = 12be-2cd\\
M_{34} = 16\cdot 3\Big(S^{1122}S^{2222}-(S^{1222})^2\Big) =
16\cdot 3\Big(\frac{ce}{6}-\frac{d^2}{16}\Big) = 8ce-3d^2
\end{array}
$$
which are minors of rectangular $2\times 4$ matrix
$$
\left(\begin{array}{c}
\partial_x S \\ \partial_y S
\end{array}\right) =
\left(\begin{array}{cccc}
4a & 3b & 2c & d \\ b & 2c & 3d & 4e
\end{array}\right)
$$
Three shifted copies of this rectangular matrix form
the $6\times 6$ matrix (\ref{ordresdetrep}) for
\be
&D_{2|4}(S) = Res(\partial_x S,\partial_y S)
= \epsilon^{ijklmn}M_{ij}\tilde M_{kl}\tilde{\tilde M}_{mn}\nn\\
&=\epsilon^{ijklmn}M_{ij}M_{k-1,l-1}M_{m-2,n-2}
= M_{12}\Big(\tilde M_{34}\tilde{\tilde M}_{56}
- \tilde M_{35}\tilde{\tilde M}_{46} +
\tilde M_{45}\tilde{\tilde M}_{36}\Big) \nn\\
&-M_{13}\Big(\tilde M_{24}\tilde{\tilde M}_{56} -
\tilde M_{25}\tilde{\tilde M}_{46}\Big) +
M_{14}\Big(\tilde M_{23}\tilde{\tilde M}_{56} -
\tilde M_{25}\tilde{\tilde M}_{36}\Big) \nn\\
&= M_{12}M_{23}M_{34} - M_{12}M_{24}^2 + M_{12}M_{34}M_{14}\nn\\
&- M_{13}^2M_{34} + M_{13}M_{14}M_{24} + M_{14}M_{12}M_{34}
- M_{14}^3
\ee
In the second line we used the fact that
$M_{\cdot 5} = M_{\cdot 6} = \tilde M_{1\cdot} =
\tilde M_{\cdot 6} = \tilde{\tilde M}_{1\cdot} =
\tilde{\tilde M}_{2\cdot} = 0$.
This result differs from $\det_{3\times 3} B$,
expressed through $M$, by a Plucker identity
$M_{12}M_{34} - M_{13}M_{24} + M_{14}M_{23}\equiv 0$
identically satisfied by minors of any $2\times 4$ matrix.
When all minors $M_{ij}=0$, i.e. $corank(B)=3$,
all the four roots of $S(x,y)$ coincide,
$\lambda_1=\lambda_2=\lambda_3=\lambda_4$.
When $corank(B) = 2$, i.e. all the six different $2\times 2$
minors of $B$ vanish, then either some three roots of $S$
coincide (say, $\lambda_1=\lambda_2=\lambda_3$)
{\it or} the four roots coincide pairwise
(say, $\lambda_1=\lambda_2$ and $\lambda_3=\lambda_4$).

$\bullet$ According to (\ref{discpol}), equation $D_{2|r}(S)=0$
defines the condition for merging of a pair out of $r$ roots. The so
defined discriminantal space has codimension one in
$r$-dimensional space ${\cal M}_{2|r}$. There are also
discriminantal spaces of higher codimension, consisting of
polynomials with bigger sets of coincident roots. We label these
spaces by $D_{2|r}[k_1,k_2,\ldots]$, implying that some $k_1$
roots coincide, then some other $k_2$ roots coincide, but differ
from the first group and so on, so that \ ${\rm codim}\
D_{2|r}[k_1,k_2,\ldots] = (k_1-1)+(k_2-1) + \ldots$. Obviously
$k_1 + k_2 + \ldots = r$. We add an agreement that $k_1\geq k_2
\geq \ldots$. According to this definition, the entire ${\cal
M}_{2|r} = D_{2|r}[1,1,\ldots]$, the ordinary discriminant
vanishes on $D_{2|r}[2,1,1,\ldots]$ of codimension $1$ and so on.

\PFig{480px-JuliusPlucker}
{150,187}
{Julius Pl\"ucker (1801 -- 1868)}

$\bullet$ At the same time entire tensor algebra (diagram) routine
is applicable to the particular case of $2|r$ tensors. The
structure group is $SL(2)$, all diagrams are representations of
this algebra: singlet if diagram has no external legs and a
combination of spins from zero to $\frac{1}{2}k$ if the number of
legs is $k$. A common zero of all elements of any representation
forms invariant subspace in ${\cal M}_{2|r}$, in particular,
invariant subspaces arise if a diagram is requested to vanish.
Discriminantal spaces are examples of invariant spaces, and there
is an obvious question of what is the correspondence between
diagrams and representations on one side and discriminantal spaces
on the other. Since the moduli space ${\cal M}$ has finite
dimension, there are only a few {\it algebraically independent}
diagrams and another question is to describe algebraic
(Plucker-like) relations among them and to define a generating
basis in the space of diagrams. A simplest example of this kind of
questions is to find the generators of the ring of {\it
invariants}, consisting of diagrams without external legs. This
ring contains the discriminant $D_{2|r}$ and one can also wonder
how it is constructed from the generators and what kind of
diagrammatic representations are available for it.

$\bullet$ A simple example of a general relation between diagrams
and roots is provided by the well known Vieta formula, \be
S_{J_1\ldots J_r} \equiv \epsilon_{J_1K_1}\ldots
\epsilon_{J_rK_r}S^{K_1\ldots K_r} \cong
\frac{1}{r!}\sum_{P\in\sigma_r} \prod_{k=1}^r X^{P(k)}_{J_k}
\label{Vietaf} \ee which is shown in Fig.\,\,\ref{Vietdia} and becomes
an obvious and direct generalization of
eqs.(\ref{linsol})-(\ref{linsolinv}) for inverse matrices, see
s.\ref{CraVie}. In (\ref{Vietaf}) $X_k$ are the roots $t_k$ of the
polynomial $S(t)$, written in homogeneous coordinates, and the
sign $\cong$ denotes projective equivalence, since normalizations
of projective solutions $X_k$ is not specified.

$\bullet$
Transformation of discriminant $D_{2|r}$ under
non-linear change of variables
$$
F:\ \ \left(\begin{array}{c}x\\y\end{array}\right)
\ \longrightarrow\
\left(\begin{array}{c}f(x,y)\\g(x,y)\end{array}\right)
$$
is more involved than a similar transformation
of a resultant, described by (\ref{composres}) in s.3.1.4.
If $S(x,y) = \prod_{i=1}^r (x-\xi_i y)$ is a
homogeneous polynomial of degree $r$ with $r$ roots
$\xi_1,\ldots,\xi_r$, then the transformation $F$
with homogeneous $f(x,y)$ and $g(x,y)$ of degrees $p$
converts it into a homogeneous polynomial
$S\circ F(x,y) = S\big(f(x,y),g(x,y)\big)$ of degree $r\!p$.
The following decomposition formula holds for discriminant
of this new polynomial:
\be
D_{2|r\!p}\,(S\circ F) = \Big(D_{2|r}(S)\Big)^p
\Big(R_{2|p}\,(F)\Big)^{r(r-1)}
C_{2|r|p}\,(S,F)
\ee
and the last factor is an {\it reducible}
polynomial in coefficients of $S$
and $F$ of degrees $2(p-1)$ and $2r(p-1)$ respectively.
If expressed through the roots of $S$ it can be decomposed
into a product of $r$ resultants:
\be
C_{2|r|p}\,(S,F) =
\prod_{i=1}^p D_{2|p}\,\Big( f(x,y)-\xi_i\,g(x,y)\Big)
\label{ditradeco}
\ee
For example, if $r=2$ and $p=1$, i.e. $S(x,y) = ax^2+bxy+cy^2$ and
$F = \left(\begin{array}{c}
\alpha x+\beta y \\ \gamma x + \delta y
\end{array}\right)$,
then the $C$-factor equals unity, since it is discriminant
of a linear function, and
$$
D_{2|2}\Big(a(\alpha x+\beta y)^2 + b(\alpha x + \beta y)
(\gamma x + \delta y) + c(\gamma x + \delta y)^2\Big)
$$ $$
= \Big(2a\alpha\beta + b(\alpha\delta+\beta\gamma) +
2c\gamma\delta\Big)^2 - 4\Big(a\alpha^2+b\alpha\gamma +
c\gamma^2\Big)\Big(a\beta^2+b\beta\delta+c\delta^2\Big)
$$ $$
=(b^2-4ac)(\alpha\delta-\beta\gamma)^2 =
D_{2|2}(ax^2+bxy+cy^2)R_{2|1}\left(
\begin{array}{c} \alpha x + \beta y\\ \gamma x + \delta y
\end{array}\right)^2
$$
A slightly more interesting is example with $r=2$ and $s=2$:
\be
&D_{2|4}\Big(a(\alpha_1 x^2+\beta_1 xy + \gamma_1 y^2)^2 \nn\\
&+b(\alpha_1 x^2+\beta_1 xy + \gamma_1 y^2)
(\alpha_2 x^2+\beta_2 xy + \gamma_2 y^2)
+ c(\alpha_2 x^2+\beta_2 xy + \gamma_2 y^2)^2\Big)\nn\\
&=(b^2-4ac)^2
\Big((\alpha_1\gamma_2-\gamma_1\alpha_2)^2
+(\alpha_1\beta_2-\beta_1\alpha_2)
(\gamma_1\beta_2-\beta_1\gamma_2)\Big)^2
\nn\ee
$$
\cdot\,\Big\{
a^2(4\alpha_1\gamma_1-\beta_1^2)^2\
+\ b^2(4\alpha_1\gamma_1-\beta_1^2)(4\alpha_2\gamma_2-\beta_2^2)
\ +\ c^2(4\alpha_2\gamma_2-\beta_2^2)^2
$$ $$
+2ab\Big(
8(\alpha_1^2\gamma_1\gamma_2 + \alpha_1\alpha_2\gamma_1^2)
-4\alpha_1\beta_1\gamma_1\beta_2
-2(\beta_1^2\gamma_1\alpha_2 + \alpha_1\beta_1^2\gamma_2)
+\beta_1^3\beta_2
\Big)
$$ $$
+2ac\Big(
8(\alpha_1^2\gamma_2^2 - \alpha_1\beta_1\beta_2\gamma_2
- \gamma_1\beta_1\beta_2\alpha_2 + \gamma_1^2\alpha_2^2)
+4\beta_1^2\alpha_2\gamma_2
+4\alpha_1\gamma_1\beta_2^2
+\beta_1^2\beta_2^2\Big)
$$ $$
+2bc \Big(
8(\alpha_1\alpha_2\gamma_2^2 +\gamma_1\gamma_2\alpha_2^2)
-4\beta_1\alpha_2\beta_2\gamma_2
-2(\alpha_1\beta_2^2\gamma_2+\gamma_2\beta_2^2\gamma_1)
+\beta_1\beta_2^3\Big)
\Big\}
$$
\be
\hspace{-6mm}= \Big(D_{2|2}(ax^2+bxy+cy^2)\Big)^2R_{2|2}\left(
\begin{array}{c} \alpha_1 x^2+\beta_1 xy + \gamma_1 y^2\\
\alpha_2 x^2+\beta_2 xy + \gamma_2 y^2
\end{array}\right)^2 C_{2|2|2}
\label{ditra22exa}
\ee
On the other hand, if we express the polynomial
$S(x,y) = ax^2+bxy+cy^2 = a(x-y\xi_+)(x-y\xi_i)$
through roots, then the same formula converts into
\be
&D_{2|4}\Big(a
\big[(\alpha_1 x^2+\beta_1 xy + \gamma_1 y^2) -
(\alpha_2 x^2+\beta_2 xy + \gamma_2 y^2)\xi_+\big]\nn\\
&\big[(\alpha_1 x^2+\beta_1 xy + \gamma_1 y^2) -
(\alpha_2 x^2+\beta_2 xy + \gamma_2 y^2)\xi_i\big]\Big)
\nn\ee
$$= \Big\{a^4(\xi_+-\xi_-)^4\Big\}
\Big((\alpha_1\gamma_2-\gamma_1\alpha_2)^2
+(\alpha_1\beta_2-\beta_1\alpha_2)
(\gamma_1\beta_2-\beta_1\gamma_2)\Big)^2
$$ $$
\cdot\
a\Big(\xi_+^2(\beta_2^2 - 4\alpha_2\gamma_2)
+ 2\xi_+(2\alpha_1\gamma_2+2\gamma_1\alpha_2-\beta_1\beta_2)
+(\beta_1^2-4\alpha_1\gamma_1)\Big)
$$ $$
\cdot\ a\Big(\xi_-^2(\beta_2^2 - 4\alpha_2\gamma_2)
+ 2\xi_-(2\alpha_1\gamma_2+2\gamma_1\alpha_2-\beta_1\beta_2)
+(\beta_1^2-4\alpha_1\gamma_1)\Big)
$$
We see that the factor $C_{2|2|2}$, represented by a polynomial
in curved brackets in (\ref{ditra22exa}), is now decomposed
into a product of two expressions in the last two lines,
which are equal to
\be
&D_{2|2}\Big(
(\alpha_1 x^2+\beta_1 xy + \gamma_1 y^2) -
(\alpha_2 x^2+\beta_2 xy + \gamma_2 y^2)\xi\Big) \nn\\
&=a\Big(\xi^2(\beta_2^2 - 4\alpha_2\gamma_2)
+ 2\xi(2\alpha_1\gamma_2+2\gamma_1\alpha_2-\beta_1\beta_2)
+(\beta_1^2-4\alpha_1\gamma_1)\Big)
\nn\ee
with $\xi = \xi_\pm$, according to (\ref{ditradeco}).
The origin of this decomposition is also obvious in this example:
discriminant of a product of polynomials is naturally decomposed
into a product of discriminants and resultants.

$\bullet$ Since discriminant $D_{2|r}(S)$ and other objects of
interest for us are $SL(2)$ invariant, and diagrams with external
legs form $SL(2)$ representations, it is important to know how
$SL(2)$ acts in ${\cal M}_{2|r}$. Originally $SL(2)$ acts as a
linear transformation of the vector space $V_2 = \{x_1,x_2\}$, and
three independent generators of infinitesimal transformations,
associated with
$$
\left( \begin{array}{c} x_1 \\ x_2 \end{array}\right)
\longrightarrow \left(\begin{array}{c} x_1 \\ x_2
\end{array}\right) + \varepsilon_1\left(\begin{array}{c} x_2 \\ 0
\end{array}\right) + \varepsilon_2\left(\begin{array}{c} x_1 \\
-x_2 \end{array}\right) + \varepsilon_3 \left(\begin{array}{c} 0
\\ x_1 \end{array}\right),
$$
act on the coefficients of $S_r(xi)$ by vector fields:
\be
&\hat\nabla_1 = \sum_{k=0}^{r-1}
(k+1)s_{k+1}\frac{\partial}{\partial s_k}, \ \ \ \hat\nabla_2 =
\sum_{k=0}^r (2k-r) s_k\frac{\partial}{\partial s_k}, \ \ \nn\\
&\hat\nabla_3 = \sum_{k=1}^r (r+1-k)
s_{k-1}\frac{\partial}{\partial s_k} \label{SL2gens} \ee
Invariants\ ${\rm Inv}(S)$\ are annihilated by these operators:
$\hat\nabla {\rm Inv}(S) = 0$, while the $2J+1$ elements of an
irreducible representation with spin $J$ are transformed through
themselves.

\subsubsection{The $n|r = 2|2$ case \label{sypo22}}

This example was already considered in s.\ref{matr|2}, now we add
some more details.

$\bullet$ The relevant polynomial $S_{2|2}(\xi) = S^{11}\xi^2 +
2S^{12}\xi + S^{22} = s_2\xi^2 + s_1\xi + s_0$ is quadratic,
projective moduli space ${\cal M}_{2|2}$ is $2$-dimensional (the
three coefficients $S$ are homogeneous coordinates in it),
discriminant
$$
D_{2|2} = \frac{1}{4}s_1^2 - s_2s_0 = \left(\begin{array}{cc}
S^{11} & S^{12} \\ S^{12} & S^{22} \end{array}\right),
$$
and in discriminantal subspace $D_{2|2}(S)=0$ of codimension one
in ${\cal M}_{2|2}$ the two roots $\xi_\pm =
\frac{1}{S^{11}}(-S^{12} \pm \sqrt{D})$ coincide. In homogeneous
coordinates these two roots can be represented as elements of
$V_2$:
$$
\left(\begin{array}{c} X_1^\pm \\ X_2^\pm \end{array}\right) =
 \left(\begin{array}{c} -S^{12} \pm \sqrt{D} \\ S^{11}
 \end{array}\right)\ \ \ \ {\rm or,\ alternatively}\ \ \
\left(\begin{array}{c} X_1^\pm \\ X_2^\pm \end{array}\right) =
 \left(\begin{array}{c} S^{22} \\ -S^{12} \mp \sqrt{D} \end{array}\right)
$$
the ratios $\xi_\pm = X_1^\pm/X_2^\pm$ are the same in both cases.

$\bullet$ The structure group $SL(2)$ acts on homogeneous
coordinates, i.e. in the space $V_2$, as follows:
$$
\left(\begin{array}{c} X_1 \\ X_2 \end{array}\right)
\longrightarrow \left(\begin{array}{c} X^\prime_1 \\ X^\prime_2
\end{array}\right) =
\left(\begin{array}{cc} e^{i\alpha} & 0 \\ 0 & e^{-i\alpha}
\end{array}\right)
\left(\begin{array}{cc} \cos\frac{\beta}{2} &
\sin\frac{\beta}{2} \\
-\sin\frac{\beta}{2} & \cos\frac{\beta}{2} \end{array}\right)
\left(\begin{array}{cc} e^{i\gamma} & 0 \\ 0 & e^{-i\gamma}
\end{array}\right)
$$
\be = \left(\begin{array}{c} e^{i(\alpha-\gamma)}
\cos\frac{\beta}{2} X_1 +
 e^{i(\alpha+\gamma)}\sin\frac{\beta}{2} X_2 \\
 e^{-i(\alpha+\gamma)}\sin\frac{\beta}{2} X_1 +
 e^{-i(\alpha-\gamma)}\cos\frac{\beta}{2} X_2  \end{array}\right)
\label{Xtrans} \ee i.e. $(X_1,X_2)$ form a spinor representation
(spin $1/2$). If $\alpha,\beta,\gamma$ are real, this formula
defines the action of the maximal compact subgroup $SU(2) \subset
SL(2|C)$. The action on ${\cal M}_{2|2}$ is defined from
invariance of the tensor $S(X) = S^{11}X_1^2 + 2S^{12}X_1X_2 +
S^{22}X_2^2$, so that covariant symmetric tensor transforms as
representation of spin $1$. It is a little more practical to
define inverse transformation from $(S^\prime)^{ij} X_iX_j =
S^{ij} X^{\prime}_i X^{\prime}_j$, i.e. $(S^\prime)^{ij} =
S^{kl}U_k^iU_l^j$ if $X^{\prime}_k = U_k^iX_i$.

$\bullet$ Invariant measure on the maximal compact subgroup
$SU(2)$ is $d\Omega = \sin\beta d\alpha\wedge d\beta\wedge
d\gamma$. With the help of this measure we can build invariant of
the structure group by taking $SU(2)$ average of any $S$-dependent
expression:
$${\rm Inv}_f(S) = \int f\Big(S^\prime(\alpha,\beta,\gamma)\Big)
d\Omega(\alpha,\beta,\gamma)$$ Of course, for arbitrary choice of
$f$ this integral vanishes. Actually, integrals over $\gamma$ and
$\alpha$ select monomials, which have equal number of $'1'$ and
$'2'$ indices in $f(S)$ and $f(S^\prime)$ respectively.

For example, at level two the integral of $S^{ij}S^{kl}$ is
non-vanishing only for $S^{11}S^{22}$ and $(S^{12})^2$ due to
$\gamma$-integration, while $\alpha$-integral picks up only terms
with $S^{11}S^{22}$ and $(S^{12})^2$ from $(S^{11})^\prime
(S^{22})^\prime(\alpha,\beta,\gamma)$ and
$\Big((S^{12})^\prime\Big)^2(\alpha,\beta,\gamma)$. Remaining
$\beta$-integration is performed with the help of \be \int
\cos^{2k}\frac{\beta}{2} \sin^{2l}\frac{\beta}{2} \ \sin\beta\
d\beta = \frac{(k+l)!}{k!\ l!} \ee

In fact, all non-vanishing averages in the $2|2$ case are
functions of discriminant $D_{2|2}$: the ring of invariants is
generated by a single generator $D_{2|2}$ and this is in intimate
relation with the fact that there is a single discriminantal space
in this case.

$\bullet$ Now we switch to classification of diagrams, see
Fig.\,\,\ref{Vieta22dia}. According to Vieta formula
(Fig.\,\,\ref{Vieta22dia}.A) \be S_{ij} = \epsilon_{ii'}\epsilon_{jj'}
S^{i'j'} = \frac{1}{2}(X^+_iX^-_j + X^-_iX^+_j) \label{S2|2} \ee
Then (Fig.\,\,\ref{Vieta22dia}.B) \be S_{ik}S^{jk} =
\frac{1}{4}(X^-_iX_+^k - X^+_iX_-^k)(X^+_jX_-^j) =
\frac{1}{2}(X^-_iX_+^k - X^+_iX_-^k)\sqrt{D}, \label{SS2|2} \nn\\\ee
where the scalar in brackets $\epsilon^{ij}X_i^+X_j^- =
2\sqrt{D}$. The next diagram, Fig.\,\,\ref{Vieta22dia}.C is given by
$$
S_{ij}S^{jk}S_{kl} = DS_{il}
$$
i.e. coincides with disconnected one in Fig.\,\,\ref{Vieta22dia}.D
Closed loop of length $2k$ is $D^{k}$, all such loop vanish on
discriminantal subspace. This completes classification of
non-trivial diagrams in $2|2$ case.

\Fig{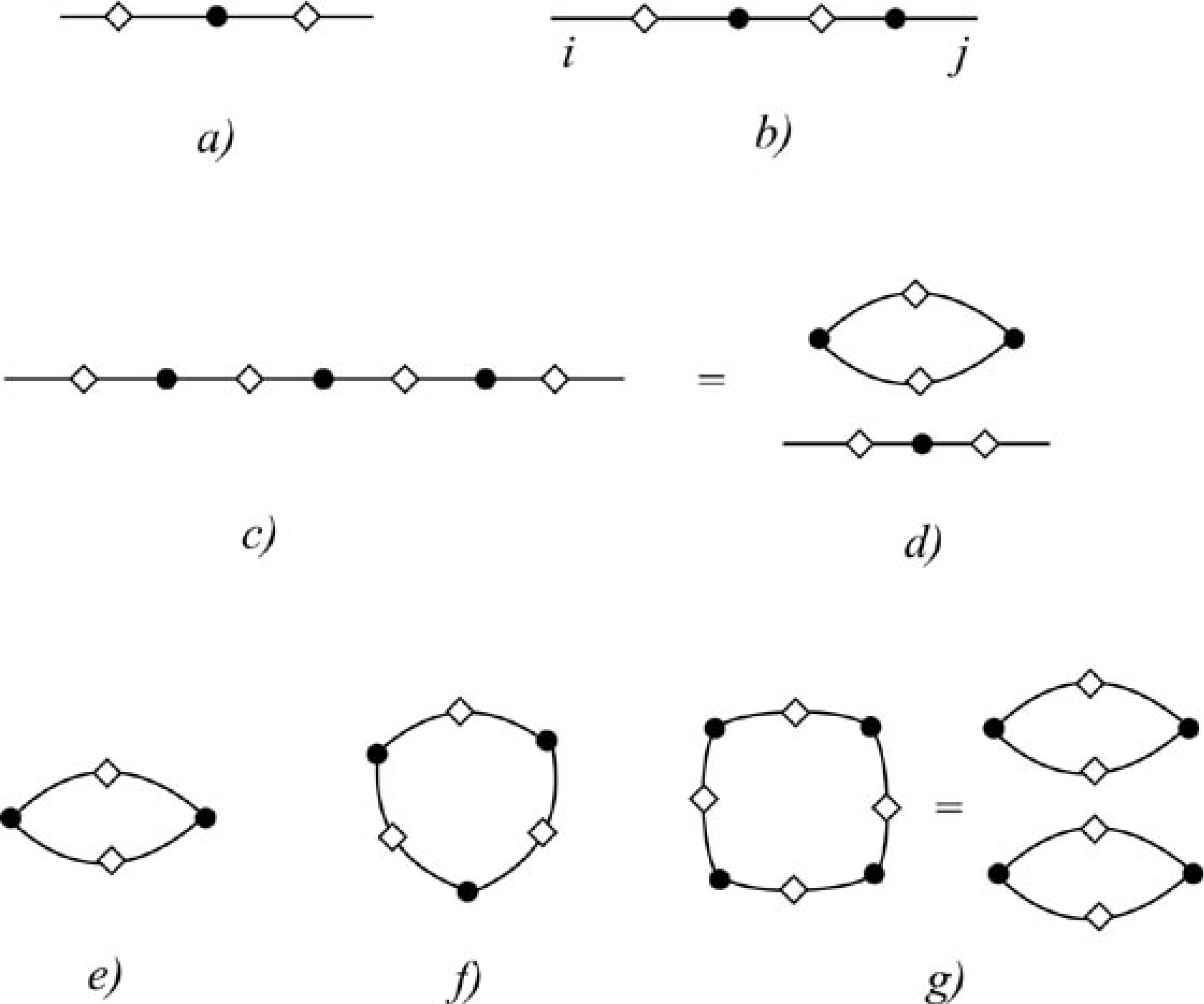}
{320,267}
{The ${\cal T}(T)$ algebra in the $n|s=2|2$ case: it has three
algebraically independent elements, shown in {\bf A}, {\bf B} and
{\bf E}. Other possible diagrams are expressed through these
three. Diagram {\bf F} and all other circles of odd length are
vanishing.}

$\bullet$ The three differential equations for invariants of the
structure group $SL(2)$, can be written down explicitly with the
help of (\ref{SL2gens}): \be
\left(\begin{array}{ccc} 0 & 2s_2 & s_1 \\
s_2 & 0 & -s_0 \\  s_1 & 2s_0 & 0 \end{array}\right)
\left(\begin{array}{c} \frac{\partial{\rm Inv}}{\partial s_2} \\ \\
\frac{\partial{\rm Inv}}{\partial s_1} \\ \\
\frac{\partial{\rm Inv}}{\partial s_0} \end{array} \right) = 0
\label{inveq22} \ee Determinant of the square matrix is zero, its
rank is $2$, equation for\ ${\rm Inv}(S)$\ reduces to
$\Big(2s_2\partial_1 + s_1\partial_0\Big){\rm Inv} = 0$, which is
solved by the method of characteristics: solution to
$\Big(v_x(x,y)\partial_x + v_y(x,y)\partial_y\Big)C(x,y) = 0$ is
provided by integration constant $C$ in solution of the ordinary
differential equation $dy/dx = -v_y(x,y)/v_x(x,y)$. Therefore
${\rm Inv}(S) = {\rm any\ function\ of}\ (s_1^2-4s_0s_2)$, i.e.
the ring of invariants is generated by discriminant $D_{2|2}(S) =
s_0s_2 - \frac{1}{4}s_1^2 = S^{11}S^{22} - (S^{12})^2$.

$\bullet$ A list of invariants and their relations in the case of
$n|r=2|2$ is collected in the table. Columns are labeled by
different types of covariant tensors (at rank $r=2$ there are
three: generic $T^{ij}$, symmetric $S^{ij} = S^{ji}$ and
antisymmetric $C^{ij} = -C^{ji}$) and rows -- by degree of
invariant : Tensors are represented by $2\times 2$ matrix with
entries $\left(\begin{array}{cc} a_{11} & a_{12} \\ a_{21} &
a_{22}
\end{array}\right)$, in symmetric case $a_{12} = a_{21}$,
in antisymmetric case $a_{11}=a_{22}=0$ and $a_{12} = -a_{21}$. In
this particular case of\ $n|r=2|2$\ entire invariant's ring for
all types of tensors is generated by discriminants.

\begin{tabular}{|c||c||c|c|}
\hline
&&&\\
$n|r = 2|2$ & polylinear & symmetric & antisymmetric \\
&&&\\
\hline
&&&\\
1 & -- & -- & Pfaffian = $a_{12}-a_{21}$ \\
&&&\\
2 & determinant = $a_{11}a_{22}-a_{12}a_{21}$ & determinant =
$a_{11}a_{22} - a_{12}^2$ &
${\rm determinant}\ =\ \frac{1}{4}{\rm Pfaff}^2$\\
&&&\\
3 & -- & -- & ${\rm Pfaff}^3$ \\
&&&\\
4 & ${\rm determinant}^2$ & ${\rm determinant}^2$ &
${\rm Pfaff}^4$ \\
&&&\\
5 & -- & -- & ${\rm Pfaff}^5$ \\
&&&\\
   &&\ldots & \\
&&&\\
\hline
\end{tabular}

$\bullet$
A resultant of two quadratic polynomials,
$A = \sum_{i,j=1}^2 A^{ij}x_i x_j$ and
$B = \sum_{i,j=1}^2 B^{ij}x_i x_j$ is proportional to the
difference of two circular diagrams
Fig.\,\,\ref{resu2quad}.C and \ref{resu2quad}.C
\be
R_{2|2}\{A,B\} \sim
A^{ij}\epsilon_{jk}B^{kl}\epsilon_{lm}A^{mn}
\epsilon_{np}B^{pq}\epsilon_{qi} -
A^{ij}\epsilon_{jk}A^{kl}\epsilon_{lm}B^{mn}
\epsilon_{np}B^{pq}\epsilon_{qi}\nn\\
\label{res2quad}
\ee
The second of these diagrams is twice the product of
two determinants, $2\det A\det B$, see Fig.\,\,\ref{resu2quad}.D.
When $A=B$ the two diagrams coincide, and their difference
vanishes, as should happen for the resultant of two
identical polynomials.

Two other pictures,
Fig.\,\,\ref{resu2quad}.A and Fig.\,\,\ref{resu2quad}.B,
show respectively the resultants of two linear polynomials,
$A = \sum_{i=1}^2 A^ix_i$, $B = \sum_{i=1}^2 B^ix_i$,
then
\be
R_{2|1}\{A,B\} = \epsilon_{ij} A^iB^j,
\label{res2lin}
\ee
and of one linear, $A = \sum_{i=1}^2 A^ix_i$,
another quadratic $B = \sum_{i,j=1}^2 B^{ij}x_i x_j$, then
\be
R_{1\times 2}\{A,B\} = \epsilon_{ik}\epsilon_{jl} A^iA^jB^{kl},
\label{res2linquad}
\ee

\Fig{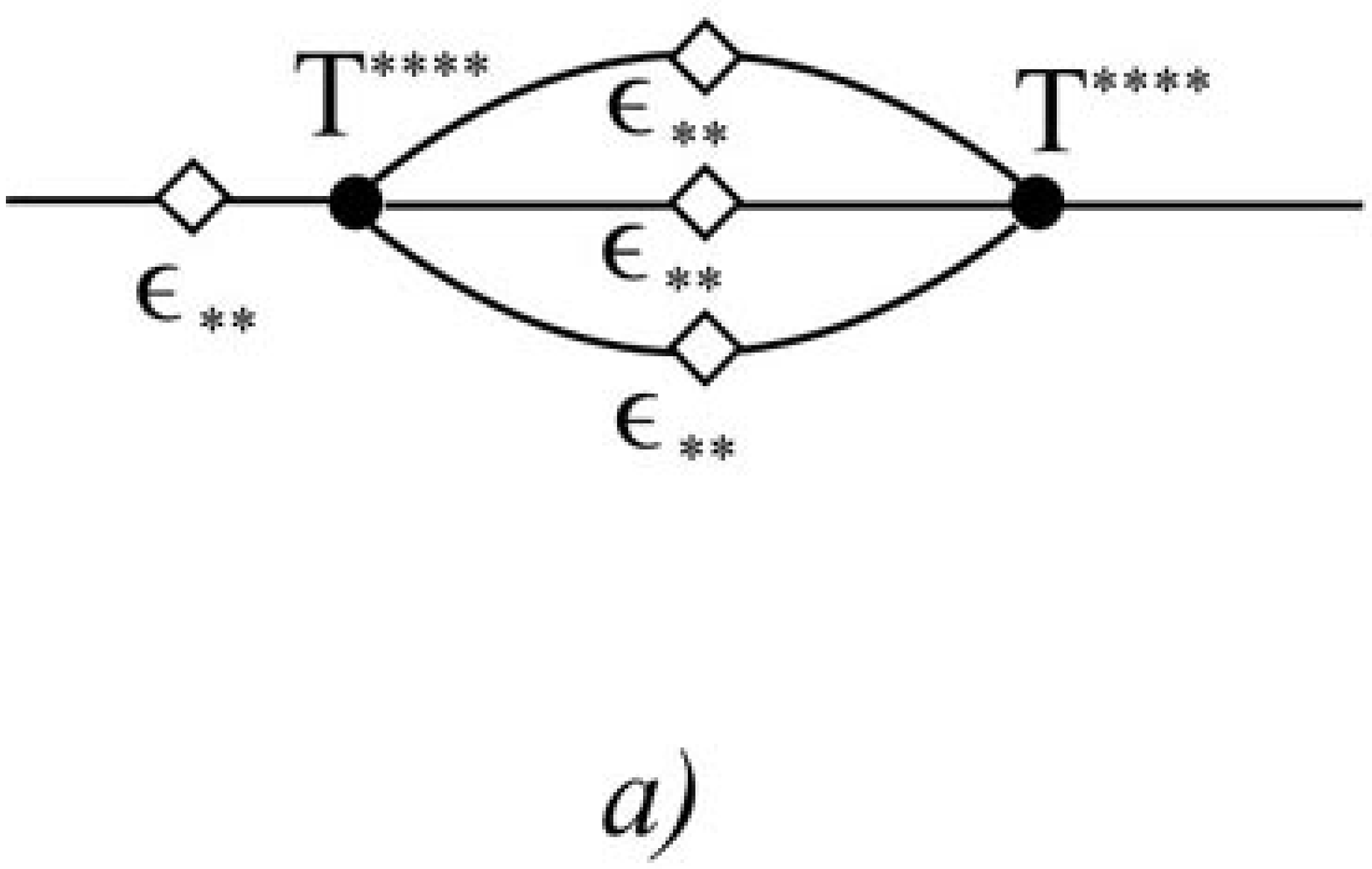}
{200,200}
{Diagrammatic representation of the simplest resultants
of two polynomials $A$ and $B$.
{\bf A.} Both $A$ and $B$ are linear, eq.\,\,(\ref{res2lin}).
{\bf B.} $A$ is linear and $B$ is quadratic,
eq.\,\,(\ref{res2linquad}).
{\bf C, D.} Two diagrams, contributing when both
$A$ and $B$ are quadratic, eq.\,\,(\ref{res2quad}).
The second diagram is actually twice a product of two
determinants.}

\subsubsection{The $n|r = 2|3$ case
\label{23resulta}}

We began consideration of this case in s.\ref{cayh}, now we add
some more details.

$\bullet$ From Vieta formula \be S_{ijk} =
\epsilon_{ii'}\epsilon_{jj'}\epsilon_{kk'}S^{i'j'k'} =
\frac{1}{6}\sum_{P\in\sigma_3}X_i^{P(1)}X_j^{P(2)}X_k^{P(3)}, \ee
where $\vec X^1,\vec X^2,\vec X^3$ are the three roots of cubic
polynomial $S(t) = S^{111}t^3 + 3S^{112}t^2 + 3S^{122}t + S^{222}$
in homogeneous coordinates (superscripts are not powers!). Then
(Fig.\ref{Vieta22dia}) \be B_{ij} = S_{imn}S_j^{mn} =
-\frac{1}{18} \Big\{X^1_iX^1_j (X^{23})^2 + X^2_iX^2_j (X^{13})^2
+ X^3_iX^3_j (X^{12})^2 + \nn \\ + (X^1_iX^2_j + X^2_iX^1_j)
X^{13}X^{23} + (X^2_iX^3_j + X^2_iX^3_j) X^{12}X^{13} +
(X^1_iX^3_j + X^3_iX^1_j) X^{12}X^{32}\Big\} \label{Bij23} \ee
where $X^{ab} = \epsilon^{ij}X^a_iX^b_j = -X^{ba}$. If two roots
coincide, say, $\vec X^2 = \vec X^3$, then $X^{23}=0$, $X^{12} =
X^{13}$ and $B_{ij} = -\frac{2}{9}X^2_iX^2_j(X^{12})^2$. When all
the three roots coincide, $\vec X^1 = \vec X^2 = \vec X^3$,
$B_{ij} = 0$. The square $B^{ij}B_{ij} \sim
(X^{12})^2(X^{13})^2(X^{23})^2 = D_{2|3}$ is equal to discriminant
of $S$ and vanishes whenever any two roots of $S(t)$ coincide.

$\bullet$ The analogue of (\ref{inveq22}) now is a little more
complicated: \be
\left(\begin{array}{cccc} 0 & 3s_3 & 2s_2 & s_1 \\
3s_3 & s_2 & -s_1 & -3s_0 \\  s_2 & 2s_1 & 3s_0 & 0
\end{array}\right)
\left(\begin{array}{c} \frac{\partial{\rm Inv}}{\partial s_3} \\ \\
 \frac{\partial{\rm Inv}}{\partial s_2} \\ \\
\frac{\partial{\rm Inv}}{\partial s_1} \\ \\
\frac{\partial{\rm Inv}}{\partial s_0} \end{array} \right) = 0
\label{inveq23} \ee Rectangular matrix has rank $3$ and there is
$4-3=1$ independent $SL(2)$ invariant -- discriminant, again the
ring of invariants is generated by $D_{2|3}$ -- symmetric version
(\ref{DiscS3}) of Cayley hyperdeterminant.




\subsubsection{The $n|r = 2|4$ case
\label{24resulta}}

\noindent

$\bullet$ The analogue of (\ref{inveq22}) and (\ref{inveq23}) is
now \be
\left(\begin{array}{ccccc} 0 & 4a_4 & 3s_3 & 2s_2 & s_1 \\
4a_4 & 2s_3 & 0 & -2s_1 & -4s_0 \\  s_3 & 2s_2 & 3s_1 & 4s_0 & 0
\end{array}\right)
\left(\begin{array}{c} \frac{\partial{\rm Inv}}{\partial s_4} \\ \\
\frac{\partial{\rm Inv}}{\partial s_3} \\ \\
\frac{\partial{\rm Inv}}{\partial s_2} \\ \\
\frac{\partial{\rm Inv}}{\partial s_1} \\ \\
\frac{\partial{\rm Inv}}{\partial s_0} \end{array} \right) = 0 \ee
Rectangular matrix has rank $3$ and there are $5-3=2$ independent
$SL(2)$ invariants: \be {\rm Inv}_2(S) =
S^{ijkl}S^{i'j'k'l'}\epsilon_{ii'}\epsilon_{jj'}
\epsilon_{kk'}\epsilon_{ll'} = 2 S^{1111}S^{2222} -
8S^{1112}S^{1222} + 6(S^{1122})^2 = 2\Big(s_0s_4 -
\frac{1}{2}s_1s_3 + \frac{1}{12}s_2^2\Big), \ee
$${\rm Inv}_3(S) = S_{ijkl}S_{i'j'mn}S_{k'l'm'n'}
\epsilon^{ii'}\epsilon^{jj'}
\epsilon^{kk'}\epsilon^{ll'}\epsilon^{ll'}\epsilon^{nn'} = $$ \be
= 6\Big(S^{1111}S^{1122}S^{2222} + 2S^{1112}S^{1122}S^{1222} -
S^{1111}(S^{1222})^2 - (S^{1112})^2S^{2222} -(S_{1122})^3\Big)-\nn\\
- 8\Big(S^{1111}(S^{1222})^2 + (S^{1112})^2S^{2222} -
2S^{1112}S^{1122}S^{1222}\Big) = s_0s_2s_4 + \frac{1}{4}s_1s_2s_3
- \frac{7}{8}s_0s_3^2 - \frac{7}{8}s_1^2s_4 - \frac{1}{36}s_2^3
\ee These are now the two generators of the ring of invariants.
Discriminant has power $6$ and is a difference \be D_{2|4} =
8\Big( \left( {\rm Inv}_2(S)\right)^3 - 6\left({\rm
Inv}_3(S)\right)^2\Big) \ee This time is not represented by a
single diagram, it is rather a sum of at least two. Note that,
since $S$ is symmetric, the lines {\it sorts} do not make sense,
and $\epsilon$'s are allowed to mix them. Accordingly, while ${\rm
Inv}_2(S)$ is a full $SL(2)^4$ invariant, the ${\rm Inv}_3(S)$ is
preserved by $SL(2)^2$ only. See Fig.\ref{Vieta24dia} for some
vacuum diagrams which describe invariants. Discriminant $D_{2|4}$
has $SL(2)^2$ invariance (though naively it is only $SL(2)$).

\Fig{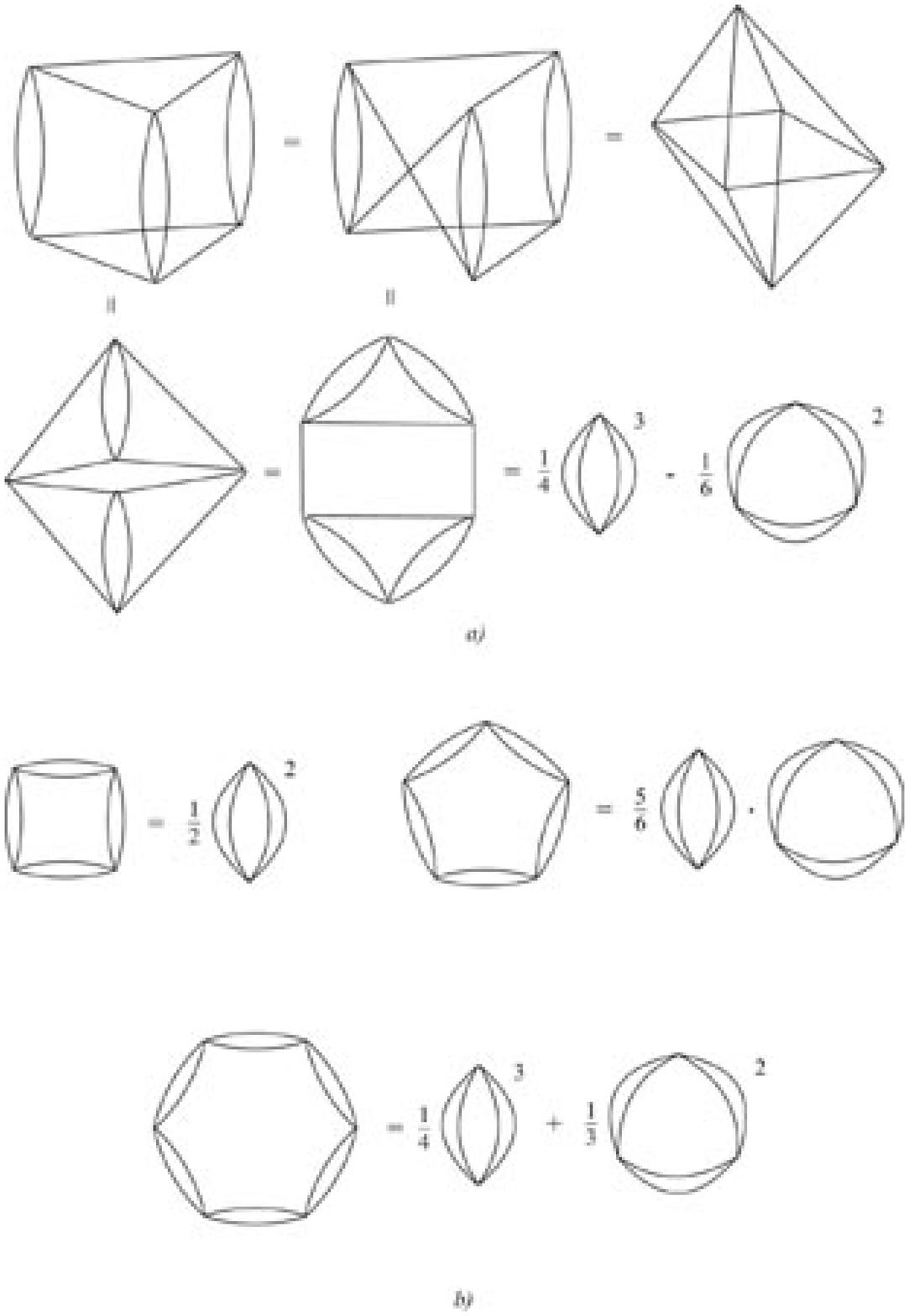} {491,577} {Examples of relations between vacuum
diagrams (relations in the ring of invariants) in the $n|s = 2|4$
case. Covariant rank-two $\epsilon$-tensors (white diamonds) in
the middles of all edges are not explicitly shown.\ {\bf A.} Many
$6$-vertex diagrams are equal, despite this does not follow from
symmetry properties alone: the linear space of invariants of
degree $6$ is $2$-dimensional.\ {\bf B.} A sequence of "sausage
diagrams" $I_m$, like all invariants they are polynomials of two
generators of invariant's ring, $I_2$ and $I_3$.}

$\bullet$ For {\it diagonal} tensor $S^{ijkl} =
a_i\delta^{ij}\delta^{ik}\delta^{il}$ $I_2={\rm Inv}_2(S) =
2a_1a_2$, $I_3={\rm Inv}_3(S) = 0$ and $D_{2|4} \sim (a_1a_2)^3$.
"Sausage diagrams" in Fig.\ref{Vieta24dia}.B are reduced to
$I_{2k}|_{{\rm diag}} = 2(a_1a_2)^k = 2^{1-k}I_2^k$,
$I_{2k+1}|_{{\rm diag}}=0$.

More informative is an intermediate case, when $S^{1112}=
s^{1222}=0$, but  $S^{1122}=\frac{1}{6}s_2\equiv B$ is
non-vanishing, along with $a_1=S^{1111}=s_0$ and
$a_2=S^{2222}=s_4$. The sausage diagrams are easily expressed
through $B$ and $A=S^{1111}S^{2222}$:
$$
\left.I_m\right|_{S^{1112}=S^{1222}=0} = 2\Big(B^m + C^2_m
AB^{m-2} + C^4_mA^2B^{m-4} + \ldots\Big) + \Big(-2B\Big)^m =
$$
\be = \left(B+\sqrt{A}\right)^m + \left(B-\sqrt{A}\right)^m +
\Big(-2B\Big)^m \label{saudred} \ee The first sum describes
diagrams with two lines in each sausage carrying coincident
indices $1$ or $2$ (Fig.\ref{24reddia}.A), the last term is given
by diagrams, where these indices are different
(Fig.\ref{24reddia}.B). Formula (\ref{saudred}) is enough to check
the relations between sausage diagrams in Fig.\ref{Vieta24dia}.B.
Discriminant
$$
\left.D_{2|4}\right|_{s_1=s_3=0} =
16s_0s_4\Big(4s_0s_4-s_2^2\Big)^2 = 64A\Big(A-9B^2\Big)^2
$$

\Fig{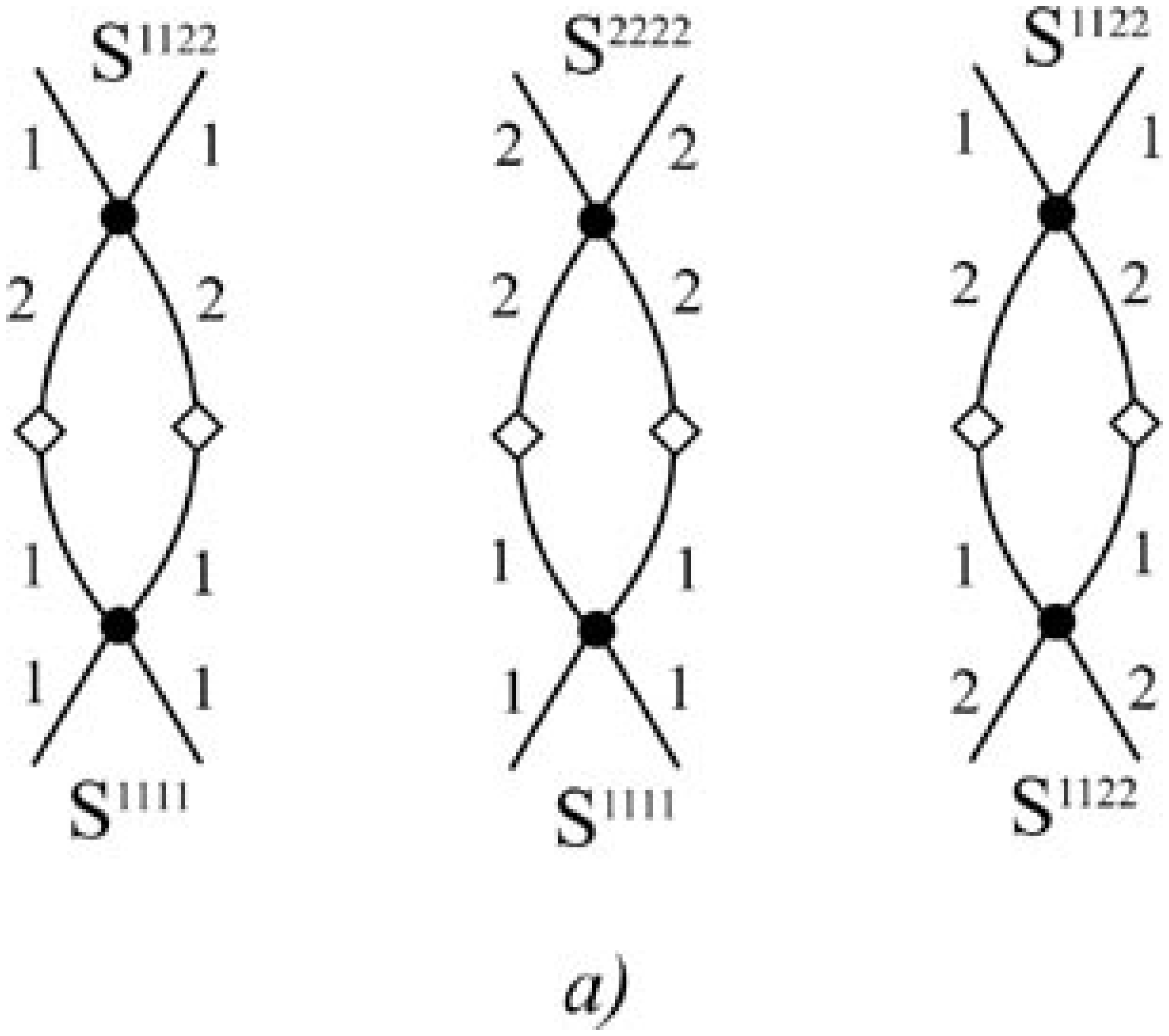} {450,248} {If $S^{1112}=S^{1222}=0$, all sausages
in a sausage diagram belong two one of two types (for
non-vanishing $S^{1112}$ and $S^{1222}$ both types of sausages can
show up in the same diagram).\ {\bf A.} Two lines in a sausage
carry the same indices. In this case vertices can contain any of
non-vanishing parameters $S^{1111}$, $S^{1122}$ and $S^{2222}$.
Diagrams of this type give rise to the first term in
eq.(\ref{saudred}).\ {\bf B.} Two lines in a sausage carry
different indices. In this case vertices can contain only
$S^{1122}$. Diagrams of this type give rise to the second term in
eq.(\ref{saudred}).}

$\bullet$ The ring of invariants for $n|r=2|4$ is isomorphic to
commutative ring of modular functions \cite{modfun}, which is also
generated by two generators of degrees $2$ and $3$: Eisenstein
series $E_2$ and $E_3$. We remind that \be E_{k}(\tau) =
\frac{1}{2\zeta(2k)}{\sum_{m,n\in Z}}'\frac{1}{(m+n\tau)^{2k}} = 1
- \frac{4k}{B_{2k}} \sum_{n=1}\sigma_{2k-1}(n)q^n, \ee where
$q=e^{2\pi i\tau}$, $\sigma_{2k-1}(n) = \sum_{d|n}d^{2k-1}$ is a
sum over divisors of $n$, Bernoulli numbers are the coefficients
in expansion\footnote{ The first Bernoulli numbers are:
$$B_2 = \frac{1}{6},\ B_4 = -\frac{1}{30},\
B_6 = \frac{1}{42},\ B_8 = -\frac{1}{30},\ B_{10}=\frac{5}{66},\
B_{12} = -\frac{691}{2730},\ B_{14} = \frac{7}{6},\ B_{16} =
-\frac{3617}{510},\ B_{18} = \frac{43867}{798},\ B_{20} =
-\frac{176611}{330}, \ldots $$ } $\frac{x}{e^x-1} =
\sum_{k=0}^\infty B_k\frac{x^k}{k!} = 1 -\frac{x}{2} +
\sum_{k=1}^\infty B_{2k}\frac{x^{2k}}{(2k)!}$, and for even $k$
the Riemann's zeta-function is $\zeta(2k) = \sum_{n=1}^\infty
n^{-2k} = \frac{(-)^{k-1}(2\pi)^{2k}}{2(2k)!}B_{2k}$. It is
amusing to compare polynomial relations among Eisenstein series
and those among sausage diagrams Fig.\ref{Vieta24dia}.B. When
invariant is unique, for $k=4,5,7,\ldots$, there is a clear
correspondence $I_{k} \leftrightarrow c_kE_k$, but it breaks
already for $I_6$ and $I_8$, when the space of invariants gets
two-dimensional (the last column contains expressions for $I_k$ at
$S^{1112}=S^{1222}=0$ from (\ref{saudred})):

\bigskip

$$
\hspace{-0.5cm}
\begin{array}{|c|c|c|}
\hline
&\ \ \ \ \ \ &\\
{\rm  modular\ forms} && {\rm sausage\ diagrams}\\
&&\\
\hline
&&\\
E_2 = 1+240q+\ldots &\leftrightarrow & I_2 \rightarrow 2(A+3B^2)\\
&&\\
E_3 = 1-504q+\ldots&\leftrightarrow& I_3 \rightarrow 6B(A-B^2)\\
&&\\
E_4 = 1+480q+\ldots = E_2^2 &\leftrightarrow&
I_4 = \frac{1}{2}I_2^2 \rightarrow 2\Big(A^2+6AB^2+9B^4\Big)\\
&&\\
E_{5} = 1-264q+\ldots = E_2E_3 &\leftrightarrow &
I_5 = \frac{5}{6}I_2I_3 \rightarrow 10B\Big(A^2+2AB^2-3B^4\Big)\\
&&\\
E_{6} = 1 + \frac{65520}{691}q+\ldots = \frac{441E_2^3 +
250E_3^2}{691} &{\rm but} & I_6 = \frac{1}{4}I_2^3 +
\frac{1}{3}I_3^2 \rightarrow
2\Big(A^3+15A^2B^2 + 15AB^4 + 33B^6\Big) \\
&&\\
E_{7} = 1-24q+\ldots = E_2^2E_3 &\leftrightarrow & I_7 =
\frac{7}{12}I_2^2I_3 \rightarrow
14B\Big(A^3 + 5A^2B + 3AB^2 - 9B^3\Big) \\
&& \\
E_{8} = 1+\frac{16320}{3617}q+\ldots = E_2\frac{1617E_2^3 + 2000
E_3^2}{3617} &{\rm but} & I_8 =
I_2\left(\frac{1}{8}I_2^3+\frac{4}{9}I_3^2\right) \rightarrow
2\Big(A^4 + 28A^3B^2 + 70A^2 B^4 + 28AB^4
+ 129B^8\Big)\\
&&\\
\ldots &&\\
&&\\
\hline
\end{array}
$$

\bigskip

The correspondence -- even when it exists -- implies certain
relations between the coefficients $c_k$, say, $c_4 =
\frac{1}{2}c_2^2$ etc. The table seems to suggest a seemingly nice
choice: $c_{2k} = 2$, $c_{2k+1} = 2k+1$. However, it converts
discriminant $D_{2|4} \sim I_2^3 - 6I_3^2$ into $\sim 4E_2^3 - 27
E_3^2$. If one wants instead that discriminant is converted into
$$\eta^{24}(\tau) = \frac{(2\pi)^{12}}{1728}(E_4^3-E_6^2)
\sim \prod_{m,n} (m+n\tau)^{12} \sim Det^{12} \bar\partial,$$ one
should rather choose $c_2=c_3=6$.

$\bullet$ Operator description of invariants -- the analogue of
(\ref{oper23.1})-(\ref{oper23.5}) -- is as follows.   Operator \be
\hat S = \left(\begin{array}{cc|cc}
S_{11}^{11} & S_{11}^{12} & S_{11}^{21} & S_{11}^{22} \\
S_{12}^{11} & S_{12}^{12} & S_{12}^{21} & S_{12}^{22} \\
\hline
S_{21}^{11} & S_{21}^{12} & S_{21}^{21} & S_{21}^{22} \\
S_{22}^{11} & S_{22}^{12} & S_{22}^{21} & S_{22}^{22}
\end{array}\right) =
\left(\begin{array}{cc|cc}
s_2 & s_1 & s_1 & s_0 \\
-s_3 & -s_2 & -s_2 & -s_1 \\
\hline
-s_3 & -s_2 & -s_2 & -s_1 \\
s_4 & s_3 & s_2 & s_1
\end{array}\right)
\ee is traceless, also all four blocks are traceless themselves
(since polynomial is irreducible representation).

For $\hat B = B_{ij}^{kl} = S_{im}^{kn} S_{jn}^{lm}$ we have: \be
\hat B = \left(\begin{array}{cccc}
2(s_2^2-s_1s_3) & s_1s_2 - s_0s_3 & s_1s_2-s_0s_3 & 2(s_1^2-s_0s_2)\\
s_1s_4 - s_2s_3 & 2(s_1s_3-s_2^2) & s_2^2-2s_1s_3 + s_0s_4 &
                  s_0s_3-s_1s_2 \\
s_1s_4 - s_2s_3 &  s_2^2-2s_1s_3 + s_0s_4 & 2(s_1s_3-s_2^2) &
                  s_0s_3-s_1s_2 \\
2(s_3^2-s_2s_4) & s_2s_3-s_1s_4 & s_2s_3 - s_1s_4 &
2(s_2^2-s_1s_3)
\end{array}\right)
\ee $\hat B$ is also traceless: ${\rm Tr} B = 0$

The two operators $\hat S$ and $\hat B$ commute: $\left[ \hat S,
\hat B\right] = 0$.

\be \hat B + \hat S^2 = \left(\begin{array}{cccc}
I_2 & 0 & 0 & 0 \\
0 & 0 & I_2 & 0 \\
0 & I_2 & 0 & 0 \\
0 & 0 & 0 & I_2
\end{array}\right)
\ee with $I_2 = 3s_2^2 + s_0s^4 - 4s_1s_3 = \frac{1}{2}{\rm Tr}
\hat S^2$ has a characteristic
$R$-matrix structure. 


$\bullet$ The common zero of invariants ${\rm Inv}_2(S)$ and ${\rm
Inv}_3(S)$ -- and thus of discriminants $D_{2|4}$ -- occurs when
three out of four roots of the polynomial coincide (pairwise
coinsidence occurs at zeroes of discriminant alone). The two other
higher discriminantal varieties -- where two pairs of the four
roots coincide and where all the four roots coincide -- are
associated with non-singlet representations of the structure group
(like the triplet condition $B_{ij}=0$, eq.(\ref{Bij23}), in the
$2|3$ case). See \cite{Sha} for more details.

\subsection{Functional integral (\ref{fint}) and its
analogues in the $n=2$ case \label{fintsec}}

\subsubsection{Direct evaluation of $Z(T)$ \label{direva}}

\noindent

$\bullet$ Partition function $Z(T)$ depends on domain of
integration for the $t$-variables. In this section we assume that
it consists of $\#(t)=p$ points on a line (a discrete
1-dimensional lattice). If $\#(t)< n$, the $\epsilon$-dependent
terms do not exist, and $Z(T)$ is trivial.

$\bullet$ For $n=2$ partition function $Z(T)$ is non-trivial
already when $t$ takes just two different values, $\#(t)=2$:
\be Z_{2^{\times r}}\Big(T\Big|\#(t)=2\Big) =
\left\{\prod_{k=1}^r \exp\left(\epsilon_{ij}
\frac{\partial^2}{\partial x_{ki}\partial
x_{kj}'}\right)\right\}\nn\\
\cdot\left.\exp \left\{T^{i_1\ldots i_r}\Big(x_{1i_1}\ldots x_{ri_r} +
x'_{1i_1}\ldots x'_{ri_r}\Big)\right\}\right|_{x,x'=0}
\label{opint2} \ee

We begin evaluation of $Z\Big(T\Big|\#(t)=2\Big)$
from the $r=2$ case. Actually, we perform the calculation for
diagonal $T^{ij}$: for $T_{diag}^{ij}x_iy_j = ax_1y_1 + bx_2y_2$.
Since for $r=2$ partition function depends on a single invariant
$T^{ij}T_{ij} = ab - T^{12}T^{21}$, evaluation of the
$ab$-dependence is sufficient. With the usual switch of notation,
$x_i(t) = x_{1i}(t)$, $y_i(t)=x_{2i}(t)$, we have for
(\ref{opint2})
\be
Z_{2\times 2}\Big(T_{diag}\Big|\#(t)=2\Big) =
\exp\left(\frac{\partial^2}{\partial x_1\partial x'_2}-
\frac{\partial^2}{\partial x_2\partial x'_1}\right)\nn\ee
\be\cdot\exp\left(\frac{\partial^2}{\partial y_1\partial y'_2}-
\frac{\partial^2}{\partial y_2\partial y'_1}\right)
\left.\exp\Big(T^{ij}(x_iy_j +
x'_iy'_j)\Big)\right|_{x,x',y,y'=0}\nn\ee
\be=\sum_{m=0}^\infty \frac{1}{m!}
\left(\frac{\partial^2}{\partial x_1\partial x'_2}-
\frac{\partial^2}{\partial x_2\partial x'_1}\right)^m
\frac{1}{m!}\left(\frac{\partial^2}{\partial y_1\partial y'_2}-
\frac{\partial^2}{\partial y_2\partial y'_1}\right)^m\nn\ee
\be\frac{1}{(2m)!}\Big(a(x_1y_1+x'_1y'_1) +
b(x_2y_2+x'_2y'_2)\Big)^m= \sum_{m=0}^\infty  \sum_{s=0}^m
\frac{1}{m!}\frac{m!}{s!(m-s)!}\nn\ee
\be\cdot(-)^{m-s}
\left(\frac{\partial^2}{\partial x_1\partial x'_2}\right)^s
\left(\frac{\partial^2}{\partial x_2\partial x'_1}\right)^{m-s}
\frac{1}{m!}\frac{m!}{s!(m-s)!}\nn\ee
\be(-)^{m-s}
\left(\frac{\partial^2}{\partial y_1\partial y'_2}\right)^s
\left(\frac{\partial^2}{\partial y_2\partial y'_1}\right)^{m-s}
\frac{1}{(2m)!}\frac{(2m)!}{m!m!}\ a^mb^m\nn\ee
\be
\Big(x_1y_1+x'_1y'_1\Big)^m\Big(x_2y_2+x'_2y'_2\Big)^m
= \sum_{m=0}^\infty  a^mb^m \sum_{s=0}^m
\frac{1}{(m!s!(m-s)!)^2}\nn\ee
\be
\left(\frac{\partial^2}{\partial x_1\partial x'_2}\right)^s
\left(\frac{\partial^2}{\partial x_2\partial x'_1}\right)^{m-s}
\left(\frac{\partial^2}{\partial y_1\partial y'_2}\right)^s
\left(\frac{\partial^2}{\partial y_2\partial
y'_1}\right)^{m-s}\nn\ee
\be
\cdot \frac{m!}{s!(m-s)!}(x_1y_1)^s(x'_1y'_1)^{m-s}
\frac{m!}{s!(m-s)!}(x_2y_2)^{m-s}(x'_2y'_2)^{s}\nn\ee
\be
= \sum_{m=0}^\infty  a^mb^m \left(\sum_{s=0}^m 1\right) =
\sum_{m=0}^\infty  (m+1)(ab)^m = \Big(1-ab\Big)^{-2} \label{ZT222}
\ee
This calculation is so simple because of the obvious selection
rules, leaving a sum over just two  parameters $m$ and $s$, which
are the same in all binomial expansions. This will no longer be
the case for $\#(t)>2$, but for $\#(t)=2$ this property persists
for any rank $r$.

Accordingly,
\be
Z_{2^{\times r}}\Big(T_{diag}\Big|\#(t)=2\Big) =
\prod_{k=1}^r \exp\left( \frac{\partial^2}{\partial x_{k1}\partial
x'_{k2}}- \frac{\partial^2}{\partial x_{k2}\partial
x'_{k1}}\right) \nn\ee
\be\cdot\left.\exp\Big(T^{i_1\ldots i_r}(x_{1i_1}\ldots x_{ri_r}
+ x'_{1i_1}\ldots x'_{ri_r})\Big)\right|_{\ {\rm all}\
x,x'=0}\nn\ee
\be= \sum_{m=0}^\infty  a^mb^m \sum_{s=0}^m  \prod_{k=1}^r
\left\{\frac{(-)^{m-s}}{s!(m-s)!} \left(\frac{\partial^2}{\partial
x_{k1}\partial x'_{k2}}\right)^s \left(\frac{\partial^2}{\partial
x_{k2}\partial x'_{k1}}\right)^{m-s} \right\} \nn\ee
\be\cdot
\frac{1}{\Big(s!(m-s)!\Big)^2} \Big(\prod_{k=1}^r x_{k1}\Big)^s
\Big(\prod_{k=1}^r x'_{k1}\Big)^{m-s} \Big(\prod_{k=1}^r
x_{k2}\Big)^{m-s} \Big(\prod_{k=1}^r x'_{k2}\Big)^s \nn\ee
\be = \sum_{m=0}^\infty  a^mb^m \left(\sum_{s=0}^m (-)^{r(m-s)}
\Big(s!(m-s)!\Big)^{r-2}\right)
\label{ZT2r2} \ee

For even $r$ all $m$ contribute, including $m=1$. This is not a
surprise, because for even $r$ (and $n=2$) there is always a
sort-respecting invariant of order $2$: ${\rm Inv}_2(T) =
T^{i_1\ldots i_r}T_{i_1\ldots i_r}$, which is equal to $ab$ for
diagonal $T$.

For odd $r\geq 3$ only even $m$ contribute to the sum, including
$m=2$. This implies that for odd $r\geq 3$ there is no
non-vanishing ${\rm Inv}_2(T)$ (it indeed is identical zero for
odd $r$), but there always is a non-trivial invariant ${\rm
Inv}_4(T)$, which reduces to $(ab)^2$ for diagonal $T$. For $r=3$
${\rm Inv}_4(T)$ is just the Cayley hyperdeterminant $D_{2\times
2\times 2}(T)$.

$\bullet$ We now list a few particular examples of (\ref{ZT2r2}).

For $r=1$ we get:
\be &Z_{2|1}^{}\Big(T\Big|\#(t)=2\Big) =
\sum_{m=0} (ab)^m \left(\sum_{s=0}^m \frac{(-)^s}{s!(m-s)!}\right)\nn\\
&= \sum_{m=0} \frac{(ab)^m}{m!}(1-1)^m = 1 \label{Z21} \ee

For $r=2$ we reproduce (\ref{ZT222}):
\be &Z_{2\times
2}\Big(T\Big|\#(t)=2\Big) = \sum_{m=0} (ab)^m \left(\sum_{s=0}^m 1
\right) = \sum_{m=0} (m+1)(ab)^m \nn\\
&= (1-ab)^{-2} = \Big(1-D_{2\times
2}(T)\Big)^{-2} \label{Z22} \ee
Here we use the expression
$D_{2\times 2}(T) = \frac{1}{2}{\rm Inv}_2(T) =
\frac{1}{2}T^{ij}T_{ij} =
\frac{1}{2}T^{ij}T^{kl}\epsilon_{ik}\epsilon_{jl}$ of discriminant
(appropriately normalized, so that $D_{2\times 2}(T_{diag})= ab$)
through the only non-trivial sort-respecting invariant of ${\cal
T}(T)$ in the $n|r=2|2$ case (comp.with Fig.\,\,\ref{Vieta22dia}.G; as
to the Pfaffian ${\rm Pfaff}(T) = T^i_i = \epsilon_{ij}T^{ij}$, it
does not respect sorts of lines and can not appear in
(\ref{Z22})). In quasiclassical limit, see eq.\,\,(\ref{eqmfint}) and
discussion afterwards, for generic (non-degenerate) $T$
discriminant $D_{2\times 2}(T)\gg 1$, and $Z_{2\times 2}(T)$
becomes just a power of $D_{2\times 2}(T)$.

For $r=3$ we get:
\be
&Z_{2\times 2\times 2}\Big(T\Big|\#(t)=2\Big) = \sum_{m=0}
(ab)^{2m} \left(\sum_{s=0}^{2m} (-)^s s!(2m-s)!\right) \nn\\
&= 1 + 3D_{2\times 2\times 2} + 40D_{2\times 2\times 2}^2 + 1260
D_{2\times 2\times 2}^3 + 72576 D_{2\times 2\times 2}^4 +
\ldots\nn\\
&=\sum_{m=0}^\infty
\frac{(2m+1)!}{m+1}D_{2\times 2\times 2}^m =\sum_{m=0}^\infty
\frac{m!}{m+1}
\frac{\Gamma\Big(m+\frac{3}{2}\Big)}{\Gamma\Big(\frac{3}{2}\Big)}
\Big(4D_{2\times 2\times 2}\Big)^m \nn\\
&= \phantom._4
F_1\left(\frac{3}{2},0,0,0;1;4D\right), \label{Z23} \ee
where hypergeometric series is defined as
\be&\phantom._rF_s(\alpha_1,\ldots,\alpha_r;\beta_1,\ldots,\beta_s;\
x) = \sum_{m=0}^\infty
\frac{(\alpha_1)_m\cdot\ldots\cdot(\alpha_r)_m}
{\beta_1)_m\cdot\ldots\cdot(\beta_s)_m}\frac{x^m}{m!} \ \nn\\
&{\rm with} \ \ \ \ (\alpha)_m = \frac{\Gamma(\alpha+m)}{\Gamma(\alpha)}
\ee We assume in (\ref{Z23}) that hyperdeterminant is normalized
so that $D_{2\times 2\times 2}(T_{diag}) = (ab)^2$. Since
(\ref{fint}) and (\ref{fint22}) respect {\it sorts} of lines, no
contractions between indices of the same tensor are allowed --
this means that the diagram Fig.\,\,\ref{Cayotherdia}.B can not
contribute, while Fig.\,\,\ref{Cayotherdia}.A vanishes and therefore
$T^2$-term can not appear at the r.h.s. of (\ref{Z23}):
$Z_{2\times 2\times 2}(T) = 1 + O(T^4)$. The large-$m$ asymptote
of the coefficients $b_m = \sum_{s=0}^{2m} (-)^s s!(2m-s)!$
defines the quasiclassical asymptotics of $Z_{2\times 2\times
2}(T)$, when $D_{2\times 2\times 2}(T) \gg 1$. For $r\geq 3$
series diverge and can be summed, say, by Borel method (perhaps,
multiply applied): $\sum a_mx^m \rightarrow \int_0^\infty
\Big(\sum \frac{a_m (xu)^m}{m!}\Big)e^{-u}du$.

For $r=4$ we get: \vspace{0.2cm}
\be
&Z_{2^{\times 4}}\Big(T\Big|\#(t)=2\Big) =
\sum_{m=0} (ab)^m \left(\sum_{s=0}^m
\big(s!(m-s)!\big)^2\right)\nn\\
&= 1 + 2(ab) + 9(ab)^2 + 80(ab)^3 +
1240(ab)^4 + 30240 (ab)^5 +\ldots \label{Z24} \ \ \ \ee
This time
information is insufficient to extend the r.h.s. (starting from
the $(ab)^3$ term) to non-diagonal $T$: there is now two
independent invariants: ${\rm Inv}_2(T)$ and the degree-$6$
discriminant $D_{2^{\times 4}}(T)$ (note that the third-order
invariant ${\rm Inv}_3(S)$ relevant for {\it symmetric} reduction
of $T$ and considered in s.\ref{24resulta}, does not respect {\it
sorts} and can not appear in the sort-respecting $Z(T)$).
Resolution of $Z_{2^{\times 4}}$ dependencies on different
invariants remains a challenging problem. As explained in
s.\ref{difrodi}, it can be expected that the large-$T$
(quasiclassical) asymptotics of $Z(T)$ will be defined by $D(T)$
alone -- and this would provide one more way to evaluate
$D_{2^{\times 4}}(T)$. However, explicit check of this hypothesis
remains to be made.

$\bullet$ In the case of symmetric $T=S$ there is only one vector
$x_i$ instead of $r$ different $x_{ki}$, $k=1,\ldots,r$, and one
can substitute (\ref{opint2}) by
\be
Z_{2|r}\Big(S\Big|\#(t)=2\Big) =  \exp\left(\epsilon_{ij}
\frac{\partial^2}{\partial x_{i}\partial x_{j}'}\right) \nn\ee
\be\left.\exp
\left\{S^{i_1\ldots i_r}\Big(x_{i_1}\ldots x_{i_r} +
x'_{i_1}\ldots x'_{i_r}\Big)\right\}\right|_{x,x'=0} =
\sum_{k,l,m,n=0}^\infty \frac{(-)^l}{k!l!m!n!}\nn\\
\left.\left(\frac{\partial^2}{\partial x_1\partial x'_2}\right)^k
\left(\frac{\partial^2}{\partial x_2\partial x'_1}\right)^l
\Big(Sx\ldots x\Big)^m\Big(Sx'\ldots x'\Big)^n\right|_{x,x'=0}
\label{opint2S} \ee

Consider first the case of $r=2$. The simplest option is to put
$S^{11}=S^{22}=0$ and $S^{12}=S^{21}=s$. Since the answer depends
on $D_{2|2}(S) = S^{11}S^{22}-(S^{12})^2$ only, this calculation
will be enough to obtain the answer. Then (\ref{opint2S}) becomes:
\be
&Z_{2|2}\Big(S\Big|\#(t)=2\Big) = \sum_{k,l,m,n=0}^\infty
\frac{(-)^l}{k!l!m!n!} \nn\\
&\cdot\left.\left(\frac{\partial^2}{\partial
x_1\partial x'_2}\right)^k \left(\frac{\partial^2}{\partial
x_2\partial x'_1}\right)^l
\Big(2sx_1x_2\Big)^m\Big(2sx'_1x'_2\Big)^n\right|_{x,x'=0}
\ee
and it is obvious that only terms with $k=l=m=n$ contribute, so
that the quadruple sum reduces to a single one: \be
Z_{2|2}\Big(S\Big|\#(t)=2\Big) = \sum_{k=0}^\infty (-)^k(2s)^k =
\frac{1}{1+4s^2} = \Big(1-4D_{2|2}(S)\Big)^{-1} \label{fins2}\ \ \ \ee
Comparing this with (\ref{Z22}) we see that the answer in
symmetric case is a square root of that for bilinear integral (see
also more familiar (\ref{ordintT}) and (\ref{ordintS}) below). The
factor $4$ in (\ref{fins2}) can be absorbed in rescaling of
$\epsilon$-containing term in (\ref{opint2S}).

As a next step we consider a slightly more sophisticated
derivation of the same answer (\ref{fins2}), when in
(\ref{opint2S}) we put $S^{12}=0$ and $S^{11}=a$, $S^{22}=b$. This
time we obtain even more sums:
\be
&Z_{2|2}\Big(S\Big|\#(t)=2\Big)= \sum_{k,l,m_1,m_2,n_1,n_2=0}^\infty
\frac{(-)^l}{k!l!m_1!m_2!n_1!n_2!}\nn\\
&\left.\left(\frac{\partial^2}{\partial x_1\partial x'_2}\right)^k
\left(\frac{\partial^2}{\partial x_2\partial x'_1}\right)^l
\Big(ax_1^2\Big)^{m_1}\Big(bx_2^2\Big)^{m_2}
\Big(a{x'_1}^2\Big)^{n_1}\Big(b{x'_2}^2\Big)^{n_2}\right|_{x,x'=0}
\label{fins2'}\ \ \ \ee
and it is clear that contributing are terms
with $k=2m_1=2n_2$ and $l = 2m_2=2n_1$, so that the double sum
remains:
\be &Z_{2|2}\Big(S\Big|\#(t)=2\Big) =
\sum_{m_1,n_1=0}^\infty
\frac{(2m_1)!(2n_1)!}{(m_1!)^2(n_1!)^2}(ab)^{m_1+n_1} \nn\\
&=\left\{\sum_{m=0}^\infty \frac{(2m)!}{(m!)^2}(ab)^m\right\}^2 =
\left\{\frac{1}{\sqrt{1-4D_{2|2}(S)
\phantom{5^{5^5}}\hspace{-0.5cm}}}\right\}^2
\label{fins22} \ee
-- in accordance with (\ref{fins2}). At the
last step we used that $(1-x)^{-\alpha} = \sum_{m=0}^\infty
\frac{\Gamma(m+\alpha)}{m!\Gamma(\alpha)}x^m =
\phantom._1F_0(\alpha;x)$ and $(2m)! = 4^m m!
\frac{\Gamma(m+1/2)}{\Gamma(1/2)}$.

\bigskip

Proceed now to the case of $r=3$. As in trilinear case (\ref{Z23})
we take diagonal $S$: $S^{111}=a$, $S^{222}=b$, all other
components zero. The answer depends on discriminant $D_{2|3}(S)$
only and is easily extractable from such reduction when
$D_{2|3}(S) = (ab)^2$. Now we have instead of (\ref{fins2'}):
\be
&Z_{2|3}\Big(S\Big|\#(t)=2\Big) = \sum_{k,l,m_1,m_2,n_1,n_2=0}^\infty
\frac{(-)^l}{k!l!m_1!m_2!n_1!n_2!}\nn\\
&\left.
\left(\frac{\partial^2}{\partial x_1\partial x'_2}\right)^k
\left(\frac{\partial^2}{\partial x_2\partial x'_1}\right)^l\Big(ax_1^3\Big)^{m_1}\Big(bx_2^3\Big)^{m_2}
\Big(a{x'_1}^3\Big)^{n_1}\Big(b{x'_2}^3\Big)^{n_2}\right|_{x,x'=0}\hspace{2mm}
\ee
The only difference is that the powers of $x$ are increased from
$2$ to $3$. This time selection rules are $k=3m_1=3n_2$ and $l =
3m_2=3n_1$, so that the remaining double is:
\be
&Z_{2|3}\Big(S\Big|\#(t)=2\Big) = \sum_{m_1,n_1=0}^\infty (-)^{n_1}
\frac{(3m_1)!(3n_1)!}{(m_1!)^2(n_1!)^2}(ab)^{m_1+n_1} \nn\\
&=\left\{\sum_{m=0}^\infty \frac{(3m)!}{(m!)^2} (ab)^m
\right\} \left\{\sum_{m=0}^\infty \frac{(3m)!}{(m!)^2}(-ab)^m
\right\}
\label{fins23} \ee
The r.h.s. is a product of two hypergeometric
function
\be Z_{2|3}\Big(S\Big|\#(t)=2\Big) = \hspace{-0.14cm}
\phantom._2F_0\Big(\frac{1}{3},\frac{2}{3};\
27\sqrt{D_{2|3}(S)}\Big)
\phantom._2F_0\Big(\frac{1}{3},\frac{2}{3};\
-27\sqrt{D_{2|3}(S)}\Big) \nn\\
=\hspace{-1mm}1 +\frac{27\cdot(27-1)}{2}D_{2|3}(S) + 27\cdot
(81-1)\cdot(27^2+27-1)\Big(D_{2|3}(S)\Big)^2\hspace{-1mm}+ \ldots
\label{ZS23}\nn\\\ee
Note that the product of two $\hspace{-0.14cm}\phantom._2F_0$
functions in (\ref{ZS23}) have the same number of factorials $m!$
in the numerator as a single $\hspace{-0.14cm}\phantom._4F_1$ in
(\ref{Z23}): $2\times (2-1) = 4-2$ (the coefficients of
$\hspace{-0.14cm}\phantom._rF_s$ grow roughly as $(m!)^{r-s-1}$;
for $r>s+1$ such series are everywhere divergent, but can be
handled, say, by Borel method).

Similarly for {\it diagonal} $S$ of rank $r$
$$
Z_{2|r}\Big(S_{diag}\Big|\#(t)=2\Big) = 
\left\{\sum_{m=0}^\infty \frac{(rm)!}{(m!)^2}(ab)^m\right\}
\left\{\sum_{m=0}^\infty \frac{(rm)!}{(m!)^2}
\left((-)^rab\right)^m\right\}
$$
\be = \hspace{-0.1cm}
\phantom._{r-1}F_0\Big(\frac{1}{r},\frac{2}{r},\ldots,\frac{r-1}{r};\
r^rab \Big)
\phantom._{r-1}F_0\Big(\frac{1}{r},\frac{2}{r},\ldots,\frac{r-1}{r};\
(-r)^rab \Big)
\label{fins2r} \nn\\\ee
However, for $r\geq 4$ consideration of
diagonal $S$ is not sufficient for restoring the full $S$
dependence. When invariant's ring has only two generators, it is
sufficient to add to (\ref{fins2r}) an analogue of calculation
(\ref{fins2}) for off-diagonal~$S$.

For example, if $r=4$, we can put $S^{1122}=S$ and all other
components vanishing, so that
\be
&Z_{2|4}\Big(S_{off-diag}\Big|\#(t)=2\Big) =
\sum_{k,l,m,n=0}^\infty \frac{(-)^l}{k!l!m!n!}\nn\\
&\cdot\left.\left(\frac{\partial^2}{\partial x_1\partial x'_2}\right)^k
\left(\frac{\partial^2}{\partial x_2\partial x'_1}\right)^l
\Big(2Sx_1^2x_2^2\Big)^m\Big(2S{x'_1}^2{x'_2}^2\Big)^n
\right|_{x,x'=0}
\ee
and only terms with $k=l=2m=2n$ contribute, so that quadruple sum
reduces to a single one: \be
Z_{2|4}\Big(S_{off-diag}\Big|\#(t)=2\Big) = \sum_{m=0}^\infty
\left(\frac{(2m)!}{m!}\right)^2(2S)^m\ =
\phantom._{3}F_0\Big(\frac{1}{2},\frac{1}{2}, 0 ;\ 32S \Big)
\label{fins24'} \nn\\\ee
The two formulas, (\ref{fins2r}) for $r=4$ and
(\ref{fins24'}) are sufficient to restore the dependence
$Z_{2|4}\Big(S\Big|\#(t)=2\Big)$ on two invariants $I_2={\rm
Inv}_2(S)= S^{ijkl}S_{ijkl}$ and $I_3={\rm Inv}_3(S) =
S^{ij}_{kl}S^{kl}_{mn}S^{mn}_{ij}$. For diagonal $S$ these
invariants are, see (\ref{saudred}): $I_2(S_{diag}) = 2ab$ and
$I_3(S_{diag})=0$, while for the off-diagonal $S$ they are:
$I_2(S_{off-diag})=6S^2$ and $I_3(S_{off-diag})= -6S^3$.

$\bullet$ Formulas like (\ref{ZT2r2}) and (\ref{fins2r}) can look
a little less exotic, if one considers their toy analogues for
{\it numbers} (rather than $n$-vectors) $x_1,\ldots,x_k$:
\be &\exp
\left(\frac{\partial^r}{\partial x_1\ldots \partial x_r} \right)
e^{Tx_1\ldots x_r} \nn\\
&=\sum_{m=0}^\infty (m!)^{r-2}T^m =
\left\{\begin{array}{cl}
{\rm for}\ r=1: & e^T \\
{\rm for}\ r=2: & (1-T)^{-1} \\
{\rm for}\ r=3: & \sum_{m=0}^\infty m!\ T^m = \int_0^\infty
\frac{e^{-u}du}{1-uT} \\
&\ldots
\end{array}\right.\ \ \ \
\ee
and
\be &\exp \left(\frac{\partial^r}{\partial x^r}\right)
e^{Sx^r} = \sum_{m=0}^\infty \frac{(rm)!}{(m!)^2}S^m \nn\\
&=\hspace{-0.1cm}
\phantom._{r-1}F_0\Big(\frac{1}{r},\frac{2}{r},\ldots,\frac{r-1}{r};\
r^rS \Big) = \left\{\begin{array}{cl}
{\rm for}\ r=1: & e^S \\
{\rm for}\ r=2: & (1-S)^{-1/2} \\
{\rm for}\ r=3: & \sum_{m=0}^\infty \frac{(3m)!}{(m!)^2} S^m\\
&\ldots
\end{array}\right.
\ee All these relations can be considered as various
tensor-algebra generalizations of Taylor formula \be
e^{a\frac{\partial}{\partial x}} f(x) = f(x+a) \ee One more
example of generalized shift operator appears in realization of
the elementary non-commutative algebra: \be {\rm if}\ \ yx = xy +
c,\ \ {\rm then}\ \ y^nx^m = e^{c\frac{\partial^2}{\partial
x\partial y}} x^my^n \ee

\subsubsection{Gaussian integrations: specifics of cases
$n=2$ and $r=2$}

\noindent

$\bullet$ As mentioned in comments after eq.\,\,(\ref{ZT222}), above
calculation becomes far more sophisticated for $\#(t)>2$. However,
in the $n|r=2|2$ case one can instead evaluate the integral
(\ref{fint}) directly: it is Gaussian in all variables and can be
calculated exactly. We split this calculation in two steps, first
valid for all $n=2$ examples, and second exploiting that $r=2$.

$\bullet$ First, for $n=2$ case the integral (\ref{fint})
\be
&Z_{2^{\times r}}(T) = \left\{\prod_{k=1}^r \left(\prod_{i=1}^{2}
\int{\cal D}x_{ki}(t){\cal D}\bar x^i_k(t)
e^{\int x_{ki}(t)\bar x_k^i(t)dt}
\right)\right.\nn\\
&\cdot\left.
\exp {{\int\int}_{t<t'}\epsilon_{ij}
\bar x^{i}_k(t)\bar x^{j}_k(t')
dtdt'} \right\} \exp {\int T^{i_1\ldots i_r}x_{1i_1}(t)\ldots
x_{ri_r}(t)dt} \label{fint5} \nn\\\ee
is Gaussian in $\bar x$ variables
and they can be explicitly integrated out.

For one particular pair $(x_{ki},\bar x_k^i = (x_i,\bar x^i)$ we
have (we assume that $\#(t)=p$ and integrals are taken along
imaginary axes to provide $\delta$-functions, factors of $\pi$ are
systematically omitted -- assumed included into the definition of
integrals):
\be
&\int d\bar x(1)\ldots d\bar x(p) \exp \left\{\Big(x(1)\bar x(1) +
\ldots + x(p-1)\bar x(p-1) + x(p)\bar x(p) \Big) \right.\nn\\
&\left.  + \Big(\bar x(1) + \ldots + \bar x(p-1)\Big)\epsilon\bar
x(p) + \Big(\bar x(1) + \ldots + \bar x(p-2)\Big)\epsilon\bar
x(p-1) + \ldots \right.\nn\\
&\left.+ \bar x(1)\epsilon \bar x(2)\right\}
= \int d\bar x(1)\ldots d\bar x(p-1) \delta \left\{x(p) +
\epsilon\Big(\bar x(1) + \ldots + \bar x(p-1)\Big)\right\} \nn\\
&\cdot\exp
\left\{\Big(x(1)\bar x(1) + \ldots + x(p-1)\bar x(p-1)\Big)
+ \Big(\bar x(1) + \ldots + \bar
x(p-2)\Big)\right.\nn\\
&\cdot\left.\epsilon\bar x(p-1) + \ldots + \bar x(1)\epsilon \bar
x(2)\right\}
= \exp \Big\{x(p-1)\epsilon x(p)\Big\}\nn\\
&\cdot \int d\bar x(1)\ldots
d\bar x(p-2) \exp\left\{ \Big( [x(1) + x(p-1) + x(p)]\bar x(1) +
\ldots \right.  \nn\\
&\left.+ [x(p-2) + x(p-1) + x(p)]\bar
x(p-2)\Big) + \Big(\bar x(1) + \ldots + \bar
x(p-3)\Big)\right.\nn\\
&\cdot\left.\epsilon\bar x(p-2) + \ldots + \bar x(1)\epsilon \bar
x(2)\right\}
\ee
Thus, after integrating over two out of $p$ ($2$-component)
variables $\bar x$, we obtain the same integral with two
differences: a prefactor $\exp \Big\{x(p-1)\epsilon x(p)\Big\}$
($\epsilon$ tensor here is contravariant, inverse of the original
covariant $\epsilon$) and shift of all remaining $x$ variables by
$x(p-1) + x(p)$. This means that the next pair of integrations --
over $\bar x(p-2)$ and $\bar x(p-3)$ will provide a new prefactor,
equal to
$$
\exp \Big\{[x(p-3)+x(p-1)+x(p)]\epsilon
[x(p-2)+x(p-1)+x(p)]\Big\}
$$
$$
= \exp \Big\{x(p-3)\epsilon \Big(x(p-2)+ x(p-1) + x(p)\Big)+
x(p-2)\epsilon \Big(x(p-1) + x(p)\Big)\Big\}
$$
Repeating this procedure again and again, we finally obtain from
$\bar x$ integrations $\exp\Big\{\sum_{t<t'} x_i(t)\epsilon^{ij}
x_j(t')\Big\}$, i.e. integration over $\bar x$-variables converts
(\ref{fint5}) into
\be
&Z_{2^{\times r}}(T) = \left\{\prod_{k=1}^r
\left(\prod_{i=1}^{2} \int{\cal D}x_{ki}(t)\right) \exp
{\int\int_{t<t'}\epsilon^{ij} x_{ki}(t)x_{kj}(t')dtdt'}
\right\}\nn\\
&\cdot\exp {\int T^{i_1\ldots i_r}x_{1i_1}(t)\ldots x_{ri_r}(t)dt}
\label{fint5'} \ee
This integral, obtained after elimination of
$\bar x$-variables, is no longer associated with any diagram
technique, instead sometime it can be evaluated explicitly

$\bullet$ In particular, for $r=2$ the integral (\ref{fint5'}) is
still Gaussian and the second step of exact calculation can be
made: \be
&Z_{2\times 2}(T) = \left\{
\int{\cal D}x_{i}(t)Dy_i(t) \exp
\int\int_{t<t'}\epsilon^{ij}
\Big(x_{i}(t)x_{j}(t') \right.\nn\\
&\left.+ y_i(t)y_j(t')\Big)dtdt' \right\}\
\exp
{\int T^{ij}x_{i}(t)y_j(t)dt} \label{fint52'} \ee
Performing
remaining integrations, we obtain \be Z_{2\times 2}(T) = \det_{ij}
\Big(\epsilon^{ij} - T^{ik}\epsilon_{kl}T^{jl}\Big)^{\mp\#(t)/2} =
\Big(1-D_{2\times 2}(T)\Big)^{\mp\#(t)} \label{fint52"} \ee The
sign in the power depends on whether $x$, $\bar x$, $y$, $\bar y$
were bosonic or Grassmannian variables. In bosonic case and for
$\#(t)=2$ eq.\,\,(\ref{fint52"}) reproduces eq.\,\,(\ref{Z22}).

\subsubsection{Alternative partition functions}

\noindent

$\bullet$ We now return to direct operator calculation from
s.\ref{direva} and use it to demonstrate the differences between
various possible partition functions $Z(T)$. We restrict examples
to the simplest case $n|r=2|2$ and $\#(t)\leq 2$. Our goal is
modest: to show the difference between the use of
\be &e^{\oplus T}
= \otimes e^T = \exp \left\{T^{ij}\left(\sum_{t}^{\#(t)}
x_i(t)y_j(t)\right) \right\}\ \nn\\
&\stackrel{\#(t)=2}{=}\ \exp
\left\{T^{ij}\left(x_iy_j + x'_iy'_j\right)\right\} \ee
and
\be
&e_\otimes(T) = \sum_{m=0}^\infty \frac{T^{\otimes m}}{m!} \nn\\
&=\sum_{m=0}^{\#(t)} \frac{1}{m!}T^{i_1j_1}\ldots T^{i_mj_m}
\sum_{t_1<\ldots <t_m}^{\#(t)} x_{i_1}(t_1) y_{j_1}(t_1)\ldots
x_{i_m}(t_m)y_{j_m}(t_m)\nn\\
&\stackrel{\#(t)=2}{=}\  1 + T^{ij}(x_iy_j+x'_iy'_j) +
\frac{1}{2}T^{ij}T^{kl}x_iy_j x'_ky'_l \ee
and between
sort-respecting and sort-mixing diagrams, generated by operators
\be
&\hat E = \prod_{t<t'}^{\#(t)} \exp \left\{\epsilon_{ij}
\left(\frac{\partial^2}{\partial x_i(t) \partial x_j(t')} +
\frac{\partial^2}{\partial y_i(t) \partial
y_j(t')}\right)\right\}\nn\\
&\stackrel{\#(t)=2}{=}\ \exp \left\{\epsilon_{ij}
\left(\frac{\partial^2}{\partial x_i \partial x'_j} +
\frac{\partial^2}{\partial y_i \partial y'_j}\right)\right\} \ee
and
\be
&\hat{\cal E} = \prod_t^{\#(t)} \exp \left\{\epsilon_{ij}
\frac{\partial^2}{\partial x_i(t) \partial
y_j(t)}\right\}\nn\\
&\prod_{t<t'}^{\#(t)} \exp \left\{\epsilon_{ij}
\left(\frac{\partial}{\partial x_i(t)}+ \frac{\partial}{\partial
y_i(t)}\right)\left(\frac{\partial}{\partial x_i(t')}+
\frac{\partial}{\partial y_i(t')}\right)\right\}\nn\ee

{\footnotesize\be= \left\{\begin{array}{cc} {\rm for}\ \#(t)=1 & \exp
\left\{\epsilon_{ij}
\frac{\partial^2}{\partial x_i\partial y_j}\right\} \\
{\rm for}\ \#(t)=2\ & \exp \left\{\epsilon_{ij}
\left(\frac{\partial}{\partial x_i} \left[\frac{\partial}{\partial
y_j}+\frac{\partial}{\partial x'_j}+ \frac{\partial}{\partial
y'_j}\right] + \frac{\partial}{\partial y_i}
\left[\frac{\partial}{\partial x'_j} + \frac{\partial}{\partial
y'_j}\right] + \frac{\partial}{\partial
x'_j}\frac{\partial}{\partial y'_j} \right)\right\}
\end{array}\right.
 \nn\ee}
 \be\label{calEop}\ee
respectively. By no means this exhausts all
possible types of operators: for example, one can consider a pure
sort-mixing one (with sort-preserving $\epsilon$-vertices
completely excluded) and, say, apply one more kind of ordering
(involving sum over permutations $P$ of indices): \be \hat {\rm
E}_\otimes = \sum_{m=0}^\infty \frac{1}{m!} \sum_{P\in \sigma(m)}
\epsilon_{i_1j_1}\ldots\epsilon_{i_mj_m}
\frac{\partial^2}{\partial x_{i_1}\partial y_{P(j_1)}}\ldots
\frac{\partial^2}{\partial x_{i_m}\partial y_{P(j_m)}}\ \ \ \ee

$\bullet$ Accordingly one can consider different partition
functions:

-- our familiar $Z(T) = \hat E e^{\oplus T}$,

-- alternative sort-respecting (\ref{opinttilde}): $\tilde Z(T) =
\hat E e_\otimes(T)$,

-- their sort-mixing analogues ${\cal Z}(T) = \hat {\cal E}
e^{\oplus T}$ and $\tilde{\cal Z}(T) = \hat {\cal E}
e_\otimes(T)$,

-- pure sort-mixing partition functions like ${\rm Z}(T) = \hat
{\rm E}_\otimes e^{\oplus T}$ and $\tilde{\rm Z}(T) = \hat {\rm
E}_\otimes e_\otimes(T)$.

Still another class of examples will be considered in a separate
section \ref{combina} below.

As an example, consider ${\cal Z}(T)$. In variance with $Z(T)$,
this quantity is non-trivial already for $\#(t)=1$ (see
Fig.\,\,\ref{nt1dia} for $\#(t)=1$ and Fig.\,\,\ref{nt2dia} for
$\#(t)=2$). However, because it contains dependence on ${\rm
Pfaff}(T)$ along with that on $D_{2\times 2}(T)$ (i.e. invariant's
ring in this {\it sort}-mixing case has two independent
generators), we can not restrict consideration to diagonal $T$.
\be
&{\cal Z}_{2|2}\Big(T\Big|\#(t)=1\Big) = \exp \left(
\epsilon_{ij}\frac{\partial^2}{\partial x_i\partial y_j}\right)
\exp \Big( T^{ij}x_iy_j\Big)\nn\\
&= \sum_{m=0}^\infty \frac{1}{m!}\left(
\frac{\partial^2}{\partial x_1\partial y_2}-
\frac{\partial^2}{\partial x_2\partial y_1}\right)^m \frac{1}{m!}
\Big( T^{11}x_1y_1 + T^{12}x_1y_2 \nn\\
&+ T^{21}x_2y_1 +T^{22}x_2y_2\Big)^m = 1 +
\Big(T^{12}-T^{21}\Big) + \Big((T^{12})^2 - T^{11}T^{22}-T^{12}T^{21} \nn\\
&+ (T^{21})^2\Big) + \ldots = \frac{1}{\det_{2\times
2}(I - \hat T)} = \frac{1}{1 - {\rm Pfaff}(T) +
D(T)\phantom{5^{5^5}}\hspace{-0.5cm}}
\label{fint52"tilde}\ee
where operator $\hat T$ is given by $2\times 2$ matrix
$\epsilon_{ik}T^{jk}$, so that ${\rm Tr}\ \hat T =
\epsilon_{ik}T^{ik} = T^{12}-T^{21} = {\rm Pfaff} (T)$ and $\det\
\hat T = T^{11}T^{22}-T^{12}T^{21} = D_{2\times 2}(T) = D(T)$.

$\bullet$ In Figs \ref{nt1dia} and \ref{nt2dia} important
relations are used, see Fig.\,\,\ref{Dvsopdia}, expressing all traces
${\rm tr} \hat T^k$ through two independent invariants ${\rm
Pfaff}(T) = {\rm tr} \hat T = T^{12}-T^{21}$ and $D_{2\times 2}(T)
= \det_{2\times 2} T = T^{11}T^{22} - T^{12}T^{21}$. Relations
follow from the chain of identities:
\be &1 - {\rm
tr}\hat T + \det T = {\det}_{2\times 2}(I - \hat T) = e^{{\rm
tr}\log (I-\hat T)} = \exp \Big(-\sum_{k=1}^\infty \frac{1}{k}{\rm
tr}\ \hat T^k\Big)\nn\\
&= 1 - {\rm tr}\ \hat T + \frac{1}{2}\Big[ \Big({\rm tr}\ \hat
T\Big)^2 - {\rm tr}\ \hat T^2\Big] \nn\\
&+ \Big[-\frac{1}{6}\Big({\rm
tr}\ \hat T\Big)^3
 + \frac{1}{2}{\rm tr}\ \hat T {\rm tr}\ \hat T^2
 -\frac{1}{3}{\rm tr}\ \hat T^3 \Big]
+ \ldots
\label{Dvsop}\ee

$\bullet$ Like in the case of $Z(T)$, the large-$T$ behavior of
${\cal Z}(T)$ in (\ref{fint52"tilde}) is fully controlled by
discriminant. In this limit the corresponding integral turns into
just \be \int e^{T^{ij}x_iy_j}d^2x d^2y = \frac{1}{{\det}\, T}
\label{ordintT} \ee Vice versa, eq.\,\,(\ref{fint52"tilde}) can be
considered as a deformation of (\ref{ordintT}), allowing
perturbative expansion in positive powers of $T$, which is not
{\it a priori} obvious at the level of (\ref{ordintT}). For
generalization of (\ref{ordintT}) from $r=2$ to the first
non-trivial case of $r=3$ see eq.\,\,(\ref{intcay}) below.

Evaluation of the bilinear counterpart of the integral
(\ref{ordintT}), $I(S) = \int e^{S^{ij}x_ix_j}dx_i$ with {\it
symmetric} tensor $T=S$ is reduced to {\ref{ordintT}} by the usual
trick:
$$
I(S)I(-S) = \int e^{S^{ij}(x'_ix'_j - x''_ix''_j)} d^2x d^2x'' =
\frac{1}{4}\int e^{S^{ij}x_iy_j}d^2x d^2y = \frac{1}{4\, {\det} S}
$$
with $x_i = x'_i+x''_j$, $y_i = x'_i - x''_i$ (symmetricity of
$S$, $S^{ij} = S^{ji}$ is essential for this change of variables
to work), and since $I(S) = (-)^nI(-S) = I(-S)$, we get \be I(S) =
\int e^{S^{ij}x_ix_j}dx_i = \frac{1}{2\sqrt{{\det}
S\phantom{5^{5^5}}\hspace{-0.5cm}}} \label{ordintS} \ee (to avoid
possible confusion we remind that $\pi$-factors are included into
the definitions of integrals). Deformation like
(\ref{fint52"tilde}) for $I(S)$ with the help of an
$\epsilon$-tensor is provided by still another important
construction to which we devote a separate subsection
\ref{combina}.

\Fig{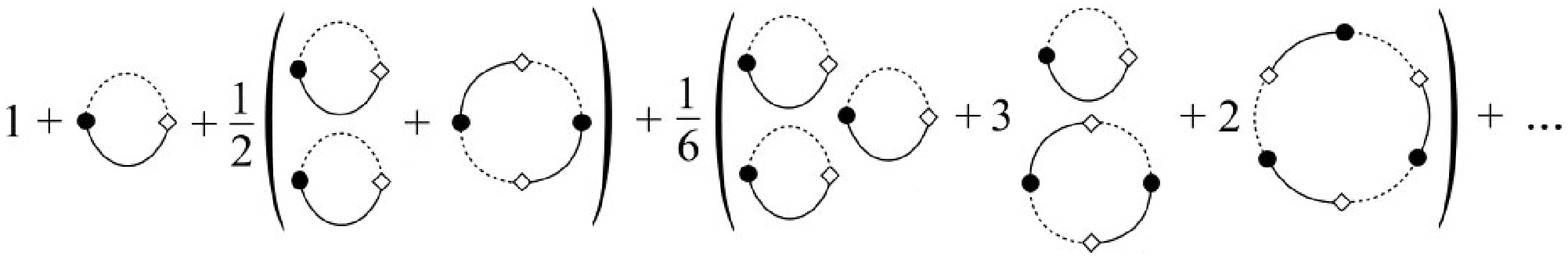}
{320,51}
{The first few diagrams for partition
function ${\cal Z}_{2|2}\Big(T\Big|\#(t)=1\Big) =
\exp\Big(\epsilon_{ij} \frac{\partial^2}{\partial x_i\partial
y_j}\Big)\ e^{T^{ij}x_iy_j} = 1 + \epsilon_{ij}T^{ij} +
\frac{2\epsilon_{ij}\epsilon_{kl}}{(2!)^2}
\Big(T^{ik}T^{jl} - T^{il}T^{jk} 
\Big) + \ldots$, see (\ref{fint52"tilde}). In this case
$\epsilon$-vertices are allowed to mix {\it sorts} of the lines,
no vertices without mixing are present because $\#(t)=1$ is too
small. In transformations relations (\ref{Dvsop}) are used, see
Fig.\,\,\ref{Dvsopdia}. }

\Fig{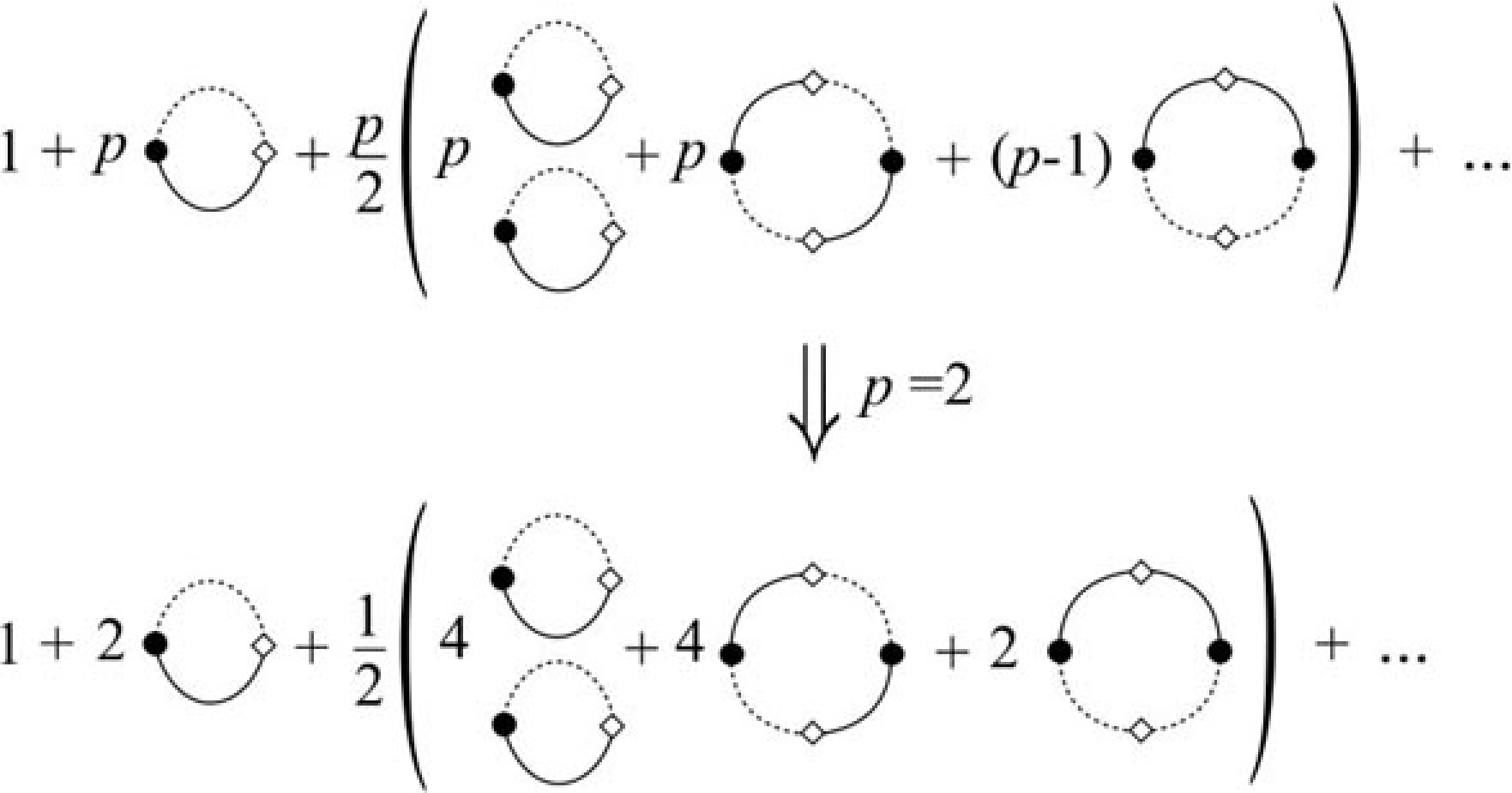}
{320,168}
{The first few diagrams for partition
function ${\cal Z}_{2|2}\Big(T\Big|\#(t)=2\Big)$,
Eq.\,\,(\ref{calEop}). Both sort-mixing and sort-respecting
$\epsilon$-vertices are contributing in this case for $m\geq 2$.
Here $p=\#(t)=2$. For $p=1$ we reproduce Fig.\,\,\ref{nt1dia}. }

\Fig{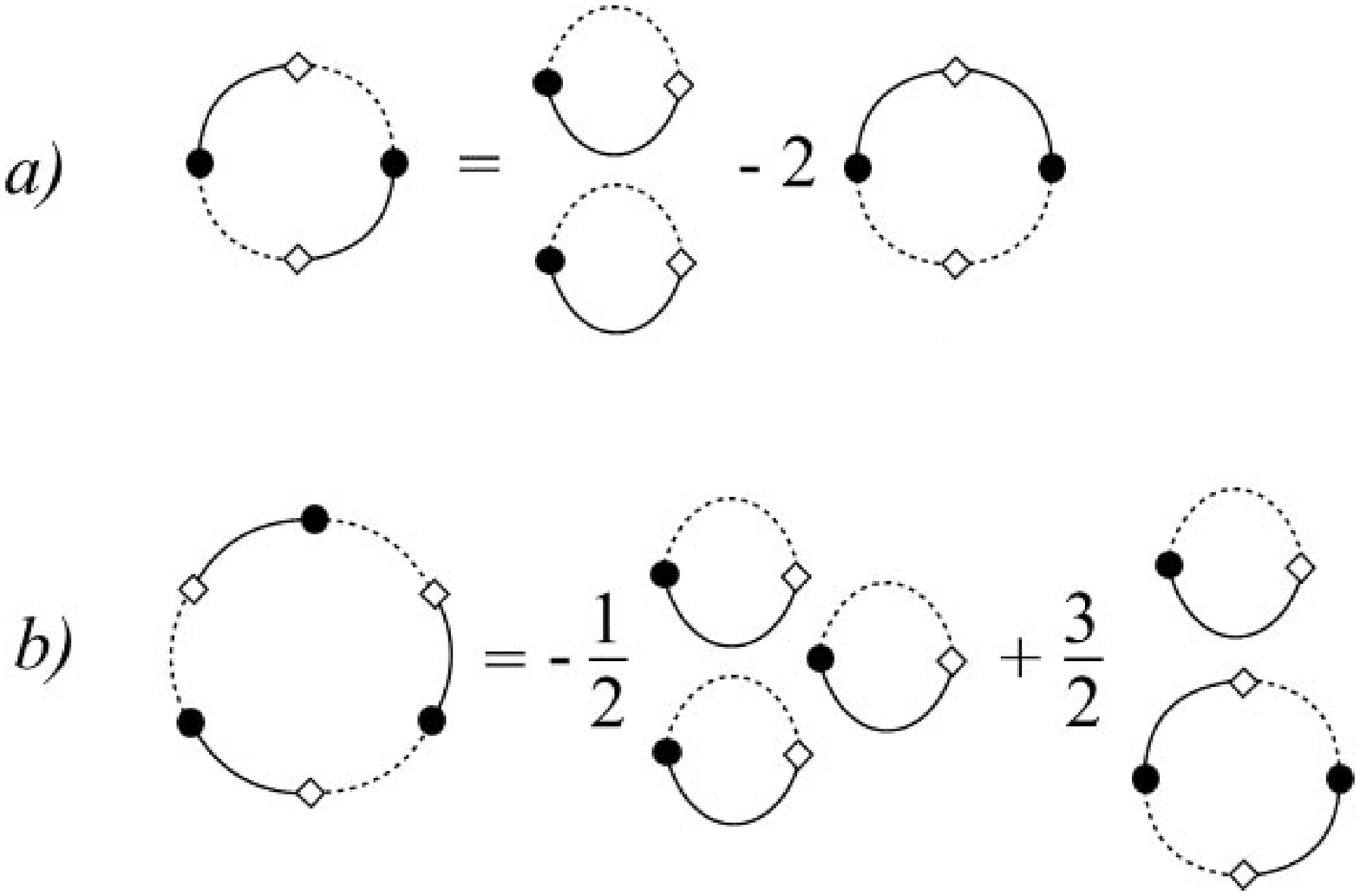} {288,189} {The first two of relations (\ref{Dvsop})
among invariants of the tensor algebra ${\cal T}(T)$ in the
$n|r=2|2$ case. \ \ {\bf A.}\ ${\rm tr} \hat T^2 = \Big({\rm
Pfaff}(T)\Big)^2 - \ 2D_{2\times 2}(T)$ or simply
$\big(T^{12}\big)^2 - 2T^{11}T^{22} + \big(T^{21}\big)^2 =
\big(T^{12}-T^{21}\big)^2 - 2
\big(T^{11}T^{22}-T^{12}T^{21}\big)$.\ \ \ {\bf B.}\ ${\rm tr}\
\hat T^3 = -\frac{1}{2}\Big({\rm tr}\ \hat T\Big)^3 +
\frac{3}{2}\, {\rm tr}\ \hat T \, {\rm tr}\ \hat T^2$. }

\subsubsection{Pure tensor-algebra (combinatorial)
partition functions \label{combina}}

\noindent

$\bullet$ From diagrammatic (tensor algebra) perspective the most
natural generating function would be just a sum over all possible
diagrams, without any reference to a functional integral, like
(\ref{fint}). It can seem that this is exactly what was done in
the previous examples, but actually the simplest
pure-tensor-algebra example has not been considered yet, and there
is an important difference: the size of the "space-time" $\#(t)$
is no longer a free parameter, moreover it needs to be made
$m$-dependent -- and this gives rise to a new class of partition
functions. Such {\it combinatorial} partition functions still
depend on the choice of the structure group (we distinguish
between the {\it sort}-respecting and {\it sort}-mixing cases) and
also on the choice of contributing diagrams (here the two simplest
options are {\it all} and {\it all connected} diagrams). Of
course, the weights also matter (this is where the
$\#(t)$-dependence is actually contained), we shall keep $1/m!$
weights in the {\it all-diagram} case, in order to keep contact
with exponentials, and to preserve the usual relation \be {\cal
Z}(T) = e^{{\cal F}(T)} \label{ZFrel} \ee between the {\it
all-diagram} and {\it connected-diagram} partition functions.

$\bullet$ In the {\it sort}-mixing case (i.e. when the {\it
structure group} is small, $G=SU(n)$) in the $m$-th order in $T$
we have $rm$ lines attached to $T$-vertices, and their other ends
are attached to $\frac{rm}{n}$ $\epsilon$-vertices of valence $n$.
The total number of diagrams is equal to the number of ways the
$rm$ lines can be split into $\frac{rm}{n}$\ \ $n$-ples, i.e. to
\be N_{n|r}(m) = \frac{(mr)!}{(n!)^{rm/n}} \label{Nsm} \ee Allowed
are only $m$, such that $rm$ is divisible by $n$, otherwise
$N_{n|r}(m) = 0$.

In the {\it sort}-respecting case (when the {\it structure group}
is maximally big, $G=SU(n_1)\times SU(n_r)$), all $r$ lines,
attached to $T$ are different, and in the $m$-th order in $T$
there are $r$ sets of $m$ lines, attached to $T$-vertices, and
their other ends are attached to $\frac{m}{n_k}$
$\epsilon$-vertices of sort $k$ and valence $n_k$. Thus the total
number of diagrams is \be N_{n_1\times\ldots\times n_r}(m) =
\prod_{k=1}^r \frac{(m)!}{(n_k!)^{m/n_k}} \label{Nsr} \ee Again
only $m$, divisible by all $n_k$, are allowed, otherwise
$N_{n_1\times\ldots\times n_r}(m) = 0$.

The numbers (\ref{Nsm}) and (\ref{Nsr}) count {\it all} diagrams,
not obligatory connected. Enumerating of {\it connected} diagrams
and their further topological classification is somewhat trickier
combinatorial problem, with complexity increasing with growing $r$
and $n$'s.

The pure combinatorial partition functions are defined as \be
{\cal Z}^{com}(T) = \sum_{m=0}^\infty \frac{1}{m!}\ \Big(\sum {\rm
over\ all}\ N(m)\ {\rm diagrams\ of\ order}\ m\Big)
\nn \\
{\cal F}^{com}(T) = \sum_{m=1}^\infty \frac{1}{m}\ \Big(\sum {\rm
over\ all}\ n(m)\ {\rm connected\ diagrams\ of\ order}\ m\Big)
\label{ZFdef} \ee Of course, one can introduce extra weights
("coupling constants") in internal sum, distinguishing between
different topologies, this is the route deep into the field of
matrix models and topological field theories
\cite{UFN2,UFN3,Berk,amm}.

$\bullet$ In operator language ${\cal Z}^{com}(T)$ is very similar
to $\tilde{\cal Z}(T) = \hat{\cal E} e_\otimes(T)$, but with
$\#(t)=m$, i.e. with $t$-lattice size adjusted to the number of
$T$-vertices. To demonstrate the differences between combinatorial
and functional-integral partition functions, we write them down
together. For $n|r=2|2$ we have for the sort-mixing combinatorial
partition function (Fig.\ref{comdia}): \be \hspace{-7.5cm} {\cal
Z}^{com}_{2|2}(T) = \hspace{7.5cm} \label{Zcomsm} \ee
\vspace{-0.35cm}
$$
=\sum_{m=0}^\infty \frac{1}{(m!)^2} \left\{ \epsilon_{ij}
\sum_{t=1}^m \frac{\partial^2}{\partial x_i(t)\partial y_i(t)} +
\epsilon_{ij} \sum_{t<t'}^m \left(\frac{\partial}{\partial x_i(t)}
+ \frac{\partial}{\partial y_i(t)}\right)
\left(\frac{\partial}{\partial x_j(t')} + \frac{\partial}{\partial
y_j(t')}\right)\right\}^m \prod_{t=1}^m
\Big(T^{ij}x_i(t)y_j(t)\Big)
$$ $$ = 1 + \epsilon_{ij}T^{ij} +
\ldots
$$
while its simplest functional-integral counterparts with the
smallest $t$-lattices $\#(t)=1$ (Fig.\ref{nt1dia}) and $\#(t)=2$
(Fig.\ref{nt2dia}) are:
$$
{\cal Z}_{2|2}\Big(T\Big|\#(t)=1\Big)
\stackrel{(\ref{fint52"tilde})}{=}
\sum_{m=0}^\infty \frac{1}{(m!)^2}
\left\{\epsilon_{ij}\frac{\partial^2} {\partial x_i\partial
y_j}\right\}^m \Big(T^{ij}x_iy_j\Big)^m =
1 + \epsilon_{ij}T^{ij} +
\frac{2\epsilon_{ij}\epsilon_{kl}}{(2!)^2}
\Big(T^{ik}T^{jl} - T^{il}T^{jk} 
\Big) + \ldots
$$
$$
\hspace{-6cm} {\cal Z}_{2|2}\Big(T\Big|\#(t)=2\Big)\
\stackrel{(\ref{calEop})}{=}\ \hspace{6cm}
$$
$$ =
\sum_{m=0}^\infty \frac{1}{(m!)^2} \left\{\epsilon_{ij}
\left(\frac{\partial}{\partial x_i} \left[\frac{\partial}{\partial
y_j}+\frac{\partial}{\partial x'_j}+ \frac{\partial}{\partial
y'_j}\right] + \frac{\partial}{\partial y_i}
\left[\frac{\partial}{\partial x'_j} + \frac{\partial}{\partial
y'_j}\right] + \frac{\partial}{\partial
x'_j}\frac{\partial}{\partial y'_j} \right)\right\}^m
\Big(T^{ij}\left[x_iy_j+x'_iy'_j\right]\Big)^m =
$$
$$
= \sum_{m=0}^\infty \frac{1}{(m!)^2} \left\{ \epsilon_{ij}
\sum_{t=1}^2 \frac{\partial^2}{\partial x_i(t)\partial y_i(t)} +
\epsilon_{ij} \sum_{t<t'}^2 \left(\frac{\partial}{\partial x_i(t)}
+ \frac{\partial}{\partial y_i(t)}\right)
\left(\frac{\partial}{\partial x_j(t')} + \frac{\partial}{\partial
y_j(t')}\right)\right\}^m \left(T^{ij}\left[\sum_{t=1}^2
x_i(t)y_j(t)\right]\right)^m =
$$  $$ = 1 + 2\epsilon_{ij}T^{ij} +
\ldots
$$

\Fig{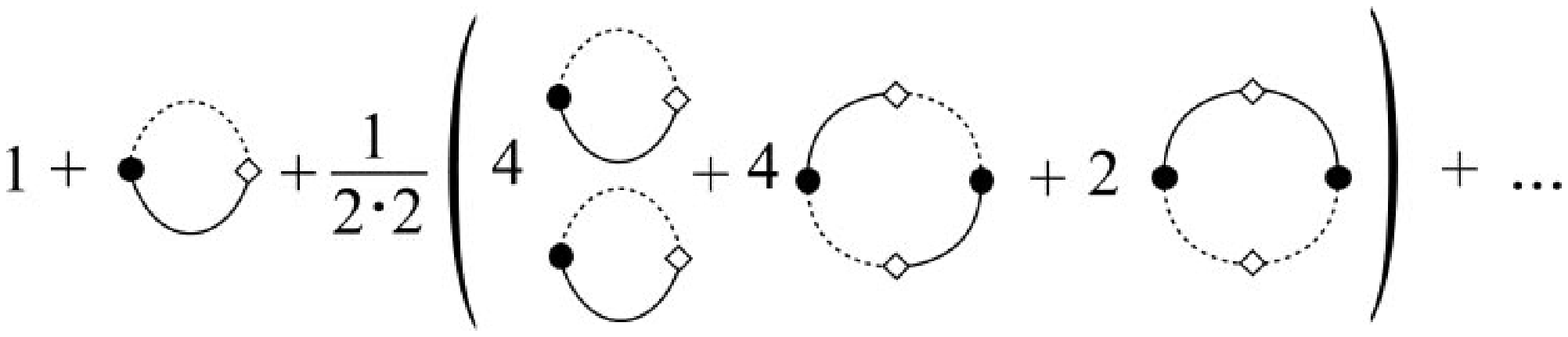} {372,78} {The first few diagrams for the sort-mixing
combinatorial partition function ${\cal Z}_{2|2}^{com}(T)$, see
(\ref{Zcomsm}). The first non-trivial diagram is taken from
Fig.\ref{nt1dia} with $\#(t)=1$, the second -- from
Fig.\ref{nt2dia} with $\#(t)=2$ -- in accordance with the rule
$p=\#(t)=m$. The contribution from Fig.\ref{nt2dia} is further
divided by $2$ because of the difference between
$\Big(T(xy+x'y')\Big)^2$ and $(Txy)(Tx'y')$. }

$\bullet$ Similarly, in the sort-respecting case the combinatorial
partition function (Fig.\ref{comspdia}) is
$$
{Z}^{com}_{2\times 2}(T) = \sum_{m=0}^\infty \frac{1}{(m!)^2}
\left\{\epsilon_{ij} \sum_{t<t'}^m
\left(\frac{\partial^2}{\partial x_i(t)\partial x_j(t')} +
\frac{\partial^2}{\partial y_i(t)\partial y_j(t')}
\right)\right\}^m \prod_{t=1}^m \Big(T^{ij}x_i(t)y_j(t)\Big) =
$$
\vspace{-0.25cm} \be = \sum_{m=0}^\infty \frac{1}{(m!)^2(2m)!}
\left(\epsilon_{ij} \sum_{t<t'}^{2m} \frac{\partial^2}{\partial
x_i(t)\partial x_j(t')} \right)^m \left(\epsilon_{ij}
\sum_{t<t'}^{2m} \frac{\partial^2}{\partial y_i(t)\partial
y_j(t')} \right)^m\ \prod_{t=1}^{2m} \Big(T^{ij}x_i(t)y_j(t)\Big)
\label{Zcomsr} \ee while its simplest functional-integral
counterpart (Fig.\ref{nt2spdia}) is
$$
Z_{2\times 2}\Big(T\Big|\#(t)=2\Big) = \sum_{m=0}^\infty
\frac{1}{[(2m)!]^2} \left\{\epsilon_{ij} \sum_{t<t'}^{\#(t)=2}
\left(\frac{\partial^2}{\partial x_i(t)\partial x_j(t')} +
\frac{\partial^2}{\partial y_i(t)\partial y_j(t')}
\right)\right\}^{2m}\!\! \left(T^{ij}\left[\sum_{t=1}^{\#(t)=2}
x_i(t)y_j(t)\right] \right)^{2m}\!\!\!\! = $$ $$ =
\sum_{m=0}^\infty \frac{1}{[(2m)!]^2} \left\{\epsilon_{ij}
\left(\frac{\partial^2}{\partial x_i\partial x'_j} +
\frac{\partial^2}{\partial y_i\partial y'_j}
\right)\right\}^{2m}\!\!\!
\Big(T^{ij}(x_iy_j+x'_iy'_j)\Big)^{2m}\!\! =
$$ \vspace{-0.25cm}
\be = \sum_{m=0}^\infty \frac{1}{(m!)^2(2m)!} \left(\epsilon_{ij}
\frac{\partial^2}{\partial x_i\partial x'_j} \right)^m\!\!
\left(\epsilon_{ij} \frac{\partial^2}{\partial y_i\partial y'_j}
\right)^m\!\! \Big(T^{ij}(x_iy_j+x'_iy'_j)\Big)^{2m}
\label{Znt2sp} \ee

\Fig{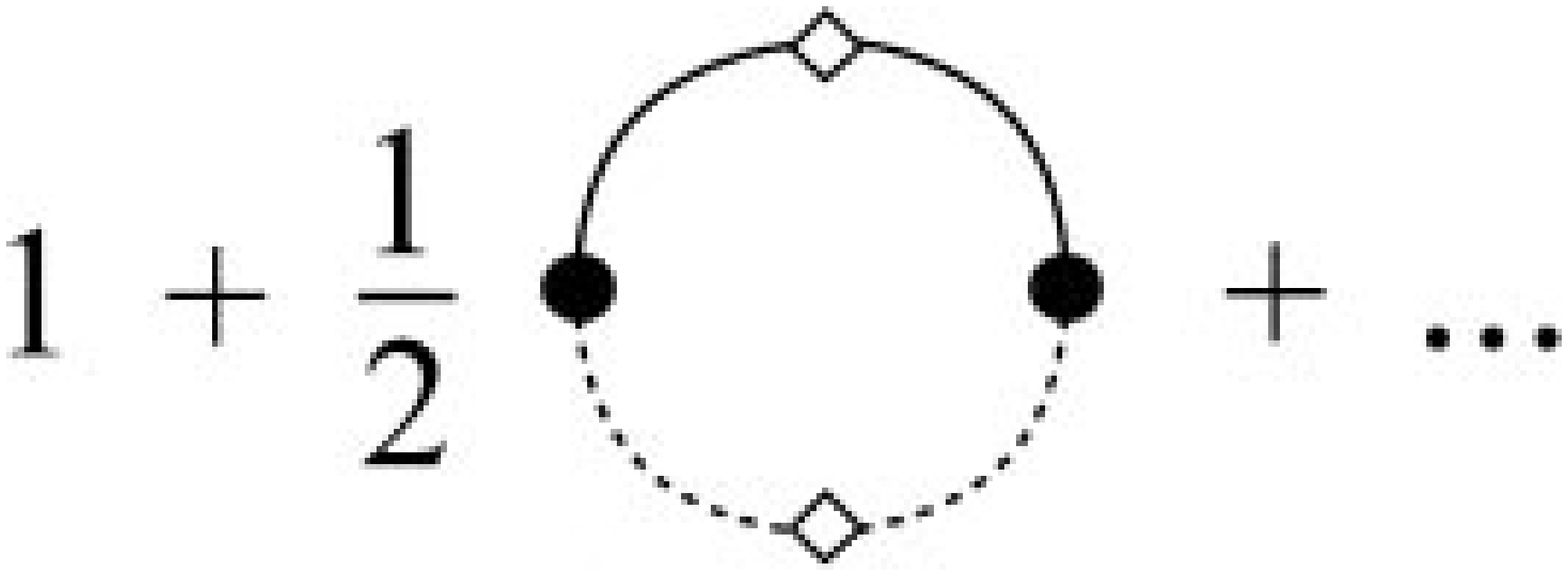} {133,48} {The first few diagrams for the
sort-respecting combinatorial partition function
${Z}_{2|2}^{com}(T)$, see (\ref{Zcomsr}). Now $\epsilon$-vertices
are not allowed to mix lines of different {\it sorts}, there are
$r=2$ different kinds of valence $n=2$ $\epsilon$-vertices. Only
diagrams with even number of $T$-vertices are non-vanishing. }

\Fig{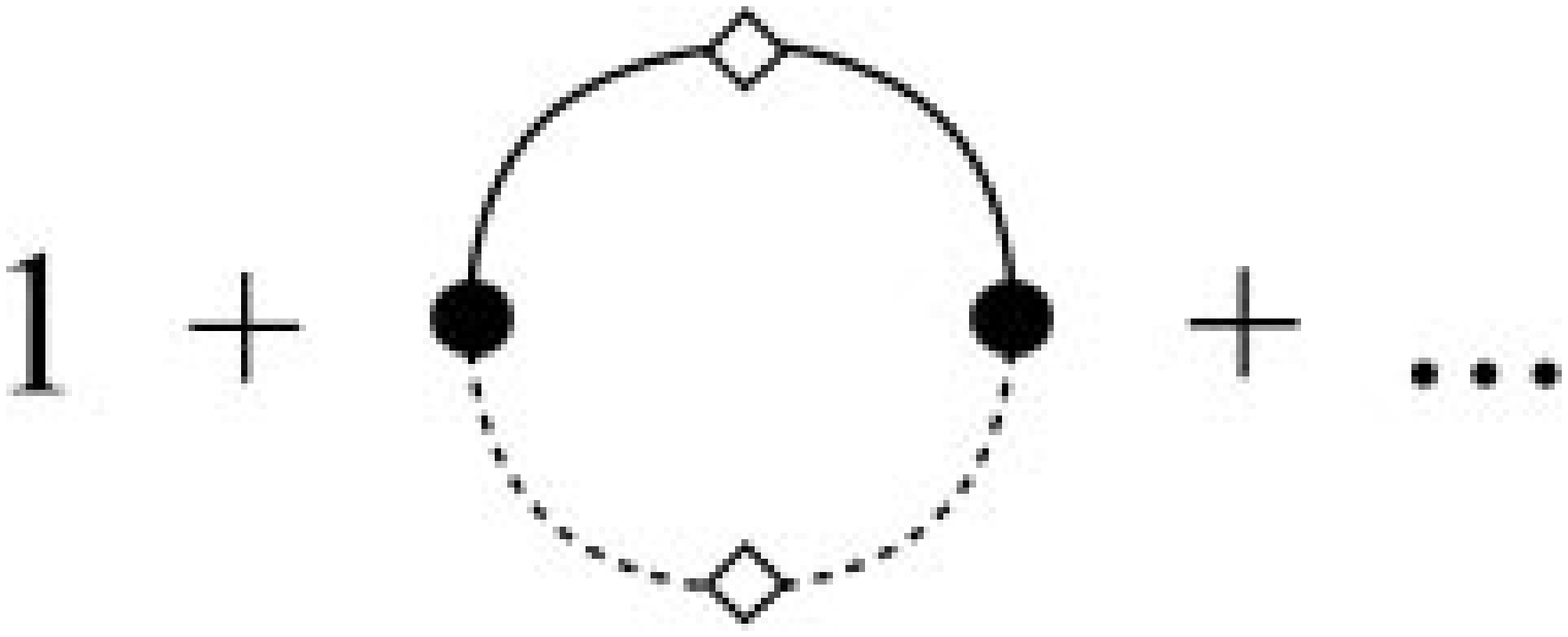} {120,48} {The first few diagrams for the
sort-respecting partition function ${Z}_{2\times
2}\Big(T\Big|\#(t)=2\Big)$, see (\ref{Znt2sp}). }

$\bullet$ Because of the lattice-size ("space-time" volume)
dependence on the order of perturbation theory, ${\cal Z}^{com}$
and $Z^{com}$ do not have immediate functional integral
representation {\it a la} (\ref{fint}) and are somewhat more
difficult to calculate, even for $n|r=2|2$.

In the $n|r=2|2$ case the set of {\it connected} vacuum diagrams
is very simple: these are circles with $m$ $T$-vertices of valence
$r=2$ and $m$ $\epsilon$-vertices of valence $n=2$. The only
reason for existence of different diagrams arises in the
sort-mixing case, when there are sort-respecting and sort-mixing
$\epsilon$-vertices.

Thus in the sort-respecting case the "prepotential" (see
Fig.\ref{prepsrdia}) is
$$
F^{com}_{2\times 2}(T) = \sum_{m=1}^\infty \frac{1}{2m}
\Big(\epsilon_{i_{2m}i_1}T^{i_1j_1}\varepsilon_{j_1j_2}T^{i_2j_2}
\epsilon_{i_2i_3}T^{i_3j_3}\varepsilon_{j_3j_4}\ldots
\varepsilon_{j_{2m-1}j_{2m}}T^{i_{2m}j_{2m}}\Big) =
$$
\vspace{-0.25cm} \be = \sum_{m=1}^\infty \frac{1}{2m}{\rm tr}
\Big(\hat T \hat{\tilde T}\Big)^{m} = -\frac{1}{2}\log
\Big\{{\det}_{2\times 2} (I - \hat T\hat{\tilde T})\Big\}
\label{prepsr} \ee \vspace{0.1cm} where $(\hat T)^i_j =
\varepsilon_{jk}T^{ik}$, i.e. \ $\hat T = \left(\begin{array}{cc}
T^{12} & T^{22} \\ -T^{11} & -T^{21}
\end{array}\right)$,\ while
$(\hat{\tilde T})^i_j = \epsilon_{jk}T^{ki}$, i.e. \ $\hat{\tilde
T}= \left(\begin{array}{cc} T^{21} & T^{22} \\ -T^{11} & -T^{12}
\end{array}\right)$, so that
$\hat T\hat{\tilde T} =$ $ \left(\begin{array}{cc} -\det T & 0 \\
0 & -\det T \end{array}\right)  = - D_{2\times 2}(T) \cdot I$,\
and \be Z^{com}_{2\times 2}(T)\ \stackrel{(\ref{ZFrel})}{=}\ \exp
\Big(F^{com}_{2\times 2}(T)\Big) = \frac{1}{\sqrt{{\det}_{2\times
2} \Big(I-\hat T\hat{\tilde T}\Big)}}= \frac{1}{1 + D_{2\times
2}(T)\phantom{5^{5^5}}\hspace{-0.5cm}} \ee Bilinear tensor $T$
maps a pair ($r=2$) of $2d$ ($n=2$) vector spaces $U\times V
\rightarrow C$. Operators $\hat T$ and $\hat{\tilde T}$
interchange the two spaces $\ \hat T:\  U \rightarrow V$, \ \
$\hat{\tilde T}:\ V \rightarrow U$,\ while \ $\hat T\hat{\tilde
T}:\  V\rightarrow V$\ and \ $\hat{\tilde T}\hat T:\  U\rightarrow
U$. In other words, $Z^{com}_{2\times 2}(T)$ is associated with an
operator \be \left(\begin{array}{cc} 0 & \hat T \\ \hat{\tilde T}
& 0
\end{array}\right) \ \ \ {\rm on\ the\ space} \ \ \
U\oplus V 
\label{oper22} \ee

It is instructive to look at the matching of coefficients in
(\ref{ZFrel}). In the sort-respecting case this relation
symbolically (with $\circled{l}$ denoting a single circle diagram
of the length $l$) states: \be Z^{com}_{2\times 2}(T) = \exp
\Big(F^{com}_{2\times 2}(T)\Big) = \exp \left(\sum_m \frac{1}{2m}
\circled{2m}\right) = \ee
$$
\begin{array}{ccccccc}
= 1 & +\Big(\frac{1}{2}\circled{2} &+ \frac{1}{4}\circled{4} &+
\frac{1}{6}\circled{6} &+ \frac{1}{8}\circled{8} &+
\frac{1}{10}\circled{10} &+ \ldots\Big) +
\\
&&+\frac{1}{2}\Big(\frac{1}{4}{\circled{2}}^2 &+
\frac{1}{4}\circled{2}\circled{4} &+ \frac{1}{16}{\circled{4}}^2 +
\frac{1}{6}\circled{2}\circled{6} &+
\frac{1}{8}\circled{2}\circled{8} +
\frac{1}{12}\circled{4}\circled{6}&+ \ldots\Big)\ +
\\
&&&+\frac{1}{6}\Big(\frac{1}{8}{\circled{2}}^3 &+
\frac{3}{16}{\circled{2}}^2\circled{4} &+
\frac{3}{32}\circled{2}{\circled{4}}^2 +
\frac{1}{8}{\circled{2}}^2\circled{6} &+ \ldots\Big)\ +
\\
&&&&+\frac{1}{24}\Big(\frac{1}{16}{\circled{2}}^4 &+
\frac{1}{8}{\circled{2}}^3\circled{4} & + \ldots\Big)\ +
\\
&&&&&+\frac{1}{120}\Big(\frac{1}{32}{\circled{2}}^5 &+
\ldots\Big)\ +
\\
&&&&&&+ \ldots\\
&&&&&&\\
\hline
&&&&&&\\
1 & \frac{1}{2} & \frac{1}{4}+\frac{1}{8} &
\frac{1}{6}+\frac{1}{8} + \frac{1}{48} & \frac{1}{8}+\frac{1}{32}
+ \frac{1}{12} +
&\frac{1}{10} + \frac{1}{16} + \frac{1}{24} + \frac{1}{64}+
\frac{1}{48} + & \\
&&&& +\frac{1}{32} + \frac{1}{16\cdot 24} &
+ \frac{1}{8\cdot 24} + \frac{1}{120\cdot 32}& \ldots \\
&&&&&&\\
= 1 & = \frac{1}{2} & =\frac{3}{8} & =\frac{5}{16} &
=\frac{35}{128} & = \frac{63}{256} & \ldots \\
&&&&&&
\end{array}
$$
In the bottom of the table we sum up the coefficients in front of
all diagrams of the given order $2m$ (i.e. in a given column), and
they indeed coincide with
$$
\frac{N_{2\times 2}(2m)}{(m!)^2(2m!)} \ \stackrel{(\ref{Nsr})}{=}
\ \frac{\Big((2m)!/2^m\Big)^2}{(m!)^2(2m!)} = \frac{(2m)!}{4^m
(m!)^2} = \frac{\Gamma\left(m+\frac{1}{2}\right)}
{m!\Gamma\left(\frac{1}{2}\right)}
$$

\Fig{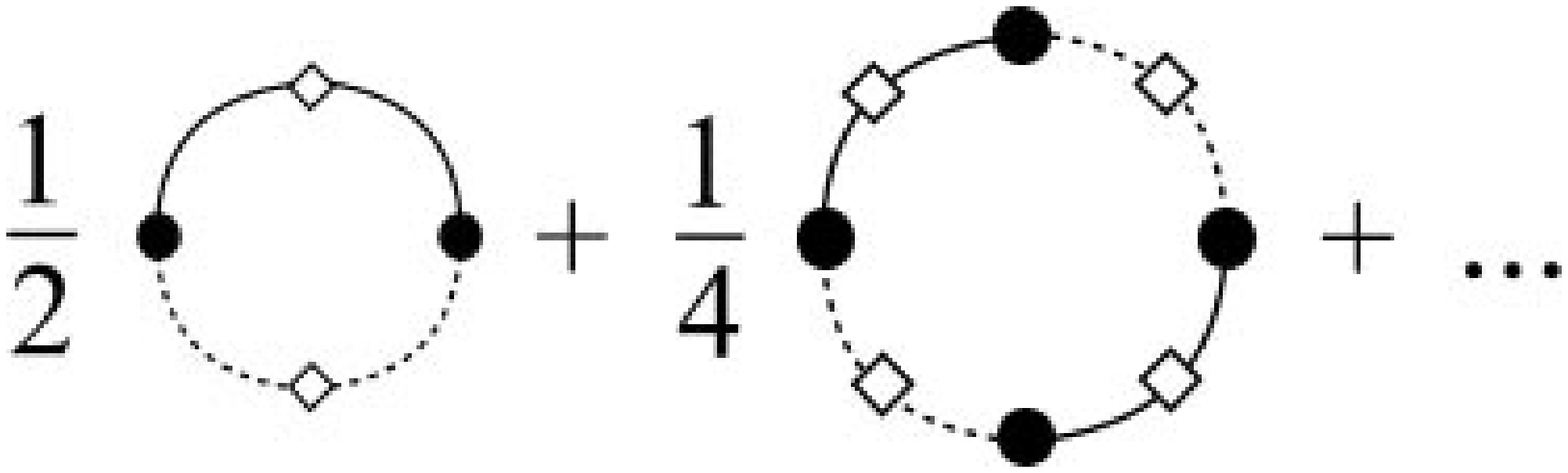} {186,55} {The first few diagrams for the
sort-respecting combinatorial prepotential $F_{2\times
2}^{com}(T)$, see (\ref{prepsr}). }

$\bullet$ In the {\it pure}-sort-mixing prepotential ${\rm F}(T)$
there is also no ambiguity in topology of diagrams, moreover now
only operators $\hat T$ arise, no $\hat{\tilde T}$ are needed, see
Fig.\ref{preppsmdia}: \be {\rm F}_{2|2}^{com}(T) =
\sum_{m=1}^\infty \frac{1}{2m} \ 2^m {\rm tr}\, \hat T^m =
-\frac{1}{2}\log \left\{{\det}_{2\times 2} (I-2\hat T)\right\}
\label{preppsm} \ee and \be {\rm Z}_{2|2}^{com}(T) = \exp\Big({\rm
F}_{2|2}^{com}(T)\Big) = \frac{1}{\sqrt{{\det}_{2\times 2}(I -
2\hat T)}} = \frac{1}{\sqrt{1-2{\rm Pfaff}(T) + 4D_{2\times 2}(T)
\phantom{5^{5^5}}\hspace{-0.5cm}}} \label{compsm22} \ee For
symmetric $T=S$ the trace\ ${\rm tr}\, \hat T = T^{12}-T^{21} =
{\rm Pfaff}(T) = 0$, so that eq.(\ref{compsm22}) turns into \be
{\cal Z}^{com}_{2|2}(S) =
\frac{1}{\sqrt{1+4D_{2|2}(S)\phantom{5^{5^5}}\hspace{-0.5cm}}} =
\frac{1}{\sqrt{1 + 4 \det S\phantom{5^{5^5}}\hspace{-0.5cm}}} \ee
and provides the tensor-algebra deformation of the integral
(\ref{ordintS}).

\Fig{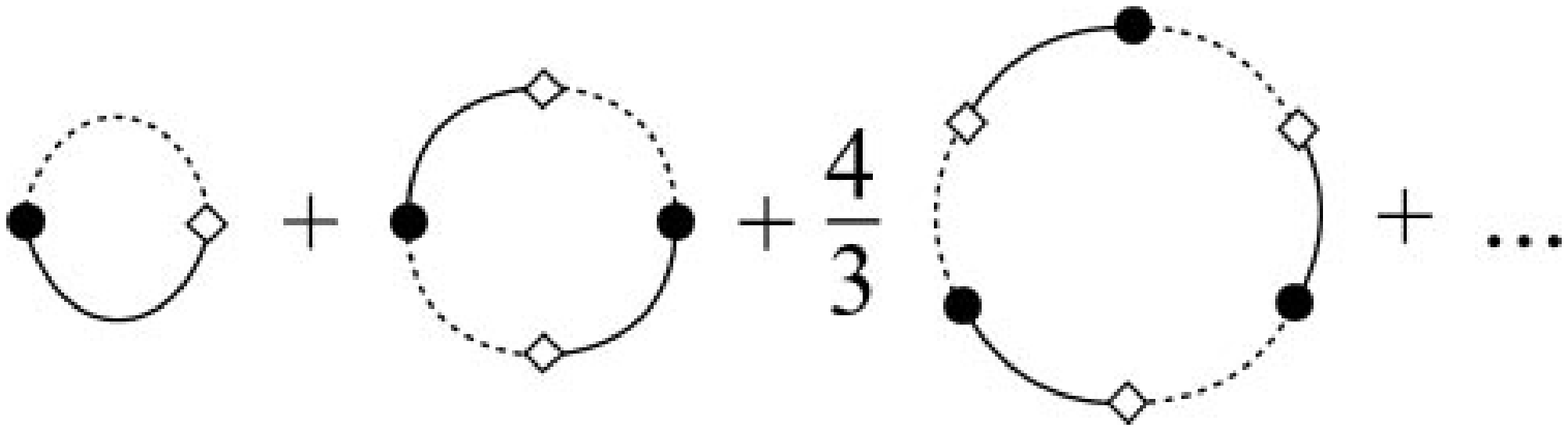} {242,65} {The first few diagrams for the {\it
pure} sort-mixing combinatorial prepotential ${\rm
F}_{2|2}^{com}(T)$, see (\ref{preppsm}). }

$\bullet$ In the ordinary sort-mixing case connected cyclic
diagrams can still differ by the sequences of sort-preserving and
sort-mixing $\epsilon$-vertices: each sort-preserving $\epsilon$
switches between sequences of operators $\hat T$ and $\hat{\tilde
T}$, see Fig.\ref{prepsmdia}.

Matching of coefficients now works as follows: \be {\cal
Z}^{com}_{2|2}(T) = \exp \Big({\cal F}^{com}_{2|2}(T)\Big) = \exp
\left(\sum_m \frac{1}{2m} 2^m\circled{m}\right) = \ee
$$
\begin{array}{ccccccc}
= 1 & +\Big(\circled{1} &+ \circled{2} &+ \frac{4}{3}\circled{3}
&+ 2\circled{4} &+ \frac{16}{5}\circled{5} &+ \ldots\Big) +
\\
&&+\frac{1}{2}\Big({\circled{1}}^2 &+ 2\circled{1}\circled{2} &+
\frac{8}{3}\circled{1}\circled{3} + {\circled{2}}^2  &+
4\circled{1}\circled{4} + \frac{8}{3}\circled{2}\circled{3}&+
\ldots\Big)\ +
\\
&&&+\frac{1}{6}\Big({\circled{1}}^3 &+ 3{\circled{1}}^2\circled{2}
&+ 3\circled{1}{\circled{2}}^2 + 4{\circled{1}}^2\circled{3} &+
\ldots\Big)\ +
\\
&&&&+\frac{1}{24}\Big({\circled{1}}^4 &+
4{\circled{1}}^3\circled{2} & + \ldots\Big)\ +
\\
&&&&&+\frac{1}{120}\Big({\circled{1}}^5 &+ \ldots\Big)\ +
\\
&&&&&&+ \ldots\\
&&&&&&\\
\hline
&&&&&&\\
1 & 1 & 1+\frac{1}{2} & \frac{4}{3}+1 + \frac{1}{6} & 2 +
\frac{4}{3}+\frac{1}{2} + \frac{1}{2} + \frac{1}{24}
&\frac{16}{5} + 2+ \frac{4}{3} + \frac{2}{3} + \frac{1}{2}+
\frac{1}{6} + \frac{1}{120}& \ldots \\
&&&&&&\\
= 1 & = 1 & =\frac{3}{2} & =\frac{5}{2} &
=\frac{35}{8} & = \frac{63}{8} & \ldots \\
&&&&&&
\end{array}
$$
Again sums of the coefficients in each column coincide with
$$
\frac{N_{2|2}(m)}{(m!)^2} \ \stackrel{(\ref{Nsm})}{=} \
\frac{(2m)!/2^m}{(m!)^2} =
2^m\frac{\Gamma\left(m+\frac{1}{2}\right)}
{m!\Gamma\left(\frac{1}{2}\right)}
$$

\Fig{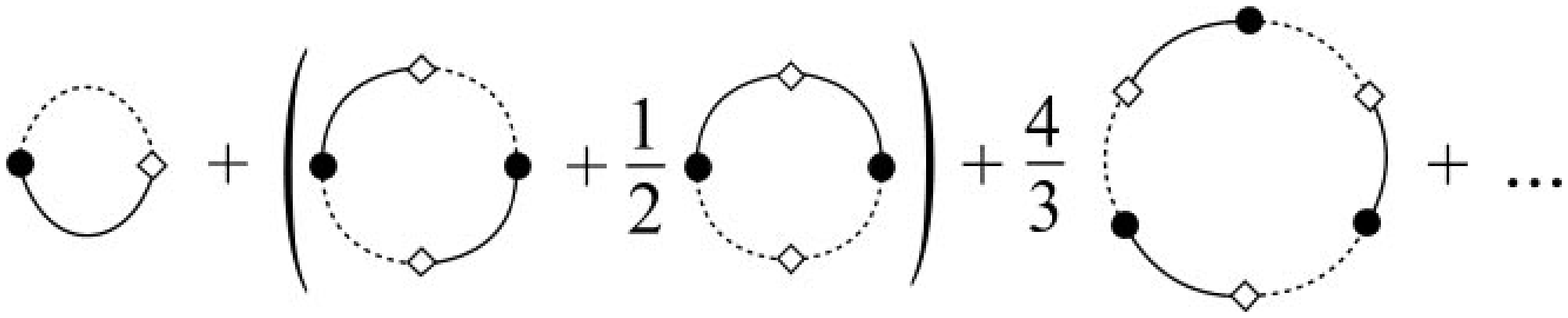} {332,65} {The first few diagrams for the
sort-mixing combinatorial prepotential ${\cal F}_{2|2}^{com}(T)$,
see~(5.108). }

$\bullet$ For higher $r>2$ the cyclic diagrams (glued segments)
are substituted by surfaces, glued from $r$-angles. In particular
case of $n|r=2|3$ with trilinear tensors the role of connected
diagrams is played by {\it dessins d'enfants} \cite{desen}.

Trilinear tensors are associated with a triple ($r=3$) of $2d$
($n=2$) vector spaces: \ $T: \ U\times V\times W \rightarrow C$.
The analogue of operator (\ref{oper22}) is given \cite{ChD} by a
pair of operators \ $U \rightarrow V\otimes W$\ and \ $V\otimes W
\rightarrow U$, i.e. \be \hat{\cal T} =
\left(\begin{array}{cc} 0 & T_i^{\cdot jk} \\
T^i_{\cdot jk} & 0 \end{array} \right) =
\left(\begin{array}{cccccc} 0 & 0 & T_1^{\cdot 11} & T_1^{\cdot
12} &
T_1^{\cdot 21} & T_1^{\cdot 22} \\
0 & 0 & T_2^{\cdot 11} & T_2^{\cdot 12} &
T_2^{\cdot 21} & T_2^{\cdot 22} \\
T^1_{\cdot 11} & T^2_{\cdot 11} & 0& 0& 0& 0 \\
T^1_{\cdot 12} & T^2_{\cdot 12} & 0& 0& 0& 0 \\
T^1_{\cdot 21} & T^2_{\cdot 21} & 0& 0& 0& 0 \\
T^1_{\cdot 22} & T^2_{\cdot 22} & 0& 0& 0& 0
\end{array}\right)
\ \ \ {\rm on} \ \ \ U\oplus (V\otimes W) \ee In fact this
operator is a member of a triple, permuted by the discrete
$\sigma_3$ factor of the structure group: the two other operators
$\hat{\cal T}'$ and $\hat{\cal T}''$ act on $V \oplus (W\otimes
U)$ and on $W \oplus (U\otimes V)$ respectively. We remind, that
indices of $T^{ijk}$ are lowered with the help of the $\epsilon$
tensor of rank $n=2$. It is easy to check that ${\rm Tr}_{6\times
6} \hat{\cal T}^p$ is non-vanishing only for $p = 4m$, also
${\det}_{6\times 6} \hat{\cal T} = 0$. The first non-vanishing
trace is equal to Cayley hyperdeterminant (normalized so that
$D(T_{diag}) = (ab)^2$):
$$
\frac{1}{4}{\rm Tr}_{6\times 6} \hat{\cal T}^4 = D_{2\times
2\times 2}(T),
$$
and higher traces are its powers, according to the analogue of
(\ref{Dvsop}): \be \vspace{-0.2cm} 1 - \frac{1}{4}{\rm
Tr}_{6\times 6} \hat{\cal T}^4 = {\det}_{6\times 6}\Big(I -
\hat{\cal T}\Big) = e^{{\rm Tr}\log (I-\hat {\cal T})} = \exp
\Big(-\sum_{m=1}^\infty \frac{1}{4m}{\rm Tr}\ \hat {\cal
T}^{4m}\Big) = \label{Dvsop3} \ee \vspace{-0.35cm}
$$
= 1 - \frac{1}{4}{\rm Tr}\ \hat {\cal T}^4 + \frac{1}{32}\Big[
\Big({\rm Tr}\ \hat {\cal T}\Big)^4 - 4{\rm Tr}\ \hat{\cal
T}^8\Big]+ \Big[-\frac{1}{6\cdot 4^3}\Big({\rm Tr}\ \hat {\cal
T}^4\Big)^3
 + \frac{1}{32}{\rm Tr}\ \hat{\cal T}^4 {\rm Tr}\ \hat{\cal T}^8
 -\frac{1}{12}{\rm Tr}\ \hat {\cal T}^{12} \Big]
+ \ldots
$$
The simplest way to resolve these equations is to consider
{diagonal} $T$:
$$
\hat{\cal T}_{diag} = \left(\begin{array}{cccccc}
0&0&0&0&0& b \\
0&0& -a &0&0&0\\
0& b &0&0&0&0\\
0&0&0&0&0&0\\
0&0&0&0&0&0\\
a &0&0&0&0&0\\
\end{array}\right)
\ \ \ {\rm and} \ \ \ \hat{\cal T}_{diag}^2 =
\left(\begin{array}{cccccc}
ab&0&0&0&0&0 \\
0&-ab&0&0&0&0\\
0&0&0&-ab&0&0\\
0&0&0&0&0&0\\
0&0&0&0&0&0\\
0&0&0&0&0&ab\\
\end{array}\right)
$$
and actually ${\rm Tr}_{6\times 6} \hat{\cal T}^{4m} = 4D_{2\times
2\times 2}^m(T)$, so that eq.(\ref{Z23}) can be also rewritten as
$$
Z_{2\times 2\times 2}\Big(T\Big|\#(t)=2\Big) = \sum_{m=0}
(ab)^{2m} \left(\sum_{s=0}^{2m} (-)^s s!(2m-s)!\right) =
$$
\vspace{-0.25cm} \be = 1 + \frac{3}{4}{\rm Tr}\ \hat{\cal T}^4 +
\frac{5}{2}{\rm Tr}\ \hat{\cal T}^8 + \frac{5\cdot 7\cdot
9}{16}{\rm Tr}\ \hat{\cal T}^{12} + \frac{7\cdot 9^2}{2}{\rm Tr}\
\hat{\cal T}^{16} + \ldots = \sum_{m=0}^\infty
\frac{(2m+1)!}{4^m(m+1)} {\rm Tr}\ \hat{\cal T}^{4m} \label{Z23'}
\ee

\subsection{Tensorial exponent}

Combinatorial partition functions are actually defined in terms of
the following formal operations:

\subsubsection{Oriented contraction}
For a pair of tensors $a\in T^nV,\alpha\in T^kV^*$ take an
ordering on the corresponding simplices $\set{1,\dots,n}$ and
$\set{1,\dots,k}$. Then we can define a contraction
$$\partial_\alpha a:=\langle\alpha,a\rangle:=\sum_{\sigma\in\tx{Hom}(k,n)} \alpha\cdot_\sigma a$$
where the sum is taken over monotone embeddings
$\sigma:\set{k}\hookrightarrow\set{n}$. \EXz{}{ For
$a_{i_1i_2i_3}\in T_3,\alpha\in T^2$ we have
$$\langle\alpha,a\rangle_k=\alpha^{ij}a_{ijk}+\alpha^{ij}a_{ikj}+\alpha^{ij}a_{kij}$$}

\subsubsection{Generating operation ("exponent")}
For a pair $\alpha,\beta\in TV^*$ we have
$$\langle\beta,\langle\alpha,a\rangle\rangle=\langle\alpha,\langle\beta,a\rangle\rangle=:\langle\alpha\beta,a\rangle$$
i.e.
$\partial_\alpha\partial_\beta=\partial_\beta\partial_\alpha$.
Then we have a well-defined operation
$$\exp\alpha:=1+\langle\alpha,\cdot\rangle+\frac{1}{2!}\langle\alpha,\cdot\rangle+\frac{1}{3!}\langle\alpha^3,\cdot\rangle+\dots$$
\EXz{}{ For $\alpha\in V^*,a=\sum_K a_Kx_1^{k_1}\dots x_N^{k_N}\in
SV$ we have
$$\exp\alpha(a)=\sum_K a_K(x_1+\langle\alpha,x_1\rangle)^{k_1}\dots (x_N+\langle\alpha,x_N\rangle)^{k_N}$$
known as \Ait{Taylor formula}.}

\subsection{Beyond $n=2$}

Explicit expressions for the resultants and discriminants for
$n>2$ are quite lengthy and we do not go into details about these
examples here. Explicit formulas for resultants are actually
written down in s.\ref{Koshul} as determinants of huge matrices.
It was also shown how much simpler expressions, like
(\ref{abc+e3}), can be deduced for tensors from particular
few-parametric families. Some more examples can be found in
\cite{GKZ}.

In the near future a systematic list of examples should be
compiled and convenient notation introduced for various
intermediate {\it building blocks} in different approaches (since
expressions in terms of primary entries -- the tensor coefficients
-- take pages already for $n=3$). Of principal importance would be
development of adequate functional integral language, associated
topological theories, {\it tensor}-models and tensorial
$\tau$-functions. {\bf This looks like a challenging and promising
field to work in.}

In the following sections we briefly touch the even more advanced
subjects, which grow from the resultant-\-discriminant theory and
can be considered as its applications.\footnote{ In these advanced
sections we no longer bother about confusing superscripts with
powers and -- in order to simplify some formulas -- we freely
interchange covariant and contravariant indices as compared to the
previous sections. } The most important is generalization of
eigenvector chapter of linear algebra: the theory of iterated maps
and their orbits, the concept of map's exponentiation,
renormalization groups, Julia and Mandelbrot sets.

\subsubsection{$D_{3|3}$, $D_{3|4}$ and $D_{4|3}$ through determinants
}

Discriminantal conditions for a symmetric tensor
$S(\vec z) = S^{i_1\ldots i_r}z_{i_1}\ldots z_{i_r}$
of the type $n|r$ describe resolvability of the system
$dS = 0$ of $n$ equations
\be
\sum_{I=1}^{N_{n|r-1}} S^{iI}z^I = 0, \ \ \ i = 1,\ldots,n
\label{Sha331}
\ee
where $I$ is symmetric multi-index $I = \{i_1\ldots i_{r-1}\}$
taking $N_{n|r-1} = \frac{(n+r-2)!}{(n-1)!(r-1)!}$ different values
and $z^I = z_{i_1}\ldots z_{i_{r-1}}$.

The system (\ref{Sha331}) is defined by a rectangular
$n\times N_{n|r-1}$ matrix $S^{iI}$
and as a {\it linear} system
it would always have solutions if $N_{n|r-1}>n$,
without any constraints on the coefficients of $S(\vec z)$.
However, $z^I$ in \ref{Sha331} are not {\it free} (independent)
variables, and these solutions are not necessarily lifted to
solutions for $\vec z$, i.e. to solutions of original system
$dS=0$.

Still, if (\ref{Sha331}) could be somehow substituted
by a system of the form
\be
\sum_{J=1}^{N_{n|p}} H^{IJ}z^J = 0, \ \ \ I = 1,\ldots,N_{n|p}
\label{Sha332}
\ee
where each equation holds whenever $dS=0$ and
with a {\it square} matrix $H^{IJ}$,
the $\det H = 0$ would be a {\it necessary} condition
for resolvability of $dS=0$, i.e.
\be
D_{n|r}(S) = {\bf irf}\Big\{\mid H\mid \Big\}
\ee
If, further, $\det H$ has the right degree
\be
d_{n|r} = n(r-1)^{n-1}
\ee
in the coefficients of $S(\vec z)$, we obtain an exact
relation -- determinantal formula
\be
D_{n|r}(S) = {\rm det}_{N_{n|p}\times N_{n|p}} H
\label{Sha333}
\ee
which could serve as a generalization of (\ref{e545}).
Remarkably, such {\it square} matrices $H$ can often be
designed \cite{Sham}.

First of all, if $r=2$, then $N_{n|r-1} = N_{n|1} = n$
and we can simply take $H^{ij} = S^{ij}$, reproducing once
again the familiar relation
$$D_{n|2}(S) = {\rm det}_{n\times n} S$$
More complicated examples exploit the result of s.4.2.1,
applied to the case of symmetric tensor $T=S$:
$dS=0$ implies $dQ=0$ with
\be
Q(\vec z) = {\rm det}_{n\times n}
\frac{\partial^2 S(\vec z)}{\partial z_i\partial z_j}
\ee
This allows to use
\be
Q^i(\vec z) = \frac{\partial Q}{\partial z_i} =
\sum_{K=1}^{N_{n|p}}Q^{iK}z^K
\ee
along with
\be
S^i(\vec z) = \frac{\partial S}{\partial z_i} =
\sum_{I=1}^{N_{n|r-1}}S^{iI}z^I
\ee
in the construction of $H$.
Quantities $Q_i(\vec z)$ have degree $n$ in the coefficients
$S$ of $S(\vec z)$ and degree $p_{n|r} \equiv n(r-2) -1$
in the $z$-variables.

In the first non-trivial example $n|r=3|3$ there is a coincidence:
$p_{3|3} = 2 = r-1$ allowing to form a matrix from $S^i$ and $Q^i$
$$
H = \left(\begin{array}{c} S_I^{i} \\ Q_I^{i} \end{array} \right)
$$
with $\#(I) = N_{3|2} = 6$ columns.
Remarkably (compare to \cite{GKZ}, Chapter 8) this is already a square matrix: the number of rows
is also $\ 2\times \# (i) = 2n = 6$.
Furthermore, $\det H$ has degree
$3\times 1 + 3\times 3 = 12 = d_{3|3}$ in the coefficients $S$,
and we conclude that
\be
D_{3|3}(S) = {\rm det}_{6\times 6}
\left(\begin{array}{c} S_I^{i} \\ Q_I^{i} \end{array} \right) =
{\rm det}_{6\times 6}
\left(\begin{array}{c} \partial_i S
\\ \partial_i Q \end{array} \right)\nn\\
=\sum_{IJK}^{C_6^3=20}(-)^{P(IJK)}S_{IJK}Q(S)_{\widetilde{IJK}},
\label{Sha334}
\ee
where $S_{IJK}=\epsilon_{ijk}S_I^iS_J^jS_K^k$ and
$Q_{IJK}=\epsilon_{ijk}Q_I^iQ_J^jQ_K^k$
are $3\times 3$ minors of rectangular matrices $S_I^i$
and $Q_I^i$, while $\widetilde{IJK}$ is
a triple, complementary to $IJK$, e.g.
$\widetilde{123}=456$ or $\widetilde{11,22,33}=12,13,23$.
For example, for $S=\frac{1}{3}(ax^3+by^3+cy^3) + 2\epsilon xyz$
from s.4.2.6 eq.\,(\ref{Sha334}) implies:
$$Q = \det \left(\begin{array}{ccc}
2ax & 2\epsilon z & 2\epsilon y \\
2\epsilon z & 2by & 2\epsilon x \\
2\epsilon y & 2\epsilon x & 2cz \end{array}\right)=
-8\Big((ax^3+by^3+cz^3)\epsilon^2 - (abc+2\epsilon^3)xyz\Big)$$
and
$$
D_{3|3}(S) = -8^3\cdot
\det\left(\begin{array}{c} ax^2+2\epsilon yz \\
by^2+2\epsilon xz \\ cz^2+2\epsilon xy \\
3a\epsilon^2x^2-(abc+2\epsilon^3)yz \\
3b\epsilon^2y^2-(abc+2\epsilon^3)xz \\
3c\epsilon^2z^2-(abc+2\epsilon^3)xy \end{array}\right)
\sim $$ $$
\sim {\rm det}_{6\times 6}\left(\begin{array}{cccccc}
a & 2\epsilon & & & & \\
3a\epsilon^2 & -(abc+2\epsilon^3) &&\\
&&b & 2\epsilon &&  \\
&&3b\epsilon^2 & -(abc+2\epsilon^3) &&\\
&&&&c & 2\epsilon \\
&&&&3c\epsilon^2 & -(abc+2\epsilon^3)
\end{array}\right) =
$$ $$
= abc(abc+8\epsilon^3)^3
$$
in accordance with (\ref{R32exa1}).

Note, that instead of (\ref{Sha334}) one can also write \cite{Sham}
$$D_{3|3}(S)=det_{10\times 10}\left(\begin{array}{c}
z_i\partial_jS\\
Q
\end{array}\right)$$

In general $p_{n|r}\neq r-1$, but in devising the matrix $H$
we still have a freedom to multiply by powers of $z$:
if all $S^i=0$ then not only all $Q^i=0$, but also any
$z^LS^i = 0$ and $z^M Q^i = 0$ with any multi-indices
$L$ and $M$.
This additional freedom allows to construct square matrices
$H$ form $dS$ and $dQ$ alone in the next non-trivial
examples of $D_{3|4}$ and $D_{4|3}$:
\be
D_{3|4}(S) =
{\rm det}_{21\times 21}
\left(\begin{array}{c} z_iz_j\partial_k S
\\ \partial_i Q \end{array} \right)
\label{Sha335}
\ee
\be
D_{4|3}(S) =
{\rm det}_{20\times 20}
\left(\begin{array}{c} z_i\partial_j S
\\ \partial_i Q \end{array} \right)
\label{Sha336}
\ee
These determinants have correct degrees  in $S$:
$\ N_{3|2}\cdot 3 \times 1 + 3\times 3 = 6\times 3 + 3\times 3 =
27 = 3\cdot 3^2 = d_{3|4}$
and
$4^2 \times 1 + 4\times 4 = 32 = 4\cdot 2^3 = d_{4|3}$ respectively.
The sizes of matrices in determinants are equal respectively to
$\ 21=N_{3|5}=3\cdot N_{3|2} + 3 = 3\cdot 6 +3\ $
and
$\ 20 = N_{4|3} = 4\cdot 4 + 4$.

It is also instructive to consider another familiar case:
$n=2$, i.e. discriminant of an ordinary homogeneous polynomial.
In terms of the present section
$$D_{2|r}(S) = \det z^q \partial S$$
where $q(r)$ is defined from the condition that the matrix
is square: $N_{2|r+q-1} = 2N_{2|q}$, i.e. $r+q = 2(q+1)$
or simply $q=r-2$. Thus
\be
D_{2|r}(S) = {\det}_{2(r-1)\times 2(r-1)}\
z_{i_1}\ldots z_{i_{r-2}} \partial_j S
\ee
It is easy to recognize in this formula the familiar expression
$D_{2|r}(S) = R_{2|r-1}(\partial_1 S,\partial_2 S)$
with resultant given by conventional formula (\ref{convres}).
It is enough to label the matrix columns by monomials
$z_1^kz_2^{r-2-k}$ so that each multiplication by $z_1$
shifts a row one step to the right.

\subsubsection{Generalization: Example of non-Koszul description of generic symmetric discriminants}

Consider now the same matrix $\ z^q\partial S\ $
for $n>2$.
Solutions $q_{n|r}$ of the equation $N_{n|r+q-1} = nN_{n|q}$,
are rarely integer, but even if they are, $q_{n|r}\geq r-1$.
However, in this case some rows of the matrix become linear
dependent, and should be excluded if we want
to have a non-vanishing determinant.
Denote by $K_{n|q}$ the linear space of homogeneous polynomials
of degree $q$ of $n$ variables, with dimension
${\rm dim}(K_{n|q}) = N_{n|q}$.
The reason for linear dependence is that
the spaces $\partial_i S \otimes K_{n|q}$ with different $i$
necessarily intersect if $q\geq r-1={\rm deg}(\partial S)$.
The {\it pair}wise intersection space is
$\partial_iS \partial_jS \otimes K_{n|q-(r-1)}$ and
its dimension is $N_{n|q-(r-1)}$.
The number of pair-wise intersections is $C^2_n=\frac{n(n-1)}{2}$,
but they themselves can intersect if $q\geq 2(r-1)$:
then $C^3_n$ triple intersections occur.
Each triple-intersection space is
$\partial_iS \partial_jS\partial_kS \otimes K_{n|q-(r-1)}$
with dimension  $N_{n|q-2(r-1)}$ and so on.
The number of linearly independent rows is therefore equal to
the alternated sum
$$
nN_{n|q} - \frac{n(n-1)}{2}N_{n|q-(r-1)} +
\frac{n(n-1)(n-2)}{6}N_{n|q-2(r-1)} - \ldots =
$$
$$ =
\sum_{j=1}^{n} (-)^{j-1}C^j_n N_{n|q-(j-1)(r-1)} =
n \sum_{j=1}^{n}
\frac{(-)^{j-1}\big(q+n-(j-1)(r-1)-1\big)!}
{j!(n-j)!\big(q-(j-1)(r-1)\big)!}
$$
\be
\label{lidesu}
\ee
For sufficiently big $q\geq (n-1)(r-1)$ all terms
in this sum are non-vanishing (no factorials of negative
integers in denominator) and the sum is equal to
$N_{n|q+r-1}$ (as a corollary of the
elementary identities $\sum_{j=0}^n(-)^jC^j_n (j-1)^m
= 0$ for $0\leq m \leq n-1$).
This means that for such $q$ rectangular matrix
$\Big(z^q\partial S\Big)$ with $N_{n|q+r-1}$ columns
and $nN_{n|q}$ rows, actually has exactly $N_{n|q+r-1}$
linearly independent rows and the corresponding principal
minor is non-vanishing for generic $S$.
Of course, this minor vanishes at discriminantal variety,
because all the equations, represented by rows of the matrix,
possess a common non-trivial solution, and $D_{n|r}(S)$
is an irreducible factor and common divisor of all such
minors with different $q\geq (n-1)(r-1)$.
In other words, the image of operator $z^q dS$, which
maps the space of degree-$q$ polynomials of $n$ variables
into the space of degree-$q+r-1$ polynomials, drops down
at discriminantal variety.
Thus phenomenon is very similar to what happens in
the case of Koszul-complex construction.
Its slight generalization, however, involves a
somewhat less trivial operator.

For the special value of $q = n(r-2)-1-(r-1) = (n-1)(r-1)-(n+1)$
the last term with $j=n$ does not contribute to the sum
(\ref{lidesu}) and it gets equal to
$$
N_{n|q+r-1} - n
$$
In this special case $q+r-1=n(r-2)-1$ is exactly the degree
of the polynomial $dQ$ and $n$ is exactly the number of
components of this vector.
This means that the matrix
$$
\left(\begin{array}{c} z^q\partial S
\\ \partial Q \end{array} \right)
$$
though rectangular for generic $n|r$,
becomes square after elimination of all linearly-dependent
rows, or, in other words, has exactly one non-vanishing
maximal minor.
This minor has a size $N_{n|n(r-2)-1}$ and degree
$N_{n|n(r-2)-1+(n-1)n}$ in $S$.
This degree is typically larger than $d_{n|r}$:
for big enough $n$ and $r$
$$N_{n|n(r-2)-1}+(n-1)n =
\frac{\big(n(r-2)-1+(n-1)\big)!}{(n-1)!
\big(n(r-2)-1\big)!} + n(n-1) \sim
$$ $$ \sim \frac{(nr)^{n-1}}{(n-1)!}
> n(r-1)^{n-1} = d_{n|r}
$$
It follows, that discriminant $D_{n|r}(S)$ is an irreducible
factor of our minor, but in general does not coincide with it.
Exact dimensions of these extra factors are:
$$\hspace{-1mm}
\begin{array}{|c|c|c|c|c|c|}
\hline
&&&&&\\
n & {\rm extra\ dimension} &r=3&r=4&r=5&r=6\\
&&&&&\\
\hline
&&&&&\\
3& \frac{3}{2}(r-3)(r-4) & 0&0& 10& 90\\
4& \frac{4}{3}(r-3)(5r^2-18r+10) & 0& 24& 330& 2940\\
5&\frac{505}{24}r^4-\frac{1885}{12}r^3+
\frac{10055}{24}r^2-\frac{5795}{12}r+225
&3& 120& 1800& 20220\\
6& \frac{294}{5}r^5-510r^4+1731r^3-2895r^2+\frac{11976}{5}r-756
& 9& 328& 5750& 79560\\
& \ldots &&&&\\
\hline
\end{array}
$$
As we already know, the extra factors are absent in three
non-trivial examples $n|r = 3|3,\ 3|4,\ 4|3$.
Since the choice of linearly-dependent is not necessarily
done in the $SL(n)$ symmetric way, the extra factors are
not, generically, $SL(3)$-invariant.

To make this construction a full-scale alternative to Koszul complexes,
one still needs to provide explicit description of the asymmetric factors.

For more examples and further discussion along these lines
see \cite{Sham}.

\setcounter{equation}{0}

\section{Eigenspaces, eigenvalues and resultants
\label{eige}}

\subsection{
From linear to non-linear case}

The theory of eigenvalues, eigenvectors and Jordanian cells is one
of the most important chapters of linear algebra. Eigenvactors
$\vec e$ and eigenvalues $\lambda$ are defined from the condition
$A^i_j e^j = \lambda e^i$, so that eigenvectors are roots of the
{\it characteristic equations}: \be \det (A-\lambda I) =
\det_{n\times n} \Big(A^i_{j} - \lambda \delta^i_{j}\Big) =
(-)^n\prod_{i=1}^n \Big(\lambda - \lambda_i(A)\Big) = 0
\label{chareqmat} \ee In fact, the $n\times n$  matrix can be
fully parameterized by its eigenvectors and eigenvalues: each
eigenvector has $n-1$ projectively independent components plus one
eigenvalue and there are $n$ independent eigenvectors, so that
$n^2  =  n \Big( (n-1) + 1 \Big)$. Let us label eigenvectors
$e_{i\mu}$ and the corresponding eigenvalues $\lambda_\mu$ by
index $\mu = 1,\ldots,n$. Then \be A^i_{j} = \sum_{\mu =1}^n
e^i_{\mu} \lambda_\mu  e^*_{\mu j} \label{matvseig} \ee The matrix
$e^i_{\mu}$ matrix is $n\times n$ square, and $e^*_{\mu j}$ is its
inverse, $\sum_{\mu=1}^n e^i_{\mu}e^*_{\mu j} = \delta^i_{j}$. If
all $\lambda_\mu=1$ we get $A^i_{j} = \delta^i_{j}$, and for
arbitrary $\lambda_\mu$ the determinant \be {\det}_{ij} A^i_{j} =
\prod_{\mu=1}^n \lambda_\mu \label{detprodeig} \ee

\bigskip

Above counting can not survive transition to non-linear case $s>1$
many things without changes. This is obvious already from
examination of the simplest numerical characteristics. The map of
degree $s$ has $nM_{n|s} = n \frac{(n+s-1)!}{(n-1)!s!}$ parameters
(instead of $n^2$ in the linear case of $s=1$). Each eigenvector
still has $n-1$ projectively independent components, but
"eigenvalue" is now a polynomial of degree $s-1$ with $M_{n|s-1} =
\frac{(n+s-2)!}{(n-1)!(s-1)!}$ parameters, see below. Naive
counting, as in the linear case, would imply that the ratio
$$  \frac{nM_{n|s}}{ n-1 + M_{n|s-1} } $$
should count the number of eigenvectors. But it is not integer!
Even in the simplest case $n=2$ $s=2$ (two quadratic polynomials),
this ratio is  $2$, while actually there will be $3$ eigenvectors.
This signals that eigenvectors and eigenvalues can not be chosen
independently, actually there are \be c_{n|s} = \frac{s^n-1}{s-1}
\ee of them, and this number can exceed $M_{n|s}$ and even
$nM_{n|s}$ (see table in s.\ref{cvsM} below). Furthemore, $c_{n|s}
\neq n$ for $s>1$, and $e^i_{\mu}$ with $\mu = 1,\ldots, c_{n|s}$
is no longer a square matrix: it is rectangular $n\times c_{n|s}$
and there is no inverse to play the role of $e^*$ in
(\ref{matvseig}). The proper generalization of (\ref{matvseig}) is
given by more sophisticated expressions like (\ref{tensexp}) or
(\ref{tensexp''}).

\subsection{Eigenstate (fixed point) problem and
characteristic equation}

\subsubsection{Generalities}

\noindent

$\bullet$ For $(n|s)$-map $\vec A(\vec z)$, described by $n$
homogeneous polynomials of degree $s$, the eigenstates are
invariant points in $P^{n-1}$, satisfying \be A^i(\vec z) =
\lambda(\vec z) z^i, \ \ \ i=1,\ldots,n \label{eist} \ee where
$\lambda(\vec z)$ is some homogeneous polynomial of degree $s-1$.
For non-trivial eigenstates to exist, $\lambda(\vec z)$ in
(\ref{eist}) should satisfy characteristic equation \be
R_{n|s}(\lambda|A) \equiv {\cal R}_{n|s}\Big\{A^i(\vec z) -
\lambda(\vec z)z^i\Big\} = 0 \label{chareq} \ee which imposes a
{\it single} constraint on the $M_{n|s-1} =
\frac{(n+s-2)!}{(n-1)!(s-1)!}$ coefficients of $\lambda(\vec z)$.

$\bullet$ Eigenvalue $\lambda(z)$ can be excluded from the system
(\ref{eist}), then eigenvectors satisfy a closed system of
equations \be z^i A^j(\vec z) = z^j A^i(\vec z) \ee This system is
over-defined: one can fix $j$ and get $n-1$ independent equations
for $i\neq j$, all other will follow.

$\bullet$ The map is defined by the tensor $A^i_\alpha$, where $i$
takes $n$ values and $\alpha = \{j_1,\ldots, j_s\}$ is a {\it
symmetric}\ multi-index, i.e. with all values, differing by
permutations of indices, like $\{j_1,\ldots, j_k,\ldots,
j_l,\ldots, j_s\}$ and $\{j_1,\ldots, j_l,\ldots, j_k,\ldots,
j_s\}$, identified. After such identification the multi-index
$\alpha$ takes $M_{n|s} = \frac{(n+s-1)!}{(n-1)!s!}$ different
values.

The polynomial $R_{n|s}(\lambda|A)$ has degree $d_{n|s} =
ns^{n-1}$ in the coefficients of $A$, but only degree
$c_{n|s}=\frac{s^n-1}{s-1}$ in the coefficients of $\lambda$. This
is because $\lambda(\vec z)\vec z$ is a rather special kind of a
homogeneous map. For example, for $n=2$, when ${\cal R}_{2|s} =
{\rm Res}_\xi\Big(y^{-s}A_1(\xi y,y), y^{-s}A_2(\xi y,y)\Big)$ is
expressed through an ordinary resultant of two polynomials in
$\xi$, these two degrees are easy to find: $d_{2|s} = 2s$ and
$c_{2|s} = s+1$.

$\bullet$ This implies that the eigenstate problem (\ref{eist})
and (\ref{chareq}) has $c_{n|s}$ different solutions. We preserve
notation $\vec e_\mu = \{e^i_\mu\}$ for these {\it eigenvectors},
but since $i=1,\ldots,n$,\ $\mu = 1,\ldots, c_{n|s}$ and
$c_{n|s}>n$ for $s\geq 2$, these eigenvectors are {\it linearly}
dependent: form an {\it overfilled} basis in generic
non-degenerate situation. Moreover, since $c_{n|s} > M_{n|s}$ for
most values of $n$ and $s$ (see Table in s.\ref{cvsM}), {\it
linear} dependencies exist even among their $s$-fold products, see
examples in s.\ref{eigexam} below.

$\bullet$ As a functional of $\lambda(\vec z)$ the characteristic
polynomial $R_{n|s}(\lambda|A)$ factorizes into a product of
$c_{n|s}$ functions, which are at most {\it linear} in the
coefficients of $\lambda(\vec z)$: \be {\cal
R}_{n|s}\left\{A_i(\vec z) - \lambda(\vec z)z_i\right\} = \prod_{k
= 1}^{c_{n|s}} L_k(\lambda|A) \label{chareqfact}\label{68eq} \ee
This is a highly non-trivial property for the case of many
variables: there is nothing like Besout theorem for arbitrary
polynomial of many variables. Still characteristic polynomial made
from a resultant possesses such decomposition! Each factor $L_\mu$
is associated with particular eigenstate of $\vec A(\vec z)$, and
decomposition (\ref{chareqfact}) is direct generalization of
(\ref{chareqmat}) from matrices to generic non-linear maps.

$\bullet$ Putting $\lambda(\vec z)=0$ in (\ref{chareqfact}) we
obtain a peculiar decomposition of the resultant: \be {\cal
R}_{n|s}\{A\} = \prod_{k=1}^{c_{n|s}} L_k(0|A), \label{resfact}
\ee which is a non-linear analogue of determinant decomposition
(\ref{detprodeig}). However, the separation of individual factors
at the r.h.s. is not uniquely defined, also the factors are
non-polynomial in the coefficients of $A$, so that the resultant
for generic map is {\it algebraically} irreducible, and its
factorization into $A$-polynomial quantities for $A$ of a special
type (for example, diagonal or with special symmetries) remains a
well-defined phenomenon.

\subsubsection{Number of eigenvectors $c_{n|s}$ as compared to
the dimension $M_{n|s}$ of the space of symmetric functions
\label{cvsM}}

The simplest way to find $c_{n|s}$ and get first impression about
the structure of characteristic equation is through consideration
of the simplest {\it diagonal} map \be A_i(\vec z) = a_iz_i^s,\ \
\ {\rm i.e.}\ \ A_{i\alpha} = \delta(\alpha,ii\ldots i) \ee This
is one of the cases when eigenvalue problem can be solved
explicitly: from \be a_iz_i^s = \Lambda z_i,\ \ \ \Lambda =
\lambda(\vec z),\ \ \ i=1,\ldots,n \ee each $z_i$ can take one of
$s$ values, either $0$ or $(\Lambda/a_i)^{1/(s-1)}$, multiplied by
any of the $s-1$ roots of unity $\omega_{s-1}^\mu$,\ $\mu =
1,\ldots,s-1$,\ $\omega_{s-1}^{s-1}=1$. The number $c_{n|s}$ of
projectively independent solutions satisfies the obvious
recurrence relation: \be c_{n+1|s} = sc_{n|s} + 1 \ee Indeed, each
of the $c_{n|s}$ solutions for $z_1,\ldots,z_n$ can be
supplemented by one of the $s$ choices for $z_{n+1}$, and there is
one essentially new solution: all $z_1=\ldots=z_n = 0$, $z_{n+1}
\neq 0$. For $n=1$ there is exactly one non-vanishing solution
modulo rescalings, so that $c_{1|s} = 1$ and therefore $c_{n|s} =
1 + s + s^2 + \ldots + s^{n-1} = \frac{s^n-1}{s-1}$. Note, that
generic resultant's degree $d_{n|s} = ns^{n-1}$ is a similar sum
of $n$ items, but all equal to $s^{n-1}$. Another remark is that
$c_{n|s}$ does not coincide with the number $M_{n|s}$ of values
the symmetric multi-index $\alpha$ in $A^i_\alpha(\vec z)$ can
take: in general \be \frac{s^n-1}{s-1} = c_{n|s} \geq M_{n|s} =
\frac{(s+n-1)!}{s!(n-1)!} \label{clM} \ee -- for fixed $s$ the
l.h.s. grows exponentially with $n$, while the r.h.s. is just a
power. More details about these two important parameters are seen
in the table:

\bigskip

{\footnotesize
$$
\hspace{-0.5cm}
\begin{array}{|c|c||c|c|c|c|c|c|}
\hline
&&&&&&&\\
&s&1&2&3&4&5&\ldots\\
&&&&&&&\\
\hline
&&&&&&&\\
&&&c_{n|2}=2^n-1&c_{n|3}=\frac{3^n-1}{2}&
c_{n|4}=\frac{4^n-1}{3}&c_{n|5} =\frac{5^n-1}{4} &\\
n &&c_{n|1}=M_{n|1}=n&&&&&\\
&&&M_{n|2}=&M_{n|3}=&M_{n|4}=&M_{n|5}=&\\
&&&=\frac{n(n+1)}{2}&\frac{n(n+1)(n+2)}{6}&
=\frac{n(n+1)(n+2)(n+3)}{24}&
=\frac{n(n+1)(n+2)(n+3)(n+4)}{120}&\\
&&&&&&&\\
\hline\hline
&&&&&&&\\
1&c_{1|s}=M_{1|s} = 1 &1&1&1&1&1&\\
&&&&&&&\\
\hline
&&&&&&&\\
2&c_{2|s} = M_{2|s} = s+1 &2&3&4&5&6&\\
&&&&&&&\\
\hline
&&&&&&&\\
&c_{3|s}=s^2+s+1&&7&13&21&31&\\
3&&3&&&&&\\
&M_{3|s}=\frac{(s+1)(s+2)}{2}&&6&10&15&21&\\
&&&&&&&\\
\hline
&&&&&&&\\
&c_{4|s}=s^3+s^2+s+1&&15&40&85&156&\\
4&&4&&&&&\\
&M_{4|s}=\frac{(s+1)(s+2)(s+3)}{6}&&10&20&35&56&\\
&&&&&&&\\
\hline
&&&&&&&\\
&c_{5|s}=s^4 + s^3+s^2+s+1&&31&121&341&781&\\
5&&5&&&&&\\
&M_{5|s}=\frac{(s+1)(s+2)(s+3)(s+4)}{24}&&15&35&70&126&\\
&&&&&&&\\
\hline
&&&&&&&\\
\ldots &&&&&&&\\
&&&&&&&\\
\hline
\end{array}
\hspace{0.5cm}
$$
}

\subsubsection{Decomposition (\ref{chareqfact}) of characteristic
equation: example of diagonal map}

Diagonal maps are also a good starting point for discussion of
decomposition (\ref{chareqfact}) of characteristic equation: some
crucial properties of this decomposition are well seen at the
level of explicit formulas.

\bigskip

{\bf The case of $n|s = 2|s$:}

\bigskip

For $n=2$ and \be \lambda(z) = \lambda_{(1)} x^{s-1} +
\lambda_{(2)} x^{s-2}y + \ldots + \lambda_{(s)} y^{s-1}
\label{lambdadiag2s} \ee we have: \be R_{2|2}(\lambda|A) =
(\lambda_{(1)}-a)(\lambda_{(2)}-b) \Big(ab - \lambda_{(1)}b -
\lambda_{(2)}a\Big), \label{r22la} \ee \be R_{2|3}(\lambda|A) =
(\lambda_{(1)}-a)(\lambda_{(3)}-b)
\Big(ab-\lambda_{(1)}b-\lambda_{(2)}\sqrt{ab}-\lambda_{(3)}a\Big)
\Big(ab-\lambda_{(1)}b+\lambda_{(2)}\sqrt{ab}-\lambda_{(3)}a\Big),
\label{r23la} \ee \be R_{2|4}(\lambda|A) =
(\lambda_{(1)}-a)(\lambda_{(4)}-b) \Big(ab - \lambda_{(1)}b  -
\lambda_{(2)} a^{1/3}b^{2/3}  - \lambda_{(3)} a^{2/3}b^{1/3} -
\lambda_{(4)} a \Big) \cdot \label{r24la} \ee
$$\cdot
\Big(ab - \lambda_{(1)}b  -
\omega_{3}\lambda_{(2)}{a^{1/3}b^{2/3}} -
\omega_3^2\lambda_{(3)}{a^{2/3}b^{1/3}} - \lambda_{(4)} a\Big)
\Big(ab - \lambda_{(1)}b  -
\omega_3^2\lambda_{(2)}{a^{1/3}b^{2/3}} -
\omega_3\lambda_{(3)}{a^{2/3}b^{1/3}} - \lambda_{(4)} a\Big),
$$
and, generically, for $n=2$ \be R_{2|s}(\lambda|A) =
(\lambda_{(1)} - a)(\lambda_{(s)} - b) \prod_{m = 1}^{s-1}\left(ab
- b \sum_{j=1}^{s}\lambda_{(j)}
\left(\frac{a}{b}\right)^{\frac{j-1}{s-1}}
\omega_{s-1}^{m(j-1)}\right) \label{r2sla} \ee The\ $c_{2|s}=s+1$\
factors at the r.h.s. are associated with the eigenvectors \be
\left(\begin{array}{c} 1 \\ 0 \end{array}\right),\ \
\left(\begin{array}{c} 0 \\ 1 \end{array}\right),\ \ {\rm and} \ \
\left(\begin{array}{c} a^{-1/(s-1)} \\
b^{-1/(s-1)}\omega_{s-1}^\nu
\end{array}\right)
\label{evect2sla} \ee where $\omega_{s-1} = e^{2\pi i/(s-1)}$ is
the primitive root of unity of degree $s-1$, and $\nu =
1,\ldots,s-1$ labels all other roots. The total degree of
$R_{2|s}$ in the coefficients $a$, $b$ of diagonal $A^i(\vec z)$
is $2s$, as required by \be {\rm deg}_AR_{n|s}(\lambda|A) = {\rm
deg}_A{\cal R}_{n|s} = d_{n|s} = ns^{n-1} \label{degcharpol} \ee

\bigskip

{\bf The case of $n|s = n|2$:}

\bigskip

For $s=2$ and $\lambda(z)$ is a linear function: \be \lambda(z) =
\lambda_1 z_1 + \lambda_2 z_2 + \ldots + \lambda_n z_n
\label{lambdadiagn2} \ee (we do not raise {\it contravariant}
indices of $z$-variables to emphasize that they are not powers).
Parameters $\lambda_i$ in here are coefficients of the linear
function and have nothing to do with polynomial coefficients
$\lambda_{(j)}$ in (\ref{lambdadiag2s}): in variance with index
$i$ in $\lambda_i$, $(j)$ in $\lambda_{(j)}$ is actually a
symmetric multi-index $(j) = (j_1\ldots j_{s-1})$, just for $n=2$
it can be parametrized by a single number $j$ -- that of unities
among $j_1,\ldots,j_{s-1}$.

For characteristic resultant in the case of $s=2$ and {\it
diagonal} map $A_i(\vec z) = a_iz_i^2$ we have: \be
R_{2|2}(\lambda|A) =
(a_1-\lambda_1)(a_2-\lambda_2)(a_1a_2-\lambda_1a_2 + \lambda_2a_1)
\label{r22la'} \ee Of course, this is the same as (\ref{r22la}).
\be R_{3|2}(\lambda|A) =
(a_1-\lambda_1)(a_2-\lambda_2)(\lambda_3-a_3) (a_1a_2 -
\lambda_1a_2 - \lambda_2a_1) (a_1a_3 - \lambda_1a_3 -
\lambda_3a_1) (a_2a_3 - \lambda_2a_3 - \lambda_3a_2) \cdot \nn \\
\cdot (a_1a_2a_3 - \lambda_1a_2a_3 -
\lambda_2a_1a_3-\lambda_3a_1a_2) \label{r32la} \ee and generically
for $n|s = n|2$: \be R_{n|2}(\lambda|A) = \prod_{i=1}^n
(a_i-\lambda_i) \prod_{i<j}^n (a_ia_j - \lambda_ia_j -
\lambda_ja_i) \prod_{i<j<k}^n (a_ia_ja_k - \lambda_ia_ja_k -
\lambda_ja_ia_k-\lambda_ka_ia_j)
\ldots = \nn \\
= \prod_{k=1}^n \left\{ \prod_{1\leq i_1<\ldots<i_k\leq n}
\Big(a_{i_1}\ldots a_{i_k}\Big) \left(1 -
\sum_{l=1}^k\frac{\lambda_{i_l}}{a_{i_l}}\right)\right\}
\label{rn2la} \ee There are \be c_{n|2} = n + \frac{n(n-1)}{2} +
\frac{n(n-1)(n-2)}{6} + \ldots = C^n_1 + C^n_2 + \ldots + C^n_n =
2^n-1 = \left.\frac{s^n-1}{s-1}\right|_{s=2} \label{cn2la} \ee
terms in this product and the total $A$-degree is \be d_{n|2} = n
+ 2\frac{n(n-1)}{2} + 3\frac{n(n-1)(n-2)}{6} + \ldots = C^n_1 +
2C^n_2 + \ldots + nC^n_n = \sum_{k=1}^n kC^n_k = 2^{n-1}n =
\left.ns^{n-1}\right|_{s=2} \label{dn2la} \ee Each elementary
factor at the r.h.s. of (\ref{rn2la}) has a simple interpretation:
\be 1 - \sum_{l=1}^k\frac{\lambda_{i_l}}{a_{i_l}} = 1 -
\lambda(\vec e_\mu), \label{rn2corrla} \ee where $\lambda(\vec
e_\mu)$ is the value of function $\lambda(\vec z)$ at the
appropriately normalized eigenvector $\vec e_\mu$. Here $\mu =
1,\ldots,c_{n|2}$ parametrizes the data
$$
\Big\{{\rm set\ of}\ k\ {\rm ordered\ indices}\Big\} = \Big\{
1\leq j_1 < \ldots \leq j_k \leq n\Big\}
$$
and associated $\vec e_\mu = \{e^j_\mu\}$ is given by \be
e^j_{\{j_1\ldots j_k\}} = \sum_{l=1}^k a_{j}^{-1}\delta^j_{j_l}
\label{evectn2la} \ee i.e. has non-vanishing components
$a_{j_l}^{-1}$ and zeroes elsewhere: compare with
(\ref{evect2sla}).

\bigskip

{\bf The case of generic $n|s$:}

\bigskip

Now $\lambda(\vec z) = \sum_{j_1\ldots j_{s-1}}^n
\lambda_{j_1\ldots j_{s-1}} z_{j_1}\ldots z_{j_{s-1}}$ is a
function of degree $s-1$ of $n$ variables, but resultant
$R_{n|s}(\lambda|A)$ can be written down explicitly without direct
reference to the coefficients $\lambda_{j_1\ldots j_{s-1}}$. The
relevant expression is direct generalization of the r.h.s. of
(\ref{rn2la}) interpreted as in (\ref{rn2corrla}), only now each
term of degree $k$ in the product is substituted by additional
product of $(s-1)^{k-1}$ factors: \be R_{n|s}(\lambda|A) =
\prod_{\mu=1}^{c_{n|s}} A_\mu\cdot \Big(1 - \lambda(\vec
e_\mu)\Big) = {\cal R}_{n|s}(A) \prod_{\mu=1}^{c_{n|s}} \Big(1 -
\lambda(\vec e_\mu)\Big) \label{rnsla} \ee Here $A_\mu =
\prod_{l=1}^k a_{i_l}$, for diagonal $A$ the resultant \be {\cal
R}_{n|s}(A) = \prod_{\mu=1}^{c_{n|s}}A_\mu = (a_1\cdot\ldots\cdot
a_n)^{s^{n-1}}, \label{diagRns} \ee and $\lambda(\vec e_\mu)$ is
the value of function $\lambda(\vec z)$ at the appropriately
normalized eigenvector $\vec e_\mu$. Eigenvectors differ from
(\ref{evectn2la}) because for $s>2$ they depend on the
$(s-1)$-order roots of $a_i$: \be e^j_\mu =
e^j_{\{j_1,\nu_{j_1};\ldots; j_k,\nu_{j_k}\}} = \sum_{l=1}^k
\delta^j_{j_l} \omega_{s-1}^{\nu_j} a_{j}^{-1/(s-1)}
\label{evectnsla} \ee compare with (\ref{evect2sla}). Additional
parameters $\nu_j = 1,\ldots,s-1$ label different choices of
roots. One of each $k$ of these parameters is unobservable, i.e.
in (\ref{evectnsla}) $\nu_{j_k} \equiv 1$, and there are only
$(s-1)^{k-1}$ free $\nu$-parameters.

This means that $\mu = 1,\ldots,c_{n|s}$ labels the data
$$
\Big\{{\rm set\ of}\ k\ {\rm ordered\ indices\ with\ additional}\
\nu-{\rm parameters}\Big\} = $$ $$ = \Big\{ 1\leq j_1 < \ldots
\leq j_k \leq n\ |\ \nu_{j_1},\ldots,\nu_{j_{k-1}} = 1,\ldots,s-1
\Big\}
$$
As generalization of (\ref{cn2la}), \be c_{n|s} = C^n_1 +
C^n_2(s-1) + \ldots + C^n_n(s-1)^{n-1} = \frac{\Big(1 +
(s-1)\Big)^{n-1} - 1}{s-1} = \frac{s^n-1}{s-1} \label{cnsla} \ee
$A$-degree of $R_{n|s}(\lambda|A)$ in (\ref{rnsla}) is -- as
generalization of (\ref{dn2la}) -- equal to \be d_{n|s} = C^n_1 +
2C^n_2(s-1) + \ldots + nC^n_n(s-1)^{n-1} = \sum_{k=1}^n
kC^n_k(s-1)^{k-1} = n\Big(1 + (s-1)\Big)^{n-1} = ns^{n-1}
\label{dnsla} \ee This formula explains also the power-counting in
(\ref{diagRns}).

\bigskip

Two {\bf obvious problems} arise when one wants to extend explicit
decompositions (\ref{rnsla}) and (\ref{diagRns}) to non-diagonal
maps.

The first is {\it appropriate normalization} of eigenvectors $\vec
e_\mu$ in (\ref{rnsla}). It is clear "from analysis of
dimensions", that ${\rm deg}_A \vec e_\mu = -\frac{1}{s-1}$, since
coefficients of $\lambda(\vec z)$ in (\ref{eist}) should be
considered as having the same degree as those of $A^i(\vec z)$,
and $\lambda(\vec e_\mu)$ in (\ref{rnsla}) should have degree
zero.

The second is invariant meaning of the coefficients $A_\mu$. In
(\ref{rnsla}) and (\ref{diagRns}) these coefficients are taken in
highly asymmetric form: for example, their $A$-degrees depend on
$\mu$ -- this is despite all eigenvectors (all branches) seem to
be on the same footing in general sutuation. In diagonal case our
choice of $A_\mu$ was essentially implied by polynomiality of
decomposition (\ref{rn2la}) -- which is lost already in diagonal
case for $s>2$ and completely
disappears in generic situation. 
In order to clarify this issue we consider a non-diagonal example.

\subsubsection{Decomposition (\ref{chareqfact}) of characteristic
equation: non-diagonal example for $n|s = 2|2$}

\noindent

{\bf Polinomial case:}

\bigskip

For non-diagonal map \be A:\ \ \ \left(\begin{array}{c} x \\ y
\end{array}\right) \longrightarrow \left(\begin{array}{c}
a_{11}x^2 + a_{12}xy \\ b_{12}xy + b_{22}y^2
\end{array}\right)
\label{nondiapol22} \ee the three $\lambda$-linear multiplicative
constituents $L_\mu(\lambda|A)$ of characteristic equation remain
polynomial in the coefficients\ $a=a_{11},\ b = b_{22},\ a_{12},\
b_{12}$: \be R_{2|2}(\lambda|A) =
(a_{11}-\lambda_1)(b_{22}-\lambda_2)
\Big((a_{11}b_{22}-a_{12}b_{12}) + \lambda_1(a_{12}-b_{22}) +
\lambda_2(b_{12}-a_{11})\Big) = \nn \\ =
(a-\tilde\lambda_1)(b-\tilde\lambda_2)\Big(ab -
\tilde\lambda_1b-\tilde\lambda_2a\Big), \label{r22larat} \ee where
\be a = a_{11}-b_{12}, \ b=b_{22}-a_{12}, \ \tilde\lambda_1 =
\lambda_1-b_{12}, \ \tilde\lambda_2 = \lambda_1-a_{12}
\label{shiftla} \ee

\bigskip

{\bf Generic non-polynomial case:}

\bigskip

For generic non-diagonal map \be A:\ \ \ \left(\begin{array}{c} x
\\ y \end{array}\right) \longrightarrow
\left(\begin{array}{c} a_{11}x^2 + a_{12}xy + a_{22}y^2\\
b_{11}x^2 + b_{12}xy + b_{22}y^2
\end{array}\right)
\ee it is convenient to eliminate $a_{12}$ and $b_{12}$ by shifts
(\ref{shiftla}), we also put $\alpha = a_{22},\ \beta = b_{11}$.
In these variables the resultant is relatively simple: \be
R_{2|2}(\lambda|A) = - \Big(\alpha\tilde\lambda_1^3 +
b\tilde\lambda_1^2\tilde\lambda_2 +
a\tilde\lambda_1\tilde\lambda_2^2 + \beta\tilde\lambda_2^3\Big)
+\Big( (b^2+a\alpha)\tilde\lambda_1^2 +
3(ab-\alpha\beta)\tilde\lambda_1\tilde\lambda_2
+ (a^2+b\beta)\tilde\lambda_2^2\Big) - \nn \\
- 2(ab-\alpha\beta)\Big(b\tilde\lambda_1 + a\tilde\lambda_2\Big)
 + (ab-\alpha\beta)^2
\label{r22lagen} \ee In accordance with (\ref{chareqfact}), it
decomposes into a product of three $\lambda$-linear factors: \be
R_{2|2}(\lambda|A) = \alpha
\Big(\eta_1-\tilde\lambda_1-\xi_1\tilde\lambda_2\Big)
\Big(\eta_2-\tilde\lambda_1-\xi_2\tilde\lambda_2\Big)
\Big(\eta_3-\tilde\lambda_1-\xi_3\tilde\lambda_2\Big)
\label{r22ladeco} \ee where $\xi$'s are the three roots of the
equation \be \alpha \xi^3 - b \xi^2 + a\xi - \beta = 0
\label{xieq22} \ee and $\eta$'s -- those of \be \alpha\eta^3 -
(b^2+a\alpha)\eta^2 + 2b(ab-\alpha\beta)\eta -(ab-\alpha\beta)^2 =
0 \label{etaeq22} \ee The choice of roots in (\ref{etaeq22})
should be correlated with that in (\ref{xieq22}): only $3$ out of
$3\times 3 = 9$ possible choices are consistent with
(\ref{r22lagen}). If $\beta = 0$ and in the limit of $\alpha = 0$
equations (\ref{xieq22}) and (\ref{etaeq22}) have simple
solutions: \be
\begin{array}{ccccc}
\xi = 0 &\ \ \ & \xi = a/b &\ \ \ & \xi \sim b/\alpha \\
\eta = a && \eta = a && \eta \sim b^2/\alpha
\end{array}
\ee and substitution of these values into (\ref{r22ladeco})
reproduces (\ref{r22larat}).

To avoid possible confusion we emphasize once again that
consistency between (\ref{r22lagen}) and (\ref{r22ladeco}) is
non-trivial: decomposition like (\ref{r22ladeco}) exists only
because of the special shape of (\ref{r22lagen}), and we see that
resultant indeed has this special shape. We also see that the
resultant decomposition \be {\cal R}_{2|2}(A) = R_{2|2}(0|A) =
\alpha\prod_{i=1}^3 \Big(\eta_i + b_{12} + \xi_ia_{12}\Big) \ee
continues to make sense, but individual factors are no longer
polynomial in the coefficients of $A$ (are made from the roots of
algebraic equations), and this explains the consistency of such
decomposition with algebraic irreducibility of the resultant
${\cal R}(A)$.

\subsubsection{Numerical examples of decomposition
(\ref{chareqfact}) for $n>2$}

If some two coefficients $\lambda$ and $\mu$ are left in the
function $\lambda(\vec z)$, and one is defined through another
$\lambda(\mu)$ from the characteristic equation
$R_{n|s}(\lambda|A) = 0$, then decomposition (\ref{chareqfact})
implies that $\lambda(\mu)$ is a set of {\it linear} functions,
i.e. straight lines (not obligatory passing through zero) on the
$(\lambda,\mu)$ plane. We present some numerically calculated
examples in Fig.\ref{decompex}: this provides additional evidence
that decomposition (\ref{chareqfact}) is indeed true for $n>2$.

\Fig{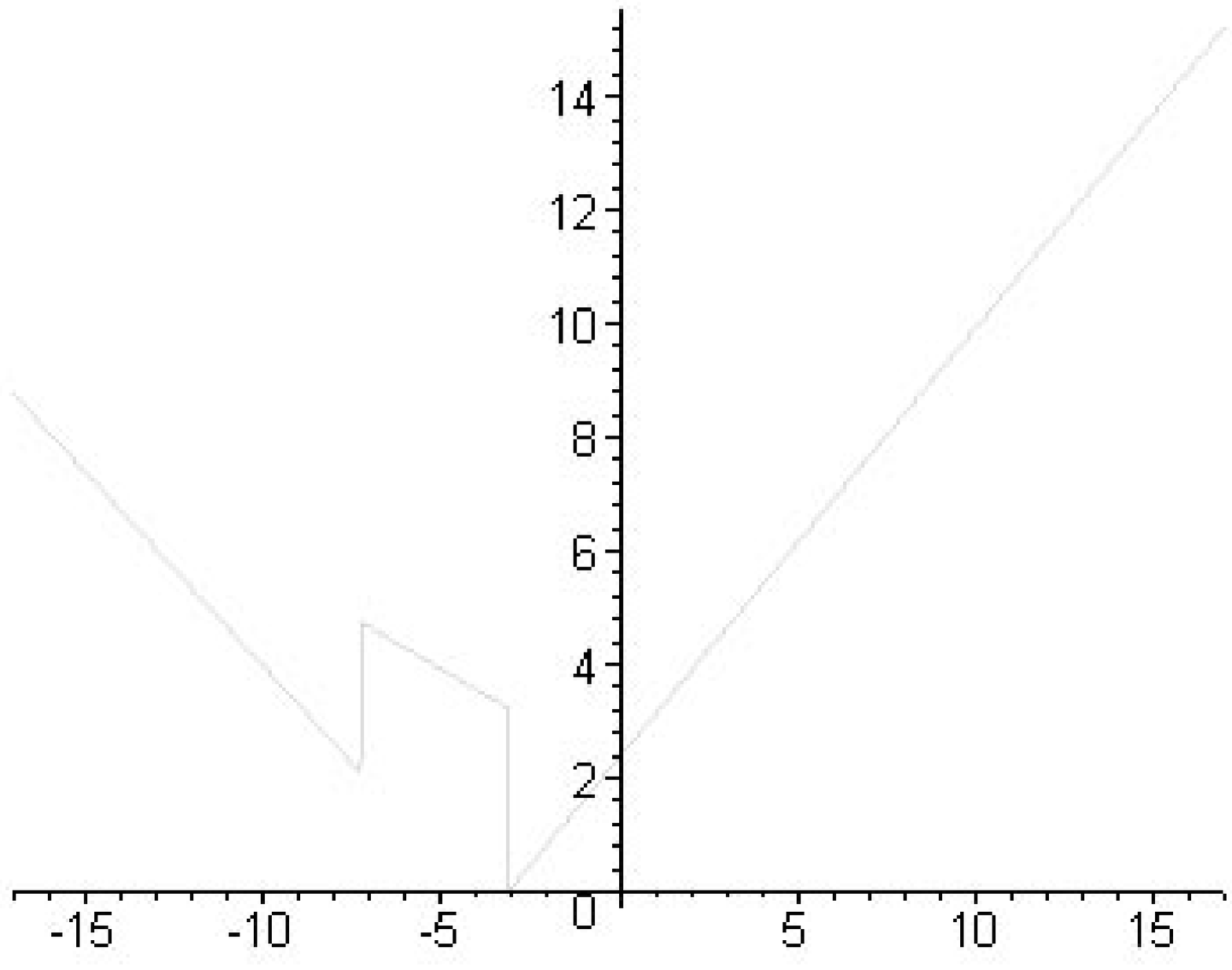} {291,224} {Two-dimensional $\lambda,\mu$ real
section of the space of $\lambda(z)$ for $n|s=3|2$. The lines are
straight, in accordance with (\ref{68eq}). Vertical segments are
irrelevant: they are due to Maple's continuous extrapolation
between different linear branches in a given domain of values of
$\lambda$.}

\subsection{Eigenvalue representation of non-linear map}

\subsubsection{Generalities}

Introduce $E^\alpha_\mu \equiv E^{j_1\ldots j_s}_{\mu} =
e^{j_1}_{\mu}e^{j_2}_{\mu}\ldots e^{j_s}_{\mu}$. The multi-index
$\alpha$ takes $M_{n|s}$ values, while multi-index $\mu$ takes
$c_{n|s}\geq M_{n|s}$ values, so that $E^{\alpha}_\mu$ is
rectangular $M_{n|s}\times c_{n|s}$ matrix. Let us further pick
some $M_{n|s}\leq c_{n|s}$ of indices $\mu$ and restrict
$E^\alpha_\mu \rightarrow \check E^\alpha_\mu$ to this subset.
Define $E^*_{\mu\alpha} = \check E^{-1}_{\mu\alpha}$ as inverse of
$\check E^\alpha_\mu$, i.e. of a  minor of size $M_{n|s}\leq
c_{n|s}$ of the original $M_{n|s}\times c_{n|s}$ matrix
$E^\alpha_\mu$.

It is in these terms that we can write a correct version of
representation (\ref{matvseig}): \be A^i_{\alpha} =
\sum_{\mu=1}^{M_{n|s}} e^i_{\mu}\Lambda_\mu E^*_{\mu\alpha} =
\sum_{\mu=1}^{M_{n|s}} \check E^{(i\vec\gamma)}_\mu
\lambda_{\mu,\vec\gamma} E^*_{\mu\alpha} \label{tensexp} \ee or,
in more detail: \be A^i_{j_1\ldots j_s} = \sum_{\mu=1}^{M_{n|s}}
\check E^{i\ l_1\ldots l_{s-1}}_\mu \lambda_{\mu(l_1\ldots
l_{s-1})} E^*_{\mu, j_1\ldots j_s}, \label{tensexp''} \ee i.e.
$\alpha = \{j_1,\ldots, j_s\}$, $\gamma = \{l_1,\ldots,
l_{s-1}\}$, and indices $l_1,\ldots, l_{s-1}$ are converted
(summed over) with the free indices of $\lambda_\mu$ (which is
actually a polynomial of degree $s-1$ and thus is symmetric tensor
of rank $s-1$).

As we shall see in examples below, representation (\ref{tensexp})
is consistent with
$$
A^i_{j_1\ldots j_s} e^{j_1}_\nu\ldots e^{j_s}_\nu = \Lambda_\nu
e^i_{\nu}
$$
or
$$
A^i_{\alpha}E^{\alpha}_{\nu} = \Lambda_\nu e^i_{\nu}
$$
for all the $c_{n|s}$ values of $\nu$, despite the sum in
(\ref{tensexp}) is only over a subset of $M_{n|s}\leq c_{n|s}$
values of $\mu$.

\subsubsection{Eigenvalue representation of Plukker coordinates}

\be I[\alpha_1\ldots\alpha_n] = \det_{1\leq i,j \leq n}
A_{i\alpha_j} = \sum_{\{\gamma_i\}} \prod_{i=1}^n
E_{\gamma_i\mu_i}\lambda_{\mu_i} \det_{ij} E^*_{\mu_i\alpha_j}
\sim \nn \\ \sim \sum_{\{\gamma_i\}} \det_{ij} E_{\gamma_i\mu_j}
\prod_{i=1}^n \lambda_{\mu_i} \det_{jk} E^*_{\mu_j\alpha_k} \ee
Determinants at the r.h.s. are $n\times n$ minors in
$M_{n|s}\times c_{n|s}$ matrix $E^\alpha_{\mu}$.

If all $\prod_{i=1}^n \lambda_{\mu_i} = 1$ we would get
$I_{\alpha_1\ldots\alpha_n} = \delta_{\{\gamma\},\{\alpha\}}$, but
it would depend on the choice of $\gamma_i$! In other words,
different choices of polynomials $\lambda_\mu(z)$ correspond to
nullification of all Plukker parameters $I$ but one
and this remaininhg one depends on the choice of polynomial.

\bigskip

Plukker parameters satisfy Plukker equations: $\forall \
\{\alpha_1,\ldots,\alpha_{2n}\}$ sets of $2n$ multi-indices (every
$\alpha$ consists of symmetrized $s$ indices, each taking $n$
values, so that there are $M_{n|s}$ different choices of $\alpha$)
we have \be \sum_{P \in \sigma_{2n}} (-)^P I[\alpha_{P(1)}\ldots
\alpha_{P(n)}]\ I[\alpha_{P(n+1)}\ldots \alpha_{P(2n)}] \equiv 0
\ee

\subsubsection{Examples for diagonal maps}

The resultant of original system: \be {\cal R}_{n|s}(a_1z_1^s,
a_2z_2^s, \ldots, a_n z_n^s) = \left(\prod_{i=1}^n a_i\right)^s
\ee

The system of equations for eigenvectors: \be a_i z_i^s = z_i,\ \
\ i=1,\ldots,n \ee Solutions are made from $z_i^{(0)} = 0$ and
$z_{i}^{(0)} = a_i^{-1/(s-1)}\omega$, where $\omega$'s are roots
of unity of degree $s-1$.

For $n>2$ the number of projectively non-equivalent solutions
actually exceeds $M_{n|s}$ and for $E_{\alpha\mu}$ one can take
any {\it square} matrix, which is a non-degenerate minor of
originally {\it rectangular} $E_{\alpha\mu}$ (originally $\#\mu =
c_{n|s} \geq \#\alpha = M_{n|s}$).

{\bf The simplest example: $n = 2$}

For $a_1=a$, $a_2=b$, $z_1 = x$, $z_2=y$ we have:

the resultant is $a^sb^s$,

eigenvectors (solutions of the system $ax^s = x,\ by^s = y$) are:
$x=0$  or $x = a^{-1/(s-1)}\equiv \alpha$, $y=0$ or $y =
b^{-1/(s-1)}\omega^j\equiv \beta \omega^j$, $j=0,\ldots,s-2$.
There are exactly $s+1 = M_{2|s}$ projectively inequivalent
combinations: \be e_\mu = \left(\begin{array}{c} \alpha \\ \beta
\omega^{\mu-1}
\end{array}\right), \ \ \mu = 1,\ldots,s-1;\ \ \
e_{s} = \left(\begin{array}{c} \alpha \\ 0 \end{array}\right); \ \
e_{s+1} = \left(\begin{array}{c} 0 \\ \beta \end{array}\right) \ee

{\bf The case of $s=2$:}

In this case $\alpha = a^{-1}$, $\beta = b^{-1}$ and \be e_1 =
\left(\begin{array}{c} \alpha \\ \beta \end{array}\right); \ \
e_{2} = \left(\begin{array}{c} \alpha \\ 0 \end{array}\right); \ \
e_{3} = \left(\begin{array}{c} 0 \\ \beta \end{array}\right) \ee
Then \be E^\alpha_\mu = \left(\begin{array}{ccc}
\alpha^2 & \alpha^2 & 0 \\
\alpha\beta & 0 & 0 \\
\beta^2 & 0 & \beta^2
\end{array}\right), \ \ \
E^*_{\mu\alpha} = E_{\mu\alpha}^{-1} = \left(\begin{array}{ccc}
0 & ab & 0 \\
a^2 & -ab & 0 \\
0 & -ab & b^2
\end{array}\right)
\ee and \be A^1_{\alpha} = \sum_{\mu=1}^3
e^1_{\mu}E^{*}_{\mu\alpha} = \alpha(0,ab,0) + \alpha(a^2,-ab,0) +
0\cdot(0.-ab,b^2) =
(a,0,0), \nn \\
A^2_{\alpha} = \sum_{\mu=1}^3 e^2_{\mu}E^{*}_{\mu\alpha} =
\beta(0,ab,0) + 0\cdot(a^2,-ab,0) + \beta(0.-ab,b^2) = (0,0,b) \ee
i.e. the two maps indeed are $A^1 = ax^2$ and $A^2=by^2$.

Characteristic equation for the full eigenvector problem \be
ax^2 = \lambda(x,y)x = (px+qy)x, \nn \\
by^2 = \lambda(x,y)y = (px+qy)y \ee is \be {\cal R}_{2|2}\Big(ax^2
- \lambda(x,y)x,\ by^2-\lambda(x,y)y\Big) = (a-p)(b-q)\Big(ab -
(bp+aq)\Big) = \nn \\ = a^2b^2 (1 - p\alpha)
(1-q\beta)\Big(1-(p\alpha + q\beta)\Big) =
\Big(1-\lambda(e_2)\Big)\Big(1 - \lambda(e_3)\Big)
\Big(1-\lambda(e_1)\Big){\cal R}(\lambda=0) \ee

{\bf The case of $s=3$:}

In this case $\alpha = 1/\sqrt{a}$, $\beta = 1/\sqrt{b}$ and \be
e_1 = \left(\begin{array}{c} \alpha \\ \beta \end{array}\right); \
\ e_2 = \left(\begin{array}{c} \alpha \\ -\beta
\end{array}\right);\ \ e_{3} = \left(\begin{array}{c} \alpha \\ 0
\end{array}\right); \ \ e_{4} = \left(\begin{array}{c} 0 \\ \beta
\end{array}\right) \ee Then \be E^\alpha_\mu = \begin{array}{c}
111 \\ 112 \\ 122 \\ 222 \end{array} \left(\begin{array}{cccc}
\alpha^3 & \alpha^3 & \alpha^3 & 0 \\
\alpha^2\beta & -\alpha^2\beta & 0 & 0 \\
\alpha\beta^2 & \alpha\beta^2 & 0 & \alpha\beta^2 \\
\beta^3 & -\beta^3 & 0 & \beta^3
\end{array}\right), \ \ \
E^*_{\mu\alpha} = E_{\mu\alpha}^{-1} = \left(\begin{array}{cccc}
0 & \frac{1}{2}a\sqrt{b} & \frac{1}{2}\sqrt{a}b & 0 \\
0 & -\frac{1}{2}a\sqrt{b} & \frac{1}{2}\sqrt{a}b & 0 \\
a^{3/2} & 0 & -\sqrt{a}b & 0 \\
0 & -a\sqrt{b} & 0 & b^{3/2}
\end{array}\right)
\ee and \be A^1_{\alpha} = \sum_{\mu=1}^4
e^1_{\mu}E^{*}_{\mu\alpha} = \alpha(0, \frac{1}{2}a\sqrt{b},
\frac{1}{2}\sqrt{a}b, 0) + \alpha(0, -\frac{1}{2}a\sqrt{b},
\frac{1}{2}\sqrt{a}b, 0)+\nn\\ + \alpha(a^{3/2}, 0, -\sqrt{a}b, 0)
+
0\cdot(0, -a\sqrt{b}, 0, b^{3/2}) = (a,0,0,0), \\
A^2_{\alpha} = \sum_{\mu=1}^4 e^2_{\mu}E^{*}_{\mu\alpha} =
\beta(0, \frac{1}{2}a\sqrt{b}, \frac{1}{2}\sqrt{a}b, 0) - \beta(0,
-\frac{1}{2}a\sqrt{b}, \frac{1}{2}\sqrt{a}b, 0)+\nn\\ +
0\cdot(a^{3/2}, 0, -\sqrt{a}b, 0) + \beta(0, -a\sqrt{b}, 0,
b^{3/2}) = (0,0,0,b) \ee i.e. the two maps indeed are $A^1 = ax^3$
and $A^2=by^3$.

{\bf The simplest example with $c_{n|s} > M_{n|s}$: $n = 3$}

In this case $c_{3|2}=7\ >\ M_{3|2}=6$. For $a_1=a$, $a_2=b$,
$a_3=c$, $z_1 = x$, $z_2=y$, $z_3=z$ the resultant is $(abc)^s$.

{\bf The case of $s=2$:}

In this case $\alpha = 1/{a}$, $\beta = 1/{b}$, $\gamma = 1/{c}$,
and there are\ $c_{3|2}=7$\  projectively non-equivalent
solutions:
$$
e_{1} = \left(\begin{array}{c} \alpha \\ 0 \\ 0
\end{array}\right); \ e_{2} = \left(\begin{array}{c} 0 \\ \beta \\
0 \end{array}\right); \ e_{3} = \left(\begin{array}{c} 0 \\ 0 \\
\gamma \end{array}\right); \
e_{4} = \left(\begin{array}{c} \alpha \\ \beta \\
       0 \end{array}\right); \
e_{5} = \left(\begin{array}{c} \alpha \\ 0 \\
       \gamma \end{array}\right);  \
e_{6} = \left(\begin{array}{c} 0 \\ \beta \\ \gamma
       \end{array}\right);  \
e_{7} = \left(\begin{array}{c} \alpha \\ \beta \\ \gamma
       \end{array}\right),
$$
forming a $3\times 7$ matrix \be e^i_{\mu} =
\left(\begin{array}{cccccc|c}
\alpha & 0 & 0 & \alpha & \alpha & 0 & \alpha \\
0 & \beta & 0 & \beta & 0 & \beta & \beta \\
0 & 0 & \gamma & 0 & \gamma & \gamma & \gamma
\end{array}\right)
\ee The corresponding $E_{\alpha\mu}$ is {\it rectangular}
$6\times 7$ matrix: \be E^\alpha_\mu = \begin{array}{c} 11 \\ 22
\\ 33 \\ 12 \\ 13 \\ 23
\end{array}
\left(\begin{array}{cccccc|c}
\alpha^2 & 0 & 0 & \alpha^2 & \alpha^2 & 0 & \alpha^2 \\
0 & \beta^2 & 0 & \beta^2 & 0 & \beta^2 & \beta^2 \\
0 & 0 & \gamma^2 & 0 & \gamma^2 & \gamma^2 & \gamma^2 \\
0 & 0 & 0 & \alpha\beta & 0 & 0 & \alpha\beta \\
0 & 0 & 0 & 0 & \alpha\gamma & 0 & \alpha\gamma \\
0 & 0 & 0 & 0 & 0 & \beta\gamma & \beta\gamma
\end{array}\right)
\ee The seven columns of this matrix are linearly dependent. A
non-degenerate (for non-vanishing $\alpha,\beta,\gamma$) square
matrix $\check E^{\alpha}_{\mu}$ can be obtained, for example, by
neglecting the last column. It is inverse of this square matrix
that we call \be E_{\mu\alpha}^{*} = \check E^{-1}_{\mu\alpha} =
\left(\begin{array}{cccccc}
a^2 & 0 & 0 & -ab & -ac & 0 \\
0 & b^2 & 0 & -ab & 0 & -bc \\
0 & 0 & c^2 & 0 & -ac & -bc \\
0 & 0 & 0 & ab & 0 & 0 \\
0 & 0 & 0 & 0 & ac & 0 \\
0 & 0 & 0 & 0 & 0 & bc
\end{array}\right)
\ee It can be now multiplied from the left by truncated $3\times
6$ matrix $\check e_{i\mu}$, obtained from $3\times 7$ $e_{i\mu}$
by omition of the last column, and we obtain: \be A^i_\alpha =
\sum_{\mu = 1}^6 \check e^i_{\mu} \check E^{*}_{\mu\alpha} =
\left(\begin{array}{cccccc}
a & 0 & 0 & 0 & 0 & 0 \\
0 & b & 0 & 0 & 0 & 0 \\
0 & 0 & c & 0 & 0 & 0
\end{array}\right)
\ee i.e. the three maps indeed are $A^1 = ax^2$, $A^2 = by^2$ and
$A^3=cz^2$. Note that this is true, despite the product
$$
\sum_{\alpha=1}^6 E^*_{\mu\alpha}E^\alpha_\nu =
\left(\begin{array}{cccccc|c}
1 & 0 & 0 & 0 & 0 & 0 & -1 \\
0 & 1 & 0 & 0 & 0 & 0 & -1 \\
0 & 0 & 1 & 0 & 0 & 0 & -1 \\
0 & 0 & 0 & 1 & 0 & 0 & 1 \\
0 & 0 & 0 & 0 & 1 & 0 & 1 \\
0 & 0 & 0 & 0 & 0 & 1 & 1
\end{array}\right)
$$
differs from the unit matrix by addition of the last column,
emerging because $c_{3|2}=7$ exceeds $M_{3|2}=6$ by one.

\bigskip

\subsubsection{The map $f(x) = x^2 + c$:}

This map is very popular in non-linear theory, it is the standard
example in textbooks on Julia and Mandelbrot sets. See s.5 of
ref.\cite{DM} for details of its algebro-geometrical description.

In homogeneous coordinates the map is
$$
\left(\begin{array}{c} x \\ y \end{array}\right) \ \longrightarrow
\ \left(\begin{array}{c} x^2 + cy^2 \\ y^2 \end{array}\right)
$$
with $(n|s) = (2|2)$. The three eigenvectors are:
$$
\left(\begin{array}{c} x \\ y \end{array}\right) =
\left(\begin{array}{c} 1 \\ 0 \end{array}\right),
\left(\begin{array}{c} x_+ \\ 1 \end{array}\right),
\left(\begin{array}{c} x_- \\ 1 \end{array}\right)
$$
with $x_\pm = \frac{1}{2}(1\pm \sqrt{1-4c})$. Thus
\be &e^i_{\mu} =
\left(\begin{array}{ccc} 1&x_+&x_- \\ 0&1&1 \end{array}\right),\ \
E^\alpha_\mu = \left(\begin{array}{ccc} 1 & x_+^2 & x_-^2 \\ 0 &
x_+ & x_- \\ 0 & 1 & 1
\end{array}\right),\nn \\
&E^*_{\mu\alpha} = E^{-1}_{\mu\alpha} = \frac{1}{x_+-x_-}
\left(\begin{array}{ccc}
x_+-x_- & -x_+^2 + x_-^2 & x_+x_-(x_+-x_-) \\
0 & 1 & -x_- \\ 0 & -1 & x_+ \end{array}\right)\ \ \ee
and
\be
&A^i_{\alpha} = \sum_{\mu=1}^3 e^i_{\mu}E^{*}_{\mu\alpha} =
\left(\begin{array}{ccc} 1&x_+&x_- \\ 0&1&1 \end{array}\right)
\frac{1}{x_+-x_-} \nn\\
&\cdot\left(\begin{array}{ccc}
x_+-x_- & -x_+^2 + x_-^2 & x_+x_-(x_+-x_-) \\
0 & 1 & -x_- \\ 0 & -1 & x_+ \end{array}\right) =
\left(\begin{array}{ccc} 1& 0 & c \\ 0&0&1 \end{array}\right)
\ee

\subsubsection{Map from its eigenvectors: the case of $n|s = 2|2$
\label{eigexam}}

Since $c_{2|2} = M_{2|2} = 3$, in this case the three eigenvectors
can be chosen independently and have $nc_{n|s} = 6$ components --
exactly the same number as the tensor $A^i_\alpha$ ($nM_{n|s} =
6$). This means that we can reconstruct $A^i_\alpha$ from the
eigenvectors. In general case the same procedure is applicable,
but only $nM_{n|s}$ out of $nc_{n|s}$ components $e^i_\alpha$ can
be chosen independently, however explicit form of the
$c_{n|s}-M_{n|s}$ constraints on $e^i_\mu$ remains to be found.

Let \be
e^i_{\mu} = \left(\begin{array}{c} x_\mu \\
y_\mu \end{array}\right) \sim \left(\begin{array}{c} u_\mu \\
v_\mu \end{array}\right). \label{eigev22} \ee Then for fixed $\mu$
the $n\times n = 2\times 2$ symmetric $ij$-matrix can be rewritten
as an $M_{n|s}=3$-component $\alpha$-vector:
$$
e^i_{\mu}e^j_{\mu} =
\left(\begin{array}{cc} u_\mu^2 & u_\mu v_\mu \\
u_\mu v_\mu & v_\mu^2 \end{array}\right) \rightarrow
\left(\begin{array}{c} u_\mu^2\\ u_\mu v_\mu \\ v_\mu^2
\end{array}\right),
$$
so that the $M_{n|s}\times M_{n|s} = 3\times 3$ matrix
$$
E^\alpha_\mu = e^{i(\alpha)}_\mu e^{j(\alpha)}_\mu =
\left(\begin{array}{ccc} u_1^2 & u_2^2 & u_3^2 \\
u_1v_1 & u_2v_2 & u_3v_3 \\ v_1^2 & v_2^2 & v_3^2
\end{array}\right)
$$
has determinant
$$
\det_{\alpha\mu} E^\alpha_\mu =
(u_1v_2-u_2v_1)(u_2v_3-u_3v_2)(u_3v_1-u_1v_3) = \Delta(u|v),
$$
is non-degenerate unless some two eigenvectors coincide
($u_iv_j=u_jv_i$), and inverse matrix
\be
&E^*_{\mu\alpha}= -\frac{1}{\Delta(u|v)} \nn\\
&\left(\begin{array}{ccc}
v_2v_3(u_2v_3-u_3v_2) & -(u_2v_3+u_3v_2)(u_2v_3-u_3v_2) &
u_2u_3(u_2v_3-u_3v_2)\\
v_3v_1(u_3v_1-u_1v_3) & -(u_3v_1+u_1v_3)(u_3v_1-u_1v_3) &
u_3u_1(u_3v_1-u_1v_3)\\
v_1v_2(u_1v_2-u_2v_1) & -(u_1v_2+u_2v_1)(u_1v_2-u_2v_1) &
u_1u_2(u_1v_2-u_2v_1)
\end{array}\right)
\nn\\\ee
Then at the r.h.s. of (\ref{tensexp''}) we have:
$$ \hspace{-1.5cm}
\begin{array}{c} il = 11 \\ il = 12 = 21 \\ il = 22 \end{array}\
\left(\begin{array}{ccc} u_1^2 & u_2^2 & u_3^2 \\
u_1v_1 & u_2v_2 & u_3v_3 \\ v_1^2 & v_2^2 & v_3^2
\end{array}\right) \times \frac{1}{\Delta(u|v)}$$
{\footnotesize$$ \times
\left(\begin{array}{ccc} v_2v_3(u_2v_3-u_3v_2)\lambda_{1}^{(l)} &
-(u_2v_3+u_3v_2)(u_2v_3-u_3v_2)\lambda_{1}^{(l)} &
u_2u_3(u_2v_3-u_3v_2)\lambda_{1}^{(l)} \\
v_3v_1(u_3v_1-u_1v_3)\lambda_{2}^{(l)} &
-(u_3v_1+u_1v_3)(u_3v_1-u_1v_3)\lambda_{2}^{(l)} &
u_3u_1(u_3v_1-u_1v_3)\lambda_{2}^{(l)} \\
v_1v_2(u_1v_2-u_2v_1)\lambda_{3}^{(l)} &
-(u_1v_2+u_2v_1)(u_1v_2-u_2v_1)\lambda_{3}^{(l)} &
u_1u_2(u_1v_2-u_2v_1)\lambda_{3}^{(l)}
\end{array}\right)
$$}
\vskip1mm
Denoting this matrix through
$$
\left(\begin{array}{ccc} A_{11}(\lambda_{\cdot l})_{11} &
A_{11}(\lambda_{\cdot l})_{12} &
A_{11}(\lambda_{\cdot l})_{22} \\
A_{12}(\lambda_{\cdot l})_{11} & A_{12}(\lambda_{\cdot l})_{12} &
A_{12}(\lambda_{\cdot l})_{22} \\
A_{22}(\lambda_{\cdot l})_{11} & A_{22}(\lambda_{\cdot l})_{12} &
A_{22}(\lambda_{\cdot l})_{22}
\end{array}\right)
$$ {\footnotesize$$\hskip1mm =
\left(\begin{array}{cc} u_1^2v_2v_3\xi_1^{(l)} +
u_2^2v_1v_3\xi_2^{(l)} + & -u_1^2(u_2v_3+u_3v_2)\xi_1^{(l)} -
u_2^2(u_3v_1+u_1v_3)\xi_2^{(l)}-\\
+ u_3^2v_1v_2\xi_3^{(l)} & -u_3^2(u_3v_1+u_1v_3)\xi_3^{(l)} \\
 &\\
u_1v_1v_2v_3\xi_1^{(l)} + u_2v_2v_1v_3\xi_2^{(l)} +&
-u_1v_1(u_2v_3+u_3v_2)\xi_1^{(l)}-u_2v_2(u_3v_1+u_1v_3)\xi_2^{(l)}-\\
+  u_3v_3v_1v_2\xi_3^{(l)} & -u_3v_3(u_3v_1+u_1v_3)\xi_3^{(l)} \\
 &\\
v_1^2v_2v_3\xi_1^{(l)} + v_2^2v_1v_3\xi_2^{(l)} + &
-v_1^2(u_2v_3+u_3v_2)\xi_1^{(l)} -
v_2^2(u_3v_1+u_1v_3)\xi_2^{(l)}- \\
+ v_3^2v_1v_2\xi_3^{(l)} & - v_3^2(u_3v_1+u_1v_3)\xi_3^{(l)}
\end{array}\right.
$$}
{\footnotesize$$\left.\begin{array}{c}
u_1^2u_2u_3\xi_1^{(l)} + u_2^2u_1u_3\xi_2^{(l)} + u_3^2u_1u_2\xi_3^{(l)}\\
\\
u_1v_1u_2u_3\xi_1^{(l)} + u_2v_2u_1v_u\xi_2^{(l)} +u_3v_3u_1u_2\xi_3^{(l)}\\
\\
v_1^2u_2u_3\xi_1^{(l)} + v_2^2u_1u_3\xi_2^{(l)} + v_3^2u_1u_2\xi_3^{(l)}
\end{array}\right)$$}
\vskip2mm\noindent
with $\xi_1^{(l)} = \lambda_1^{(l)} (u_2v_3-u_3v_2)$, $\xi_2^{(l)}
= \lambda_2^{(l)} (u_3v_1-u_1v_3)$, $\xi_3^{(l)} = \lambda_3^{(l)}
(u_1v_2-u_2v_1)$, we get: \be A^1_{jk} = A_{11}(\lambda_{\cdot
1})_{jk} + A_{12}(\lambda_{\cdot 2})_{jk},
\nn \\
A^2_{jk} = A_{12}(\lambda_{\cdot 1})_{jk} + A_{22}(\lambda_{\cdot
2})_{jk} \ee i.e.
$$
A^1_{\alpha} = \frac{1}{\Delta(u|v)}\left(\begin{array}{c}
u_1v_2v_3\Xi_1 + u_2v_1v_3\Xi_2 +\\
+ u_3v_1v_2\Xi_3
\end{array}\right.$$
$$\left.\begin{array}{|c|c}
-u_1(u_2v_3+u_3v_2)\Xi_1 - u_2(u_3v_1+u_1v_3)\Xi_2 -
&u_1u_2u_3(\Xi_1+\Xi_2+\Xi_3)\\
-u_3(u_1v_2+u_2v_1)\Xi_3 &
\end{array}\right)$$

$$
A^2_{\alpha} = \frac{1}{\Delta(u|v)}\left(\begin{array}{c}
v_1v_2v_3(\Xi_1 + \Xi_2 + \Xi_3)\\
\
\end{array}\right.$$\vskip3mm
$$\left.\begin{array}{|c|c}
-v_1(u_2v_3+u_3v_2)\Xi_1 -
v_2(u_3v_1+u_1v_3)\Xi_2 -
&v_1u_2u_3\Xi_1+ v_2u_1u_3\Xi_2+\\
-v_3(u_1v_2+u_2v_1)\Xi_3 & + v_3u_1u_2\Xi_3
\end{array}\right)$$\vskip2mm\noindent
where $\Xi_1 = (u_2v_3-u_3v_2)\Lambda_1$, $\Xi_2 =
(u_3v_1-u_1v_3)\Lambda_2$, $\Xi_3 = (u_1v_2-u_2v_1)\Lambda_3$ and
$\Lambda_\mu = u_\mu \lambda_\mu^{(1)} + v_\mu \lambda_\mu^{(2)}$.
Note that $\lambda_\mu^{(1)}$ and $\lambda_\mu^{(2)}$ enter
expressions for $A$ only in peculiar combinations $\Lambda_\mu$.
The $6$ independent components of tensor $A$ can be parametrized
by $3$ $u$'s and by $3$ $v$'s or by $3$ $u$'s and by $3$
$\Lambda$'s. Above equations can be also rewritten as \be
&A^1_{\nu} = \frac{1}{\Delta(u|v)}\sum_{\alpha = 11,12,22} u_\alpha
\Xi_\alpha \Big(v_\beta v_\gamma,\ - (u_\beta v_\gamma + u_\gamma
v_\beta),\
u_\beta u_\gamma\Big), \nn\\
&A^2_{\nu} = \frac{1}{\Delta(u|v)}\sum_{\alpha = 11,12,22} v_\alpha
\Xi_\alpha \Big(v_\beta v_\gamma,\ - (u_\beta v_\gamma + u_\gamma
v_\beta),\ u_\beta u_\gamma\Big) \label{A22throughev} \ee
($\alpha,\beta,\gamma$ stand for cyclic permutations of $(1,2,3) =
(11,12,22)$: if $\alpha = 11$, then $\beta = 12$, $\gamma = 22$,
if $\alpha = 12$, then $\beta = 22$, $\gamma = 11$, if $\alpha =
22$, then $\beta = 11$, $\gamma = 12$). It is easy to check from
(\ref{eigev22}) and (\ref{A22throughev}) that $A^1(\vec e_\mu) =
A^1_\alpha E^\alpha_\mu = \Lambda_\mu u_\mu$ and $A^2(\vec e_\mu)
= A^2_\alpha E^\alpha_\mu = \Lambda_\mu v_\mu$, i.e. that \be \vec
A\Big(\vec e_\mu\Big) = \Lambda_\mu \vec e_\mu \ee

\bigskip

{\bf Plucker elementary determinants} can be expressed through the
eigenvectors:
\be
&\left(\begin{array}{c} I_1 \\ I_2 \\ I_3 \end{array}\right) =
\left(\begin{array}{c} I[12,22] \\ I[22,11] \\
I[11,12] \end{array}\right) =
\left(\begin{array}{c} I[2,3] \\ I[3,1] \\
I[1,2] \end{array}\right) =\nn\\
&= -\frac{1}{\Delta(u|v)}\left\{
\Lambda_1\Lambda_2(u_1v_2 - u_2v_1)
\left(\begin{array}{c} u_3^2 \\ u_3v_3 \\
v_3^2 \end{array}\right) \right.\nn\\
&+\left. \Lambda_2\Lambda_3(u_2v_3 - u_3v_2)
\left(\begin{array}{c} u_1^2 \\ u_1v_1 \\
v_1^2 \end{array}\right) + \Lambda_3\Lambda_1(u_3v_1 - u_1v_3)
\left(\begin{array}{c} u_2^2 \\ u_2v_2 \\
v_2^2 \end{array}\right)\right\}
\nn\ee
The {\bf resultant}
\be &{\cal R}_{2|2} = I_1I_3 - I_2^2 =
\frac{\Lambda_1\Lambda_2\Lambda_3}{\Delta(u|v)}
\Big(\Lambda_1(u_2v_3-u_3v_2) \nn\\
&+ \Lambda_2(u_3v_1-u_1v_3) +
\Lambda_3(u_1v_2-u_2v_1)\Big) \label{r22eige} \ee
This formula is
especially convenient when $\Lambda$'s are taken for $3$
independent parameters (along with a triple of either $u$'s or
$v$'s -- then another triple is kept fixed).

\subsubsection{Appropriately normalized eigenvectors and
elimination of $\Lambda$-parameters}

Alternatively, parameters $\Lambda$ can be absorbed into rescaling
of $u$'s and $v$'s: \be u_\mu = \Lambda_\mu^{1/(s-1)}\check u_\mu
=
\Lambda_\mu\check u_\mu,\nn \\
v_\mu = \Lambda_\mu^{1/(s-1)} \check v_\mu = \Lambda_\mu \check
v_\mu \label{apron} \ee Such {\it appropriately normalized}
eigenvectors $\check{\vec e_\mu}$ satisfy \be \vec
A\Big(\check{\vec e_\mu}\Big) = \check{\vec e_\mu} \ee

{\bf The case of $n|s=2|2$:}

\bigskip

If expressed in terms of $\check{\vec e_\mu}$,
(\ref{A22throughev}) becomes: \be \vec A =
\sum_{(\mu\nu\lambda)=(123),(231),(312)} \check{\vec e_\mu}
\frac{\Big(\check{\vec e_\nu}\times \ldots\Big) \Big(\check{\vec
e_\lambda}\times\ldots\Big)} {\Big(\check{\vec
e_\nu}\times\check{\vec e_\mu}\Big) \Big(\check{\vec
e_\lambda}\times\check{\vec e_\mu}\Big)} \ee (orthogonality
properties are explicit in this formula) or, in more detail, \be
A^i_{jk} = \sum_{(\mu\nu\lambda)=(123),(231),(312)} \frac{\check
e^i_\mu \Big(\check e_{\nu j}\check e_{\lambda k} + \check e_{\nu
k}\check e_{\lambda j}\Big)} {2\Big(\check e_{\nu l}\check
e^l_\mu\Big) \Big(\check e_{\lambda m}\check e^m_\mu\Big)}
\label{Athroughnorei22} \ee where $\check e_{\nu j} =
\epsilon_{jk}\check e^k_\nu$. Then the resultant (\ref{r22eige})
acquires the form: \be {\cal R}_{2|2} = \frac{\check{\vec
e_1}\times\check{\vec e_2} + \check{\vec e_1}\times\check{\vec
e_2} + \check{\vec e_1}\times\check{\vec e_2}} {\Big(\check{\vec
e_1}\times\check{\vec e_2}\Big) \Big(\check{\vec
e_1}\times\check{\vec e_2}\Big) \Big(\check{\vec
e_1}\times\check{\vec e_2}\Big)} \ee The three $2$-component
vectors are linearly dependent: \be p_1\check{\vec e_1} +
p_2\check{\vec e_2} + p_3\check{\vec e_3} = 0 \ee and \be {\cal
R}_{2|2} = \frac{p_1+p_2+p_3}{p_1p_2p_3}
\left(\frac{p_1}{\check{\vec e_2}\times\check{\vec e_3}} \right)^2
= \frac{p_1+p_2+p_3}{p_1p_2p_3} \left(\frac{p_2}{\check{\vec
e_3}\times\check{\vec e_1}} \right)^2 =
\frac{p_1+p_2+p_3}{p_1p_2p_3} \left(\frac{p_3}{\check{\vec
e_1}\times\check{\vec e_2}} \right)^2 \ee Thus the resultant
vanishes either when $p_1+p_2+p_3 = 0$ or when at least two out of
three {\it normalized} vectors $\check{\vec e_\mu}$ grow
infinitely. However, one should keep in mind that $p_1$ and $p_2$
are {\it not} free parameters: $-\frac{p_1 \check{\vec e_1} +
p_2\check{\vec e_2}}{p_1+p_2}$ can be a normalized eigenvector
only for special values of $p_1$ and $p_2$: for example, in the
{\it polynomial} $n|s=2|2$ case (\ref{nondiapol22}) we have \be
p_1 = -\frac{1-\tilde\beta}{1-\tilde\alpha\tilde\beta}p_3,\ \ \
p_2 = -\frac{1-\tilde\alpha}{1-\tilde\alpha\tilde\beta}p_3 \ee
with $\tilde\alpha = b_{12}/a_{11}$, $\tilde\beta =
a_{12}/b_{22}$, and $p_1+p_2+p_3 =
-\frac{(1-\tilde\alpha)(1-\tilde\beta)}
{1-\tilde\alpha\tilde\beta}p_3$ vanishes only when either $p_1=0$
or $p_2=0$ or $p_3=0$ (and these are singular points, where more
carefull analysis is needed).

\bigskip

{\bf The case of $n|s = 2|s$:}

\bigskip

In this case we have $c_{2|s} = s+1$ $n=2$-component eigenvectors
and direct generalization of (\ref{Athroughnorei22}) is: \be
A^i_{j_1\ldots j_s} = {\rm symm}\left(
\sum_{\mu=1}^{s+1}\check e^i_\mu \frac{\check e_{\nu_1 j_1} \ldots
\check e_{\nu_s j_s}} {\Big(\check e_{\nu_1 l}\check
e^l_\mu\Big)\ldots \Big(\check e_{\nu_s l}\check e^l_\mu\Big)}
\right) \label{Athroughnorei2s} \ee As in (\ref{Athroughnorei22})
the set $\nu_1 \neq \nu_2 \neq \ldots \neq \nu_s \neq \mu$ i.e.
all eigenvectors appear in each term of the sum, symmetrization is
over $j_1,\ldots,j_s$ or $\nu_1,\ldots,\nu_s$, the indices are
lowered with the help of the $\epsilon$-tensor $\epsilon_{ij}$,
and the $\mu$-th term in the sum obviously annihilates all
eigenvectors except for $\vec e_\mu$.

\bigskip

{\bf The case of $n|s = n|1$ (linear maps):}

\bigskip

This another simple case. Still it deserves mentioning, because
now $n$ can exceed $2$ and the rank-$n$ $\epsilon$-tensor can not
be used to low index of a single vector: a group of $n-1$ vectors
is needed. There are exactly $c_{n|1}=n$ eigenvectors. Specifics
of the linear ($s=1$) case is that $\Lambda$-parameters can {\it
not} be eliminated and there is no distinguished normalization of
eigenvectors. Accordingly the analogue of (\ref{Athroughnorei22})
is: \be A^i_{j} = \sum_{\mu=1}^n \Lambda_\mu e^i_\mu\frac{
\epsilon^{\mu\mu_1\ldots\mu_{n-1}} \epsilon_{jl_1\ldots l_{n-1}}
e_{\mu_1}^{l_1}\ldots e_{\mu_{n-1}}^{l_{n-1}}} {\det_{j\nu}
e^j_\nu} \label{Athroughnorein1} \ee Determinant in denominator is
nothing but the trace of the numerator:
$$
{\det}_{j\nu} e^j_\nu = e^j_\nu \epsilon^{\nu\mu_1\ldots\mu_{n-1}}
\epsilon_{jl_1\ldots l_{n-1}}
e_{\mu_1}^{l_1}\ldots e_{\mu_{n-1}}^{l_{n-1}}
$$
Of course, (\ref{Athroughnorein1}) is just a minor reformulation
of (\ref{matvseig}).

\bigskip

{\bf Arbitrary $n|s$:}

\bigskip

In general one expects a synthesis of eqs.(\ref{Athroughnorei2s})
and (\ref{Athroughnorein1}): \be A^i_{j_1\ldots j_s} = {\rm
symm}\left( \sum_{\mu=1}^{M_{n|s}}\check e^i_\mu \prod_{k=1}^s
\frac{\epsilon_{j_kl_1\ldots l_{n-1}} \check e_{\nu_k,1}^{l_1}
\ldots \check e_{\nu_k,n-1}^{l_{n-1}}} {\epsilon_{jl_1\ldots
l_{n-1}} \check e_\mu^j \check e_{\nu_k,1}^{l_1} \ldots \check
e_{\nu_k,n-1}^{l_{n-1}}} \right) \label{Athroughnoreins} \ee
However, this formula should be used with care. Not only there is
a sum over only $M_{n|s}$ out of $c_{n|s}\geq M_{n|s}$
eigenvectors $\check {vec e}_\mu$, the set of only $s(n-1)$ out of
$c_{n|s}-1 = s+\ldots + s^{n-1}$ eigenvectors $\check {\vec
e}_{\nu_k l_m}$ should be specified for every $\check {vec
e}_\mu$.

\subsection{Eigenvector problem and unit operators
\label{unopei}}

Eigenstate problem 
can be reformulated to look even more similar to the
linear-algebra case if it is written in terms of {\it unit
operators}. Namely, rewrite (\ref{eist}) as \be \vec A(\vec z) =
\vec I_{\lambda}(\vec z) \label{eistunop} \ee Then (\ref{chareq})
turns into \be R_{n|s}(\lambda|A) = {\cal R}_{n|s}\Big(A -
I_\lambda\Big) = 0 \label{charequnop} \ee The map
$I^i_{\lambda}(\vec z)$ is {\it unit} in the sense that it leaves
{\it all} points of {\it projective} space $P^{n-1}$ intact:
$I^i_\lambda(\vec z) = \lambda(\vec z)z^i$ and the common
rescaling $\lambda(\vec z)$ is projectively unobservable. Of
course, from the point of view of  affine space $C^n$ with
homogeneous coordinates $\vec z$, the map $z^i \rightarrow
I^i(\vec z)$ is not unit. Moreover, maps with different functions
$\lambda(\vec z)$ are all different: the space of unit maps is
that of polynomials of degree $s-1$ of $n$ variables and have
dimension $M_{n|s-1} = \frac{(n+s-2)!}{(n-1)!(s-1)!}$, which is
bigger than one for $s>1$. Instead for $s>1$ all maps $I_\lambda$
are degenerate: \be {\cal R}_{n|s}(I_\lambda) = 0\ \ \ \forall
\lambda\ \ \ {\rm if}\ \ s>1 \ee Unit operators and their
application to construction of exponentiated maps (and thus to the
definition of non-linear analogue of Lie algebras)    will be
discussed in a little more detail in s.\ref{unop} below.

A problem of unit operators has the following interesting
generalization. For a pair of operators $A,B:\ V\rightarrow W$,
both  acting from one space $V$ into another $W$, one can ask, if
there are any elements $z\in V$, which have collinear (i.e.
projectively equivalent)
images: 
\be \vec A(\vec z) = \lambda(\vec z)\vec B(\vec z) \ee with some
function (scalar) $\lambda(\vec z)$. Note, that in variance with
the eigenspace problem, the target space $W$ does not need to
coincide with original $V$. Existence of such $\vec z$ imposes a
constraint on the the pair of maps $A,B$. Similarly, for a triple
of operators one can ask that the images of a single vector $z\in
V$ are linearly dependent: \be \vec A(\vec z) = \lambda_B(\vec
z)\vec B(\vec z) + \lambda_C(\vec z)\vec C(\vec z). \ee One can
also ask for collinearity of the three images: \be \vec A(\vec z)
= \mu_B(\vec z)\vec B(\vec z) = \mu_C(\vec z)\vec C(\vec z) \ee --
this imposes {\it two} constraints on the triple $A,B,C$. A
hierarchy of such problems involves any number of operators.

\setcounter{equation}{0}

\section{Iterated maps}

Eigenvectors, considered in the previous section \ref{eige}, are
{\it orbits} of the map $\vec A(\vec z)$ of order one (fixed
points in projective space $P^{n-1}$). As we saw in s.\ref{eige},
already {\it eigenvector} theory is somewhat non-trivial in
non-linear situation. Far more interesting is the theory of
higher-order orbits. In linear case the problem of higher order
orbits does not add much new, see a short discussion in s.6.2 of
ref.\cite{DM}. This is essentially because after the matrix-map is
diagonalized, it is a simple task to study its integer and
rational powers (roots): root of diagonal matrix is not
generically diagonal, but still rather simple to describe and
classify. A little bit trickier is consideration of matrices with
Jordan cells, but they can also be studied by elementary methods.
Anyhow, the theory of higher-order maps -- somewhat unjustly --
does not attract considerable attention in the linear-algebra
courses. The story is much less trivial and can not be neglected
in non-linear case (in addition it plays important role in
physical applications).

The theory of iterated maps of a single variable (i.e. of two
homogeneous variables, $n=2$) is developed in \cite{DM}. In this
section we make the first steps of its generalization to
non-linear maps of many variables. As everywhere in this paper we
work with projectivized spaces, and considerations of
ref.\cite{DM} are reproduced in particular case of $n=2$.

Zeroes of characteristic polynomials
$R_{n|s}(\lambda^{(m)}|A^{\circ m})$ are associated with the
periodic orbits of degree $m$ of the map $A$, and are direct
counterparts of discriminants ${\rm Disc}(F_m)$ of the functions
$F_m(x) = f^{\circ m}(x)-x$ of one variable. The first purpose of
iteration-map studies is to find multi-dimensional analogues of
irreducible functions $G_m(x)$ (describing orbits of {\it exact}
order $m$, with all divisors of $m$ excluded), the corresponding
discriminants $D_m = {\rm Disc}(G_m)$, resultants $R_{m,k} = {\rm
Res}(G_m,G_k)$ and their zeroes, forming the Mandelbrot set in the
space of all functions. The presentation below  is restricted to a
few basic definitions and elementary first steps.

\subsection{Relation between
$R_{n|s^2}(\lambda^{s+1}|A^{\circ 2})$ and $R_{n|s}(\lambda|A)$}

From (\ref{composres}) we deduce: \be {\cal R}_{n|s^2} (A\circ A)
=
\Big({\cal R}_{n|s}(A)\Big)^{s^{n-1}(1+s)}, \nn \\
{\cal R}_{n|s^3} (A\circ A\circ A) =
\Big({\cal R}_{n|s}(A)\Big)^{s^{2(n-1)}(1+s+s^2)}, \nn \\
\ldots \nn \\
{\cal R}_{n|s^m}(A^{\circ m}) = \Big({\cal
R}_{n|s}(A)\Big)^{s^{(m-1)(n-1)}\frac{s^m-1}{s-1}} \ee The powers
at the r.h.s. can also be written as $s^{(m-1)(n-1)}c_{m|s}$.

If $\vec A(\vec z) = \lambda(\vec z)\vec z$, then $\vec A^{\circ
2}(\vec z) = \lambda(\vec z)^{s+1}\vec z$,\ $\vec A^{\circ 3}(\vec
z) = \lambda(\vec z)^{s^2+s+1}\vec z$ and $\vec A^{\circ m}(z) =
\lambda(\vec z)^{\frac{s^m-1}{s-1}}\vec z$. This implies that the
characteristic resultant of the iterated map with appropriately
adjusted "eigenvalue parameter" $\lambda$ is divisible by ${\cal
R}_{n|s}(\lambda|A)$ and by all other characteristic resultants
with iteration degrees, which are divisors of $m$: \be {\cal
R}_{n|s^m}\Big(\left.\lambda^{\frac{s^m-1}{s-1}}\right| A^{\circ
m}\Big) = \prod_{{\rm divisors}\ k\ {\rm of}\ m} {\cal
R}_{n|s^k}\Big(\left.\lambda^{\frac{s^k-1}{s-1}}\right| A^{\circ
k}\Big)\ldots \ee Actually this does not exhaust all divisors,
there are more, denoted by $\ldots$, accounting for "correlations"
between the intermediate resultants. The last irreducible factor
is characteristic of the order-$m$ orbits {\it per se}: it
vanishes when such orbits merge, see \cite{DM} for details.

\bigskip

{\bf Examples of diagonal maps for $n=2$}

\bigskip

Consider {\it diagonal} map $P^1\rightarrow P^1$ ($n=2$) of degree
$s$ \be A: \  \left(\begin{array}{c} x\\ y \end{array}\right)
\longrightarrow \left(\begin{array}{c} ax^s\\ by^s
\end{array}\right), \ \ \ \ A^{\circ 2}: \ \left(\begin{array}{c}
x\\ y \end{array}\right) \longrightarrow \left(\begin{array}{c}
a^{s+1}x^{s^2}\\ b^{s+1}y^{s^2} \end{array}\right) \ee

{\bf $s=2$:}

From (\ref{r2sla}) we get:
$$
R_{2|4}(\mu|A^{\circ 2}) = (\mu_1-a^3)(\mu_4-b^3) (a^3b^3 -
\mu_1b^3  - \mu_2{ab^2}  - \mu_3{a^{2}b} - \mu_4 a^3 ) \cdot $$ $$
\cdot (a^3b^3 - \mu_1b^3  - \omega_3\mu_2{ab^2}  -
\omega_3^2\mu_3{a^{2}b} - \mu_4 a^3 ) (a^3b^3 - \mu_1b^3  -
\omega_3^2\mu_2{ab^2}  - \omega_3\mu_3{a^{2}b} - \mu_4 a^3 )
$$
Substituting $\mu(\vec z) = \lambda^3(\vec z)$, what for $(n|s) =
(2|2)$ menas that $\mu_1 = \lambda_1^3$, $\mu_2 =
3\lambda_1^2\lambda_2$, $\mu_3=3\lambda_1\lambda_2^2$ and
$\mu_4=\lambda_2^3$ we obtain
$$
R_{2|4}(\lambda^3|A^{\circ 2}) =
(\lambda_1^3-a^3)(\lambda_2^3-b^3) (a^3b^3 - \lambda_1^3b^3  -
3\lambda_1^2\lambda_2 ab^2  - 3\lambda_1\lambda_2^2 a^2b -
\lambda_2^3 a^3 ) \cdot $$ $$ \cdot (a^3b^3 - \lambda_1^3b^3  -
3\omega_3\lambda_1^2\lambda_2 ab^2  -
3\omega_3^2\lambda_1\lambda_2^2 a^2b - \lambda_2^3 a^3 ) (a^3b^3 -
\lambda_1^3b^3  - 3\omega_3^2\lambda_1^2\lambda_2 ab^2  -
3\omega_3\lambda_1\lambda_2^2 a^2b - \lambda_2^3 a^3 ) =
$$ $$
(\lambda_1^3-a^3)(\lambda_2^3-b^3) \Big((ab)^3 -
(\lambda_1b+\lambda_2a)^3\Big) \Big((ab)^3 -
(\lambda_1b+\omega_3\lambda_2a)^3\Big) \Big((ab)^3
-(\lambda_1b+\omega_3^2\lambda_2a)^3\Big)
$$
Now we use decomposition $p^{s+1}-q^{s+1} = \prod_{k=1}^{s+1}(p -
\omega_{s+1}^kq)$ to rewrite this as
$$
R_{2|4}(\lambda^3|A^{\circ 2}) = \prod_{k=1}^3
\left\{(\lambda_1\omega_3^k-a)(\lambda_2\omega_3^k-b)
\prod_{l=1}^3
\Big(ab-(\lambda_1b+\omega_3^l\lambda_2a)\omega_3^k\Big) \right\},
$$
while, see (\ref{r22la}),
$$
R_{2|2}(\lambda|A) = (\lambda_1-a)(\lambda_2-b) \Big(ab -
(b\lambda_1 + a\lambda_2)\Big)
$$

{\bf $s=3$:}
$$
R_{2|9}(\mu|A^{\circ 2}) = (\mu_1-a^4)(\mu_9-b^4)\prod_{k=1}^8
\prod_{k=1}^8\Big((ab)^4 - b^4\mu_1 - \omega_8^k
b^{7/2}a^{1/2}\mu_2 - \omega_8^{2k}b^3a\mu_3 - \ldots -
a^4\mu_9\Big)
$$
Substituting $\mu(\lambda)$ from
$$
\mu_1x^8 + \mu_2x^7y + \mu_3x^6y^2 + \ldots + \mu_9y^8 =
(\lambda_1x^2 + \lambda_2xy + \lambda_3y^2)^4,
$$
we get:
$$
R_{2|9}(\lambda^4|A^{\circ 2}) = (\lambda_1^4 - a^4)(\lambda_3^4 -
b^4) \prod_{k=1}^8 \Big((ab)^4 - (b\lambda_1 +
\omega_8^k\sqrt{ab}\lambda_2 + \omega_8^{2k}a\lambda_3)^4\Big) =
$$
$$
= \prod_{m=1}^4\left\{(\lambda_1 -
\omega_4^ma)(\lambda_3-\omega_4^mb) \prod_{k=1}^8 \Big(ab -
\omega_4^m(b\lambda_1 + \omega_8^k\sqrt{ab}\lambda_2 +
\omega_8^{2k}a\lambda_3)\Big)\right\},
$$
while, see (\ref{r23la}),
$$
R_{2|3}(\lambda|A) = (\lambda_1 - a)(\lambda_3-b)\prod_{k=1}^2
\Big(ab - (b\lambda_1 + \omega_2^k\sqrt{ab}\lambda_2 +
a\lambda_3)\Big)
$$

{\bf Arbitrary $s$ (and $n=2$):}

Particular non-trivial branches (for diagonal map their are two
trivial ones: $L_{s^2} = \mu_{s^2}-b^{s+1}$ and $L_{s^2+1}=
\mu_1-a^{s+1}$) from (\ref{chareqfact})
$$
L_k(\mu|A^{\circ 2}) = (ab)^{s+1} - b^{s+1} \sum_{j=1}^{s^2}
\omega_{s^2-1}^{k(j-1)}
\left(\frac{a^{s+1}}{b^{s+1}}\right)^{\frac{j-1}{s^2-1}} \mu_{j}
$$
$$\mu(\vec z) = \sum_{j=1}^{s^2} \mu_j x^{s^2-j}y^{j-1} =
\lambda^{s+1}(\vec z) = \left(\sum_{i=1}^{s} \lambda_i
x^{s-i}y^{i-1}\right)^{s+1}
$$
$$
L_k(\lambda^{s+1}|A^{\circ 2}) = (ab)^{s+1} - \left( b\sum_{i=1}^s
\omega_{s^2-1}^{k(i-1)} \left(\frac{a}{b}\right)^{\frac{i-1}{s-1}}
\lambda_{i}
\right)^{s+1} = \prod_{m=1}^{s+1} \left( ab - \omega_{s+1}^m
\left( b\sum_{i=1}^s \omega_{s^2-1}^{k(i-1)}
\left(\frac{a}{b}\right)^{\frac{i-1}{s-1}}
\lambda_{i}
\right)\right)
$$
$$k = 1,\ldots, s^2-1$$
while
$$
L_m(\lambda|A) = \left( ab - b \sum_{i=1}^{s}
\omega_{s-1}^{m(i-1)}
\left(\frac{a}{b}\right)^{\frac{i-1}{s-1}}\lambda_i \right)
$$
$$m = 1,\ldots,s-1$$
Taking a prodruct of all branches, we obtain \be
R_{2|s^2}\left(\lambda^{s+1}|A^{\circ 2}\right) =
\prod_{k=1}^{s+1} \left\{(\lambda_1\omega_{s+1}^k - a)
(\lambda_s\omega_{s+1}^k - b) \prod_{l=1}^{s^2-1} \Big(ab -
\omega_{s+1}^k b \sum_{j=1}^{s-1}\lambda_j
\left(\frac{a}{b}\right)^{\frac{j-1}{s-1}}
\omega_{s^2-1}^{l(j-1)}\Big) \right\} \ee Note that roots of unity
of two different degrees, $s+1$ and $s^2-1$ appear in this formula
(for $s=2$ they occasionally coincide). Decomposing $l = m(s+1) +
r$ with $r = 1, \ldots, s+1$, we obtain one more representation:
\be R_{2|s^2}\left(\lambda^{s+1}|A^{\circ 2}\right) =
\prod_{k=1}^{s+1}\left\{ R_{2|s}(\lambda \omega_{s+1}^k|A)
\prod_{r=1}^{s}\frac{R_{2|s}(\lambda \omega_{s+1}^k|A_r)}
{(\lambda_1\omega_{s+1}^k - a_r) (\lambda_s\omega_{s+1}^k -
b_r)}\right\} \label{decomp2sla} \ee where $a_r$ and $b_r$ in
diagonal map $A_r$ are related to original $a$ and $b$ in $A$
through
$$
\frac{b_r}{a_r} = \frac{b}{a}\omega_{s+1}^r
$$
The l.h.s. and r.h.s. in (\ref{decomp2sla}) have degrees
$(s^2+1)(s+1)= (s+1)\Big(s+1 + s(s-1)\Big) $ and $2s^2(s+1)=
(s+1)\Big(2s + s(2s-2)\Big) $ in the coefficients of $\lambda$ and
$A$ respectively.

\subsection{Unit maps and exponential of maps:
non-linear counterpart of $algebra \ \leftrightarrow\ group$
relation \label{unop}}

\noindent

$\bullet$ We introduced {\it unit maps} in s.\ref{unopei}. By
definition they leave {\it all} points of projective space
$P^{n-1}$ invariant, i.e. satisfy \be I^i(\vec z) = \lambda_I(\vec
z)z^i \ \ \ {\rm for}\ {\bf any}\ z^i\ {\rm and\ some\ function}\
\ \lambda_I(\vec z) \label{projunitdef} \ee While for linear maps
($s=1$) the unity is projectively (i.e. up to a common factor)
unique: $I^i(\vec z) = \ {\rm const}\cdot \delta^i_jz^j$, for
non-linear maps ($s\geq 2$) there are many {\it unities}, labeled
by arbitrary polynomials $\lambda(\vec z)$ of degree $s-1$. For
example, for $s=2$ unit are the maps
$$
\left(\begin{array}{c} x \\ y \end{array}\right) \longrightarrow
\left(\begin{array}{c} x^2 \\ xy \end{array}\right) \ \ \ {\rm
and} \ \ \ \left(\begin{array}{c} x \\ y \end{array}\right)
\longrightarrow \left(\begin{array}{c} xy \\ y^2
\end{array}\right)
$$ 
and entire family
$$
\left(\begin{array}{c} x \\ y \end{array}\right) \longrightarrow
\alpha \left(\begin{array}{c} x^2 \\ xy \end{array}\right) + \beta
\left(\begin{array}{c} xy \\ y^2 \end{array}\right) = (\alpha x  +
\beta y) \left(\begin{array}{c} x \\ y \end{array}\right)
$$
-- all are non-linear unit maps, leaving any $\xi = x/y$ intact,
$\xi \longrightarrow \xi$. Of course, all unit maps commute and
any composition of unities of degree $s$ is a unity of degree
$s^2$.

$\bullet$ If $I^i(\vec z)$ is a unit map, it allows to associate
with any "nearby" map $A^i(\vec z) = I^i(\vec z) + \varphi^i(\vec
z)$ a new -- exponentiated -- map ${\cal E}_I(\varphi)$ of
projective space $P^{n-1}$. Association $\varphi \rightarrow {\cal
E}_I^\varphi$ is the non-linear counterpart of the $algebra \
\leftrightarrow\ group$ relation, now it depends on the choice of
$I$: \be A^i(\vec z) = I^i(\vec z) + \varphi^i(\vec z) \
\longrightarrow {\cal E}_I^{\varphi} = \lim_{N\rightarrow \infty}
\Big(I + \frac{1}{N}\varphi\Big)^{\circ N} \label{expomapdef} \ee
While $A^i(\vec z)$ is well defined in terms of homogeneous
coordinates $\vec z$, this is {\it not} true for ${\cal
E}_I^\varphi$. Indeed, if $A^i(\vec z)$ was of degree $s>1$, the
exponential ${\cal E}_I^\varphi$ is a limit of maps of increasing
degrees $s^N$, which is hard to define on the space of $\vec z$'s.
Remarkably, after projection to projective space the limit and
thus ${\cal E}_I^\varphi$ acquire sense and can be often found in
explicit form. For $\phi$ being a quadratic map given by an
element $\phi\in \cl{T}^2_1$ of tensor algebra, the contractions
involved in taking the limit in \ref{expomapdef} are presented by
diagrams on Fig.~\ref{exptree}.

\Fig{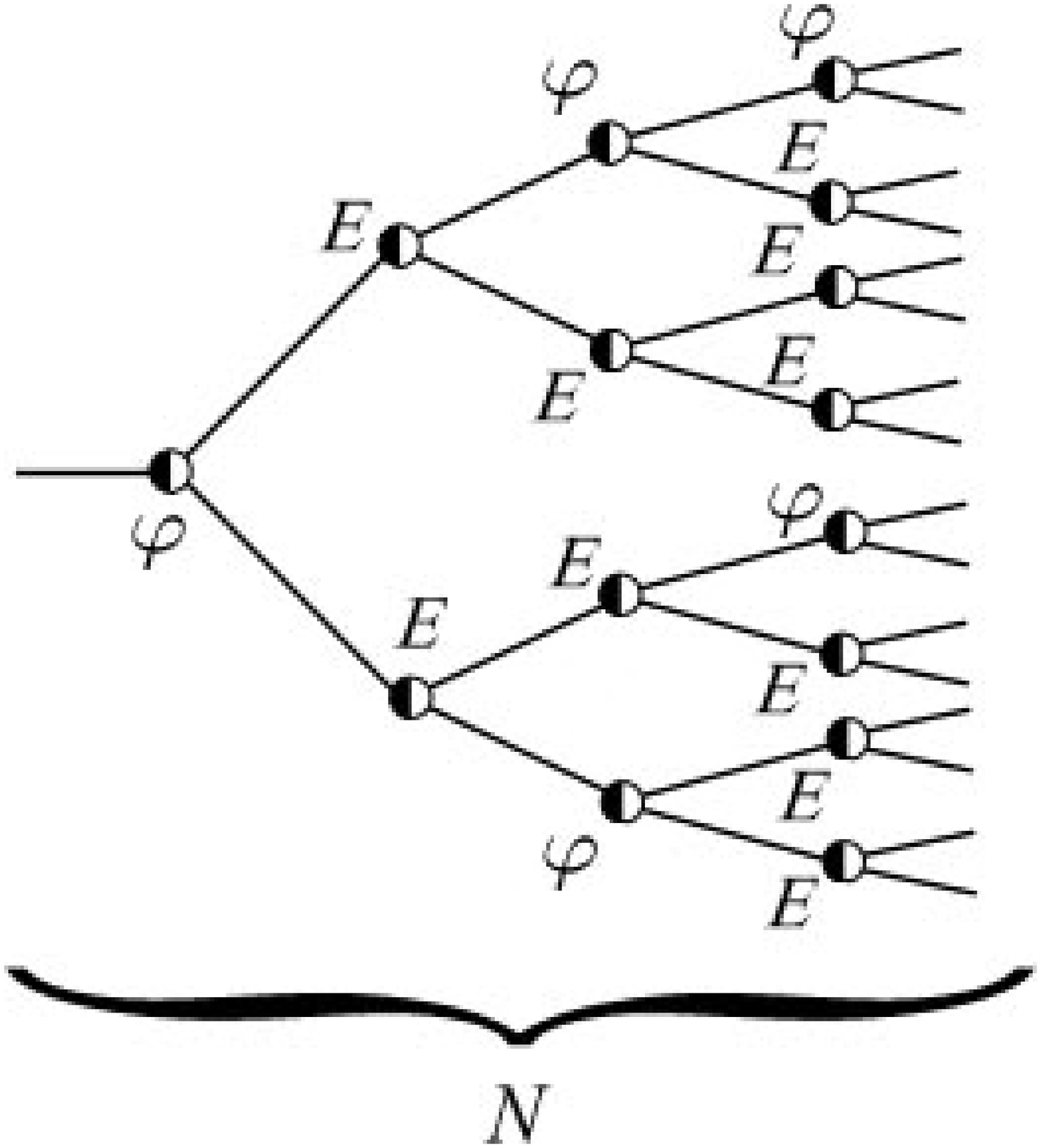} {174,192} {One of the contactions appearing in the
exponential limit procedure \ref{expomapdef} for
$\phi\in\cl{T}^2_1$, involved in computing the term of order $5$
in $\phi$ (there are $5$ $\phi$-vertices in the diagram, with the
rest being unity vertices), for iteration step $N=4$.}

$\bullet$ The role of the {\it exponentiation functor} ${\cal
E}_I^\varphi$ is to {\it stabilize} the iteration procedure.
Usually the subsequent maps of a point form an orbit, which, even
if periodic, walks in a big domain of the space. Moreover,
behaviour of different orbits is rather different. The orbits of
exponentiated maps are very different: they are all well ordered
and similar. This is achieved by actually restricting iterations
to maps in vicinity of unities: and this makes exponentiation
depending on the choice of the unity.

$\bullet$ Actually, if considered from the perspective of
exponential maps, the concept of unit map is somewhat ambiguous:
it depends on an additional structure, which can be chosen in
different ways. For the limit in (\ref{expomapdef}) to exist, $I$
should possess self-reproducing (ergodic) property: after some
number of iterations it should -- in some sense -- get close to
what it originally was. In eq.(\ref{projunitdef}) this happens
already at the first iteration step, but after projectivization.
However, in other contexts other possibilities can be used:
projectivization can be changed for another kind of
identification, stability does not require coincidence at the
first step, it may happen later and may be only approximate etc
etc. See \cite{ADM}, where the structure constants of associative
algebra are used to define a {\it unit operator} and {\it
exponents}, relevant, for example, for the study of triangulations
and string field theory.

$\bullet$ Exponential maps can be conveniently studied with the
help of differential equations. This is because for fixed $I$ and
$\varphi$ and for arbitrary $c$-numbers $t_1,t_2$ exponentiated
maps satisfy multiplicativity property \be {\cal
E}_I^{t_1\varphi}\circ {\cal E}_I^{t_2\varphi} = {\cal
E}_I^{(t_1+t_2)\varphi} \ee If we now put $\xi^i = z^i/z^1$ (so
that $\xi^1\equiv 1$), then for $\vec\xi(t) = {\cal
E}_I^{t\varphi}\Big(\vec\xi(0)\Big)$, \be \dot\xi^i(t) = \frac{d
\xi^i(t)}{dt} = \frac{\varphi^i\Big(\vec\xi(t)\Big) - \xi^i(t)
\varphi^1\Big(\vec\xi(t)\Big)}{\lambda\Big(\vec\xi(t)\Big)}
\label{diffeqforexp} \ee

$\bullet$ In the following subsection we give examples of {\it
direct} computation of the limit (\ref{expomapdef}), and compare
them with solutions of differential (RG-like) equations
(\ref{diffeqforexp}). We add also a preliminary comment on
relation between discrete and continuous dynamics -- a story of
its own value. Of considerable interest at the next stage would be
relation between resultant theory and exponential maps, but it is
left beyond the scope of the present text. The problem here is
that resultants are defined for non-linear maps in homogeneous
coordinates, but are {\it not} invariant under projective
rescalings and thus can not be simply reduced to the maps of
projective spaces. Improtant fact is that for $s>1$ the resultant
of any unity is zero, since any non-trivial root of degree-$(s-1)$
polynomial $\lambda(\vec z) = 0$ from  $I^i(\vec z) = \lambda(\vec
z) z^i$ solves $I^i(\vec z) = 0$: therefore by definition ${\cal
R}\{I\} = 0$. We will be also interested in non-linear analogues
of relation $\det (I - \varphi) = \exp \left\{-\sum_{k=1}^\infty
\frac{1}{k}{\rm tr}\varphi^k\right\}$.

\subsection{Examples of exponential maps}

\subsubsection{Exponential maps for $n|s=2|2$}

If maps are represented by $n\times M_{n|s} = 2\times 3$ matrices,
then unities with $\lambda(\vec z) = \vec\lambda\vec z =
\lambda_1z_1 + \lambda_2 z_2$ have the form \be I_{\vec\lambda} =
\lambda_1 \left(\begin{array}{ccc}1&0&0\\0&1&0\end{array}\right) +
\lambda_2 \left(\begin{array}{ccc}0&1&0\\0&0&1\end{array}\right) =
\left(\begin{array}{ccc}\lambda_1&\lambda_2&0\\
0&\lambda_1&\lambda_2\end{array}\right) \ee They are degenerate,
the system $\vec I_{\vec\lambda}(\vec z) = \lambda(\vec z)\vec z$
always has a non-vanishing solution: with $\lambda(\vec z) = 0$,
i.e. $(z_1,z_2) = (-\lambda_2,\lambda_1)$. Therefore the resultant
${\cal R}_{2|2}(I_{\vec\lambda}) =
\left|\left|\begin{array}{cccc} \lambda_1&\lambda_2&0&0\\
0& \lambda_1&\lambda_2&0\\ 0& \lambda_1&\lambda_2&0 \\
0&0& \lambda_1&\lambda_2\end{array}\right|\right| = 0$. Iterations
of $I_{\vec\lambda}$ are also unities, $I_{\vec\lambda}^{\circ
m}(\vec z) = \lambda(\vec z)^{2^m-1}\vec z$, but of degree $s^m =
2^m$, and their resultants also vanish.

Exponentiation functor ${\cal E}_{I_{\vec\lambda}}$ first puts any
map $I_{\vec\lambda} + \varphi$ into vicinity of $I_{\vec\lambda}$
and then converts it into a map ${\cal
E}_{I_{\vec\lambda}}^\varphi$ of projective spaces $P^{n-1}
\rightarrow P^{n-1}$. (Of interest is also representation of
${\cal E}_{I_{\vec\lambda}}^\varphi$ in homogeneous spaces, but
even the adequate choice of the space depends on $\varphi$.)
Convenient way to find ${\cal E}_{I_{\vec\lambda}}^\varphi$ is
through solving the differential equation (\ref{diffeqforexp}): id
$\xi = z_2/z_1$, then $\xi(t) = {\cal
E}_{I_{\vec\lambda}}^{t\varphi}(\xi)$ satisfies $\dot\xi(t) =
\frac{\varphi^2(1,\xi) - \xi\varphi^1(1,\xi)}{\lambda(1,\xi)}$
with boundary condition $\xi(t=0) = \xi$. Then $\xi(t=1) = {\cal
E}_{I_{\vec\lambda}}^\varphi(\xi)$.

\bigskip

{\bf Examples}

\bigskip

Some simple exponents for the unity $I =
\left(\begin{array}{ccc}1&0&0\\0&1&0\end{array}\right)$ are
collected in the table. In this case differential equations look
simpler if written for the variable $\xi = z_2/z_1$ because for
this unity $\lambda(\vec z) = z^1$ and $\lambda(1,\xi) = 1$. For
$\tilde I =
\left(\begin{array}{ccc}0&1&0\\0&0&1\end{array}\right)$ with
$\tilde\lambda(\vec z) = z_2$ they would have exactly the same
form in terms of
$\tilde\xi = 1/\xi = z_1/z_2$, when 
$\tilde\lambda(\tilde\xi,1)=1$. In the table parameter $\alpha$ is
absorbed into $t$: exponentiated map is obtained from the third
column at $t=\alpha$.

\bigskip

\begin{tabular}{|c|c|c|c|}
\hline &&&\\
$I + \varphi$ & $\dot\xi = \varphi_2(1,\xi) - \xi\varphi_1(1,\xi)$
& $\xi(t)$ &
$\left(\begin{array}{c} x \\ y \end{array}\right)$ \\
&&&\\
\hline
&&&\\
$\left(\begin{array}{ccc} 1+\alpha & 0 & 0 \\ 0 & 1 &
0\end{array}\right)$ & $\dot \xi = -\xi$ & $\xi \rightarrow \xi
e^{-t}$ &
$\left(\begin{array}{c} xe^t \\ y \end{array}\right)$  \\
&&&\\
$\left(\begin{array}{ccc} 1 & \alpha & 0 \\
0 & 1 & 0\end{array}\right)$ & $\dot \xi = -\xi^2$ & $\xi
\rightarrow \frac{\xi}{1+t\xi}$ &
$\left(\begin{array}{c} x + ty \\ y \end{array}\right)$\\
&&&\\
$\left(\begin{array}{ccc} 1 & 0 & \alpha \\
0 & 1 & 0\end{array}\right)$ & $\dot \xi = -\xi^3$ & $\xi
\rightarrow \frac{\xi}{\sqrt{1 + 2t\xi^2}}$ &
$\left(\begin{array}{c} \sqrt{x^2 + 2ty^2} \\ y
\end{array}\right)$ \\
&&&\\
$\left(\begin{array}{ccc} 1 & 0 & 0 \\
\alpha & 1 & 0\end{array}\right)$ & $\dot \xi = 1$ & $\xi
\rightarrow \xi+t$ &
$\left(\begin{array}{c} x \\ tx+y \end{array}\right)$\\
&&&\\
$\left(\begin{array}{ccc} 1 & 0 & 0 \\
0 & 1+\alpha & 0\end{array}\right)$ & $\dot \xi = \xi$ & $\xi
\rightarrow \xi e^{t}$ &
$\left(\begin{array}{c} x \\ ye^t \end{array}\right)$\\
&&&\\
$\left(\begin{array}{ccc} 1 & 0 & 0 \\
0 & 1 & \alpha \end{array}\right)$ & $\dot \xi = \xi^2$ & $\xi
\rightarrow \frac{\xi}{1-t\xi}$&
$\left(\begin{array}{c} y \\ x-ty \end{array}\right)$\\
&&&\\
\hline
\end{tabular}

\bigskip

Five of these maps have obvious {\it polynomial} representations
in homogeneous vector space of dimension two (where the original
map $A$ is defined), however in one case such representations does
not exist (it either contains square root or can be represented
polynomially in a linear spaces of higher dimension: of $x^2$,
$xy$ and $y^2$). In any case, homogeneous representations are not
unique: most important is that there is no obvious way to obtain
them directly from original $I + \varphi$, omitting the
intermediate projection onto projective space $C^2 \rightarrow
P^1$.

Even more interesting is exponent of generic map: it is a typical
example of multi-time $\tau$-functions. For generic map
$$
\dot \xi = -\varphi^1_{22}\xi^3 +
(\varphi^2_{22}-\varphi^1_{12})\xi^2 +
(\varphi^2_{12}-\varphi^1_{11})\xi + \varphi^2_{11} =
-c\prod_{i=1}^3(\xi-\mu_i)
$$
we get: \be \left(\frac{\xi(t)-\mu_1}{\xi-\mu_1}\right)^{\mu_{23}}
\left(\frac{\xi(t)-\mu_2}{\xi-\mu_2}\right)^{\mu_{31}}
\left(\frac{\xi(t)-\mu_3}{\xi-\mu_3}\right)^{\mu_{12}} =
e^{ct\Delta(\mu)} \ee with $\Delta(\mu) =
\mu_{12}\mu_{23}\mu_{31}$, or \be
\left(\frac{\xi(t)-\mu_1}{\xi(t)-\mu_3}\right)^{\mu_{23}}
\left(\frac{\xi(t)-\mu_2}{\xi(t)-\mu_3}\right)^{\mu_{31}} =
e^{ct\Delta(\mu)}
\left(\frac{\xi-\mu_1}{\xi-\mu_3}\right)^{\mu_{23}}
\left(\frac{\xi-\mu_2}{\xi-\mu_3}\right)^{\mu_{31}} \ee

\subsubsection{Examples of exponential maps for $2|s$}

These formulas are straightforwardly generalized to arbitrary
$2|s$: for $\dot\xi = -c\prod_{i=1}^{s+1}(\xi - \mu_i)$ the answer
is \be \prod_{i=1}^{s+1}
\left(\frac{\xi(t)-\mu_i}{\xi-\mu_i}\right)^{1/\Delta_i} = e^{-ct}
\ee with $\Delta_i = \prod_{j\neq i}^{s+1}(\mu_i-\mu_j)$.

For example, for $s=3$ and $\left(\begin{array}{c} x \\ y
\end{array}\right) \longrightarrow \left(\begin{array}{c} x^3 \\
x^2y + \alpha y^3 \end{array}\right)$
or $I + \varphi  = \left(\begin{array}{cccc} 1 & 0 & 0 & 0 \\
0 & 1 & 0 & \alpha \end{array}\right)$ we get for $\xi = y/x$:
$\dot\xi = \xi^3$ and $\xi(t) = \frac{\xi}{\sqrt{1-2t\xi^2}}$.

\subsubsection{Examples of exponential maps for $n|s=3|2$}

In this case the matrices are $n\times M_{n|s} = 3\times 6$. We
put $(z_1,z_2,z_3) = (x,y,z)$ and $\xi = y/x$, $\eta = z/x$. Some
simple exponents for the unity $I = \left(\begin{array}{cccccc}
x^2 & y^2 & z^2 & xy & xz & yz \\
\hline 1&0&0&0&0&0\\ 0&0&0&1&0&0\\ 0&0&0&0&1&0\end{array}\right)$
are collected in the table:

\bigskip
\hspace{-1.0cm}
\begin{tabular}{|c|c|c|c|}
\hline &&&\\
$I + \varphi$ & $\begin{array}{c}
\dot\xi = \varphi_2(1,\xi,\eta) - \xi\varphi_1(1,\xi,\eta)\\
\dot\eta = \varphi_3(1,\xi,\eta) - \eta\varphi_1(1,\xi,\eta)
\end{array}$
& $\left(\begin{array}{c} \xi(t) \\ \eta(t) \end{array}\right)$&
$\left(\begin{array}{c} x \\ y \\ z \end{array}\right)$
\\
&&&\\
\hline
&&&\\
$\left(\begin{array}{cccccc} 1&0&0&0&0&0\\ \alpha &0&0&1&0&0\\
0&0&0&0&1&0
\end{array}\right)$ &
$\begin{array}{c} \dot \xi = 1 \\ \dot\eta = 0 \end{array}$ & $
\left(\begin{array}{c}\xi + t \\ \eta \end{array}\right) $ &
$\left(\begin{array}{c} x \\ tx+y \\ z \end{array}\right)$ \\
&&&\\
$\left(\begin{array}{cccccc} 1&0&0&0&0&0\\ 0&\alpha &0&1&0&0\\
0&0&0&0&1&0
\end{array}\right)$ &
$\begin{array}{c}\dot\xi =\xi^2\\ \dot\eta =0\end{array}$&
$\left(\begin{array}{c}\frac{\xi}{1-t\xi}\\
\eta\end{array}\right)$&
$\left(\begin{array}{c} x\\ \frac{xy}{x-ty}\\ z\end{array}\right)$\\
&&&\\
$\left(\begin{array}{cccccc} 1&0&0&0&0&0\\ 0&0&\alpha &1&0&0\\
0&0&0&0&1&0
\end{array}\right)$ &
$\begin{array}{c}\dot\xi =\eta^2\\ \dot\eta =0\end{array}$&
$\left(\begin{array}{c}\xi + t\eta^2\\ \eta\end{array}\right)$&
$\left(\begin{array}{c} x \\ \frac{xy+tz^2}{x}\\
z \end{array}\right)$\\
&&&\\
$\left(\begin{array}{cccccc} 1&0&0&0&0&0\\ 0&0&0&1+\alpha &0&0\\
0&0&0&0&1&0
\end{array}\right)$ &
$\begin{array}{c} \dot\xi = \xi\\ \dot\eta =0\end{array}$&
$\left(\begin{array}{c}\xi e^t \\ \eta \end{array}\right) $&
$\left(\begin{array}{c} x \\ ye^t \\ z \end{array}\right)$ \\
&&&\\
$\left(\begin{array}{cccccc} 1&0&0&0&0&0\\ 0&0&0&1&\alpha&0\\
0&0&0&0&1&0
\end{array}\right)$ &
$\begin{array}{c} \dot\xi =\eta\\ \dot\eta =0\end{array}$&
$\left(\begin{array}{c}\xi + t\eta \\ \eta \end{array}\right) $&
$\left(\begin{array}{c} x \\ y+tz \\ z \end{array}\right)$ \\
&&&\\
$\left(\begin{array}{cccccc} 1&0&0&0&0&0\\ 0&0&0&1&0&\alpha \\
0&0&0&0&1&0
\end{array}\right)$ &
$\begin{array}{c}\dot\xi=\xi\eta\\ \dot\eta =0\end{array}$&
$\left(\begin{array}{c}\xi e^{\eta t} \\ \eta\end{array}\right)$&
$\left(\begin{array}{c} x \\ ye^{tz/x} \\ z \end{array}\right)$\\
&&&\\
\hline
&&&\\
$\left(\begin{array}{cccccc} 1+\alpha &0&0&0&0&0\\ 0&0&0&1&0&0 \\
0&0&0&0&1&0
\end{array}\right)$ &
$\begin{array}{c} \dot \xi = -\xi \\ \dot\eta = -\eta
\end{array}$& $\left(\begin{array}{c}\xi e^{-t} \\ \eta e^{-t}
\end{array}\right)$&
$\left(\begin{array}{c} x e^t \\ y \\ z \end{array}\right)$ \\
&$\xi/\eta = {\rm const}$&&\\
$\left(\begin{array}{cccccc} 1&0&0&\alpha &0&0\\ 0&0&0&1&0&0 \\
0&0&0&0&1&0
\end{array}\right)$ &
$\begin{array}{c} \dot \xi = -\xi^2 \\ \dot\eta = -\xi\eta
\end{array}$& $\left(\begin{array}{c}\frac{\xi}{1 + t\xi} \\
\frac{\eta}{1+t\xi}
\end{array}\right) $ &
$\left(\begin{array}{c} x+ty \\ y \\ z \end{array}\right)$ \\
&&&\\
\hline
&&&\\
$\left(\begin{array}{cccccc} 1&0&0&0&0&0\\ 0&0&0&1& \alpha &0 \\
0&0&0&-\beta &1&0
\end{array}\right)$ &
$\begin{array}{c}\dot\xi = \alpha\eta\\ \dot\eta
=-\beta\xi\end{array}$& $\left(\begin{array}{c}
\xi\cos \omega t 
+ \eta\sqrt{\frac{\alpha}{\beta}} \sin\omega t \\
\eta \cos\omega t - \xi \sqrt{\frac{\beta}{\alpha}} \sin \omega t
\end{array}\right) $ &
$\left(\begin{array}{c} x \\
y\cos \omega t + z\sqrt{\alpha/\beta}\sin\omega t
\\ z\cos \omega t - y\sqrt{\beta/\alpha}\sin\omega t \end{array}\right)$ \\
&&&\\
\hline
\end{tabular}
\hspace{1.0cm}

In the last, two-parametric {\it rotational} example, $\alpha$ is
not absorbed into $t$: one could rather absorb $\omega =
\sqrt{\alpha\beta}$.

\setcounter{equation}{0}

\section{Potential applications}

In this section we very briefly comment on several most
straightforward applications of non-linear algebra. Each of these
subject deserves a separate big chapter, and may be not a single
one, but we do not go into details. Also, potential applications
are by no means exhausted by the subjects below: we chosed {\it
these} just because they are the most obvious and direct. One can
easily extend the list.

\subsection{Solving equations}

\subsubsection{Craemer rule}

Once one learned to solve any equation of a single variable --
whatever it means, say, passed the undergraduate course of
algebra,-- it is natural to switch to solving {\it systems} of
equations of {\it many} variables. In another undergraduate course
-- of linear algebra -- one do exactly this, but only for {\it
linear} equations, and, as we tried to explain in this paper, this
restriction has absolutely no -- except for, probably, pedagogical
-- reason to be imposed. Given a system of equations of any number
of variables, one would naturally start solving them iteratively:
first solve the last equation for the last variable, express it
through the other variables, substitute this expression into the
previous equations, ans so on. Actually, in practice many people,
when they need to solve a system of {\it linear} equations, do
exactly this instead of using the Craemer rule in any of its four
versions (\ref{liCraI}) - (\ref{liCraIV}).

Of course, one can try to do the same for non-linear equations.
The problem is, however, that -- even if one has a {\it some}
favorite way to solve equation of a single variable, for example,
numerically, -- it hardly can be used for this purpose. Numerical
methods are ineffective in solving equations with free parameters
in the coefficients (the role of these parameters is played by
remaining variables). Analytical expressions rarely exist. Worse
than that: even if analytical expression, say, in radicals,
occasionally exists, its substitution into the next equation will
give one with coefficients, which include radicals (depending on
the remaining variables!), so that we get away from original
context of, say, polynomial equations. Actually, the route fastly
brings one into a under-developed theory of Galua extentions with
free parameters: a full-scale string-theory-kind of a problem,
very interesting, but yet far from being ready-for-use in
down-to-earth applications.

This makes the existence of Craemer rule (\ref{CraIII}) in
non-linear setting much more important, not only conceptually but
also practically(!), than it was in linear case. It reduces the
problem of solving a {\it system} of non-linear equations -- an
{\it a priori} non-algortithmic problem, as we just exaplained --
to that of finding roots of functions of single variables. Of
course, there is a price to pay:

(i) one should know explicit formulas for resultants (this is
somewhat tedious, but absolutely constructive procedure: see, for
example, s.\ref{Koshul} for explicit construction) and

(ii) in this single variable the equation has relatively high
degree in the single unknown variable: up to order $c_{n|s} =
\frac{s^n-1}{s-1}$, i.e. somewhat less than degree $d_{n|s} =
ns^{n-1}$ of the resultant in the {\it coefficients} of original
system (original tensor), but still pretty big (moreover, it grows
exponentially with the number of variables $n$).

\subsubsection{Number of solutions}

Above mentioned iterative procedure -- though {\it inpractical}
and substituted in applications by a far more effective Craemer
rule (\ref{CraIII}) -- is {\it conceptually} adequate and can be
also used in theoretical considerations. For example, it allows to
enumerate solutions. Imagine that we solved the non-homogeneous
system and -- in the spirit of iterative procedure -- assume that
the first equation is used to define the first variable, the
second equation -- the second variable and so on. Then, if the
$i$-th equations has degree $s_i$, we see that if all other
variables are fixed, the $i$-th one satisfies a polynomial
equation of degree $s_i$ and thus can take $s_i$ different values.
This gives for the number of different solutions to
non-homogeneous system\footnote {Of course, among these solutions
can occasionally be zero-vectors: like in the case of
non-homogeneous system
$\left\{\begin{array}{c}ax^s=x\\by^2=y\end{array}\right.$ this
zero is occasional, because a minor deformation, say, to $ax^2 = x
+ \delta$, moves the root away from zero. Also some roots can
occasionally be projectively equivalent: like it happens in
eigenvalue problems, considered in s.\ref{eige}, where after
subtraction of zero-vector and projectively equivalent solutions
the total number $s^n$ reduced to $c_{n|s} = \frac{s^n-1}{s-1}$. }
\be \#_{s_1,\ldots,s_n} = \prod_{i=1}^n s_i, \ \ \ \ {\rm in\
particular} \ \ \ \ \#_{n|s} = s^n \ee

For a homogeneous system the situation is a little different.
Roughly speaking, when after $n-1$ iterations we come down to the
first equation for a single remaining variable (and all other
variable expressed through it with the help of all other
equations), this first equation is either automatically satisfied
or not. If resultant is non-vanishing, i.e. the homogeneous system
is inconsistent, the answer is "no" (we do not count identically
vanishing solution which always exists for the homogeneous
system). However, if the resultant {\it is} vanishing, the answer
can still be "no": it is guaranteed only that for {\it some}
expression of the first variable through the other ones the first
equation is resolvable, but not for {\it all}. Only {\it one} of
the $\prod_{i\neq 1}^n s_i$ branches is guaranteed to provide a
root of the first equation -- and only {\it one} of its $s_1$
roots. This means, that even if its resultant vanishes,
homogeneous system can have a single solution (of course, since
the system is {\it homogeneous}m it will be a single point in
projective space $P^{n-1}$, in original space $C^n$ of homogeneous
variables it is entire one-dimensional sub-space). In special
cases the number of solutions can occasionally be higher: the
number lies between $1$ and $\#_{s_1,\ldots,s_n}$.

\bigskip

{\bf Example:}

\bigskip

Take for the homogeneous ${n|s}={2|2}$ system \be \left\{
\begin{array}{c} a^2x^2-y^2 = 0 \\ cxy-y^2 = 0
\end{array}\right.
\ee The first equation gives: $y = \pm ax$. The second one gives:
$y=0$ and $y = cx$. For $c\neq \pm a$ there are no non-vanishig
solutions: this is in accordance with the fact that the resultant
of this system is ${\footnotesize\left|\left|\begin{array}{cccc}
a&0&-1&0\\0&a&0&-1\\0&c&-1&0\\0&0&c&-1\end{array}\right|\right|}
=a^2-c^2$. If $c=a$, there is a single (one-parametric) solution
$y=ax$, if $c=-a$, there is another single solution $y=-ax$.

Let us now make the system non-homogeneous: \be \left\{
\begin{array}{c} a^2x^2-y^2 = p \\ cxy-y^2 = q
\end{array}\right.
\ee The system describes two hyperbolas, see Fig.\ref{hyperbo}.A,
which -- without any relation between $c$ and $a$ -- intersect at
$s^n=4$ points (probably, complex and even infinite: this is even
more obvious for two ellipses, Fig.\ref{hyperbo}.B). In the
homogeneous limit of $p=q=0$ hyperbolas turn into straight lines,
Fig.\ref{hyperbo}.C, which either intersect at zero only, if
$c^2\neq a^2$, or coincide, if $c=\pm a$ (but coincidence occurs
only for one pair of the lines).

\Fig{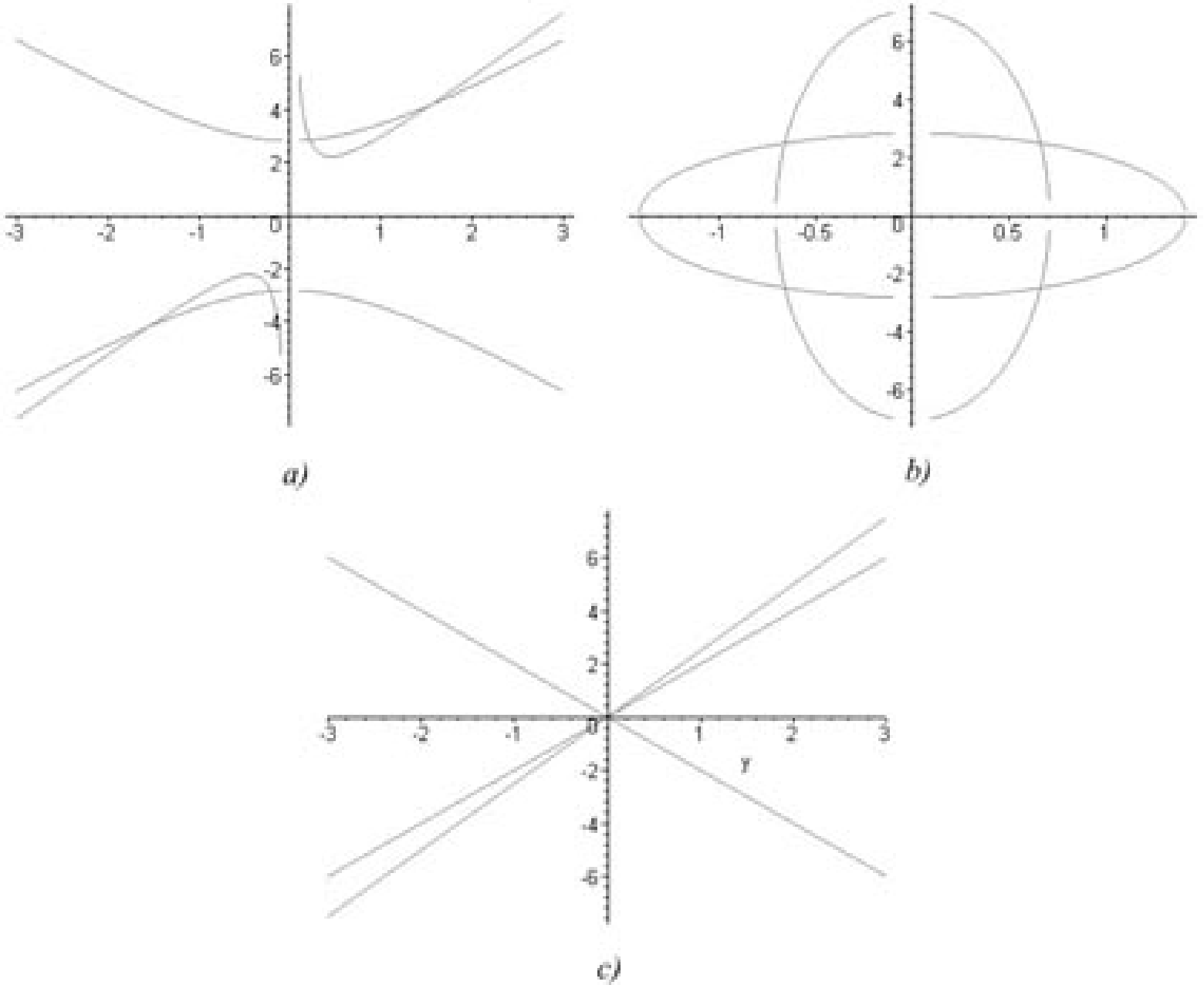} {402,329} {a) hyperbolic case: $a=1/2, p=2, c=2/5,
q=1/5$, \ \ b) elliptic case (
$x^2/2+y^2=2,\ x^2/10+y^2=1/2$), \ \ c) degenerate case $p=q=0$
(the fourth straight line, $y=0$, coincides with coordinate
$x$-axis)}

One can similarly analyze the system \be
\left\{ \begin{array}{c} (y-\xi_1 x)(y-\xi_2 x) = 0 \\
(y-\xi'_1x)(y-\xi'_2 x) = 0
\end{array}\right.
\ee with the resultant
$(\xi_1-\xi'_1)(\xi_1-\xi'_2)(\xi_2-\xi'_1)(\xi_2-\xi'_2)=0$.

\subsubsection{Index of projective map
\label{iprom}}

As a map of projective spaces $P^{n-1} \longrightarrow P^{n-1}$,
$\vec A(\vec z)$ is characterized by the {\it index} $i(A)$, i.e.
the number of stable pre-images of any point $\vec a$ in the
target space $P^{n-1}$. These pre-images are solutions of the
system \be A^i(\vec z) = t\mu(\vec z) a^i, \ee where $\mu(\vec z)$
can be arbitrary function of $\vec z$ of degree $s = {\rm deg}_z
A(\vec z)$. Whenever the image of the homogeneous map $\vec A(\vec
z)$ crosses the given (complex) direction $\vec a$, there exists a
non-vanishing solution $t\neq 0$ of \be r_{n|s}(t) = {\cal
R}_{n|s}\Big\{A^i(\vec z) - t\mu(\vec z)a^i \Big\} = 0
\label{criindex} \ee The vanishing roots $t=0$ exist only if the
free term of $r(t)$ polynomial is zero, and this term is nothing
but the resultant ${\cal R}_{n|s}\{A\}$. Thus at discriminantal
points in moduli space, where the map $A$ degenerates and
$r_{n|s}(t=0) = {\cal R}_{n|s}\{A\}$ vanishes, the index drops by
one (and further drops down at the sub-varieties of higher
codimension).

\bigskip

{\bf Examples}

\bigskip

{\bf Linear maps ($s=1$):}

\bigskip

In this case $\mu(\vec z) = \mu_jz_j$, and \be r_{n|1}(t) =
\det_{ij} \Big(A_{ij} - t a_i\mu_j\Big) = t\Big(\mu_j\check
A_{ji}a_i\Big) + \det A \ee is linear in $t$ and for
non-degenerate $A$, \ $\det A \neq 0$, the index $i(A) = i_{n|1} =
1$, i.e. the map is one-to-one (note that the combination
$\mu_j\check A_{ji}a_i \neq 0$ with $\vec a \neq 0$ and $\vec\mu
\neq 0$ if $\det\check A = \det A^{n-1}\neq 0$). If the map is
degenerate, $\det A = 0$, then index $i(A)=0$, i.e. generic point
in $P^{n-1}$ target space has {\it no} pre-image. Pre-images exist
for a sub-variety of directions which actually has codimension \
${\rm corank}(A)$\ in $P^{n-1}$. However, for ${\rm corank}(A)>1$,
the minor $\check A_{ji}$ and thus $r_{n|1}(t)$ vanish
identically, so that more sensitive characteristics of the map $A$
should be used to investigate the situation.

This example shows that when the index drops from $i(A)=1$ to
$i(A)=0$ this results in the decrease of {\it dimension} of the
image (see Fig.\ref{mapinddia}), in other cases degeneracy of the
map only decreases the number of times the target space is covered
by the image.

\Fig{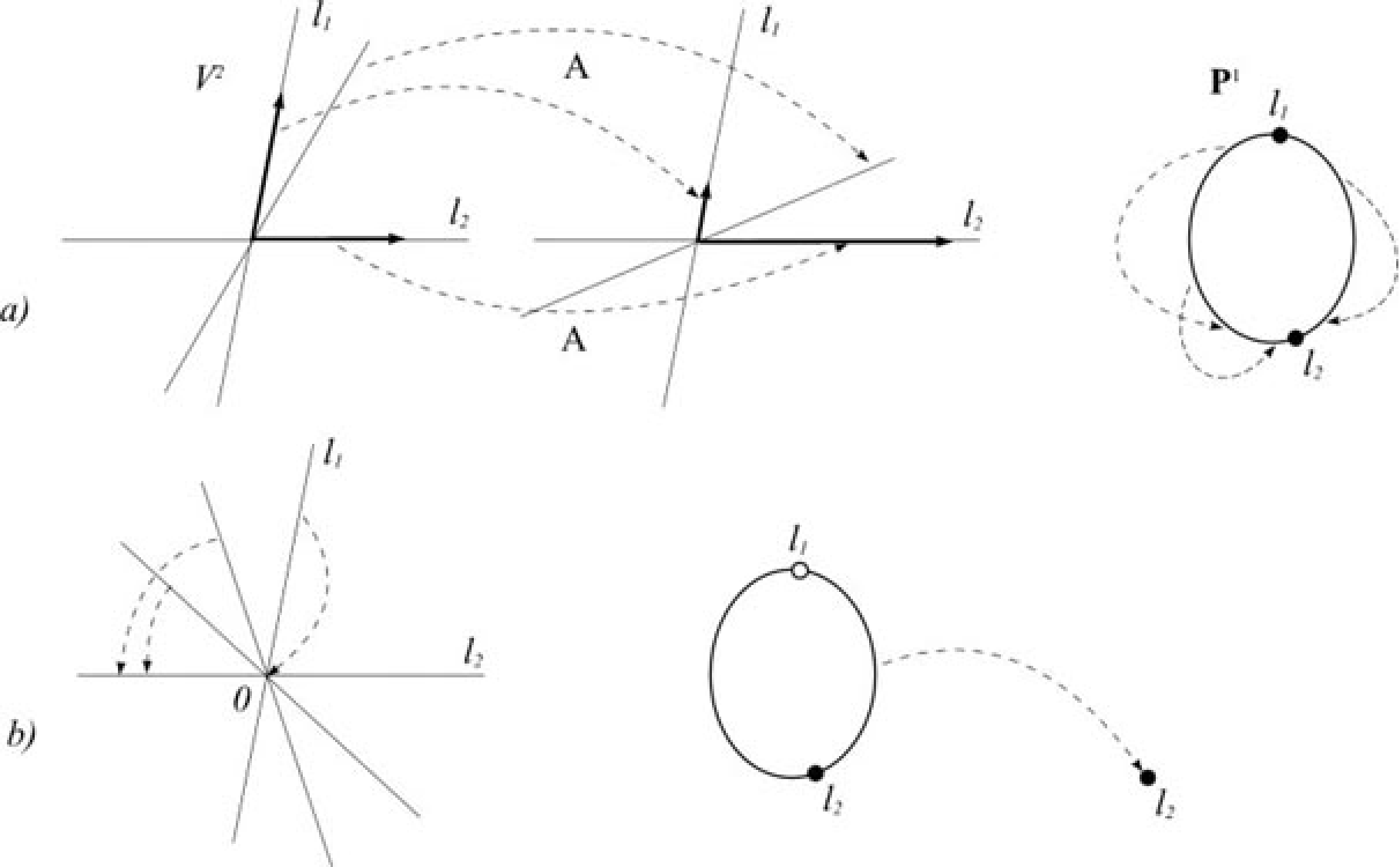} {403,250} {a) a non-degenerate operator on $V^2$
defines a map of $\bb{P}^1=\bb{P}V^2$ into itself with stationary
points corresponding to eigenspaces. As one of eigenvalues tend to
$0$ (operator "degenerates") the images of almost all points of
$\bb{P}^1$ tend to the point corresponding to the other eigenspace
b) a degenerate operator on $V^2$ is the projection along its
kernel $l_1$, which corresponds to the map of $\bb{P}^1$
(punctured at $l_1$) to a point (the projective line "bursts" at
the kernel point), which is a covering having index $0$ almost
everywhere}

\bigskip

{\bf Maps of $P^1$ ($n=2$):}

\bigskip

In this case $r_{2|s}(t)$ is the ordinary resultant of two
polynomials of degree $s$. If, for example, we take $\mu(\vec z) =
\mu y^n$ (nothing actually depends on this choice), then $t$
appears only in the free term of both polynomials and therefore
can contribute at most $t^s$ to determinant in
(\ref{ordresdetrep}). Thus for non-degenerate maps the index
$i_{2|s}=s$.

In this example, instead of introducing $t$ and counting its
powers in the resultant, one could simply solve the equation \be
\frac{P_1(x/y)}{P_2(x/y)} = \frac{a_1}{a_2} = {\rm const} \ee This
is a polynomial equation of degree $s$ for $\xi = x/y$ with $s$
solutions, i.e. we reproduce the index $i_{2|s} = s$. Note, that
the {\it eigenvector} problem, \be \frac{P_1(x/y)}{P_2(x/y)} =
\frac{x}{y} \ee is equivalent to a polynomial equation of degree
$s+1$ with $c_{2|s} = s+1$ solutions. Similarly, generic
non-homogeneous system \be \frac{P_1(x/y)}{P_2(x/y)} =
\frac{p_1(x/y)}{p_2(x/y)} \ee with polynomials $p_{1,2}(x/y)$ of
degree $s-1$ posesses $s(s-1)$ solutions for $\xi = x/y$,
belonging to the $s$ branches of solution for the $y$-variable:
this gives a total of $s(s-1)+s = s^2 = \#_{2|s}$ solutions to
non-homogheneous system.

\subsubsection{Perturbative (iterative) solutions
\label{peso}}

A special place among solutions to non-linear equations is
occupied by ``explicit formulas" like \be x_\pm = \frac{-b \pm
\sqrt{b^2-4ac}}{2a} \label{soqueq1} \ee for the roots of
non-homogeneous quadratic equation \be ax^2 + bx + c = 0.
\label{queq} \ee One of these two roots, $x_+$, has a remarkable
expansion:
$$
x_+ = \frac{b}{2a}\left(\sqrt{1-\frac{4ac}{b^2}}-1\right)
= -\sum_{k=0}^\infty \frac{2^{k+1}(2k-1)!!}{(k+1)!}
\frac{a^kc^{k+1}}{b^{2k+1}} =
$$
\be
= -\frac{c}{b} - \frac{ac^2}{b^3} - 2\frac{a^2c^3}{b^5}
- 5\frac{a^3c^4}{b^7} - 14\frac{a^4c^5}{b^9}
- 42\frac{a^5c^6}{b^{11}} - \ldots
\label{soqueq}
\ee
Integer-valued(!) coefficients in this series are the
celebrated Euler-Catalan numbers $\alpha_k^{(2)} =
\frac{(4k-2)!!!!}{(k+1)!} = \frac{(2m)!}{m!(m+1)!}$,
appearing in numerous combinatorial problems, including
description of Gaussian phases in matrix models
\cite{amm}.

\PFig{Catalan}
{200,276}
{Eugene Catalan (1814 -- 1894)}

Of course, expansion (\ref{soqueq}) is nothing but
an iterative solution to (\ref{queq}):
$$
x = -\frac{c}{b} - \frac{a}{b}x^2 =
-\frac{c}{b} - \frac{a}{b}
\left(-\frac{c}{b} - \frac{a}{b}x^2\right)^2
= -\frac{c}{b} - \frac{ac^2}{b^3} -
2\frac{a^2c}{b^3}x^2 - \frac{a^3}{b^3}x^4 =
$$
\be
= -\frac{c}{b} - \frac{ac^2}{b^3} -
2\frac{a^2c}{b^3}\left(-\frac{c}{b} - \frac{a}{b}x^2\right)^2
- \frac{a^3}{b^3}\left(-\frac{c}{b} - \frac{a}{b}x^2\right)^4
= \ldots
\ee
and therefore possesses a straightforward representation
in terms of {\it tree} Feynman diagrams:
\newpage
$$
x_+ = \ \sum {\rm rooted\ trees} \ \ =
\ \ \ \
\begin{picture}(25,10)
\put(-3,3){\circle{6}}
\put(20,3){\vector(-1,0){20}}
\put(23,3){\circle*{6}}
\end{picture}
\ + \ \ \
\begin{picture}(38,20)
\put(-3,3){\circle{6}}
\put(20,3){\vector(-1,0){20}}
\put(35,18){\vector(-1,-1){15}}
\put(37,20){\circle*{6}}
\put(35,-12){\vector(-1,1){15}}
\put(37,-14){\circle*{6}}
\end{picture} \ \
$$
\be
+ \ \ \left( \ \
\overbrace{
\begin{picture}(45,40)
\put(-3,3){\circle{6}}
\put(20,3){\vector(-1,0){20}}
\put(35,18){\vector(-1,-1){15}}
\put(35,-12){\vector(-1,1){15}}
\put(37,-14){\circle*{6}}
\put(50,33){\vector(-1,-1){15}}
\put(52,35){\circle*{6}}
\put(50,3){\vector(-1,1){15}}
\put(52,1){\circle*{6}}
\end{picture}
\ \ + \ \
\begin{picture}(50,40)
\put(-3,3){\circle{6}}
\put(20,3){\vector(-1,0){20}}
\put(35,18){\vector(-1,-1){15}}
\put(37,20){\circle*{6}}
\put(35,-12){\vector(-1,1){15}}
%
\put(50,3){\vector(-1,-1){15}}
\put(52,5){\circle*{6}}
\put(50,-27){\vector(-1,1){15}}
\put(52,-29){\circle*{6}}
\end{picture}
}^2
\ \right) \ \ + \ \ \left(\ \
\begin{picture}(50,60)
\put(-3,3){\circle{6}}
\put(20,3){\vector(-1,0){20}}
\put(35,18){\vector(-1,-1){15}}
\put(35,-12){\vector(-1,1){15}}
%
\put(50,33){\vector(-1,-1){15}}
\put(52,35){\circle*{6}}
\put(50,13){\vector(-3,1){15}}
\put(52,12){\circle*{6}}
\put(50,-7){\vector(-3,-1){15}}
\put(52,-6){\circle*{6}}
\put(50,-27){\vector(-1,1){15}}
\put(52,-29){\circle*{6}}
\end{picture}
\ \ \right.
\label{diaqueq}
\ee
\be
&+ \left. \
\overbrace{
\begin{picture}(60,55)
\put(2,3){\circle{6}}
\put(20,3){\vector(-1,0){15}}
\put(35,18){\vector(-1,-1){15}}
\put(35,-12){\vector(-1,1){15}}
\put(37,-14){\circle*{6}}
\put(50,33){\vector(-1,-1){15}}
\put(50,3){\vector(-1,1){15}}
\put(52,1){\circle*{6}}
\put(65,48){\vector(-1,-1){15}}
\put(67,50){\circle*{6}}
\put(65,18){\vector(-1,1){15}}
\put(67,16){\circle*{6}}
\end{picture}
+ \
\begin{picture}(60,50)
\put(2,3){\circle{6}}
\put(20,3){\vector(-1,0){15}}
\put(35,18){\vector(-1,-1){15}}
\put(35,-12){\vector(-1,1){15}}
\put(37,-14){\circle*{6}}
\put(50,33){\vector(-1,-1){15}}
\put(52,35){\circle*{6}}
\put(50,3){\vector(-1,1){15}}
%
\put(65,18){\vector(-1,-1){15}}
\put(67,20){\circle*{6}}
\put(65,-12){\vector(-1,1){15}}
\put(67,-14){\circle*{6}}
\end{picture}
+ \
\begin{picture}(60,40)
\put(2,3){\circle{6}}
\put(20,3){\vector(-1,0){15}}
\put(35,18){\vector(-1,-1){15}}
\put(37,20){\circle*{6}}
\put(35,-12){\vector(-1,1){15}}
%
\put(50,3){\vector(-1,-1){15}}
\put(50,-27){\vector(-1,1){15}}
\put(52,-29){\circle*{6}}
\put(65,18){\vector(-1,-1){15}}
\put(67,20){\circle*{6}}
\put(65,-12){\vector(-1,1){15}}
\put(67,-14){\circle*{6}}
\end{picture}
+ \
\begin{picture}(65,40)
\put(2,3){\circle{6}}
\put(20,3){\vector(-1,0){15}}
\put(35,18){\vector(-1,-1){15}}
\put(37,20){\circle*{6}}
\put(35,-12){\vector(-1,1){15}}
%
\put(50,3){\vector(-1,-1){15}}
\put(52,5){\circle*{6}}
\put(50,-27){\vector(-1,1){15}}
%
\put(65,-12){\vector(-1,-1){15}}
\put(67,-10){\circle*{6}}
\put(65,-42){\vector(-1,1){15}}
\put(67,-44){\circle*{6}}
\end{picture}
}^4 \ \right) \ \nn\\
 &+ \ \ \ldots
\nn\ee
The white circle at the root of the tree represents
the root $x_+$ which we are looking for,
black circles at the ends of the branches are valence-one
vertices with the charges $c$, triple vertices carry charges $a$,
and propagators are inverse $-b$'s.
It is now clear why Catalan numbers $\alpha_k^{(2)}$
in (\ref{soqueq}) are integers:
they simply count the number of different diagrams
of the given order in (\ref{diaqueq}).

In this formula we no longer need to consider $a,b,c$ as
numbers: they can be appropriate tensors, of ranks
$3$, $2$ and $1$ respectively, and (\ref{diaqueq})
actually provides a solution to the {\it system} of
$n$ non-homogeneous quadratic equations,
\be
\sum_{j,k=1}^n A_i^{jk}x_jx_k + \sum_{j=1}^n B_i^j x_j + C_i = 0,
\ \ \ i = 1,\ldots, n,
\label{sysqueq}
\ee
namely,
\be
x_i = - (B^{-1})_i^jC_j - (B^{-1})_i^jA_j^{kl}(B^{-1})_k^mC_m
(B^{-1})_l^nC_n + \ldots
\label{sosysqueq}
\ee
(sums over repeated indices are implied).
One can consider this series as a definition of tensorial
square root, in the spirit of (\ref{soqueq1}).
Note that several different tensorial structures can arise
in the given order of perturbation theory: this happens
starting from the forth term in (\ref{soqueq}), the coefficient
$5$ actually splits into two different contributions
and should be rather written as $5=4+1$,
since four over-braced diagrams are equal but do not
coincide with the fifth one.

As usual for perturbative methods in quantum field theory,
in this approach quadratic {\it matrices} -- tensors of rank $2$,
like $B$ -- play a special role: they can be
easily inverted with the help of {\it determinant} theory.
Also typical for perturbative method is restriction to
{\it one out of many} possible solutions: above formulas
nicely describe only $x_+$, but fully ignore the other,
non-perturbative, solution $x_-$.
The situation is even worse
for the {\it system} (\ref{sysqueq}): only one out of $s^n=2^n$
solutions is given by (\ref{sosysqueq}).
The reason for these restrictions is that perturbative solution
describes the only root, which remains non-singular when
$a \rightarrow 0$, or, generically, when the resultant
${\cal R}_{n|2}(A)$ of the rank-$3$ tensor $A$
in (\ref{sysqueq}) vanishes.
Other -- non-perturbative -- solutions contain fractional
powers of ${\cal R}_{n|2}(A)$ in denominators.
Adequate description of {\it all} solutions is one of the
purposes of the future development of non-linear algebra.

\PFig{cft}
{320,146}
{}


It goes without saying that quadratic equations are in no way
distinguished.
For example, Cardano formula
\be
x =
\left(-\frac{d}{2}+\sqrt{\frac{d^2}{4}+\frac{c^3}{27}}\right)^{1/3}+
\left(-\frac{d}{2}-\sqrt{\frac{d^2}{4}+\frac{c^3}{27}}\right)^{1/3}
\label{socueq1}
\ee
for solution to cubic non-homogeneous equation,
\be
x^3+cx+d = 0
\label{cueq}
\ee
can be treated in the same way and (one of!) the roots
is given by an obvious diagram technique, which simultaneously
describes solution to the {\it system}
\be
A_i^{jkl}x_jx_kx_l + B_i^{jk}x_jx_k + C_i^{j}x_j + D_i = 0.
\ee
This time it is quadratic matrix $C$
that will appear in denominator.
Again, from this perspective
it is not a surprise that the coefficients of appropriate
expansion of the seemingly complicated (\ref{socueq1})
are actually integers(!):
\be
x = \sum_{k=0}^\infty (-)^{k+1}\alpha_k^{(3)}
\frac{d^{2k+1}}{c^{3k+1}}
= -\frac{d}{c} + \frac{d^3}{c^4} - 3\frac{d^5}{c^7} +
12\frac{d^7}{c^{10}} - 55\frac{d^9}{c^{13}}
+ \ldots\nn
\ee
\be
&=&\ \
\begin{picture}(29,20)
\put(-3,3){\circle{6}}
\put(20,3){\vector(-1,0){20}}
\put(23,3){\circle*{6}}
\end{picture}
+
\begin{picture}(33,20)
\put(2,3){\circle{6}}
\put(20,3){\vector(-1,0){15}}
\put(35,18){\vector(-1,-1){15}}
\put(37,20){\circle*{6}}
\put(35,3){\vector(-1,0){15}}
\put(37,3){\circle*{6}}
\put(35,-12){\vector(-1,1){15}}
\put(37,-14){\circle*{6}}
\end{picture}
+ 3\cdot
\begin{picture}(50,20)
\put(2,3){\circle{6}}
\put(20,3){\vector(-1,0){15}}
\put(35,18){\vector(-1,-1){15}}
\put(37,20){\circle*{6}}
\put(35,3){\vector(-1,0){15}}
\put(35,-12){\vector(-1,1){15}}
\put(37,-14){\circle*{6}}
\put(50,18){\vector(-1,-1){15}}
\put(52,20){\circle*{6}}
\put(50,3){\vector(-1,0){15}}
\put(52,3){\circle*{6}}
\put(50,-12){\vector(-1,1){15}}
\put(52,-14){\circle*{6}}
\end{picture}
\nn\ee

\be
&+&\left( 9\cdot
\begin{picture}(65,20)
\put(2,3){\circle{6}}
\put(20,3){\vector(-1,0){15}}
\put(35,18){\vector(-1,-1){15}}
\put(37,20){\circle*{6}}
\put(35,3){\vector(-1,0){15}}
\put(35,-12){\vector(-1,1){15}}
\put(37,-14){\circle*{6}}
\put(50,18){\vector(-1,-1){15}}
\put(52,20){\circle*{6}}
\put(50,3){\vector(-1,0){15}}
\put(50,-12){\vector(-1,1){15}}
\put(52,-14){\circle*{6}}
\put(65,18){\vector(-1,-1){15}}
\put(67,20){\circle*{6}}
\put(65,3){\vector(-1,0){15}}
\put(67,3){\circle*{6}}
\put(65,-12){\vector(-1,1){15}}
\put(67,-14){\circle*{6}}
\end{picture} \ \
+ 3\cdot
\begin{picture}(50,40)
\put(2,3){\circle{6}}
\put(20,3){\vector(-1,0){15}}
\put(35,18){\vector(-1,-1){15}}
\put(40,3){\vector(-1,0){20}}
\put(42,3){\circle*{6}}
\put(35,-12){\vector(-1,1){15}}
%
\put(50,33){\vector(-1,-1){15}}
\put(52,35){\circle*{6}}
\put(50,23){\vector(-3,-1){15}}
\put(52,24){\circle*{6}}
\put(50,13){\vector(-3,1){15}}
\put(52,12){\circle*{6}}
\put(50,-7){\vector(-3,-1){15}}
\put(52,-6){\circle*{6}}
\put(50,-17){\vector(-3,1){15}}
\put(52,-18){\circle*{6}}
\put(50,-27){\vector(-1,1){15}}
\put(52,-29){\circle*{6}}
\end{picture}\ \ \right)+ \ldots
\label{socueq2}\ee

Propagators in these diagrams are inverse matrices $-C$,
black circles represent valence-one vertices $D$
and we do not take valence-three vertices into account,
i.e. assume that $B=0$, like in (\ref{cueq}) -- of course
diagrams with $B$-vertices can be easily included.
Coefficients $\alpha_k^{(3)}$ in (\ref{socueq2})
-- the cubic Catalan numbers -- are equal to the
numbers of diagrams of a given order:
\be
\alpha_k^{(3)} = \frac{(3k)!}{k!(2k+1)!}
\ee
In order to extract them from Cardano formula (\ref{socueq1}),
one first rewrites (\ref{socueq1}) as
\be
x = -\frac{d}{c}\sum_{m\geq 0}^\infty
\frac{\Gamma(2m+\frac{2}{3})}{(2m+1)!\Gamma(\frac{2}{3})}
\frac{\beta^m}{(1+\beta)^{m+\frac{1}{3}}}
 \ \ \ \ \ {\rm with} \ \ \ \ \beta = \frac{27d^2}{4c^3}
\ee
and then obtains
\be
&\alpha_k^{(3)} =
\left(\frac{27}{4}\right)^{\! k}
\frac{\Gamma(k+\frac{1}{3})}{\Gamma(\frac{1}{3})}
\sum_{m=0}^k \frac{(-)^m}{(2m+1)!(k-m)!}
\frac{\Gamma(2m+\frac{2}{3})\Gamma(\frac{1}{3})}
{\Gamma(m+\frac{1}{3})\Gamma(\frac{2}{3})} \nn\\
&=\frac{(3k)!}{k!(2k+1)!}
\label{srule}
\ee
In manipulations with the sum in (\ref{srule}) one
can also rewrite it as
$$
\alpha_k^{(3)} = \left(\frac{27}{4}\right)^{\! k}
\frac{\Gamma(k+\frac{1}{3})}{\Gamma(\frac{1}{3})}
\sum_{m=0}^k \frac{(-)^m\gamma_m}{(k-m)!}
$$
where $\gamma_0 = 0$ and
$$
\gamma_{m+1} = \frac{6m+5}{3(m+1)(2m+3)}\gamma_m
$$
so that $\gamma_1 = \frac{5}{9}$,
$\gamma_2 = \frac{11}{2\cdot 27}$,
$\gamma_3 = \frac{11\cdot 17}{2\cdot 3^5\cdot 7}\ $
and so on, or simply
$$
\gamma_m = \frac{(6m-1)\overbrace{!!!!!!}^6}{3^m m!(2m+1)!!}.
$$

It is straightforward to
go further beyond quadratic and cubic equations.
Perturbative solution to
\be
ax^s + px+q = 0
\ee
with any $s$ is given by direct generalization of
eqs.(\ref{soqueq}) and (\ref{socueq2}):
\be
x = \sum_{k\geq 0}^\infty
\frac{(sk)!}{k!\big((s-1)k+1\big)!}
\frac{a^k q^{s(k-1)+1}}{(-p)^{sk+1}}
\label{solseq}
\ee
and coefficients
$\alpha_k^{(s)} = \frac{(sk)!}{k!\big((s-1)k+1\big)!}$
are integer for all natural $s$.
Again, the formula has diagrammatic representation
with vertices of valence $s+1$ and $1$.
As usual, in
this representation it is easy both to switch rom a single
equation to a system of $n$ equations for $n$ variables,
\be
A_i^{j_1\ldots j_s}x_{j_1}\ldots x_{j_s} + P_i^jx_j + Q_j=0,
\label{syseq}
\ee
and to add all omitted vertices
$x^m$ with all intermediate valencies $m+1$.
Perturbative is only one out of $s^n$ solution to (\ref{syseq}),
remaining non-perturbative solutions are singular in the limit
${\cal R}_{n|s}(A)=0$.
Formula (\ref{solseq}) does not make much difference between
different values of $s$, however the rate of coefficients growth
increases with increasing $s$ and Pade-like transforms can be
needed to improve convergence of the series.
For $s>5$ there is no way to represent Catalan numbers
$\alpha_k^{(s)} = \frac{(sk)!}{k!\big((s-1)k+1\big)!}$
as expansion coefficients of any compositions
of radicals, like (\ref{socueq1}),
no sum rule like (\ref{srule}) exists.

\subsection{Dynamical systems theory}

Despite many years of intensive developement the theory of
dynamical systems \cite{cath}, at least of their most typical and
interesting "chaotic" phases, remains far behind the modern
standards of theoretical physics. An impressive amount of
phenomenological information and infinite catalogues of incredible
beauties are available, but no universal, systematic and commonly
accepted approaches emerged so far. Non-linear algebra is a
natural framework for building such an approach, and it will be
among its primary applications in foreseable future.

\subsubsection{Bifurcations of maps, Julia and Mandelbrot sets}

Let ${\cal M}_n$ be a set of maps $f:\ X^n \rightarrow X^n$ of
$n$-dimensional space $X^n$ into itself, \be x_i \rightarrow
f_i(x) \ee Define \be J_f(x) = {\rm det}_{n\times n}
\frac{\partial f_i}{\partial x_j} \ee This $J_f$ is multiplicative
w.r.t. a composition of maps: \be J_{f\circ g}(x) = J_f(g(x))
J_g(x). \ee Zeroes of $J_f$ form the "critical hypersurface" of
$f$ in $X$: \be \Sigma_f = \left\{ x:\ J_f(x) = 0 \right\},\ \ \
\Sigma_f \subset X \ee Unification of all critical surfaces \be
Julia(f) = \bigcup_{m=1}^\infty \Sigma_{f^{\circ m}} \ \ \subset\
X \ee is called {\it Julia set} of $f$.

In the space ${\cal M}_n$ we specify the bifurcation variety
${\cal B}_n \subset {\cal M}_n$, consisting of maps $f$, where
$\Sigma_f$ is singular/reshuffled: \be {\cal B} = \left\{ f: \
\left.
\begin{array}{c} J_f(x) = 0 \\ \frac{\partial J_f(x)}{\partial x_i} = 0
\end{array}\right.\ \right\}
\ee This variety is actually a kind of a left- and right-"ideal"
in ${\cal M}$ w.r.t. the composition of maps: if $f \in {\cal B}$
then $\forall g \in {\cal M}$ both $f\circ g \in {\cal B}$ and
$g\circ f \in {\cal B}$. Variety ${\cal B}$ is called {\it
Mandelbrot set}.

\bigskip

{\bf Example:}

Let $n=1$, $X = C$ and $y=f(x)$. Then $J_f = f'(x)$ and $\Sigma_f$
consists of all critical points of $f$.

The bifurcation variety ${\cal B}_1$ consists of functions $f$
with coincident critical points, where equations $f'(x) = 0$ and
$f''(x)=0$ are compatible: \be f \in {\cal B}_1   \ \
\leftrightarrow \ \ {\rm Disc} (f') = 0 \ee

\bigskip

Numerous examples of Julia and Mandelbort sets (for various
occasional one-parametric families of maps $f$) are famous,
largely because of simplicity of their computer simulations and of
their well developed fractal structure, especially appealing for
the lovers of complexity and disorder. Still, as we know from
\cite{DM}, Julia and Mandelbrot sets are well defined
algebro-geometric structures with {\it analytically calculable}
and well controlled hierarchical structure. As usual in algebraic
geometry, somewhat more economic is description in homogeneous
coordinates, with $f$ substituted by homogeneous maps $\vec A(\vec
z)$.

\subsubsection{The universal Mandelbrot set}

The universal Mandelbrot set ${\cal B}$ was defined in
ref.\cite{DM} as a sub-variety in the moduli space ${\cal M}$ of
all maps, where {\it orbits} exchange stability.\footnote{ We
called it {\it boundary} of Mandelbrot set in \cite{DM}, to avoid
contradiction with terminology in the literature. However, as
emphasized in \cite{DM}, "Mandelbrot set" is a much better term.
Similar is terminological situation with the Julia set. } For this
to happen, the two orbits, of orders $m_1$ and $m_2$, need to
merge, and this means that the two characteristic equations,
${\cal R}\Big(\lambda^{m_1}|A^{\circ m_1}\Big) = 0$ and ${\cal
R}\Big(\lambda^{m_2}|A^{\circ m_2}\Big) = 0$ should be compatible
-- and this imposes a constraint on the map $A$. Roughly speaking,
${\cal B}$ is the union of all such constraints, over all possible
$m_1$ and $m_2$ (actually, one of these numbers should divide
another and this property is behind the pronounced hierarchical
structure of ${\cal B}$, inherited by all its {\it sections} which
are usually considered under the name of particular Mandelbrot
sets). This simple idea provides a pure algebro-geometric
description -- together with powerfull and explicit technical
tools -- of the universal Mandelbrot set: actually it becomes a
sort of the Universal Discriminant ${\cal D} \subset {\cal M}$.

At the next step one should make definitions more accurate, at
least in three respects.

First,  one should separate orbits of order $m$ from those of
lower orders, which are divisors of $m$,-- this is done by
extracting certain factors from the iterated resultants ${\cal
R}\Big(\lambda^{m}|A^{\circ m}\Big)$. This procedure is greatly
simplified by remarkable Besout decomposition (\ref{chareqfact})
of such resultants (a highly non-trivial property for polynomials
of many variables!) and is not too much different from its
single-variable ($n=2$) counterpart, considered in great detail in
\cite{DM}.

Second, for non-linear maps ($s\geq 2$) the "eigenvalues"
$\lambda$ in characteristic equations do not have direct "physical
meaning", since they can be eliminated by eigenvectors rescaling.
This problem can be easily resolved by convertion from homogeneous
to projective coordinates -- as was done in the single-variable
case ($n=2$) in \cite{DM}, -- but for $n\geq 3$ such convertion
makes resultant theory overcomplicated.

Third, one should add information about Julia sets, which form a
kind of a {\it Julia Bundle over the Universal Mandelbrot Set}.
Its structure is even more interesting and involves a mixture of
group-theoretical and Hodge-like structures, typical for the
modern theory of integrable systems and $\tau$-functions. This
emerging connection is one of the most beautifull illustrations
for the "unity of the opposite" priciple: extreme order
(integrability) and extreme disorder (chaotical dynamics) are
essentially the same.

\subsubsection{Relation between discrete and continuous dynamics:
iterated maps, RG-like equations and effective actions}

Understanding transition between continuous and discrete, or, more
accurately, identifying universality classes of discrete models
which have the same continuous limits, remains an eternal open
problem of theoretical physics. At the level of linear algebra the
problem is not seen, while non-linear algebra is able to address
it at the most fundamental level. The problem is closely related
to exponentiation of non-linear maps, briefly introduced in
s.\ref{unop}. In \cite{DM} this research direction appears in
association with {\it effective actions} \cite{MN}. Effective
actions are solutions to discrete analogue of Callan-Symanzik
equation of quantum field theory, if discrete dynamics is
interpreted as Kadanoff's discrete {\it renormalization group}
flow.

\bigskip

{\bf The problem:}\footnote{We acknowledge cooperation with
B.Turovsky and Ya.Zabolotskaya on the study of continuous
interpolations of discrete dynamics.}

\bigskip

What is the relation between discrete holomorphic dynamics of one
variable, \be z \rightarrow f(z) \label{didy} \ee and continuous
dynamics \be \dot z = \beta(z)? \label{cody} \ee Continuous
dynamics produces the function (world line) $z(t)$ and we assume
that \be z(t+1) = f\big(z(t)\big) \label{woli} \ee The {\bf
problem I} is to describe the set of functions $\beta(z)$ for
given $f(z)$ and the {\bf problems II and III} are to describe
associated {\it world lines} $z(t)$ and {\it trajectories} (curves
in the complex $z$-plane, which are images of $z(t)$).

Obviously, there can be many different $\beta(z)$ for a single
$f(z)$: different continuous evolutions can provide the same
discrete evolution and the {\bf problem IV} is to specify
additional requirements which makes the lifting \be \{f(z)\}
\longrightarrow \{\beta(z)\} \label{lift} \ee unambiguous.

\bigskip

{\bf Reformulation in terms of functional equation:}

\bigskip

Equation (\ref{cody}) has immediate solution (functional inverse
of $z(t)$) \be g(z) = \int^z \frac{dx}{\beta(x)} = t, \ee and in
terms of $g(z)$ \be \beta(z) = \frac{1}{g'(z)} \ee i.e. instead of
finding $\beta(z)$ for given $f(z)$ it is enough to find $g(z)$,
which is solution to the functional equation (\ref{woli}) \be
g\big(f(z)\big) - g(z) = 1 \label{fgeq} \ee Note that ${\cal
F}(x,\phi)$ from footnote 2 of ref.\cite{DM} is arbitrary function
of the difference $g(x)-\phi$.

\bigskip

{\bf The problem of periodic orbits:}

\bigskip

If $z_0$ is a fixed point of $f$, i.e. $f(z_0) = z_0$, the
function $g(z)$ should have a singularity at $z_0$ in order to
satisfy (\ref{fgeq}). In the vicinity of $z_0$ we have from
(\ref{fgeq}) \be g\big(z_0 + \epsilon f'(z_0) + \ldots\big) -
g\big(z_0 + \epsilon\big) = 1 \ee i.e. \be g(z) \sim
\frac{\log(z-z_0)}{\log f'(z_0)} + O(z-z_0) \label{logsing} \ee so
that \be \beta(z) = \frac{1}{g'(z)} \sim \Big(\log
f'(z_0)\Big)(z-z_0) + O\Big((z-z_0)^2\Big) \ee One could expect
that $g(z)$ is also singular in the vicinities of all points of
periodic orbits of any order, not just of the first one, i.e. on
the entire boundary of the Julia set of $f$ -- and thus $g(z)$
could seem to be a very sophisticated function. However, this is
not obligatory the case, and $g(z)$ can be rather simple.
Understanding of this phenomenon is the {\bf problem V}.

\bigskip

{\bf World lines behavior in the vicinity of fixed points:}

\bigskip

A fixed point $z_0$ of $f$, $f(z_0) = z_0$, is a zero of
$\beta(z)$ and we often have
$$\beta(z) = \sum_{k=1}^\infty \beta_k(z-z_0)^k$$
and
$$z(t)-z_0 = Ce^{\beta_1 t} +
\frac{\beta_2}{\beta_1}\left(CE^{\beta_1t}\right)^2
 + \ldots = \sum_{k=1}^\infty C_ke^{k\beta_1 t}$$
Then
$$f\big(z(t)\big)-f(z_0) = z(t+1)-z_0 =
\sum_{k=1}^\infty e^{k\beta_1}C_ke^{k\beta_1 t} = 
e^{\beta_1}\big(z(t)-z_0\big) +
e^{\beta_1}\left(e^{\beta_1}-1\right)
\frac{\beta_2}{\beta_1}\big(z(t)-z_0\big)^2 + \ldots $$ On the
other hand the l.h.s. is equal to
$$f(z(t))-f(z_0) = \sum_{k=1}^\infty
\frac{1}{k!}\partial^kf(z_0)(z-z_0)^k$$ and we have \be
\beta_1 = \log f'(z_0), \nn \\
\beta_2 = \frac{\beta_1 f''(z_0)}{2f'(z_0)\big(f'(z_0)-1\big)},
\nn \\
\ldots \label{beta12} \ee

\bigskip

{\bf EXAMPLES}

\bigskip

{\bf The case of $f(x) = x^n$: }

\bigskip

In this case the function $g(z)$ is \cite{DM}:
$$g(z) = \frac{\log(\log z)}{\log n}$$
and $$\beta(z) = 1/g'(z) = \log n ( z\log z)$$ World lines are
$$z(t) = e^{n^t}$$ and trajectories are logarithmic spirals
$$z = e^{(\lambda+i)\phi}\ \ {\rm or}\ \ \rho = |z| = e^{\lambda\phi}$$

Note that $g(z)$ is singular at the fixed points $z_0=0$ and
$z_0=1$, but not singular at other points of the unit circle
$|z|=1$ (which is the boundary of Julia set in this case).
However, the double-logarithmic $g(z)$ has sophisticated branching
structure. Asymptotic formula (\ref{logsing}) holds in the
vicinity of $z_0 = 1$ where $\beta(z) \sim z-z_0$, but not of
$z_0=0$, where $\beta(z) \sim (z-z_0)\log(z-z_0)$.

\bigskip

{\bf Important correction:}

\bigskip

For $f(z) = z^n$ we have $z(t+1) = z^n(t)$, but actually
$$\log z(t+1) = n\log z(t) + 2\pi i k$$
and $k$ is additional parameter, affecting the evolution: \be z(t)
= e^{-\frac{2\pi i k}{n-1}} \left[ z(0) e^{\frac{2\pi i
k}{n-1}}\right]^{n^t} = z^1_k\left[\frac{z(0)}{z^1_k}e^{2\pi i
m}\right]^{n^t} \ee Here $z^1_k = e^{-\frac{2\pi i k}{n-1}}$ is
actually one of the $n-1$ unbstable orbits of order one, and $m =
entier\left(\frac{k}{n-1}\right)$ counts the number of additional
(hidden) rotations in the continuous interpolation.

Evolution with a given $k$ and $m=0$ is {\it small} (local) in the
vicinity of the fixed point $z^1_k$
and there the reasoning 
leading to eqs.(\ref{beta12}) is applicable.

\bigskip

{\bf The case of $F(x) = x^n + c$:}

\bigskip

The function $g(z|c) = g_0(z) + cg_1(z) + c^2g_2(z) + \ldots$
where $g_0(z)$ is found in the previous subsection and $g_1(z)$
satisfies \be g_1(z^n) - g_1(z) = -g_0'(z^n) \sim \frac{1}{z\log
z} \ee

\bigskip

{\bf Appendix: Solving functional equations}

\bigskip

{\bf The basic finite-difference equation:}

\bigskip

The basic functional equation \be g(\xi+\hbar) - g(\xi) = 1
\label{basfe} \ee -- the discrete analogue of $g'(\xi) = 1$ with
generic solution $g(\xi) = \xi + {\rm const}$ -- has generic
solution \be g(\xi) = \frac{1}{\hbar}\xi + \theta(\xi) \ee where
$\theta$ is arbitrary periodic function, \be \theta(\xi+\hbar) =
\theta(\xi) \ee

There are two different directions of generalization
(\ref{basfe}): one can change the r.h.s. and the l.h.s.

\bigskip

{\bf Discrete integration:}

\bigskip

In the first case one substitutes (\ref{basfe}) by \be
g(\xi+\hbar) - g(\xi) = h(\xi) \label{fidieq} \ee -- the discrete
analogue of $g'(\xi) = h(\xi)$ with generic solution $g(\xi) =
\int^\xi h(\xi)$.

Similarly, one can call the solution of (\ref{fidieq}) the {\it
discrete (Jackson) integral}
$$g(\xi) = \sum_{n=0}^{N=\infty} h(\xi - n\hbar)$$
However, this is largely a symbolical notion, because the
ambiguity in discrete integration is arbitrary periodic function,
not just a single constant, what is reflected in ambiguity in
taking the large-$N$ limit. For example, even for $h=1$ it is not
easy to explain in what sense the solution $\frac{1}{\hbar}\xi +
\theta(\xi)\ \stackrel{?}{=}\ \sum_{n=0}^\infty 1$.

More useful can be a {\bf table of meaningful discrete integrals}:
$$
\begin{tabular}{|c|c|}
\hline
&\\
$h(\xi)$ & $\hbar g(\xi)$ \\
&\\
\hline
&\\
1& $\xi$\\
&\\
$\xi$& $\frac{1}{2}\xi^2 - \frac{\hbar}{2}\xi$\\
&\\
$\xi^2$& $\frac{1}{3}\xi^3 - \frac{\hbar}{2}\xi^2 + \frac{\hbar^2}{6}\xi$\\
&\\
$\xi^n$& $\frac{1}{n+1}\xi^{n+1} + \sum_{k=0}^n c_{n;k}\hbar^{k+1}\xi^{n-k}$\\
&\\
\hline
\end{tabular}
$$

\bigskip

{\bf Other discrete versions of first derivative:}

\bigskip

Discrete versions of continuous equation $g'(\xi) = h(\xi)$ can
differ not only by the choice of the r.h.s., but also by the
choice of discrete derivative at the l.h.s. and changes of
$g$-variable can be needed to bring the discrete equation to the
form (\ref{fidieq}). Several examples are collected in the next
table.

$$
\begin{tabular}{|c|c|c|}
\hline
&&\\
{\rm original equation} & {\rm transformation rule} &
{\rm transformation rule}\\
&$G \rightarrow g$ & $H \rightarrow h$ \\
&&\\
\hline
&&\\
$G(kx) - G(x) = H(x)$ & $G(x) = g\left(\hbar\frac{\log x}{\log
k}\right)$&
$H(x) = h\left(\hbar\frac{\log x}{\log k}\right)$\\
&&\\
$G(x^n) - G(x) = H(x)$ & $G(x) = g\left(\hbar\frac{\log(\log
x)}{\log n}\right)$&
$H(x) = h\left(\hbar\frac{\log(\log x)}{\log n}\right)$\\
&&\\
\hline
\end{tabular}
$$

\subsection{Jacobian problem}

This is one of the famous simply-looking but hard mathematical
problems, where diagram technique and other quantum field theory
methods are expected to work.

For {\it polynomial} map $y_i = f_i(x)$ inverse map $x_i =
\phi_i(y)$, $f\circ \phi = id$ is also {\it polynomial} whenever
Jacobian \be J_f(x) = {\det}_{ij}\frac{\partial f_i}{\partial x_j}
=1 \ee Since Jacobian for polynomial map is always polynomial, it
is clear that if $f$ and $\phi$ are both polynomial, $J_f(x)=
J_\phi^{-1}(x)$ is necessarilly independent $x$ (constant is the
only polynomial, whose {\it algebraic} inverse is a polynomial:
Jacobian problem is to generalize this statement to the {\it
functional} inverse). Non-trivial is inverse statement, that $J_f
= 1$ implies that $\phi$ is polynomial.\footnote{We are indebted
to A.Losev for comments about Jacobian problem.}

\bigskip

{\bf Example of quadratic map (with a rank-3 tensor):}

\bigskip

\be y^i = f^i(x) = x^i + A^i_{jk}x^jx^k \ee \be J_f(x) =
{\det}_{ij}\left(\delta^i_{j} + 2A^i_{jk}x^k\right) \ee \be x =
\frac{1}{2A}\Big(\sqrt{1+4Ay} - 1\Big) \ee \be
x^i = y^i -A^i_{jk}y^jy^k + 2A^i_{jk}y^k (A^j_{lm}y^ly^m) - \nn \\
-\Big(A^i_{jk}(A^j_{lm}y^ly^m) (A^k_{pq}y^py^q) + 4A^i_{jk}y^k
(A^j_{lm}y^l) (A^m_{pq}y^py^q)\Big) + \ldots \ee is given by the
obvious sum of tree diagrams. This map is polynomial when $A$ is
nilpotent, say, $A^i_{jk}A^j_{lm} = 0$. A possibility: $A^i_{jk} =
u^iv_jw_k$, $u^iv_i = 0$. In this case $J_f = 1 + A^i_{ik}x^k = 1
+ (u^iv_i)(w_kx^k) = 1$.



\subsection{Taking integrals}

In the context of quantum field theory, when tensor is used to
define the {\it action} in the integral (ordinary or functional),
discriminantal subspace is known as that of {\it classical}
configurations. However, for {\it homogeneous} actions the
vanishing of all first derivatives implies the vanishing of action
itself and, what is even more important, -- if only homegeneity
degree exceeds two, $r>2$ -- degeneracy of the second-derivatives
matrix, which stands pre-exponent in denominator (for bosonic
integrals, for Grassmannian integrals it rather appears in the
numerator and makes integral vanishing -- this is the usual avatar
of singularity in anti-commutative setting). This means that the
integral is singular in the space of theories (actions) ${\cal
M}_{n_1\times\ldots \times n_r}$ at discriminantal subspace where
${\cal D}_{n_1\times\ldots \times n_r} = 0$. Even if action is
non-homogeneous, the highest-degree part of the action by
definition homogeneous, and it is dominant at high values of
integration variables. This means that when {\it its} discriminant
vanishes, the integral changes drastically: this discriminantal
subspace is actually the {\it phase transitions} hypersurface. Of
course, terminology of quantum field theory is not necessary here,
phenomenon manifests itself already at the level of simplest
non-quadratic integrals, starting from the archetypical theory of
Airy functions.

\subsubsection{Basic example: matrix case, $n|r = n|2$}

For a square $n\times n$ matrix $T_{ij}$ \be \int \prod_{i,j=1}^n
e^{T_{ij}x^iy^j} dx^idy^j = \int \prod_i \delta\Big(T_{ij}y^j\Big)
\prod_j dy^j = \frac{1}{\det_{ij} T_{ij}} \label{intdiscr2} \ee
Integral diverges when $\det T = 0$.

\subsubsection{Basic example: polynomial case, $n|r = 2|r$}

For a homogeneous poynomial of degree $r$ of two variables $x$ and
$y$ \be
\int e^{P_r(x,y)}y^{r-2}dxdy = \int e^{y^rP_r(t)} y^{r-1}dy dt
\longrightarrow \frac{1}{\prod_{i=1}^s (t-\mu_i)} = \sum_{i=1}^s
\frac{1}{\Delta_i}\frac{1}{t-\mu_i} \ee where $\Delta_i =
\prod_{j\neq i} (\mu_j-\mu_i)$. Integral of a function $g(t)$ with
this weight provides $\sum_i \frac{g(\mu_i)}{\Delta_i}$. This sum
is singular when discriminant ${\cal D}(P_s) = 0$. When measure
does not contain the specially adjusted factor $y^{r-2}$, the
singularity at ${\cal D}(P_s) = 0$ persists, it just gets more
sophisticated: turns from a simple pole into ramification.

\subsubsection{Integrals of polylinear forms}

Following \cite{D3}, define the {\it integral discriminant}
$\tilde {\cal D}_{n_1\times \ldots\times n_r}(T)$ of the rank-r
tensor (polymatrix) $T_{i_1,\ldots,i_r}$ through an integral over
$r$ vectors $x_1,\ldots,x_r$: \be \oint
\exp\Big(T_{i_1,\ldots,i_r} x_1^{i_1}\ldots x_r^{i_r}\Big)
dx_1^{i_1}\ldots dx_r^{i_r} = \frac{const} {\tilde{\cal
D}_{n_1\times\ldots\times n_r}
(T)} \label{defD} \ee The {\it integral discriminant} $\tilde
{\cal D}_{n_1\times\ldots\times n_r}(T)$ in (\ref{defD}) is not
necessarilly a polynomial of entries of $T$, but -- as we saw in
two basic examples above -- it vanishes whenever the tensor $T$ is
degenerate, i.e. when the ordinary polynomials discriminant ${\cal
D}_{n_1\times\ldots\times n_r}(T) = 0$. This is because any
non-trivial solution to the equation $\partial T/\partial
x_k^{i_k} = 0$ defines a direction in the integration domain,
where exponent does not contribute, and therefore the integral
diverges. Therefore \be \tilde {\cal D}_{n_1\times\ldots\times
n_r}(T) = {\cal D}_{n_1\times\ldots\times
n_r}^{\gamma_{n_1\times\ldots\times n_r}}(T) \chi(T),
\label{chifun} \ee where the $\chi(T)$ is actually a function of
ratios of invariants of the structure group -- it is of
generalized hypergeometric type and is generically transcendental
and contour-dependent,-- but one naturally expects that it does
not have zeroes at finite {\it projective} values of the
coefficients of $T$: all such zeroes are controlled by the {\t
algebraic discriminant}. The power
$$\gamma_{n_1\times\ldots\times n_r} = \frac{\sum_{k=1}^r n_k}
{r\cdot {\rm deg}_T {\cal D}_{n_1\times\ldots\times n_r}(T)}$$ of
the algebraic discriminant at the r.h.s. of (\ref{chifun}) is
easily evaluated by dimension-counting (for $T$-degrees of the
algebraic discriminants see s.\ref{degdi}). Eq.(\ref{chifun})
allows one to consider algebraic discriminants as the {\it
quasiclassical asymptotics} of the integral ones, see also
\cite{GKZ}.

\bigskip

Eq.(\ref{defD}) defines $\tilde{\cal D}_{n_1\times\ldots\times
n_r}$ iteratively in $r$: \be \frac{1}{\tilde{\cal
D}_{n_1\times\ldots\times n_r}(T)} = \oint\frac{d\vec
x_r}{\tilde{\cal D}_{n_1\times\ldots \times n_{r-1}}\Big(T[\vec
x_r]\Big)} \label{intAx} \ee where integral is over a single
($n_r$-component) vector $x$ and the rank-$(r-1)$ tensor \be
\Big(T[\vec x_r]\Big)_{i_1\ldots i_{r-1}} = T_{i_1\ldots
i_r}x^{i_r} \ee Decompositions like (\ref{chareqfact}) and
(\ref{resfact}) can probably be used to cleverly handle the
remaining integrals in (\ref{intAx}).

Another usefull representation is in terms of determinant of the
matrix $\hat T[x]$: \be \frac{1}{\tilde{\cal
D}_{n_1\times\ldots\times n_r}(T)} = \oint \frac{d^{r-2}
x}{\det_{ij} T_{ij}[x]} = \oint\frac{dx_1^{k_1}\ldots
dx_{r-2}^{k_{r-2}}} {\det_{ij} \Big(T_{ijk_1\ldots
k_{r-2}}x_1^{k_1}\ldots x_{r-2}^{k_{r-2}}\Big)} \label{intdet} \ee
Actually, these representations have direct counterparts for {\it
algebraic} discriminants, above contour integrals provide a sort
of a {\it quantum deformation} of algebraic operations: compare
(\ref{intAx}) with (\ref{itediscor}) and (\ref{intdet}) with
(\ref{disres2}). Of course integrals depend also on the choice of
the integration contours: differtent choices define different
branches of integral discriminant. We do not go into details about
this subject and related Stokes phenomena, attracting much
attention all the way from the classical theory of Airy functions
and quasiclassical quantum mechanics to modern theory of
$\tau$-functions.

\subsubsection{Multiplicativity of integral discriminants}

With appropriately adjusted $A$-independent
$const(n_1,\ldots,n_r)$ in eq.(\ref{defD}) the integral
discriminants possess peculiar multiplicativity property: \be
\tilde{\cal D}_{r_1+r_2} (A) = \tilde{\cal D}_{r_1+1}(B)
\tilde{\cal D}_{r_2+1}(C)\ \ \ {\rm for} \ \ \ A=BC, \ee where
$A$, $B$ and $C$ are tensors of ranks $r = r_1+r_2$, $r_1+1$ and
$r_2+1$ respectively (for the sake of brevity we omit
$n_k$-indices, but of course, for the product to exist $n_{r_1+1}$
in $A$ should coincide with $n_{r_2+1}$ in $B$; note also that
indices of $A$ or $B$ can easily be raised at expense of the
simultaneous lowering of index of the corresponding $x$-variable
-- this affects only the size of the structure group). In more
detail, the product is defined as \be A^{(r_1+r_2)}_{i_1\ldots
i_{r_1}j_1\ldots j_{r_2}} = B^{(r_1+1)}_{i_1\ldots
i_{r_1}k}C^{(r_2+1)}_{kj_1\ldots j_{r_2}} \ee so that \be
\frac{1}{\tilde{\cal D}_r(A)} \sim \oint \exp\Big(A_{i_1\ldots
i_{r_1}j_1\ldots j_{r_2}} x_1^{i_1}\ldots
x_{r_1}^{i_{r_1}}y_1^{j_1}\ldots y_{r_2}^{j_{r_2}}\Big)
dx_1^{i_1}\ldots dx_{r_1}^{i_{r_1}}dy_1^{j_1}\ldots
dy_{r_2}^{j_{r_2}} = \ee
$$
= \oint e^{-p_kq^k} dp_kdq^k \left(\oint \exp\Big(B_{i_1\ldots
i_{r_1}k} x_1^{i_1}\ldots x_{r_1}^{i_{r_1}}q^k\Big)
dx_1^{i_1}\ldots dx_{r_1}^{i_{r_1}}\right) \left(\oint
\exp\Big(C_{j_1,\ldots,j_{r_2}}^k y_1^{j_1}\ldots
y_{r_2}^{j_{r_2}}p_k\Big) dy_1^{j_1}\ldots
dy_{r_2}^{j_{r_2}}\right) \sim
$$
\be \sim \oint \frac{e^{-p_kq^k} dp_kdq^k} {\tilde{\cal
D}_{r_1}\Big(B(q)\Big) \tilde{\cal D}_{r_2}\Big(C(p)\Big)} =
\frac{1}{\tilde{\cal D}_{r_1+1}(B) \tilde{\cal D}_{r_2+1}(C)} \ee
The first equality is the definition of ${\cal D}(A)$, in the
second one we used the obvious relation
$$ \oint dp dq e^{-p_kq^k} e^{B_k\{x\}q^k} e^{C^k\{y\}p_k}
= e^{B_k\{x\}C^k\{y\}} = e^{A\{x,y\}},$$ the third one is the
definition of ${\cal D}\Big(B(q)\Big) {\cal D}\Big(C(p)\Big)$,
while the last equality uses the fact that under the integral \be
\frac{1}{\tilde{\cal D}_r\Big(A(x)\Big)} \sim
\frac{\delta(x)}{\tilde{\cal D}_{r+1}(A)}\ \label{Axdelta} \ee
(comp. with (\ref{intAx})). The meaning of (\ref{Axdelta}) is that
for generic $A$ the only singularity in $x$ is at $x=0$.

Alternative -- and much simpler -- derivation is provided by
representation (\ref{intdet}): since $B_{ik}\{x\}C^k_j\{y\} =
A_{ij}\{x,y\}$ \be \frac{1}{\tilde{\cal D}(A)} = \oint
\frac{d^{r_1-1}xd^{r_2-1}y}{\det_{ij} A_{ij}\{x,y\}} = \oint
\frac{d^{r_1-1}xd^{r_2-1}y}{\det_{ij} \Big(B_{ik}\{x\}
C^k_{j}\{y\}\Big)} = \oint \frac{d^{r_1-1}x}{\det_{ij}
B_{ij}\{x\}} \oint \frac{d^{r_2-1}y}{\det_{ij} C_{ij}\{y\}} =
\frac{1}{\tilde{\cal D}(B)\tilde{\cal D}(C)} \ee

In particular, multiplicativity property implies that under a
transformation from the structure group (multiplication by some
$n_k\times n_k$ matrix $U_k$ -- which affects only one of the $r$
indices in $T$)
$$\tilde{\cal D}(U_kT) = \tilde{\cal D}(T) \det U_k$$
since, as we know from (\ref{intdiscr2}), for a matrix
$\tilde{\cal D}(U_k) = {\det}_{n_k\times n_k} U_k$.

Another trivial example is: $1\times\ldots\times 1$ tensors: since
integral discriminant ${\cal D}_{1\times\ldots\times 1}(T)=T$ we
have an obvious identity $AB = AB$ for {\it numbers} $A$ and $B$.

\subsubsection{Cayley $2\times 2\times 2$ hyperdeterminant
as an example of coincidence between integral and algebraic
discriminants}

In the $2\times 2 \times 2$ case, see s.\ref{cayh}, the Cayley
hyperdeterminant ${\cal D}_{2\times 2\times 2}(T)$ is the only
invariant of the structure group $SL(2)\times SL(2)\times SL(2)$,
therefore the function $\chi(T)$ in (\ref{chifun}) has nothing to
depend on and is just a $T$-independent constant. Indeed, in this
case the integral (\ref{defD}) can be evaluated exactly
\cite{D3}:\footnote{ Other examples in \cite{D3} should be treated
with caution, as we already mentioned, in most cases beyond
$2\times 2\times 2$ the integral discriminant is transcendental
function and does not coincide with the power of a polynomial
algebraic discriminant. }
$$
\oint\ldots\oint e^{T_{ijk}x_iy_jz_k} d^2x d^2y d^2z \
\stackrel{(\ref{intdet})}{\sim} \ \oint\oint
\frac{d^2z}{\det_{2\times 2} \hat T[z]} =
\oint\oint\frac{dz_1dz_2}{a(z_1-\xi_+z_2)(z_1-\xi_-z_2)} \sim
\frac{1}{a(\xi_+ - \xi_-)} \sim \frac{1}{\sqrt{b^2-4ac}}
$$
In this formula
$$
\det_{2\times 2} \hat T[z] =
\Big(T_{11i}T_{22j}-T_{12i}T_{21j}\Big)z_iz_j = az_1^2 + bz_1z_2 +
cz_2^2 = a(z_1-\xi_+z_2)(z_1-\xi_-z_2), \ \ \ \ \xi_\pm =
\frac{1}{2a}\Big(-b \pm \sqrt{b^2-4ac}\Big)
$$
and
$$
b^2-4ac  = \Big(T_{111}T_{222} + T_{112}T_{221} - T_{121}T_{212} -
T_{122}T_{211}\Big)^2 - 4\Big(T_{111}T_{221}-T_{121}T_{211}\Big)
\Big(T_{112}T_{222} - T_{122}T_{212}\Big) \
\stackrel{(\ref{Cahyp})}{=}\ {\cal D}_{2\times 2\times 2}(T)
$$
so that indeed the $2\times 2\times 2$ integral discriminant is
the square root (since $\gamma_{2\times 2\times 2} =
\frac{2+2+2}{3\times 4} = \frac{1}{2}$) of the algebraic one: \be
\tilde{\cal D}_{2\times 2\times 2}(T) \sim \sqrt{{\cal D}_{2\times
2\times 2}(T)} \label{intcay} \ee i.e.  $\chi(T)\sim 1$ in
eq.(\ref{chifun}), as expected.

\subsection{Differential equations and functional integrals}

A simple example, where non-linear algebra is needed in the theory
of partial-derivatives differential equations, is provided by
direct Shroedinger equation with non-quadratic Hamiltonian: \be
i\hbar\frac{\partial \Psi}{\partial t} =
S\Big(-i\hbar\frac{\partial}{\partial \vec x} \Big)\Psi + V(\vec
x)\Psi, \ee where $S(\vec p)$ and $V(\vec x)$ are symmetric
functions of their arguments. The problem is already non-trivial
for vanishing potential $V(\vec x)$ and for homogeneous kinetic
term $S(\vec p)$ of degree $r\geq 3$ -- provided the number of
variables $n>1$. Solution to Shroedinger equation is formally
given by functional integral \be \Psi(\vec x,t) = \int
e^{\frac{i}{\hbar}\int\Big(S(\vec p) - \vec p\dot{\vec x} + V(\vec
x)\Big)d\tau} {\cal D}p(\tau){\cal D}x(\tau) \ee If $S(\vec p)$ is
quadratic, one can easily integrate over $\vec p(t)$ and obtain a
new integral: over paths $x(\tau)$, passing through the given
point $x$ at the time $\tau = t$, with  weights, dictated by the
action $\int \Big(S^{-1}(\dot{\vec x}) - V(x)\Big) d\tau$ (certain
problems arise when discriminant $\det \partial^2 S=0$, but they
can be handled by a variety of methods). However, if $S(\vec p)$
has higher degree $r$, i.e. $S$ is a tensor of rank $r\geq 3$,
then even elimination of $p$-variables becomes an interesting
technical problem, related to integral discriminants (and thus, in
the quasiclassical limit $\hbar \rightarrow 0$, to algebraic
discriminants).

Even more interesting are generalizations to non-homogeneous
kinetic terms, inclusion of potentials, generalizations to
non-linear (in $\Psi$) differential equations and/or transition to
the limit of large $n$. Even the theory of $\Psi$-linear and
$p$-quadratic (${\rm deg}_p S(p)=2$) differential equations --
like that of Virasoro constraints in matrix models -- is highly
non-trivial: the structure of associated discriminantal spaces
(the phase structure of matrix-model partition functions) appears
very rich and interesting, see \cite{amm} and references therein.

\subsection{Renormalization and Bogolubov's recursion formula}

Renormalization theory in quantum field theory (QFT) solves the
following problem. Assume that we have a function $F_\Lambda(T)$
of some variables $T$, and the shape of this function (the
coefficients of its $T$-expansion) depends on additional parameter
$\Lambda$. The question is: can we absorb the $\Lambda$-dependence
into a shift of $T$-variables, i.e. can we construct a new
function $Q_\Lambda(T)$, such that
$F_\Lambda\Big(T+Q_\Lambda(T)\Big)$ is no longer
$\Lambda$-dependent?

In QFT applications the role of $F_\Lambda(T)$ is usually played
by partition function, obtained by functional integration over
fields where the weight -- exponentiated action -- depends on the
coupling constants $T$. The origin of the problem is that the
integrals often diverge and one needs to introduce cut-off
parameter $\Lambda$ to make them well-defined. This cut-off should
be somehow eliminated, and one is ready to substitute original
(``bare") couplings by some $\Lambda$-dependent $\tilde T_\Lambda
= T + Q_\Lambda(T)$, if this is sufficient to make the final
answer independent of $\Lambda$ (often it is enough to make it
finite in the limit when $\Lambda \rightarrow \infty$). However,
one is not ready to sacrifice (change) the {\it shape} of
partition function $F_\Lambda(T)$: it belongs to a very
distinguished class of ``integrable $\tau$-functions" and any
deformation will take it away from this class. The only thing
which one can change safely is its argument, $T \rightarrow \tilde
T$.

Thus renormalization problem includes two inputs: a given function
$F_\Lambda(T)$ and a linear projector ${\cal P}_+$ in the space of
$\Lambda$-dependent functions, selecting the ``wanted" ones among
all, say, $\Lambda$-independent or finite in the limit when
$\Lambda \rightarrow \infty$. In fact, concrete nature of
projector is inessential for solving the problem: it can be any
projector acting on the coefficients of formal $T$-series $F(T)$.
In these terms the problem is:

\bigskip

\centerline{{\bf Given $F(T)$ and ${\cal P}_+$,\ \ find $Q(T)$,\ \
satisfying}} \vspace{-0.2cm} \be {\cal
P}_-\left\{F_\Lambda\Big(T+Q(T)\Big)\right\} = 0 \label{prob1} \ee
\centerline{{\bf and}} \be {\cal P}_+\Big\{Q(T)\Big\} = 0
\label{prob2} \ee Here ${\cal P}_- = Id - {\cal P}_+$ is the
complementary projector onto the space of ``unwanted" functions.
The second condition puts $Q(T)$ fully into the space of
``unwanted" functions, thus making the problem unambiguously
defined.

\PFig{Bogolubov066}
{200,194}
{Nikolai Bogolubov (1909 -- 1992)}

In this form this is a typical problem of non-linear algebra, and
its solution -- the celebrated Bogolubov's recursion formula
\cite{Bo} -- can be considered as one of its first beautiful
results. We refer to \cite{GMS,CK,MS} for detailed description of
the subject, and consider here only the most trivial example from
\cite{MS}.

Let us take \be \left\{ \begin{array}{c}
F(T) = T + \Lambda T^2, \\
{\cal P}_-\Big\{f(\Lambda)\Big\} = f(\Lambda) - f(0),
\end{array} \right.
\ee i.e. ``unwanted" are functions, which depend on $\Lambda$ and
the goal is to eliminate $\Lambda$-dependence by
$\Lambda$-dependent shift of variables $T \rightarrow \tilde T = T
+ Q_\Lambda(T)$, so that $F(\tilde T) = \tilde T + \Lambda \tilde
T^2$ no longer depends on $\Lambda$. Imposing the constraint
(\ref{prob2}) we reduce the problem to \be \tilde T + \Lambda
\tilde T^2 = T \ee and therefore \be \tilde T = \frac{\sqrt{1 +
4\Lambda T}-1}{2\Lambda}
= T - \Lambda T^2 + 2\Lambda^2 T^3\nn\\
- 5\Lambda^3 T^4 + 14\Lambda^4 T^5 - 42 \Lambda^5 T^6 + \ldots
\label{soquad2}\ee
One easily
recognizes in this formula already familiar eq.\,\,(\ref{soqueq}) and
it is therefore not a big surprise that a diagrammatic
representation exists for this solution and for the generic
problem (\ref{prob1}) and (\ref{prob2}) as well.

Such solution is provided by Bogolubov's forest formulas, the
simpler one is
\be \hat F(\Gamma/\Gamma) \hat Q(\Gamma) = -{\cal
P}_-\left\{ \hat F(\Gamma) + \sum_{\big\{\gamma_1 \cup \ldots \cup
\gamma_k\big\}} \hat F\Big(\Gamma/\gamma_1\ldots \gamma_k\Big)
\hat Q(\gamma_1)\ldots\hat Q(\gamma_k)\right\}
\nn\ee
\be
\label{fof}\ee
where both the $F(T)$ and the counter-term $Q(T)$ can be
expanded into sums over Feynman diagrams: \be F(T) = \sum_\Gamma
\hat F(\Gamma) Z(\Gamma|T), \ \ \ \ Q(T) = \sum_\Gamma \hat
Q(\Gamma) Z(\Gamma|T) \ee with basic functions $Z(\Gamma|T)$,
capturing the topology of the graph $\Gamma$: $Z(\Gamma|T)$ is a
product of couplings $T$ at the vertices with indices (if any)
contracted along the links. The sum in (\ref{fof}) is over all
possible box-subgraphs $\big\{\gamma_1 \cup \ldots \cup
\gamma_k\big\}$ of the graph $\Gamma$, i.e. parts of $\Gamma$
lying in a collection of non-intersecting ``boxes", and
$\Gamma/\gamma_1\ldots \gamma_k$ is obtained by contracting all
boxes to points.

The diagram technique behind (\ref{fof}) is not obligatory the
same that we considered in s.\ref{peso}, actually, many different
techniques can be used. An important difference is that {\it loop}
diagrams should be included along with {\it trees} when we switch
from solutions to equations (of motion) to partition functions:
loops are responsible for {\it quantum corrections} to classical
quantities, represented by tree diagrams. See \cite{GMS,CK,MS} for
the first steps of development of this important chapter of
non-linear algebra.

\section{Acknowledgements}

This work was partly supported by the grants: RFBR 04-02-17227
(V.D.), RFBR 04-02-16880, INTAS 05-1000008-7865, the NWO project
047.011.2004.026, the ANR-05-BLAN-0029-01 project (A.M.) and by
the Russian President's grant for support of the scientific
schools LSS-8004.2006.2.

\end{document}